**Multiferroicity**

**— The coupling between magnetic and polarization orders**

K. F. Wang

*Nanjing National Laboratory of Microstructures, Nanjing University, Nanjing 210093, China*

J. –M. Liu[a]

*Nanjing National Laboratory of Microstructures, Nanjing University, Nanjing 210093, China*

*School of Physics, South China Normal University, Guangzhou 510006, China*

*International Center for Materials Physics, Chinese Academy of Sciences, Shenyang, China*

Z. F. Ren[b]

*Department of Physics, Boston College, Chestnut Hill, MA 02467, USA*

[Abstract] Multiferroics, defined for those multifunctional materials in which two or more kinds of fundamental ferroicities coexist, have become one of the hottest topics of condensed matter physics and materials science in recent years. The coexistence of several order parameters in multiferroics brings out novel physical phenomena and offers possibilities for new device functions. The revival of research activities on multiferroics is evidenced by some novel discoveries and concepts, both experimentally and theoretically. In this review article, we outline some of the progressive milestones in this stimulating field, specially for those single phase multiferroics where magnetism and ferroelectricity coexist. Firstly, we will highlight the physical concepts of multiferroicity and the current challenges to integrate the magnetism and ferroelectricity into a single-phase system. Subsequently, we will summarize various strategies used to combine the two types of orders. Special attentions to three novel

---
[a] E-mail: liujm@nju.edu.cn
[b] E-mail: renzh@bc.edu

mechanisms for multiferroicity generation: (1) the ferroelectricity induced by the spin orders such as spiral and $E$-phase antiferromagnetic spin orders, which break the spatial inversion symmetry, (2) the ferroelectricity originating from the charge ordered states, and (3) the ferrotoroidic system, will be paid. Then, we will address the elementary excitations such as electromagnons, and application potentials of multiferroics. Finally, open questions and opportunities will be prospected.

***Keywords:***

Multiferroicity, ferroelectricity, magnetism, magnetoelectric coupling, multiferroics, polarization, magnetization, time reversion symmetry breaking, spatial inversion symmetry breaking, helical spin ordered state, charge-ordered state, electromagnon, ferrotoroidicity

# Contents

## 1. Introduction

Magnetic and ferroelectric materials permeate every aspect of modern science and technology. For example, ferromagnetic materials with switchable spontaneous magnetization $M$ driven by external magnetic field $H$ have been widely used in data storage industries. The discovery of the giant magnetoresistance effect (GMR) significantly promoted magnetic memory technology and incorporated it into the eras of magnetoelectronics or spintronics. The fundamental and application issues associated with magnetic random-access memories (MRAMs) and related devices have been intensively pursued, in order to achieve high-denisty integration and also overcome the large handicap of the relatively high writing energy [1-4]. On the other hand, the sensing and actuation industry relies heavily on ferroelectric materials with spontaneous polarization $P$ reversible upon an external electric field $E$, because most ferroelectrics, especially those perovskite oxides, are high-performance ferroelastics or piezoelectrics with spontaneous strain. The coexistence of strain and polarization allows these materials to be used in broad applications in which elastic energy is converted into electric energy or vice versa [5]. In addition, there has been continuous effort along with the use of ferroelectric random-access memories (FeRAMs) [6] as novel non-volatile and high-speed memory media, and in promoting their performance superior to semiconductor flash memories.

As for the trends toward device miniaturization and high-density data storage, an integration of multifunctions into one material system has become highly desirable. Stemming from the extensive applications of magnetic and ferroelectric materials, it is natural to pursue a new generation of memories and sensing/actuating devices powered by materials that combine magnetism and ferroelectricity in effective and intrinsic manners (as shown in Fig. 1). The coexistence of several order parameters will bring out novel physical phenomena and offers possibilities for new device functions. The multiferroics addressed in this article represent one such type of materials, which do allow opportunities for human being to develop efficient control of magnetization or/and polarization by electric field or/and magnetic field (Fig. 1 and Fig. 2), and to push their multi-implications. The novel prototype devices based on multiferroic functions may offer particularly super performance for spintronics, for example, reading the spin states, and writing the polarization states to reverse the spin states by electric field, to overcome the high-writing energy in magnetic random-access memories.

Considering that little attention has been paid to multiferroicity until recently, it now offers us the opportunity to explore some important issues which have rarely been reachable. Although ferroelectricity and magnetism have been the focus of condensed matter physics and materials science since their discovery, quite a number of challenges in dealing with multiferroicity within the framework of fundamental physics and technological applications have emerged. There are, in principle, two basic issues to address in order to make multiferroicity physically understandable. The first one is the coexistence of ferroelectricity (electric dipole order) and magnetism (spin order) in one system (hereafter, composite integration strategies for the two types of functions is excluded for detail discussion except a sketched introduction in Sec.2.1), since it was once proven extremely difficult for the two orders to coexist in a single material. Even so, exploring the microscopic conditions by which the two orders can coexist intrinsically in one system as a nontrivial problem has never been given up. Second, an efficient coupling between the two orders in a multiferroic system (we always refer this coupling to the magnetoelectric coupling) seems to be even more important than their coexistence, because such a magnetoelectric coupling represents the basis for multi-control of the two orders by either electric field or magnetic field. Investigations demonstrated that a realization of such strong coupling has been even more challenging and, thus, the core of recent multiferroic researches.

It should be mentioned here that most multiferroics synthesized so far are transitional metal oxides with perovskite structures. They are typically strongly correlated electronic systems in which the correlations among spins, charges/dipoles, orbitals and lattice/phonons are significant. Therefore, intrinsic integration and strong coupling between ferroelectricity and magnetism are essentially related to the multi-latitude landscape of interactions between these orders, thus making the physics of multiferroicity extremely complicated. Nevertheless, it is also clear that multiferroicity provides a more extensive plateform to explore the novel physics of strongly correlated electronic systems, in addition to high $T_C$ superconductor and colossal magnetoresistance (CMR) manganites, etc.

Since its discovery a century ago, ferroelectricity, like superconductivity, has been linked to the ancient phenomena of magnetism. Attempts to combine the dipole and spin orders into one system started in the 1960s [7, 8], and some multiferroics, including boracites ($Ni_3B_7O_{13}I$, $Cr_3B_7O_{13}Cl$) [8], fluorides ($BaMF_4$, $M$=Mn, Fe, Co, Ni) [9, 10], magnetite $Fe_3O_4$ [11], (Y/Yb)MnO$_3$ [12], and BiFeO$_3$ [13], were identified in the following decades. However, such a combination in these multiferroics has been proven to be unexpectedly tough. Moreover, a successful combination of the two orders does not necessarily guarantee a strong magnetoelectric coupling and convenient mutual control between them. Fortunately, recent work along this line has made substantial progress by discovering/inventing some multiferroics, mainly in the category of frustrated magnets, which demonstrate the very strong and intrinsic magnetoelectric coupling. Our theoretical understanding of this breakthrough is attributed to the physical approaches from various length scales/levels. Technologically, growth and synthesis techniques for high quality single crystals and thin films become available. All of these are responsible for an upsurge of interest in this topic in last several years. In Tables I and II are collected several kinds of single phase multiferroics discovered and investigated recently [14-20].

This article intends to review the state-of-the-art breakthroughs in this stimulating research field and is organized in the following manner. In Sec. 2, the relationship and difference between the magnetoelectric coupling and multiferroicity will be addressed and the issue why the coexistence of magnetism and ferroelectricity is physically unfavored will be discussed. Sec. 3 is devoted to the theoretical and experimental efforts made so far, by which the magnetism and ferroelectricity were essentially combined and the improper ferroelectricity induced by specific magnetic and charge orders was eventually demonstrated. The elementary excitations in multiferroics — electromagnons, is clarified in Sec. 4. We highlight in Sec. 5 another way to reach strong magnetoelectric coupling: ferrotoroidical systems. The potential applications and unsolved problems associated with multiferroicity will be prospected in Sec. 6 and Sec. 7.

It should be mentioned that the authors of this article are not in a position to cover every aspect of multiferroicity and its related topics. In fact, such a task is very hard and will not be intended, not only because of the rapid advances of this field. The conclusion and perspectives are biased by the authors' point of view. We apologize for our inability to mention the work of many researchers. Surely, we are in an immediate position to take responsibility for all technical deficiencies in this article, if any.

## 2. Magnetoelectric effects and multiferroicity

### 2.1. *Magnetoelectric effects*

The magnetoelectric effect, in its most general definition, describes the coupling between electric and magnetic fields in matters (i.e. induction of magnetization ($M$) by an electric field ($E$) or polarization ($P$) generated by a magnetic field ($H$)). In 1888, Röntgen observed that a moving dielectric body placed in an electric field became magnetized, which was followed by the observation of the reverse effect: polarization generation of a moving dielectric in a magnetic field [21]. Both, however, are not the intrinsic effects of matters. In 1894, by crystal symmetry consideration, Curie predicted the possibility of an intrinsic magnetoelectric effect in some crystals [22]. Subsequently, Debye coined this kind of effect as a "magnetoelectric effect" [23]. The first successful observation of the magnetoelectric effect was realized in $Cr_2O_3$, and the magnetoelectric coupling coefficient was 4.13 ps/m [24]. Up to now, more than 100 compounds that exhibit the magnetoelectric effect have been discovered or synthesized [14-20, 25].

Thermodynamically, the magnetoelectric effect can be understood within the Landau theory framework, approached by the expansion of free energy for a magnetoelectric system, i.e.

$$F(E,H) = F_0 - P_i^s E_i - M_i^s H_i - \frac{1}{2}\varepsilon_0\varepsilon_{ij}E_iE_j - \frac{1}{2}\mu_0\mu_{ij}H_iH_j - \alpha_{ij}E_iH_j$$
$$- \frac{1}{2}\beta_{ijk}E_iH_jH_k - \frac{1}{2}\gamma_{ijk}H_iE_jE_k - \dots\dots \quad , \quad (1)$$

where $F_0$ is the ground state free energy; subscripts ($i, j, k$) refer to the three components of a variable in spatial coordinates; $E_i$ and $H_i$ the components of electric field $E$ and magnetic field $H$, respectively; $P_i^s$ and $M_i^s$ the components of spontaneous polarization $P^s$ and magnetization $M^s$; $\varepsilon_0$ and $\mu_0$ the dielectric and magnetic susceptibilities of vacuum; $\varepsilon_{ij}$ and $\mu_{ij}$ are the second-order tensors of dielectric and magnetic susceptibilities; $\beta_{ijk}$ and $\gamma_{ijk}$ are the third-order tensor coefficients; and most importantly, $\alpha_{ij}$ is the components of tensor $\alpha$ which is designated as the linear magnetoelectric effect and corresponds to the induction of polarization by a magnetic field or a magnetization by an electric field. The rest of the terms in the preceding equations correspond to the high-order magnetoelectric effects parameterized by tensors $\beta$ and $\gamma$ [25]. Then the polarization is

$$P_i(E,H) = -\frac{\partial F}{\partial E_i} = P_i^s + \varepsilon_0 \varepsilon_{ij} E_j + \alpha_{ij} H_j + \frac{1}{2} \beta_{ijk} H_j H_k + \gamma_{ijk} H_i E_j + \dots\dots, \qquad (2)$$

and the magnetization is

$$M_i(E,H) = -\frac{\partial F}{\partial H_i} = M_i^s + \mu_0 \mu_{ij} H_j + \alpha_{ij} E_j + \beta_{ijk} H_j E_i + \frac{1}{2} \gamma_{ijk} E_j E_k + \dots\dots, \qquad (3)$$

Unfortunately, usually the magnetoelectric effect in single phase compounds is too small to be practically applicable. The breakthrough in terms of the giant magnetoelectric effect was achieved in composite materials, for example, in the simplest case the multilayer structures composed of a ferromagnetic piezomagnetic layer and a ferroelectric piezoelectric layer [25-28]. Other kinds of magnetoelectric composites including co-sintered granular composites and column-structure composites were also developed [29-31]. In the composites, the magnetoelectric effect is generated as a product property of the magentostrictive and piezoelectric effects, which is a macroscopic mechanical transfer process. A linear magnetoelectric polarization is induced by a weak *a.c.* magnetic field imposed onto a *d.c.* bias magnetic field. Meanwhile, a magnetoelectric voltage coefficient up to 100 V·cm$^{-1}$Oe$^{-1}$ in the vicinity of electromechanical resonance was reported [25]. These composites are acceptable for practical applications in a number of devices such as microwave components, magnetic field sensors and magnetic memories. For example, it was recently reported that the magnetoelectric composites can be used as probes in scanning probe microscopy to develop a near-field room temperature scanning magnetic probe microscope [32]. For the complete introduction of the magnetoelectric effects in composite materials, readers are referred to the review papers by Fiebig [25] and Nan *et al* [26], and hereafter we will no longer touch the magnetoelectric composite materials.

One way to significantly enhance the magnetoelectric response in single phase compounds is to make use of strong internal electromagnetic fields in the components with large dielectric and magnetic susceptibilities. It is well known that ferroelectric/ferromagnetic materials have the largest dielectric/magnetic susceptibility, respectively. Ferroelectrics with ferromagnetism, i.e. ferroelectomagnets [33] would be prime candidates for an enhanced magnetoelectric effect. Consequently, Schmid coined the materials with two or more primary ferroic order parameters (ferroelectricity, ferromagnetism, and ferroelasticity) as multiferroics [34]. What should be mentioned is that, except for the coexistence of ferroelectricty and ferromagnetism, the materials with strong coupling between primary ferroelastic and ferromagnetic order parameters, in the class of ferromagnetic martensitic systems, were synthesized about 10 years ago, too. For a review of ferroelsatic materials, one may refer to the excellent book of Salje [35]. Since no substantial breakthrough for ferromagnetic-ferroelastic coupling has been reported, in this article, we restrict our main concern specifically to single phase multiferroic compounds exhibiting (anti)ferromagnetism and (anti)ferroelectricity simultaneously.

### 2.2. *Incompatibility between ferroelectricity and magnetism*

Given such a definition of multiferroics, the incompatibility between ferroelectricity and magnetism comes out as the first issue to address. From a point of view of symmetry consideration, ferroelectricity needs the broken spatial inverse symmetry while the time reverse symmetry can be invariant. A spontaneous polarization would not appear unless a structure distortion of the high-symmetry paraelectric (PE) phase breaks the inversion symmetry. The polarization orientation must be different from those crystallographic directions which constrains the symmetry of the point group. In contrast, the broken time-reversal symmetry is the prerequisite for magnetism and spin order, while invariant spatial inverse symmetry applies for most conventional magnetic materials in use but not a prerequisite. Among all of the 233 Shubnikov magnetic point groups, only 13 point groups, i.e., *1, 2, 2′, m, m′, 3, 3m′, 4, 4m′m′, m′m2′, m′m′2′, 6*, and *6m′m′*, allow the simultaneous appearance of spontaneous polarization and magnetization. This restriction in the crystallographic symmetry results in the fact that multiferroics in nature are rare. Even so, it is known that some compounds belonging to the above 13 point groups do not show any multiferroicity. Therefore, approaches different from simple symmetry considerations are needed.

Most technologically important ferroelectrics such as $BaTiO_3$ and $(Pb,Zr)TiO_3$ are transitional metal oxides with perovskite structure ($ABO_3$). They usually take cubic structure at high temperature with a small B-site cation at the center of an octahedral cage of oxygen ions and a large A-site cation at the unit cell corners [5, 6]. In parallel, there are a large number of magnetic oxides in a perovskite or a perovskite-like structure. Attempts to search for or synthesize multiferroics were mostly concentrated on this class of compounds. Nevertheless, in spite of hundreds of magnetic oxides and ferroelectric oxides, there is practically no overlap between them. This leads to an unfortunate but clear argument that magnetism and ferroelectricity tend to exclude each other. This is an issue that has been addressed repeatedly. So far, the overall picture suggests that all conventional ferroelectric perovskite oxides contain transition metal (TM) ions with a formal configuration $d^0$, such as $Ti^{4+}$, $Ta^{5+}$, $W^{6+}$, at B-sites (i.e., the TM ions with an empty $d$-shell). The empty $d$-shell seems to be a prerequisite for ferroelectricity generation, while it does not mean that all perovskite oxides with empty $d$-shell TM ions must exhibit ferroelectricity.

Magnetism, on the contrary, requires the TM ions at B-site with partially filled shells (always $d$- or $f$-shells), such as $Cr^{3+}$, $Mn^{3+}$, $Fe^{3+}$, because the spins of electrons occupying completely the filled shell add to zero and do not participate in magnetic ordering. The difference in filling the TM ion $d$-shells at B-site, which is required for ferroelectricity and magnetism, makes these two ordered states mutually exclusive. However, a closer look at this process reveals even more abundant physics associated with this issue.

Ferroelectrics have spontaneous polarization that can be switched by electric field. In particular, they undergo a phase transition from a high-temperature, high-symmetry PE phase that roughly behaves as ordinary dielectrics, into a low-symmetry polarized phase at low temperature accompanied by an off-centre shift of B-site TM ions, as shown in Fig.3 (structurally distorted). In fact, ionic-bond perovskite oxides are always centrosymmetric (therefore not ferroelectric-favored). This is because, for centrosymmetric structures, the short range Coulomb repulsions between electron clouds on adjacent ions are minimized. The ferroelectric stability is therefore determined by a balance between these short-range repulsions favoring the non-ferroelectric centrosymmetric structure, and additional bonding considerations which stabilize the ferroelectric phase.

Currently, two distinctly different chemical mechanisms for stabilizing the distorted structures in ferroelectric oxides are proposed in the literature. In fact, both are described as a second-order Jahn-Teller effect. In this section, we only address one of them: the ligand-field hybridization of a transition metal cation with its surrounding anions. Take $BaTiO_3$ as an example. The empty $d$-states of TM ions, like $Ti^{4+}$ in $BaTiO_3$, can be used to establish strong covalency with the surrounding oxygen anions which soften the Ti-O repulsion [17, 36]. It is favorable to shift the TM ions from the centre of $O_6$ octahedra towards one (or three) oxygen(s) to form a strong covalent bond at the expense of weakening the bonds with other oxygen ions, as shown in Fig. 4(a). The hybridization matrix element $t_{pd}$ (defined as the overlap between the wave functions of electrons in Ti and O ions) changes to $t_{pd}(1+gu)$, where $u$ is the distortion and $g$ is the coupling constant. In the linear approximation, corresponding terms in the energy $\sim(-t_{pd}^2/\Delta)$, where $\Delta$ is the charge transfer gap, cancel with each other [17]. However, the second order approximation produces an additional energy difference:

$$\delta E \cong -(t_{pd}(1+gu))^2/\Delta - (t_{pd}(1-gu))^2/\Delta + 2t_{pd}^2/\Delta = -2t_{pd}^2(gu)^2/\Delta, \qquad (4)$$

If the corresponding total energy gain $\sim u^2$ exceeds the energy loss due to the ordinary elastic energy $\sim Bu^2/2$ of the lattice distortion, such a distortion would be energetically favorable and the system would become ferroelectric. Referring to Fig. 4(b), one observes that only the bonding bands would be occupied (solid arrows) if the TM ion has an empty $d$-shell, a process that only allows for electronic energy. If there is an additional $d$-electron on the corresponding $d$-orbital (dashed arrow), this electron will occupy an antibonding hybridized state, thus suppressing the total energy gain. This seems to be one of the factors suppressing the tendency of magnetic ions to make a distorted shift associated with ferroelectricity [17, 36].

Surely, the incompatibility between ferroelectricity and magnetism has even more complicated origins than the above model. More realistic ingredients should be included in order to understand the suppression of ferroelectricity in systems with magnetic ions. For example, it has been argued that the breaking of singlet valence state $(( d_\uparrow p_\downarrow - d_\downarrow p_\uparrow ) / \sqrt{2} )$ by local spin in magnetic ions is responsible for the incompatibility [17]. This issue still deserves further attentions.

### 2.3. *Mechanisms for ferroelectric and magnetic integration*

As stated above, ferroelectric perovskite oxides need B-site TM ions with an empty *d*-shell to form ligand hybridization with surrounding anions. This type of electronic structure likely excludes magnetism. However, not all experimental and theoretical results support the argument that ferroelectricity and magnetism are absolutely incompatible, and an integration of them seems to be possible. First, the famous Maxwell equations governing the dynamics of electric field, magnetic field and electric charges, tell us that rather than being two independent phenomena, electric and magnetic fields are intrinsically and tightly coupled to each other. A varying magnetic field produces an electric field, whereas electric current, or a charge motion, generates a magnetic field. Second, the formal equivalence of the equations governing the electrostatics and magnetostatics in polarizable media explains the numerous similarities in the physics of ferroelectricity and ferromagnetism, such as their hysteresis behavior in response to the external field, anomalies at the critical temperature and domain structures. On one hand, these coupling phenomena and similarities in terms of the electric dipoles and spins in polarizable media imply the potential to integrate ferroelectricity and magnetism into single phase materials. On the other hand, the hybridization between the B-site cation and anion (i.e. the covalent bond) in ferroelectrics can be seen as the virtual hopping of electrons from the oxygen-filled shell to the empty *d*-shell of the TM ion. On the contrary, however, it is the uncompensated spin exchange interaction between adjacent magnetic ions that induces the long range spin order and macroscopic magnetization, where the spin exchange interaction can be mapped into the virtual hopping of electrons between the adjacent ions. This similarity also hints a possibility to combine these two orders into one system.

With respect to the roadmaps for integrating ferroelectricity and magnetism, we incipiently address the conceptually simplest situation: to synthesize materials which contain separate functional units. Usually, one mixes the non-centro-symmetric units, which may arouse a strong dielectric response and ferroelectricity, together with those units with magnetic ions. An alternative approach refers to perovskite oxides once more, where the A-sites are usually facilitated with cations of a $(ns)^2$ valence electron configuration, such as $Bi^{3+}$, $Pb^{3+}$, which favor the stability of ferroelectrically distorted structures. At the same time, the B-sites are facilitated with magnetic ions providing magnetism. This approach avoids the exclusion rule of ferroelectricity and magnetism at the same sites because, here, the ferroelectricity is induced by the ions at the A-sites instead of the same B-site ions for magnetism.

Nevertheless, such kind of simple approaches do allow for ferroelectricity and magnetism in one system, but may not necessarily offer strong magnetoelectric coupling, partially because the microscopic mechanisms responsible for ferroelectricity and magnetism are physically very different. The eventual solution to this paradox, if any, is to search for ferroelectricity that is intrinsically generated by special spin orders. This not only enables an effective combination of the two orders but also the spontaneous mutual control of them. Fortunately, substantial progress along this line has been achieved in recent a few years, and some novel multiferroics in which ferroelectricity is induced by a geometric distortion and a helical/conical spin order, as well as a charge-ordered structure, have been synthesized. Details of these efforts and results will be presented in next section.

## 3. Approaches to coexistence of ferroelectricity and magnetism

### 3.1. *Independent systems*

As mentioned above, the conceptually simplest approach is to synthesize multiferroics with two structural units functioning separately the ferroelectricity and magnetism. The first and well known examples are borates, such as $GdFe_3(BO_3)_4$, which contains ferroelectricity active $BO_3$ groups and magnetic ions $Fe^{3+}$ [37, 38]. Besides the multiferroicity, these materials exhibit interesting optical properties. Boracites, such as $Ni_3B_7O_{13}I$, are also in this class [8, 39]. One can cite many similar compounds, like $Fe_3B_7O_{13}Cl$ [40], $Mn_3B_7O_{13}Cl$ [41] etc, which may exhibit multiferroic behaviors, noting that they don't have a perovskite structure.

We address perovskite oxides here. The first route toward perovskite multiferroics was taken by Russian researchers. They proposed to mix both magnetic TM ions with $d$ electrons and ferroelectrically active TM ions with $d^0$ configurations at the B-sites (i.e., substituting partially the $d^0$-shell TM ions by magnetically active 3d ions while keeping the perovskite structure stabilized). It is hoped that the magnetic ions and $d^0$-shell TM ions favor separately a magnetic order and a ferroelectric order, although this may be difficult if the magnetic doping is over-concentrated. The typical (and one of the most studied) compound is $PbFe_{1/2}^{3+}Nb_{1/2}^{5+}O_3$ (PFN) in which $Nb^{5+}$ ions are ferroelectrically active and $Fe^{3+}$ ions are magnetic, respectively. While a theoretical prediction of the ferroelectric and antiferromagnetic orders respectively below certain temperatures was given, simultaneous experiments confirmed the ferroelectric Curie temperature of ~385 K and the Néel point of ~143 K [7, 42-44], noting that the two ordering temperatures are far from each other. A saturated polarization as high as ~65 $\mu C/cm^2$ in epitaxial PFN thin films was also reported, as shown in Fig. 5(a) [45], demonstrating the excellent ferroelectric property.

The coupling between magnetic order and ferroelectric order in this kind of multiferroics is, in most cases, very weak because these two orders originate from different kinds of ions. The consequent magnetoelectric coupling can be understood phenomenologically. According to the Ginzburg-Landau-Devonshire theory investigated by Kimura $et$ $al$ [46], the thermodynamic potential $\Phi$ in a multiferroic system can be expressed as:

$$\Phi = \Phi_0 + aP^2 + \frac{b}{2}P^4 - PE + a'M^2 + \frac{b'}{2}M^4 - MH + \gamma P^2 M^2 , \qquad (5)$$

where $\Phi_0$ is the reference potential, $a$, $a'$, $b$, $b'$ are related coefficients, respectively, and the term "$\gamma P^2 M^2$" is the coupling between $P$ and $M$ ( i.e. the magnetoelectric coupling term).

Surely, a variation of $M$ would influence the ferroelectricity, and eventually, the magnetic transition would result in a change of dielectric constant $\varepsilon \propto \partial^2 \Phi / \partial P^2$ around the transition point. Although this response would be quite weak because of the very small coefficient $\gamma$, one can use this response to check the validity of this theory. As an example, for PFN, the difference in dielectric constant between experimentally measured $\varepsilon(T)$ and the data extrapolated from the paramagnetic region at temperature $T>T_N$, can be denoted as $\delta\varepsilon$. By $\varepsilon \propto \partial^2 \Phi / \partial P^2$, one easily obtains $\delta\varepsilon \sim \gamma M^2$ (i.e. $\delta\varepsilon$ is proportional to the square of magnetization). Yang *et al* synthesized high quality PFN single crystals using a high-temperature flux technique and carefully studied the magnetic and dielectric properties as a function of temperature [47]. Obvious anomalies in the dielectric constant $\varepsilon$ near the Néel point ($\sim$143 K) was observed, as shown in Fig. 5(c). A linear relationship between $\delta\varepsilon$ and $M^2$ in the range of 130 K and 143 K was demonstrated, shown in Fig. 5(d), confirming the Ginzburg-Landau-Devonshire theory. This work revealed that there does exist magnetoelectric coupling between the ferroelectric order and magnetic order in PFN. Here, the low temperature magnetic order was approved by the Mössbauer spectra. In particular, the weak ferromagnetic order, as shown in Fig. 5(b), was argued to originate from the magnetoelectric coupling interaction [47].

Besides PFN, other multiferroics falling in the category of $AB_{1-x}B'_xO_3$, such as $PbFe_{1/2}^{3+}Ta_{1/2}^{5+}O_3$ [41] and $PbFe_{1/2}^{3+}W_{1/2}^{5+}O_3$ [48], were synthesized. Similar investigations performed on these materials also revealed a weak magnetoelectric coupling between the ferroelectric and spin orders. Again, it was shown that the weak magnetoelectric coupling exists because of the different and independent origins in the two types of orders. We may call these multiferroics as independent multiferroic materials.

### 3.2. *Ferroelectricity induced by lone pair electrons*

#### 3.2.1. *Mechanism for ferroelectricity induced by lone pair.* In addition to the ligand-field hybridization of a B-site TM cation by its surrounding anions, which is responsible for the ferroelectric order, the existence of $(ns)^2$ (lone pair) ions may also favor breaking the

inversion symmetry, thus inducing and stabilizing the ferroelectric order. In general, those ions with two valence electrons can participate in chemical bonds using (sp)-hybridized states such as $sp^2$ or $sp^3$. Nevertheless, this tendency may not be always true and, for some materials, these two electrons may not eventually participate in such bonding. They are called the "lone pair" electrons. $Bi^{3+}$ and $Pb^{3+}$ ions have two valence electrons in $s$-orbit, which belong to the lone pairs. The lone pair state is unstable and will invoke a mixing between the $(ns)^2$ ground state and a low-lying $(ns)^1(np)^1$ excited state, which eventually leads these ions to break the inversion symmetry [49-51]. This "stereochemical activity of the lone pair" helps to stabilize the off-center distortion and, in turn, the ferroelectricity. In typical ferroelectrics $PbTiO_3$ and $Na_{0.5}Bi_{0.5}TiO_3$, both the lone pair mechanism and the ligand-field hybridization take effect simultaneously [49].

The ions with lone pair electrons, such as $Bi^{3+}$ and $Pb^{3+}$, always locate at A-sites in an $ABO_3$ perovskite structure. This allows for magnetic TM ions to locate at B-sites so that the incompatibility for TM ions to induce both magnetism and ferroelectricity is partially avoided. The typical examples are $BiFeO_3$ and $BiMnO_3$, where the B-site ions contribute to the magnetism and the A-site ions via the lone pair mechanism lead to the ferroelectricity. In view of the origins for the two types of orders and magnetoelectric coupling, this approach shows no essential difference from the independent multiferroic materials highlighted in Sec. 3.1.

What are amazing are the intense investigations on $BiFeO_3$ and $BiMnO_3$ all over the world, which focus on the enhanced ferromagnetism and ferroelectricity. The strong magnetoelectric coupling in macroscopic sense, such as the mutual control of ferroelectric domains and antiferromagnetic domains, were revealed by recent experiments. Therefore, it may be beneficial to spend some space for addressing the two materials.

In both $BiMnO_3$ and $BiFeO_3$, $Bi^{3+}$ ions with two electrons on 6s orbit (lone pair) shift away from the centro-symmetric positions with respect to the surrounding oxygen ions, favoring the ferroelectricity. The magnetism is, of course, from $Fe^{3+}$ or $Mn^{3+}$ ions. $BiMnO_3$ is unique, in which both $M$ and $P$ are reasonably large. In fact, it is one of the very exceptional multiferroics offering both ferroelectric and ferromagnetic orders. $BiMnO_3$ has a monoclinic perovskite structure (space group $C2$) [52, 53], and shows a ferroelectric transition at $T_{ferroelectric}$~800 K accompanied by a structure transition shown in Figs. 6(a)~6(d) with the remanent polarization of ~16 $\mu C/cm^2$ [54-56], and a ferromagnetic transition at $T_{FM}$~110 K shown in Fig. 6(e) [57], below which the two orders coexist. The electron localization functions (ELFs) obtained by a first principle calculation facilitate a visualization of the bonding and long pairs in real space which, in turn, approves the "lone pair" mechanism in $BiMnO_3$ [51]. In Fig. 7(a) is presented the valence ELFs of cubic $BiMnO_3$ projected onto different lattice planes, together with the ELFs of cubic $LaMnO_3$ for comparison. The blue end of the scale bar represents the state with nearly no electron localization, while the white end represents complete localization. It is clearly shown that the ELFs on the Mn-O plane of both cubic compounds have similar patterns and even a similar spin polarization. However, large differences can be found on the Bi-O plane. The 6s "lone pairs" around Bi ions are approximately spherical, forming the orange rings of localization. This spherically distributed lone pairs form a domain of localization that is reducible and tends to be unstable. In addition, the localization tendency of the lone pairs to form a lobe pattern can be strong enough to drive a structural distortion [48-50]. The calculated ELFs for monoclinic $BiMnO_3$ are shown in Fig. 7(b). In order to adapt the traditional lone-pair geometry, the visible regions in the iso-surface correspond to the lobe-like Bi lone pairs allowed by the distorted geometry of the monoclinic structure. Further calculations reveal that the localized lone pair in the distorted structure is not only composed of the expected Bi 6s and 6p states, but also of some contribution from the 2p states on the oxygen ligands [51]. These predictions suggest that the lone pairs on the Bi ions in $BiMnO_3$ are stereochemically active and are the primary driving force for the highly distorted monoclinic structure, and thus, ferroelectricity in $BiMnO_3$ [51].

The magnetoelectric coupling between the ferroelectricity and magnetism in $BiMnO_3$ would be weak, as argued above and confirmed experimentally. The observed dielectric constant shows only a weak anomaly at $T_{FM}$ and is fairly insensitive to external magnetic field. The maximum decrease of dielectric constant $\varepsilon$ upon a field of 9 T appearing around $T_{FM}$ is ~0.6%, as shown in Fig.6 (f) [46].

### 3.2.2. *Room-temperature multiferroic BiFeO₃.* 

$BiFeO_3$ is another well known multiferroic material because it is one of the few multiferroics with both ferroelectricity and magnetism above room temperature. The rhombohedrally distorted perovskite structure can be indexed with $a=b=c$=5.633Å, $\alpha=\beta=\gamma$=59.4º and space group $R3c$ at room temperature, owing to the shift of Bi ions along the [111] direction and distortion of $FeO_6$ octahedra surrounding the [111] axis, as shown in Fig. 8(a) [58-61]. The electric polarization prefers to align along the [111] direction, as shown by the arrow. The ferroelectric Curie point is $T_C$~1103 K and the antiferromagnetic Neel point is $T_N$~643 K, while weak ferromagnetism at room temperature can be observed due to a residual moment in a canted spin structure [59, 60]. The high ferroelectric Curie point usually refers to a large polarization since other typical ferroelectrics with such Curie points have a polarization up to ~100 $\mu C/cm^2$. However, for $BiFeO_3$ single crystals, the   measured $P$ along the [001] direction at 77 K was 3.5 $\mu C/cm^2$, indicate a possible $P$ of only 6.1 $\mu C/cm^2$ along the [111] direction, as reported in earlier work [62]. For polycrystalline samples, the expected value of $P$ should be smaller. The reason for this small polarization is possibly due to the high leakage current as a result of defects and the nonstoichiometry of the test materials.

In fact, this issue has been cleared out recently. To overcome this obstacle, recent work focused on new synthesizing methods [63-66] and solid solutions of $BiFeO_3$ with other $ABO_3$ ferroelectric materials [67-73]. By improving the method for single crystal growth, high quality single crystals of $BiFeO_3$ with a polarization of ~60 $\mu C/cm^2$ was obtained [63, 64], indicating that the [111]-oriented polarization can reach up to 100 $\mu C/cm^2$, as shown in Fig. 8(b). A new sintering method for polycrystalline ceramics, the so-called liquid phase rapid sintering, was developed in the authors' laboratory with which the volatilization of Bi ions during the sintering was essentially avoided [65]. Rapid annealing of pre-sintered $BiFeO_3$ ceramics was also demonstrated in order to enhance the electric property [66]. The ferroelectricity and magnetism can also be significantly enhanced by substituting Bi ions with rare-earth ions such as $La^{3+}$ and $Pr^{3+}$, similarly due to the suppression of Bi evaporation and mixed valence of Fe ions [67-70].

While practical applications prefer high quality $BiFeO_3$ thin films in heteroepitaxial form, a large amount of effort was devoted to thin films of $BiFeO_3$ [74-80] where the crystal structure is monoclinic rather than rhombohedral as seen in bulk ceramic samples, due to the strain of substrates. Nowadays, high quality $BiFeO_3$ epitaxial films with room-temperature polarization as high as 60~80 $\mu C/cm^2$ which approaches the theoretical value, are available [74, 81]. Moreover, researches revealed that the in-plane strains in the thin films could drive a rotation of the spontaneous polarization on the (110) plane, while the polarization magnitude itself remains almost constant, which is responsible for the strong strain tunablity of the out-of-plane remanent polarization in (001)-oriented $BiFeO_3$ films [81].

BiFeO_3 has a complicated magnetic configuration. Neutron scattering experiments revealed that the antiferromagnetic spin order is not spatially homogenous but rather a spatially modulated structure [60], manifested by an incommensurate cycloid structure of a wavelength of $\lambda \sim 62$ nm, as shown in Fig. 8(c) and (d). The spiral spin propagation wave vector $q$ is along the $[10\overline{1}]$ directions and the polarization is along the [111] directions. These two directions define the $(\overline{1}2\overline{1})$ cycloidal plane where the spin rotation occurs, as shown by Fig. 8(d) and the shaded region in Fig. 8(c). Due to this feature, the antiferromagnetic vector is locked within the cycloid, averaged to zero over a scale $\sim\lambda$, and responsible for the very weak magnetization of bulk $BiFeO_3$. It is expected that this cycloid structure may be partially destroyed if the sample size is as small as the cycloid wavelength ($\sim 62$ nm), predicting enhanced magnetization and even weak ferromagnetism in nanoscale $BiFeO_3$ samples. It is this mechanism that results in the enhanced magnetization in the thin film sample [74]. Other grain reducing methods for improving the ferromagnetism of $BiFeO_3$ were also reported. For example, $BiFeO_3$ nanowires and nanoparticles do show ferromagnetism [82, 83], as shown in Fig. 9. Moreover, the optical decomposition of organic contaminants by a nanopowder of $BiFeO_3$ as a high photocatalyst was also demonstrated recently [83, 84].

Based on the same reasons for $BiMnO_3$, one may postulate that the magnetoelectric coupling in $BiFeO_3$ would be very weak, too. However, some recent studies found that the ferroelectric polarization is closely tied to the incommensurate cycloid spin structure and a significant magnetoelectric effect was observed in $BiFeO_3$. Because this situation is very similar to the ferroelectricity induced by spiral spin order, we will carefully discuss this effect in Sec.3.4.5.

Besides $BiMnO_3$ and $BiFeO_3$, attention has been given to other Bi-containing multiferroics in the same category. For example, bismuth layer-structured ferroelectrics $Bi_{4+n}Ti_3Fe_nO_{12+3n}$ ($n=1$), which is a member of the Aurivillius-type materials and has a four-layered perovskite structure, is composed of units with nominal composition $(Bi_3Ti_3FeO_{13})^{2-}$ sandwiched between two $(Bi_2O_2)^{2+}$ layers along the c-axis [85]. It has both the ferroelectric and magnetic orders below a certain temperature. In order to enhance the ferromagnetism and ferroelectricity in $BiFeO_3$, researchers focused on a theoretical prediction [86, 87] that $Bi_2(Fe,Cr)O_6$ would exhibit huge macroscopic magnetization and polarization, due to the ferromagnetic superexchange interaction between Fe and Cr ions which induces the ferromagnetic state in $La_2(Fe,Cr)O_6$ [88, 89]. However, it is challenging to synthesize materials with ordered Fe and Cr ions. Meanwhile, compared with pure $BiFeO_3$, samples with disordered Fe/Cr configuration showed no significant improvement of the multiferroicity [90, 91]. Similarly, $Bi_2NiMnO_6$ was studied carefully, too, due to the ferromagnetic superexchange interaction between Ni and Mn ions [92-93]. It is also worthy of noting that multiferroic $PbVO_3$ facilitated with another lone-pair ion, $Pb^{2+}$, was synthesized recently [95-98], which is very similar to conventional ferroelectric material, $PbTiO_3$. Furthermore, $Cu_2OSeO_3$, which is another lone pair containing materials, exhibits the coexistence of piezoelectricity and ferrimagentism but unfortunately no spontaneous polarization was measured. It exhibits significant magentocapacitance effects below the ferromagnetic Curie temperature of ~60 K [99, 100]. This is because $Cu_2OSeO_3$ is metrically cubic down to 10 K but the ferrimagnetic ordering reduces the symmetry to rhombohedral *R3* which excludes the spontaneous ferroelectric lattice distortion. Similar effects were observed in $SeCuO_3$, too [101].

### 3.3. *Geometric ferroelectricity in hexagonal manganites*

For those ferroelectrics addressed in the last two sections, the main driving force for the ferroelectric transitions comes from the structural instability toward the polar state associated with electronic pairing. They were coined as "proper" ferroelectrics. Different from this class of ferroelectrics, some other ferroelectrics have their polarization as the by-product of a complex lattice distortion. This class of materials, together with all other ferroelectrics with their polarization originating from by-product of other order configurations, were coined as "improper" ferroelectrics. Hexagonal manganites *R*MnO$_3$ with *R* the rare-earth element (Ho-Lu, or Y), fall into the latter category, and are often cited as typical examples that violate the "d$^0$-ness" rule.

**3.3.1.** *Geometric ferroelectricity and coupling effects in YMnO₃.* We take YMnO$_3$ as an example [12, 102-105]. It is a well known multiferroic system with a ferroelectric Curie temperature $T_{ferroelectric}$=950 K and an antiferromagnetic Neel temperature $T_N$=77 K. The hexagonal manganites and orthorhombic manganites, $R$MnO$_3$ where $R$ is the relative large ions such as La, Pr, Nd, etc, have very different crystal structures from those of small $R$ ions, in spite of their similar chemical formulas. The hexagonal structure adopted by YMnO$_3$ and other manganites with small $R$ ions consists of non-connected layers of MnO$_5$ trigonal bipyramids corner-linked by in-plane oxygen ions (O$_P$), with apical oxygen ions (O$_T$) which form close-paced planes separated by a layer of Y$^{3+}$ ions. The schematic views of the crystal structure are given in Fig. 10(a).

The different crystal structures are facilitated with different electronic configurations. In contrast to conventional perovskites, YMnO$_3$ has its Mn$^{3+}$ ions not inside the O$_6$ octahedra but coordinated by a 5-fold symmetry (i.e. in the centre of O$_5$ trigonal bipyramid). Similarly, $R$-ions (e.g. Y ions) are not in a 12-fold but a 7-fold coordination. Consequently, the crystal field level scheme of Mn ions in these compounds is different from the usual one in an octahedral coordination. The $d$-levels are split into two doublets and an upper singlet, instead of a triplet $t_{2g}$ and a doublet $e_g$ in orthorhombic perovskites (Fig. 10(b)). Therefore, the four $d$ electrons of Mn$^{3+}$ ions occupy the two lowest doublets, leaving no orbital degeneracy. Consequently, Mn$^{3+}$ ions in these compounds are not Jahn-Teller active.

Early work in the 1960s established YMnO$_3$ to be ferroelectric with space group *P6₃cm*, and revealed an A-type antiferromagnetic order with non-collinear Mn spins oriented in a triangular arrangement [12, 102]. The ferroelectric polarization arises from an off-center distortion of Mn ions towards one of the apical oxygen ions. However, careful structural analysis revealed that Mn ions remain very close to the center of the oxygen bipyramids and, thus, are definitely not instrumental in providing the ferroelectricity [106]. The first principle calculation also predicts that the off-center distortion of Mn ions is energetically unfavorable. The main difference between the paraelectric *P63/mmc* structure and ferroelectric *P6₃cm* one is that all ions in the paraelectric phase are restricted within the planes parallel to the *ab* plane, whereas in the ferroelectric phase, the mirror planes perpendicular to the hexagonal *c*-axis are lost, as shown in Fig. 10(a) and Fig. 11. The structural transition from the centrosysmmetric *P63/mmc* to the ferroelectric *P6₃cm* is mainly facilitated by two types of atomic displacements. First, the $MnO_5$ bipyramids buckle, resulting in a shorter *c*-axis and the $O_T$ in-plane ions are shifted towards the two longer Y-$O_P$ bonds. Second, the Y ions vertically shift away from the high-temperature mirror plane, keeping the constant distance to $O_T$ ions. Consequently, one of the two ~2.8 Å Y-$O_P$ bond length is reduced down to ~2.3 Å, and the other is elongated to 3.4 Å, leading to a net electric polarization [106]. The polarization dependent X-ray absorption spectroscopy (XAS) at O K and Mn $L_{2,3}$ edges of $YMnO_3$ demonstrated that the Y 4d states are indeed strongly hybridized with the O 2p states. This results in large anomalies in the Born effective charges on the off-centered Y and O ions [107].

The above picture suggests that the main dipole moments are contributed by the Y-O pairs instead of the Mn-O pairs. This is an additional example for the A-site ions induced ferroelectricity, but details of the mechanism for such distortion remain puzzling up to date. A close packing demand is one possible reason. To reach the close packing, the rigid $MnO_5$ trigonal bipyramids in $YMnO_3$ prefer to tilt and then lead to the loss of inversion symmetry and the ferroelectricity. Moreover, for a hexagonal $RMnO_3$, a combinatorial approach by structural characterization and electronic structure calculation, as done already, seems to devalue the role of re-hybridization and covalency in driving the ferroelectric transition, which is instead cooperatively driven by the long-range dipole-dipole interactions and oxygen rotations [106]. Interestingly, the huge Y-$O_P$ off-center displacements are quite distinct from the small displacements induced by chemical activity available for conventional ferroelectric perovskite oxides, but the induced electric polarization remains much smaller. Thus, one may argue that this is a completely different mechanism for ferroelectric distortion [108, 109].

The spin configuration of hexagonal $YMnO_3$ is frustrated, which will be addressed carefully in next section. The easy plane anisotropy of Mn spins restricts the moments strictly on the *ab* plane, which are thus dominated by the strong in-plane antiferromagnetic superexchange interaction. The inter-plane exchange between the Mn spins is two orders of magnitude lower. Therefore, $YMnO_3$ is an excellent example of a quasi two dimensional Heisenberg magnet on a triangular lattice with a spin frustration generated by geometric constraint. Accordingly, the Mn spins undergoing long range order at $T_N$ usually develop into a non-collinear configuration with a $120^\circ$ angle between neighboring spins [110-115].

For hexagonal manganites, all theoretical and experimental evidences consistently favor the Y-$d^0$ness with re-hybridization being the driving force for the ferroelectricity. This stands for a substantial new approach to ferroelectricity. In this framework, the strong coupling between ferroelectric order and magnetic order (magnetoelectric coupling) may be expected because both orders are essentially associated with the lattice structure. For example, Fiebig *et al* employed optical second harmonic generation to map the coupled magnetic and ferroelectric domains in YMnO$_3$ [113]. In this case, as proposed by the symmetry analysis, YMnO$_3$ has four types of 180° domains denoted by (+*P*, +*l*), (+*P*, -*l*), (-*P*, -*l*) and (-*P*, +*l*), respectively, where ±*P* and ±*l* are the independent components of the ferroelectric and antiferromagnetic order parameters. Any ferroelectric domain wall will be coupled with an antiferromagnetic domain wall, as shown in Fig.12, thus the sign of the product *Pl* must be conserved upon crossing a ferroelectric domain wall [113]. Moreover, a significant anomaly of the dielectric constant in response to theelectric field along the *ab* plane ($\varepsilon_{ab}$) can be observed at $T_N$, but no anomaly at $T_N$ is available when the electric field is along the *c*-axis [114]. These experiments provide fascinating evidence that supports the strong magnetoelectric coupling in YMnO$_3$.

### 3.3.2. *Magnetic phase control by electric field in HoMnO$_3$.*

For *R*MnO$_3$, such as hexagonal HoMnO$_3$, in addition to the complex Mn spin structure, usually $R^{3+}$ ions also carry their own spin (magnetic moment) that is non-collinear with the Mn spins. The ferroelectric phase of HoMnO$_3$ appears at the Curie point $T_C$=875 K, and possesses *P6$_3$cm* symmetry with a polarization *P*=5.6 μC/cm$^2$ [116-119] along the hexagonal *c*-axis. Besides the Mn$^{3+}$ ions, Ho$^{3+}$ ions with *f* electrons also contribute a nonzero magnetic moment with the easy axis anisotropy along the *c*-axis, noting that the Mn$^{3+}$ spins are restricted within the basal *ab* plane due to the anisotropy. The as-induced frustration favors four kinds of possible triangular antiferromagnetic configurations, as shown in Fig. 13(a), in which the magnetic ordered states are composed of three magnetic sublattices with Mn$^{3+}$ ($3d^3$) ions at the $6c$ positions and Ho$^{3+}$ ($4f^{10}$) ions at the $2a$ and $4b$ positions, respectively. At low temperature, the exchange coupling between $Ho^{3+}$ and $Mn^{3+}$ magnetic subsystems becomes strong enough so that additional distinct changes of magnetic structure may occur. Below $T_N \sim 76$ K, the Mn spins favor the non-collinear antiferromagnetic ordering. The coupling between the Mn spins and Ho spins drives an in-plane rotation of the Mn spins at $T_{SR} \sim 33$ K. Correspondingly, the Ho spins become magnetically polarized and a small magnetization from the antiferromagnetic sublattice was detected and enhanced as temperature fell down. In fact, the measured *c*-axis magnetic susceptibility has an abrupt decrease at $T_{SR}$, although the change is small, indicating the onset of the antiferromagnetic Ho spin order with magnetic moments aligned along the hexagonal *c*-axis. At an even lower temperature, $T_{Ho} \sim 5$ K, another spin reorientation transition associated with the Ho spins takes place, leading to a low temperature phase with *P6₃cm* magnetic symmetry and a remarkable enhancement of Ho spin moment. This configuration remains antiferromagnetic. The two $Ho^{3+}$ sublattices are assumed to be Ising-like ordered along the *z(c)*-axis, exhibiting the antiferromagnetism or ferri-/ferromagnetism.

It is important to mention that the dielectric property of $HoMnO_3$ is very sensitive in response to the subtle variation of the magnetic order [117, 118]. The dielectric constant as a function of temperature, *ε(T)*, under zero magnetic field, exhibits three distinct anomalies, shown in Fig. 13(b) and 13(c). At the Neel point, *ε(T)* shows a clear decrease due to the onset of an antiferromagnetic order with the Mn spins. This feature was confirmed in other hexagonal manganites or multiferroics and is usually viewed as a symbol of antiferromagnetic ordering. The transition into the *P6₃cm* magnetic structure at $T_{Ho} \sim 5.2$ K is accompanied by a sharp increase of *ε(T)*. The most notable anomaly of *ε(T)* is the sharpest peak at $T_{SR} \sim 32.8$ K. In addition, the dielectric constant and these anomalies exhibit evident dependence on magnetic field. A magnetic field *H,* imposed along the *c*-axis, shifts the sharpest peak at $T_R$ toward a lower temperature, and the peak at $T_{Ho}$ toward a higher temperature. Eventually, the two peaks develop similar plateaus and merge at $H \sim 33$ kOe, as shown in Fig. 13(c). Above $H \sim 40$ kOe, all anomalies associated with *ε(T)* are suppressed, leaving a small drop at $\sim 4$ K. These additional anomalies indicate the phase complexity and mark the generation of field induced reentrant novel phase due to the indirect coupling between the ferroelectric and antiferromagnetic orders [117, 118].

The most fascinating effect with hexagonal *R*MnO₃ is the magnetic phase control by an electric field, as demonstrated in HoMnO$_3$ [119]. Using an optical second harmonic generation technique, it was observed that at $T_N$, external electric field may drive HoMnO$_3$ into a magnetic state different from that under zero electric field, thereby modulating the magnetic order of the Mn$^{3+}$ sublattice, as shown in Fig. 13. Moreover, compared with YMnO$_3$, HoMnO$_3$ has an extra magnetic sublattice consisting of Ho$^{3+}$ ions, which shows an interesting response to electric field. In the presence of an electric field, the para- or antiferromagnetic state under zero field is converted into a ferromagnetic order with strong macroscopic magnetization. The proposed mechanism for this phase control is the microscopic magnetoelectric coupling originating from the interplay of the Ho$^{3+}$-Mn$^{3+}$ interactions and ferroelectric distortion [119]. The large difference in far-infrared spectroscopy regarding the antiferromagnetic resonance splitting of Mn ions between YMnO$_3$ and HoMnO$_3$ demonstrates the ferromagnetic exchange coupling between Mn ions and surrounding Ho ions [120]. However, the role of Ho$^{3+}$ ionic spins in HoMnO$_3$ remains ambitious up to now. For example, the X-ray resonant scattering experiment indicated that the magnetic structure of Ho$^{3+}$ ions remains unchanged upon an applied electric field as high as $10^7$ V/m [121], which may suggest no contribution of Ho$^{3+}$ spins to the ferromagnetic state of HoMnO$_3$ under an electric field.

Similar effects were also identified in other multiferroics in the same category, such as YbMnO$_3$ [122], InMnO$_3$ [123, 124] and (Lu/Y)CrO$_3$ [125-127]. But the detail mechanism of ferroelectricity in these compounds remains a puzzle. For example, more recently a new concept of "local noncentrosymmetry" in YCrO$_3$ has been proposed to account for the small value of polarization observed in spite of the large $A$-cation off-centering distortion [125, 126]. It is amazing that these multiferroics may possibly be prepared in a constrained manner so that a metastable phase can be maintained using special approaches. For instance, bulk TbMnO$_3$ is of an orthorhombic structure (it is ferroelectric, to be addressed in next section), but a hexagonal metastable TbMnO$_3$ can be epitaxially deposited on an in-plane hexagonal Al$_2$O$_3$ substrate [128]. With respect to the bulk phase, the hexagonal TbMnO$_3$ films may exhibit ~20 times larger remnant polarization with the ferroelectric Curie point shifting to ~60 K. Additionally, while an antiferroelectric-like phase and a clear signature of the magnetoelectric coupling were observed in hexagonal TbMnO$_3$ films, the metastable orthorhombic (Ho/Y)MnO$_3$ can be synthesized under high pressure condition [129]. In the orthorhombic HoMnO$_3$, below the antiferromagnetic Neel point, the Ho spins tilt toward the *a*-axis from their original alignment (along the *c*-axis) in the hexagonal phase, and a larger magnetoelectric coupling was detected, probably being ascribed to the *E*-phase antiferromagnetic order [130], which will be carefully discussed in Sec.3.6.

### 3.4. *Spiral spin order induced multiferrocity*

So far, we have reviewed various mechanisms for multiferrocity in several types of multiferroics. These mechanisms definitely shed light on researches on novel multiferroics. Nevertheless, it has to be noted that the perspectives of these mechanisms are somewhat disappointing. In these multiferroics, the ferroelectricity and magnetism basically originate from different ions or subsystems. In a general and macroscopic sense, one may not expect a very strong magnetoelectric coupling in these multiferroics. An exception is owed to the ferroelectricity directly induced by the spin order, meaning that an intrinsic magnetoelectric coupling occurs between the ferroelectric and magnetic order parameters. Keeping this in mind, the primary problem is how to overcome the inter-exclusion between ferroelectricity and magnetism so that any special spin order can induce ferroelectricity.

### 3.4.1. *Symmetry consideration.*

**3.4.1. *Symmetry consideration.*** The inter-exclusion between ferroelectricity and magnetism originates not only from the $d^0$-ness rule, but also from the symmetry restriction of the two types of orders. Ferroelectricity needs the broken spatial inverse symmetry and usually invariant time reverse symmetry, in which electric polarization *P* and electric field *E* change their signs upon an inversion operation of all spatial coordinates $r \rightarrow -r$ but may remain invariant upon an operation of time reversal $t \rightarrow -t$. In contrast, the broken time-reversal symmetry is the prerequisite for magnetism (spin order), in which magnetization *M* and magnetic field *H* change their signs upon time reversal and may remain invariant upon spatial inversion. Consequently, a multiferroic system that is both ferromagnetic and ferroelectric requires the simultaneous breaking of the spatial-inversion and time-reversal symmetries. The magnetoelectric coupling between polarization *P* and magnetization *M* is derived based on this general symmetry argument [131-133].

First, time reversal $t \rightarrow -t$ must leave the magnetoelectric coupling invariant. As this operation transforms $M \rightarrow -M$, and leaves $P$ invariant, the lowest order magnetoelectric coupling term has to be quadratic in $M$. However, the fourth-order term $-P^2M^2$ does not contribute to any ferroelectricity because it is compensated by the energy cost for a polar lattice distortion proportional to $-P^2$, although $-P^2M^2$ term may account for the small change in dielectric constant at a magnetic transition (as identified for $BiMnO_3$ etc.) [46]. However, given the case of a spatially inhomogeneous spin configuration (i.e. magnetization $M$ is a function of spatial coordinates), the above symmetry argument allows for the 3rd-order magnetoelectric coupling (i.e. the coupling between a homogeneous polarization and an inhomogeneous magnetization can be linear in $P$ and contains one gradient of $M$) [132].

This simple symmetry argument immediately leads to the following magnetoelectric coupling term in the Landau free energy [132, 133]:

$$\Phi_{ME}(r) = P \cdot \{ \gamma \cdot \nabla(M^2) + \gamma' [ M(\nabla \cdot M) - (M \cdot \nabla)M ] + ... \} , \qquad (6)$$

where $r$, $P$, and $M$ are vectorized spatial coordinate, polarization and magnetization; $\gamma$ and $\gamma'$ are the coupling coefficients. The first term on the right-hand side is proportional to the total derivative of the square of magnetization and would not give contribution unless $P$ is assumed to be independent of spatial coordinate $r$. By including the energy term associated with $P$, i.e. $P^2/2\chi_e$, where $\chi_e$ is the dielectric susceptibility, into the free energy, a minimization of the free energy with respect to $P$ produces:

$$P = \gamma' \chi_e [ M(\nabla \cdot M) - (M \cdot \nabla)M ] , \qquad (7)$$

This simple symmetry argument predicts the possible multiferroicity in spin frustrated systems which always prefer to have spatially inhomogeneous magnetization due to the competing interactions. For example, a one-dimensional spin chain with a ferromagnetic nearest-neighbor interaction $J<0$ has a uniform ground state with parallel-aligned spins. An additional antiferromagnetic next-nearest-neighbor interaction $J'>0$ which meets $J'/|J|>1/4$, (i.e. the Heisenberg model $H=\sum_n[J \cdot S_n \cdot S_{n+1} + J' \cdot S_n \cdot S_{n+2}]$, where $S_i$ is the Heisenberg spin moment at site $i$ referring to a spin chain), frustrates this simple spin order [134], as shown in Fig. 14(a). The frustrated ground state is characterized by a spiral spin order (spiral spin-density wave) and can be expressed as:

$$S_n = S_1 e_1 \cos Q \cdot r + S_2 e_2 \sin Q \cdot r \,, \qquad (8)$$

where the unit vectors $e_i$ ($i=1,2,3$) form an orthogonal basis, $e_3$ is the axis around which spins rotate and vector $Q$ is given by $\cos(Q/2)=-J''(4J)$. If only $S_1$ or $S_2$ are nonzero, this equation describes a sinusoidal spin-density wave, which cannot induce any ferroelectricity because it is invariant upon the spatial inversion operation $r \rightarrow -r$. Given that $S_1$ and $S_2$ are both nonzero, Eq.(8) describes a spiral spin order (spiral spin-density wave) with the spin rotation axis $e_3$. Like any other magnetic order, the spiral spin order spontaneously breaks the time reversal symmetry. In addition, it also breaks the spatial inversion symmetry because the sign reversal of all coordinates inverts the direction of the spin rotation in the spiral. Therefore, the symmetry of the spiral spin state allows for a simultaneous presence of multiferroicity. Using Eq.(7) and Eq.(8), one finds that the average polarization is transverse to both $e_3$ and $Q$:

$$\overline{P} = \frac{1}{V}\int d^3x P = \gamma' \, \chi_e S_1 S_2 [\, e_3 \times Q \,] \,, \qquad (9)$$

The above simple model can be extended to two- or three-dimensional spin systems. In general, two or more competing magnetic interactions can induce the spin frustration and the spiral (helical) spin order which, in turn, breaks the spatial inversion and time reversal symmetries simultaneously, thus establishing the ferroelectric order.

What should be mentioned here is a question that the spiral spin order (spiral spin-density wave) is a prerequisite for generating ferroelectricity remains unsure. It was theoretically predicted that the acentric dislocated spin-density wave (SDW) may also drive a ferroelectric polarization [133]. For a SDW order described by $M=M_0 \cdot \cos(q_m x + \varphi)$ where $q_m$ is the magnetic ordering wave vector and $\varphi$ is its phase, magnetization $M$ is phase-dislocated with respect to the lattice wave vector. As the spins are collinear and sinusoidal, a center of symmetry exists but no directionality is available, eventually no ferroelectricity is possible. However, for an acentric SDW system, $M^2$ falls behind with respect to polarization $P$, which is the immediate consequence of the finite phase difference $\varphi$. Thus, $M^2$ has some directionality in relation to $P$, which is a sufficient condition for a direct coupling between the two types of orders and a macroscopic polarization [133].

Surely, one may expect additional long-range and spatially inhomogeneous spin structures which can produce nonzero polarization $P$, following Eq.(7). This issue remains interesting and deserves further investigation.

**3.4.2. *Microscopic mechanism.*** In addition to the symmetry argument disclosed above, a microscopic mechanism responsible for ferroelectricity in magnetic spiral systems is required. Unfortunately, it was found that such a mechanism is very complex and a clear answer has not yet been available. Currently, three theories on the microscopic aspect of magnetoelectric coupling in magnetic spiral multiferroics have been proposed: the inverse Dzyaloshinskii-Moriya (DM) model (exchange striction approach) [135, 136], the spin current model (KNB model) [137], and the electric current cancellation model [138].

***The inverse DM model:*** A plausible microscopic mechanism for ferrroelectricity in the spin spiral system is the displacement of oxygen ions driven by the antisymmetric Dzyaloshinskii-Moriya (DM) interaction [139, 140] which is a relativistic correction to the usual superexchange interaction. In fact, it has been a long time issue whether a weak (canted) ferromagnetism can be generated by the DM interaction in some compounds such as $La_2CuO_2$. As early as 1957, Dzyaloshiskii pointed out that a "weak" ferromagnetism may be possible in antiferromagnetic compounds such as $Fe_2O_3$ but may not in the isostructural oxide $Cr_2O_3$. This prediction was made within the framework of symmetry argument. Dzyaloshiskii proposed that an invariant in the free energy expansion of the following form [141]:

$$E_{DML} = D \cdot (M \times L),$$                                                          (10)

where $D$ is the materials-specific vector coefficient, $M$ the magnetization, and $L$ the antiferromagnetic order parameter (vector), will result in appearance of the second order parameter $M$ at the antiferromagnetic ordering temperature. In other words, if the symmetry of a pure antiferromagnetic state is such that the appearance of a small magnetization does not lead to further symmetry lowering, any microscopic mechanism which favors a nonzero magnetization, even if it is rather weak, will lead to $M \neq 0$. A possible microscopic mechanism was proposed subsequently by Moriya, who pointed out that such an invariant with the required form can be realized by an antisymmetric microscopic coupling between two localized magnetic moment $S_i$ and $S_j$ [140]:

$$E_{ij}^{DM} = d_{ij} \cdot ( S_i \times S_j ),$$ (11)

where $d_{ij}$ is the prefactor. This invariant term is the so-called DM interaction, and $d_{ij}$ is the DM factor.

For a spiral spin ordered state, the classical low temperature spin structure can be described as $S_n^i = S_0^i \cos(n\theta + \alpha^i)$, where $i=(x, y, z)$. A detailed consideration for typical multiferroic TbMnO$_3$ was given by Sergienko and Dagotto [135] and is described here. For TbMnO$_3$, $S_0^x = S_0^y = S_0^z = 1.4$, $\theta = 0.28\pi$, $\alpha^i$ is a constant, but not critical to the physics. Assuming that the positions of Mn ions are fixed and oxygen ions may displace from their center positions, the isotropic superexchange interaction of a Mn-O-Mn chain in the $x$ direction (as shown in Fig. 14(c)) can be described as:

$$H_{ex} = -\sum_n [ J_0 + \frac{1}{2} J'_{\parallel} x_n^2 + \frac{1}{2} J'_{\perp} ( y_n^2 + z_n^2 )] ( S_n \cdot S_{n+1} ),$$ (12)

where $J_0$, $J'_{\perp}$, and $J'_{\parallel}$ are the exchange constants, $r_n = (x_n, y_n, z_n)$ is the displacement of oxygen ion located between the Mn spins $S_n$ and $S_{n+1}$. In an orthorhombically distorted structure, the displacement of oxygen ion can be described as $r_n = (-1)^n r_0 + \delta r_n$, where $r_0$ is a constant and $\delta r_n$ is the additional displacement associated with the incommensurate structure. Taking into account the elastic energy $H_{el} = \kappa \sum_n ( \delta x_n^2 + \delta y_n^2 + \delta z_n^2 ) / 2$ associated with the displacement, where $\kappa$ is the stiffness, the total free energy upon a minimization yields:

$$\delta z_n = (-1)^n \frac{J'_{\perp} z_0}{2\kappa} \sum_i S_0^i \{ \cos\theta + \cos[(2n+1)\theta + 2\alpha_i ] \},$$ (13)

and similar expressions for $\delta x_n$ and $\delta y_n$ can be obtained. Note that this displacement still cannot induce the ferroelectric polarization because of $\sum_n r_n = 0$.

Further consideration has to go to the antisymmetric DM interaction $D^i(r_n) \cdot (S_n \times S_{n+1})$ which will change its sign under the spatial inversion. For a perovskite structure, the DM factor $D^x(r_n) = \gamma(0, -z_n, y_n)$ and $D^y(r_n) = \gamma(z_n, 0, -x_n)$ for the Mn-O-Mn chain along the $x$ and $y$ directions, respectively. The Hamiltonian, depending on $\delta r_n$ for the Mn-O-Mn chain along the $x$ directions, respectively, can be written as:

$$\delta H_{DM} = \sum_n D^x(\delta \vec{r}_n) \cdot [\, S_n \times S_{n+1}\,] + H_{el}, \qquad (14)$$

and a minimizing of Eq.(12) with respect to $\delta \vec{r}_n$ (exchange strictive effect) yields:

$$\delta \tilde{z}_n = \frac{\gamma}{\kappa} S_0^x S_0^y \sin\theta \sin(\alpha_x - \alpha_y),$$
$$\delta \tilde{x} = \delta \tilde{y} = 0 \qquad (15)$$

Hence the DM interaction drives the oxygen ions to shift in one direction perpendicular to the spin chain, thus resulting in an electric polarization, as shown in the lower panel of Fig. 14(d). If the spin configuration is collinear, parameter $\delta \vec{r}_n$ as given by Eq.(15) vanishes, i.e. the paraelectric state. For example, in $La_2CuO_4$, the weak ferromagnetism (as shown in the upper part of Fig.14(d)) would induce alternative displacement of O atoms and then no ferroelectric polarization. This suggests that a non-collinear spin configuration is a necessary ingredient of ferroelectricity generation by the DM interaction.

Applying this conceptual picture to a realistic system, such as a perovskite manganite $R\text{MnO}_3$, one can develop a practically applicable microscopic model. Combining the orbitally degenerate double-exchange model together with the DM interaction, a microscopic Hamiltonian for orthorhombical multiferroic manganites can be described as [135]:

$$H = -\sum_{ia,\alpha\beta\gamma} t_{\alpha\beta}^a d_{i\alpha\sigma}^+ d_{i+a,\beta\sigma} - J_H \sum_i s_i \cdot S_i + J_{AF} \sum_{ia} S_i \cdot S_{i+a} +$$
$$+ \sum_{ia} D^a(\vec{r}) \cdot [\, S_i \times S_{i+a}\,] + H_{JT} + \frac{\kappa_1}{2} \sum_i (Q_{xi}^2 + Q_{yi}^2) + \frac{\kappa_2}{2} \sum_i (\sum_m Q_{mi}^2), \qquad (16)$$

where the first term on the right-hand side accounts for electron hopping (kinetic energy term), the second term is the Hund coupling, the third one is an antiferromagnetic superexchange interaction between neighbor local spins, the fourth term includes the DM interaction, the fifth refers to the Jahn-Teller term, and the last two terms come from the ferroelectric phonon modes (the displacement of O atoms). The roles of these terms are summarized in Fig. 15. A simulation based on this Hamiltonian revealed the appearance of incommensurate magnetic ferroelectric phase induced by ordered oxygen displacement, as shown in Fig. 16(a), and the simulated relative displacement of oxygen ions (i.e. ferroelectric polarization) is shown in Fig. 16(b). This model produces a phase diagram that is in excellent agreement with experiments [135].

A Monte Carlo simulation on the multiferroic behaviors of a two-dimensional $\text{MnO}_2$

lattice based on this model for multiferroic manganites was reported recently. The simulated ferroelectric polarization induced by the spiral spin ordering and its response to the external magnetic field agree with reported experimental observations [136]. Furthermore, the possible coexistence of clamped ferroelectric domains and spiral spin domains is predicted in this simulation. In short, it has been argued that the DM interaction, competing with other exchange interactions, stabilizes the helical (spiral) spin order, while the exchange striction effect favors the ferroelectric polarization.

***The KNB model:*** The spin current model to be addressed here was proposed by Katasura, Nagaosa, and Balatsky, then called the KNB model. It serves as the second microscopic explanation of multiferroicity in a spiral spin ordered system and also refers to manganites [137]. This model is very famous and has been widely utilized to explain a number of experimentally observed facts due to its clear physics and simple picture.

For a spin chain, the spin current from site $n$ to site $n+1$ can be expressed as $j_{n,n+1} \propto S_n \times S_{n+1}$, which describes the precession of spin $S_n$ in the exchange field created by spin $S_{n+1}$. The DM interaction leads to the spiral spin configuration and acts as the vector potential or gauge field to the spin current. The induced electric dipole between the site-pair is then given by $P_{n,n+1} \propto r_{n,n+1} \times j_{n,n+1}$, where $r_{n,n+1}$ is the vector pointing to site $n+1$ from site $n$. Although the model may be over-simplified, it is physically equivalent to the exchange striction approach.

***Electric current cancellation model:*** This model stems from fundamental electromagnetic principles [138]. The current operator of electrons is defined as the change in Hamiltonian with respect to the variance of vector potential of electromagnetic field, i.e.

$$J = -c \delta H / \delta A , \tag{17}$$

where $A$ is the vector potential of electromagnetic field and $c$ is the light velocity. In non-relativistic quantum mechanics, the definition of electric current includes three terms generated from three different physical origins: (1) the contribution of standard momentum, (2) the spin contribution, (3) the contribution of spin-orbital coupling. For example, we consider a single electron in a band structure described by the Hamiltonian:

$$H_e = \frac{(p - e\frac{A}{c})^2}{2m^*} + \alpha(p - e\frac{A}{c}) \cdot [\sigma \times \nabla V(r)] - \mu(\nabla \times A) \cdot \sigma, \qquad (18)$$

where $m^*$ is the effective mass of electrons, $\alpha$ the effective spin-orbital coupling parameter, $\mu = ge/2mc$ and $\sigma$ is the spin of electrons. In the absence of external electrodynamic field, i.e. $A=0$, for a given wave function $\Psi(r)$, the electric current from above equation is given by:

$$j = j_0 + \mu c\nabla \times (\Psi^*\sigma\Psi) + \alpha e(\Psi^*\sigma\Psi) \times \nabla V(r)$$
$$j_0 = \frac{ieh}{2m^*}[(\nabla \Psi^*)\Psi - \Psi^*(\nabla \Psi)] \qquad , \qquad (19)$$

where $h$ is the Planck constant, and the three terms precisely correspond to the three physical origins mentioned above. For the magnetization of electrons in the band with a simple spiral magnetic order, one has:

$$M = M_0[cos(qx/a), sin(qx/a), 0], \qquad (20)$$

where $M_0$ is the magnetic moment, $q$ is the spiral wave vector, $a$ is the lattice constant, and $x$ is the coordinate. The electric current associated with the magnetization is given by:

$$J_M = \mu c\nabla \times M = \frac{\mu cqM_0}{a}(0,0,cos\,qx/a), \qquad (21)$$

which represents the current along the $z$ direction.

In an insulator, the net electric current with such configuration must be zero, based on Kohn's proof of the insulator property. The total electric current contributed from $j_0$ in the band also vanishes since the lattice mirror symmetry in the $x$-$y$ plane is not broken for the noncollinear multiferroics in the absence of external magnetic field. Therefore, the electric current from the magnetic ordering must be counterbalanced by the electric current induced from the spin-orbit coupling. This cancellation requirement leads to:

$$\mu c\nabla \times M + \alpha eM \times \nabla V(r) = 0, \qquad (22)$$

By a simple algebraic modification and averaging over the total space, the above equation becomes:

$$\frac{\alpha e^2}{\mu c}\langle E\rangle = \left\langle \frac{(M \cdot E)M}{M_0^2}\right\rangle + \left\langle \frac{M \times (\nabla \times M)}{M_0^2}\right\rangle, \qquad (23)$$

where $\langle\dots\rangle$ refers to the space averaging and $\nabla V(r) = -eE(r)$. The first term in the right side of Eq.(23) usually vanishes when a space averaging for a spatially modulated spin density is made. The total ferroelectric polarization can then be written as:

$$P = \frac{\varepsilon_0 \mu c}{\alpha e^2} \left\langle \frac{M \times (\nabla \times M)}{M_0^2} \right\rangle,$$                             (24)

It is worthy of mention that the generated polarization $P$ is inversely proportional to the effective spin-orbital coupling parameter, a very unusual argument. Moreover, one can conclude that there is no contribution to the ferroelectricity from the completely filled bands since electrons in a fully filled band do not have magnetization response. Therefore, the contribution to the ferroelectricity only comes from the band which is partially filled, i.e. multiferroics must not be an conventional insulator but an insulator with partially filled band. The strong electron-electron coupling or spin-exchange coupling between the electrons on the band and the localized spin moment can cause an insulator with partially filled band.

The significance of this model is presented by a limitation on the ferroelectric polarization, i.e. the energy gap $\Delta_g$ in the insulator. If there is an internal electric field $E$ which is spontaneously generated, the electric field must satisfy $e\,|\,E\,|\,h\,/\sqrt{2m^*\Delta_g} < \Delta_g$ in order to maintain the validity of the insulator. A semi-quantitative estimation gives a polarization of only ~100μC/m$^2$ for typical manganites, a disappointing prediction from the point of view of technological applications.

In spite of different microscopic origins, the three models outlined above give a similar prediction: $P_{n,n+1} \propto r_{n,n+1} \times (S_n \times S_{n+1})$. Furthermore, these models are all based on the transverse spiral spin ordered state in which the spin spiral plane contains the propagation vector of spin modulation. This postulation, in fact, may not be always true. Some other spiral spin ordered states, which are not reachable by the three models, can indeed induce ferroelctricity, to be addressed below. In summary, the issue of multiferroicity as generated in spiral spin ordered systems remains to be attractive, thus making a more careful consideration necessary. However, it is now generally accepted that the spin-orbit coupling and the DM interaction do play important roles.

**3.4.3. *Experimental evidences and materials.*** The wealth of evidence that supports spiral spin order induced multiferroicity and intrinsic magnetoelectric coupling parallels the current theoretical progress. We collect some of the main results below and we will see that the theory of spiral spin order induced ferroelectricity is in principle an appropriate description although this theory does not account of all observed phenomena so far.

***1D spiral spin chain systems:*** We first deal with the one-dimensional (1D) spin systems. The 1D chain magnet with competing nearest-neighbor ferromagnetic interaction ($J$) and next-nearest-neighbor antiferromagnetic interaction ($J'$) will develop its configuration into a frustrated spiral spin order as long as $|J'/J| > 1/4$ [142], as already theoretically predicted in Fig. 14(a). Experimentally, the spin configuration of $LiCu_2O_2$ can be approximately treated as a quasi-1D spin chain system and the crystal structure is shown in Fig. 17(a), where magnetic $Cu^{2+}$ ions are blue and nonmagnetic copper ions are green with red dots for oxygen ions. The blue bonds constitute the quasi 1D triangle spin ladders, with the weaker inter-ladder interaction ($J_\perp$) than the in-ladder interactions ($J_1$ and $J_2$). Therefore, each ladder can be viewed as an independent 1D spin chain, as shown in Fig. 17(b). In fact, the picture of a quasi-1D spin spiral is also physically sound since the equivalent nearest-neighbor exchange interactions and frustration ratio estimated experimentally for $LiCu_2O_2$ are $J_1$=5.8meV and $J_2/J_1$=0.29>1/4 [142, 143]. Indeed, a noncollinear spiral spin order was identified for these quasi-1D spin ladders with a spiral propagation vector (0.5, $\xi$, 0) and $\xi$=0.174 was determined. Consequently, within the theoretical framework addressed above, the ferroelectric polarization along the *c*-axis ($P_c$) would be expected, and was experimentally evident in $LiCu_2O_2$, as shown in Fig. 17(c). The anomaly of the dielectric constant at the magnetic transition point and the spontaneous $P_c$ below this point, shown in Fig. 18, are quite obvious [142].

More exciting is the intrinsic magnetoelectric coupling between the spin order and ferroelectric order, which is evident in the response of polarization to external magnetic field [142]. The external field along the *b*-axis drives the rotation of spins within the *bc* plane (Fig. 17(c)) toward the *ab* plane, as shown in Fig. 17(d), and correspondingly, a switch of the polarization orientation from the *c*-axis to the *a*-axis, as shown in Fig. 18, was observed [142].

Nevertheless, it should be mentioned that not all of the experimental results on $LiCu_2O_2$ can be successfully explained by this one-dimensional spin chain model [144, 145], while similar copper oxide, $LiCuVO_4$, was identified as multiferroic material too [146, 147]. For example, we look at the response of polarization $P$ to external magnetic field. Whatever the magnetic field applies along the $b$-axis or $a$-axis, the spiral spin order will be transferred into a parallel aligned configuration which would no longer generates any spontaneous polarization, while experimentally the suppression of polarization along the $c$-axis is accompanied with the appearance of polarization along the $a$-axis, which is not explainable theoretically. Therefore, one may argue that additional contribution to the polarization generation is involved. Furthermore, for $LiCu_2O_2$ and $LiCuVO_4$, early neutron scattering study revealed the incommensurate magnetic structure with a modulation vector (0.5, 0.174, 0), in which the $Cu^{2+}$ magnetic moment lies in the $CuO_2$ ribbon plane (i.e. the $ab$ plane) [142]. However, according to the KNB model or the inverse-DM model, the spontaneous polarization along the $a$-axis is associated with the $ab$-plane spin spiral. This is true for $LiCuVO_4$ [146], but unfortunately for $LiCu_2O_2$ the polarization aligns along the $c$-axis [142]. A possible reason is that the KNB model and the inverse DM model were formulated for the $t_{2g}$ electron system, while for $LiCu_2O_2$ an unpaired spin resides in the $e_g$ orbital.

This issue was recently checked carefully by X-ray absorption spectroscopy and neutron scattering, and a possible $bc$-plane spin spiral was proposed [145]. Moreover, experiments revealed that the ground state of $LiCu_2O_2$ have long-range 2D-like incommensurate magnetic order rather than being a spin liquid of quantum spin-1/2 chains due to the large interchain coupling which suppresses quantum fluctuations along the spin chains. And the spin coupling along the $c$-axis is essential for generating electric polarization [148]. Nevertheless, so far no conclusive understanding has been reached.

***2D spiral spin systems:*** Two-dimensional (2D) frustrated spin system is exampled by the Kagome staircase $Ni_3V_2O_8$ which can be viewed as a quasi-2D spin structure with a frustrated spin order. Similar experiments regarding the electric polarization together with the spin structure and phase diagram are summarized in Fig. 19 [149-151].

The well known geometrically frustrated spin systems go to those 2D triangular lattices with an antiferromagnetic interaction, as shown in Fig. 14(b). While the second spin can easily align in antiparallel with the first spin due to the antiferromagnetic interaction, the third one, however, can't align in a stable way to the first and second spins simultaneously, leading to a frustrated spin structure. Surely, real systems seem far more complicated than this simple picture and the inter-spin interactions can be competitive and entangled. Given the classical Heisenberg spins, the 2D triangular lattice generally favors the 120° spiral-spin order at the ground state. Depending on the sign of anisotropy term $H=D\Sigma(S_i^z)^2$ where $S_i^z$ is the $z$-axis component of spin $S_i$, the spin spiral is confined in parallel ($D>0$, easy-plane type) to, or perpendicular ($D<0$, easy-axis type) to the triangular-lattice plane [152].

RbFe(MoO$_4$)$_2$ (RFMO) exhibits the typical easy-plane triangular lattice, which is described by space group $P\bar{3}m1$ at room temperature. At $T_0$=180 K, the symmetry is lowered to $P3$ by a lattice distortion, as shown in Fig. 20(a), in which the out-of-plane ions lead to two types of triangles: the "up triangles" with a green oxygen tetrahedron above the plane and the "down triangles" with a tetrahedron below the plane [153]. For $T<T_0$, RFMO contains perfect Fe$^{3+}$ triangular lattice planes in which spins $S$=5/2 are coupled through antiferromagnetic superexchange interactions. The magnetism is dominated by the intra-plane interactions of an energy scale of ~1.0 meV and the inter-plane interaction of at least 25 times weaker [154]. Therefore, RFMO is essentially a $XY$ antiferromagnet on a triangular lattice with a long-range magnetic ordering at $T_N$=3.8 K. The magnetic ground state is shown in Fig. 20(b). The magnetic ordering wave vector in the reciprocal lattice units is $q$=(1/3, 1/3, $q_z$) with $q_z$~0.458 at $T<T_N$ under zero magnetic field. This feature implies the absence of a mirror plane perpendicular to the $c$-axis, and experimental measurement revealed an electric polarization of ~5.5μC/cm$^2$ along the $c$-axis [153].

However, according to the KBN model or the inverse DM model, the generated local polarization is $P_{n,n+1} \propto r_{n,n+1} \times (S_n \times S_{n+1})$, and thus lies in the basal plane for RFMO. In view of the three fold rotation axis, the macroscopic polarization $P=\Sigma_n P_{n,n+1}$ vanishes. It means that neither the KNB model nor the inverse DM model can explain the origin of ferroelectricity in RFMO.

CuCrO$_2$ with the delafossite structure (as shown in Fig. 21(a)) is another typical triangular-lattice antiferromagnet with the easy-axis type, and the magnetic properties are dominated by Cr$^{3+}$ ions with $S$=3/2 spin [155]. Recent studies revealed the 120° spin structure with the easy-axis anisotropy along the $c$-axis, in which the spin spiral is in the (110) plane and the spins rotate in the plane perpendicular to the wave vector, as shown in Fig. 22. Again, the KNB model or the inverse DM model predicts that only polarization perpendicular to the spin spiral plane (along the [110] directions) is possible and the net polarization vanishes because any 120° spin structure produces the same $S_n \times S_{n,n+1}$ for all bonds in the triangular lattice. Nevertheless, experiments revealed a sharp anomaly of dielectric constant at $T_N$ and a polarization of ~20μC/m$^2$ below $T_N$ [155].

Besides RFMO and CuCrO$_2$, LiCrO$_2$ and NaCrO$_2$ also exhibit the 2D triangular-lattice structure, but they don't exhibit any electrical polarization over the whole temperature range since they are probably antiferroelectrics due to a different sock salt structure, as shown in Fig. 21(b) [155]. CuFeO$_2$ is a quasi-2D example consisting of Cu and Fe triangular layers, as shown in the inset of Fig. 23(a) [156, 157]. The complex magnetization behavior such as five *M-H* plateaus was observed, which is in physics attributed to the spin-phonon coupling [157]. For a magnetic field between 6~13 Tesla, the ground state will evolve from the collinear commensurate order into noncollinear incommensurate frustrated state. The non-zero polarization inside this magnetic field range, accompanied with remarkable dielectric anomalies at the magnetic transition point below 11K, was observed, as shown in Fig. 23. A doping at the Fe-sites with nonmagnetic ions like Al$^{3+}$ and Ga$^{3+}$ can also induce the noncollinear incommensurate spin state and then observable electric polarization [158, 159], It is revealed that the possible microscopic origin of the ferroelectricity is the variation in the metal-ligand hybridization with spin-orbit coupling [160].

***3D spiral spin systems:*** We finally highlight recent investigations on three-dimensional (3D) frustrated spin systems [161-195]. Typical examples are perovskite manganites Tb(Dy)MnO$_3$ [161-180]. We pay special attention to TbMnO$_3$ which has been extensively investigated. At room temperature, TbMnO$_3$ has an orthorhombically distorted perovskite structure (space group *Pbnm*), different from antiferromagnetic-ferroelectric hexagonal rare-earth manganites *R*MnO$_3$ (*R*=Ho, Y, etc). The $t_{2g}^3 e_g^1$ electronic configuration of the Mn$^{3+}$ site is identical to the parent compound of colossal magnetoresistance manganites LaMnO$_3$ where the staggered $d_{3x^2-r^2}/d_{3y^2-r^2}$ orbital order favors the ferromagnetic spin order in the *ab*-plane and antiferromagnetic order along the *c*-axis. A replacement of La by smaller ions, such as Tb and Dy, enhances the structural distortion and strengthens the next-nearest-neighbor antiferromagnetic exchange, compared with the nearest-neighbor ferromagnetic interaction in the *ab*-plane. Consequently, the competition between the two types of interactions frustrates the spin configuration within the *ab*-plane and then induces successive magnetic phase transitions at low temperature. Theoretical investigation predicted that the Jahn-Teller distortion, together with the relatively weak next-nearest-neighbor superexchange coupling in perovskite multiferroic manganites is shown to be essential for the spiral spin order [161].

At room temperature, the crystal symmetry of $TbMnO_3$ has an inversion center, and the system is nonpolar. Magnetic and neutron scattering experiments showed that the spin structure of $Tb(Dy)MnO_3$ favors an incommensurate collinear sinusoidal antiferromagnetic ordering of $Mn^{3+}$ spins along the *b*-axis, taking place at $T_N$=41 K with a wave vector *q*=(*0, $k_s$~0.29, 1*) in the P*bnm* orthorhombic cell, as shown in Fig. 24(a), (b) and (c) [162]. It is easily understood that the collinear sinusoidal antiferromagnetic state is paraelectric and the ferroelectric phase may not appear unless the spin order is spiral or helicoidal-like. The nonzero polarization appears only below ~30 K ($T_{lock}$) where an incommensurate-commensurate (or lock-in) transition occurs, generating a helicoidal structure with the magnetic modulation wave vector $k_s$ which is nearly temperature-independent and locked at a constant value ~0.28 [162].

It is easily predicted that the generated electric polarization $P \sim e \times k_s$ where *e* is the unit vector connecting the neighboring two spins, is parallel to the *c*-axis, because vector $k_s$ is along the *b*-axis and the spin helicoidal points to the *a*-axis. This prediction is consistent with experiments, as shown in Fig. 24(d) and (e). The dielectric constant along the *c*-axis ($\varepsilon_c$) exhibits a sharp peak at the lock-in point ($T_{lock}$), below which only the polarization along the *c*-axis is observable under a zero magnetic field. Further decrease of the temperature leads to the third anomaly of magnetization and specific heat as a function of temperature at ~7 K, at which the $Tb^{3+}$ spins initiates the long-range ordering with a propagation vector (0, ~0.42, 1). Simultaneously, the electric polarization also exhibits a small anomaly.

$TbMnO_3$ is similar to those improper ferroelectrics mentioned earlier and its polarization is a secondary order parameter induced by the lattice distortion. Because the lattice modulation (distortion) is accompanied with the spin order, the intrinsic magnetoelectric coupling between the spin and polarization may be expected. In fact, experiments confirmed the re-alignment of polarization by an external magnetic field. A magnetic field over ~5 Tesla applied along the *b*-axis significantly suppresses the polarization along the *c*-axis ($P_c$), below a temperature $T_{flop}$ which increases with increasing magnetic field, as shown in Fig. 25(c). In contrast, a finite polarization along the *a*-axis ($P_a$), is generated, with the onset point perfectly consistent with $T_{flop}$ (as shown in Fig. 25(d)). These experiments demonstrate convincingly the intrinsic magnetoelectric coupling effect characterized by a spontaneous switching of polarization from one alignment to another, as shown in Fig.25(c) and (d). DyMnO$_3$ also exhibits polarization flop from $P\|c$ to $P\|a$ by applied magnetic field. Whereas in TbMnO$_3$ the polarization flop is accompanied by a sudden change from incommensurate to commensurate wave vector modulation, in DyMnO$_3$ the wave vector varies continuously through the flop transition [175].

At the same time, the colossal magnetodielectric effect associated with the remarkable response of dielectric constant along the *c*-axis and *a*-axis, respectively, to external magnetic field, is shown in Fig. 25(a) and (b). This colossal effect was argued to be related to the softening of element excitations in these systems, just the same as the softening of phonons in normal ferroelectrics [175]. However, careful study of the dielectric spectra of DyMnO$_3$ found that this colossal effect is a phenomenon emerging only below $10^5$-$10^6$ Hz and the spectrum shape is not the resonance type but the relaxation type, indicating an origin other than the bosonic excitations [181]. It was postulated that this colossal effect may be attributed to the local electric field-driven motion of the multiferroic domain walls between the *bc*-plane spin cycloid ($P\|c$) and *ab*-plane spin cycloid ($P\|a$) domains, as shown in Fig.26. Moreover, this motion exhibits an extremely high relaxation rate of about $10^7$ s$^{-1}$ even at low temperature, indicating that the multiferroic domain wall emerging at the polarization flop is thick rather than the Ising-like thin domain wall identified in conventional ferroelectrics [181].

It should be pointed out that the effect of magnetic field on the electric polarization is orientation-dependent. This remains to be a nontrivial issue [169]. On one hand, when the magnetic field is applied along the *a*-axis or the *b*-axis, both the magnetization and polarization along the *c*-axis exhibit double metamagnetic transitions, and the polarization ($P_c$) is drastically suppressed at the second metamagnetic transition (~10 Tesla for $H\|a$ and ~4.5 Tesla for $H\|b$). This suppression is due to the flop of the electric polarization from the $c$-axis to the $a$-axis, as shown in Fig. 25(c) and Fig. 27(d) and (e), coinciding with a first order transition to a commensurate but still long-wavelength magnetically modulated state (revealed by the magnetization curves in Fig. 27(a) and (b)), with a propagation vector of (0, 1/4, 1) [154]. On the other hand, a magnetic field above ~5 Tesla applied along the $c$-axis causes a single metamagnetic transition, and suppresses the polarization along any crystallographic orientation, as shown in Fig. 27(c) and (f). This effect is related to the disappearance of the incommensurate antiferromagnetic ordering with the (0, 1, 0) magnetic Bragg reflection.

As for the mechanism of the electric polarization flop induced by external magnetic field along the $a$-axis or $b$-axis, two possible scenarios were proposed. The first and direct scenario is that the field induced phase with $P$ along the $a$-axis is also a spiral spin ordered state, corresponding to the spin rotation from the $a$-axis to the $c$-axis. However, recent neutron scattering experiment revealed that the field-induced magnetic phase is a non-spiral commensurate phase with the propagation vector (0, 1/4, 0) [168]. This spin modulation induced lattice distortion is attributed to the ferroelectric order due to the $E$-type antiferromagnetic, which will be discussed again in Sec.3.6.

The multiferroicity in systems with spiral spin order was confirmed in several other perovskite manganites. Fig.28 summaries the phase diagram by plotting temperature $T$ against the Mn-O-Mn bond angle $\phi$ which scales the rare earth ionic radius. The shaded region corresponds to the spiral spin order and, thus, the multiferroicity [163]. Those manganites with even smaller rare earth ions may exhibit geometrical ferroelectricity, as already discussed in Sec.3.3.

To the end of this subsection, we would mention that the multiferroics with spiral spin order do show the intrinsic magnetoelectric coupling, as demonstrated by careful experiments. However, their ferromagnetism seems to be very weak since essentially no spontaneous magnetization is available due to the helical or spiral spin order. An extension of this spiral spin order concept can partially avoid this problem. For example, conical spin state is also a kind of spiral spin order, in which the spontaneous component $S_\|$ (homogeneous ferromagnetic part) and spiral component of the magnetization coexist, as shown in Fig. 29(a) [196]. If the spiral component lies in the ($e_1$, $e_2$) plane, $S_∥$ points to the $e_3$-axis, one has the spin moment $S_n=S_1e_1cos(Q·r)+ S_2e_2sin(Q·r)+S_∥e_3$, where $Q$ is the wave vector and $r$ is the space coordinate. Chromite spinels, $CoCr_2O_4$ (Fig. 29(b)) [197-200] do show such exceptional conical spin structure.

In $CoCr_2O_4$, $Co^{2+}$ and $Cr^{3+}$ ions occupy the tetrahedral (A) and octahedral (B) sites respectively. Because of the nearest-neighbor and isotropic antiferromagnetic A-B and B-B exchange interaction ($J_{AA}$ and $J_{BB}$) with $J_{BB}/J_{AA}>2/3$, a conical state with the spiral wave vector $Q∼0.63$ was identified below ∼27K [110]. The ferromagnetic $M$-$H$ hysteresis and spontaneous polarization $P∼Q×[001]∼[-1,1,0]$, were identified, as seen in Fig. 29(c) and (d) [197]. A reversal of external magnetic field could trigger the switching of polarization because of the transition of ($M$, $Q$) to ($–M$, $–Q$), as seen in Fig. 29(e) and Fig. 30(c). This process is very quick, attractive for potential applications. Moreover, there is another magnetic transition at $T_L∼14K$ which is a magnetic lock-in transition and has the first-order nature. The spontaneous polarization exhibits a discontinuous jump and changes its sign without reversal of spin spiral wave vector $Q$ at this transition temperature. This fact is contrast to the above discussion, as shown in Fig.30(a) [201]. Below this temperature, although the electric polarization can be reversed, a reversal of $H$ also induces the 180º flip of $Q$ and then polarization $P$, as shown in Fig. 30(c) [201].

### 3.4.4. *Multiferroicity approaching room temperature.* All of the physics associated with multiferroicity from spiral spin structure illustrates the fact that the spiral magnetic order often arises from the competing magnetic interactions. These competing interactions usually reduce the ordering temperature of conventional spin ordered phase. Hence, it is hardly possible for the the spiral spin order (phase)-induced ferroelectricity to appear above a temperatures of ∼40K, far below room temperature required for service of most devices.

One of the possible ways to overcome this barrier is to search for those magnetic materials with very strong competing magnetic interactions, and this effort has been marked with some progress recently. In fact, it was once revealed that the magnitude and sign of the principal super-exchange interaction $J$ in low-dimensional cuprates depend remarkably on the Cu-O-Cu bond angle $φ$ [202, 203]. In cuprates with $φ ∼180º$, $J$ has an order of magnitude of

~$10^2$ meV, thus favoring ferromagnetic order. Upon decreasing $\varphi$, $J$ is monotonically suppressed and eventually becomes negative (favoring iferromagnetic order) at $\varphi \sim 95º$, as shown in Fig. 31(a). Therefore, for those cuprates with $\varphi$ deviating away from 180º, the ferromagnetic interaction ($J$) competes with the higher-order superexchange interactions, often leading to the spiral magnetic order with relatively high ordering temperature. While LiCu$_2$O$_2$ discussed above is the typical example exhibiting the spiral magnetic order and simultaneously ferroelectricity below ~25K, the relationship between parameters $J$ and $\varphi$ in cuprates allows us to tune the strength of the spiral magnetic order. For example, CuO with *C2/c* monoclinic crystal structure can be viewed as a composite of two types of zigzag Cu-O chains running along the [10$\bar{1}$] and [101] directions, respectively, with $\varphi$=146º and 109º. A Cu-O-Cu angle of 146º seems to be an intermediate value between 95º and 180º and a large magnetic super-exchange interaction is expected. A strong competition between this super-exchange interaction and the ferromagnetic one was identified, resulting in the incommensurate spiral magnetic order (AF2) which appears over the temperature range from 213K to 230K, as shown in Fig. 31(b). This argument was confirmed by a clear ferroelectric polarization measured in this temperature region, as shown in Fig. 31(c) [202].

Besides CuO, hexaferrite Ba$_{0.5}$Sr$_{1.5}$Zn$_2$Fe$_{12}$O$_{22}$ is another multiferroic system offering the spiral magnetic and ferroelectric orders at a relatively high temperature [204]. Similarly, hexaferrite Ba$_2$Mg$_2$Fe$_{12}$O$_{22}$ was also found to exhibit magnetic field induced ferroelectricity at relatively high temperature, although it does not show ferroelectricity under zero magnetic field [205]. The helical spin order with propagation vector $k_s$ along the [001] direction appears at ~200 K, and so does the as-induced ferroelectricity, under a very small magnetic field (~30 mT). More interesting here is that a magnetic field as small as ~30mT is sufficient to stimulate a transverse conical spin structure with respect to the magnetic field direction. In agreement with the inverse DM model and the KNB model, this transverse conical spin order allows a polarization $P$ to align perpendicular to both magnetic field and the propagation vector $k_s$. An oscillating or multidirectionally rotating field is able to excite the cyclic rotation of polarization $P$. For example, the rotating magnetic field with magnitude from 30 mT to 1.0 T, within the plane normal to the [001] direction, drives polarization $P$ to vary in proportion to *sinθ* where *θ* is the angle defined by the magnetic field and spiral propagation axis, as shown in Fig. 32 [205].

**3.4.5.** ***Electric field control of magnetism in spin spiral multiferroics.*** It is now believed that the ferroelectricity in those frustrated magnetic oxides originates from specific frustrated spin configuration, e.g. the spiral spin structure. Therefore, a control of polarization by magnetic field becomes quite natural and was extensively demonstrated. Because of the intrinsic magnetoelectric coupling in those materials, one may also expect an effective control of the magnetization by electric field as the counter-part of the magnetic field control of polarization. Nevertheless, so far no dynamic and macroscopically reliable evidence of this magnetization control by electric field has been available, while the microscopic identification in some multiferroics was reported very recently. One example is $TbMnO_3$, in which the magnetization switching associated with the polarization reversal was observed using spin-polarized neutron scattering [180]. This seems to be the first evidence to demonstrate that the helical spin order can be modulated by an electric field or by polarization switching. In Fig. 33(a) are shown the recorded two satellite scattering peaks at position (4, $\pm q$, *L*=1) with the neutron spins parallel (mode $I_\uparrow$) and antiparallel (mode $I_\downarrow$), respectively, to the scattering vector in the ferroelectric phase. It is seen that the intensities of the two peaks are reversed upon a switching of the polarization between the two states along the $\pm c$-axis. This suggests that the spin helicity, in clockwise or counterclockwise mode, can be controlled by reversing the polarization, as schematically shown in Fig. 33(b) [180]. Although this effect is very weak and the spin helicity reversal might not be realized at a temperature above $T_C$, a reversible magnetoelectric coupling between magnetization and polarization in $TbMnO_3$ was identified.

Furthermore, this effect was recently demonstrated also in $BiFeO_3$. Although the ferroelectricity in $BiFeO_3$ is commonly believed to originate from the lone-pair electrons of Bi ions and the magnetoelectric coupling should be weak, it was pointed out that the $Fe^{3+}$ ions are ordered antiferromagnetically (G-type) and their moment alignment constitutes a cycloid with a period of ~62 nm [206]. Because of the rhombohedral symmetry, there are three equivalent propagation vectors for the cycloidal rotation: $k_1$=[δ, 0, -δ], $k_2$=[0, δ, -δ], and $k_3$=[-δ, *δ, 0*] with *δ*~0.0045. This allows one to argue the possibility that such a cycloidal spin structure may also contribute to the ferroelectricity in BiFeO$_3$, or one may expect intrinsic coupling between the ferroelectricity and the cycloidal spin structure. This idea was indirectly confirmed experimentally by a successful observation of the intimate link between the cycloidal magnetic structure of Fe$^{3+}$ ions and the polarization vector [206-209].

The coupling between ferroelectric domains and antiferromagnetic domains in BiFeO$_3$ provides direct evidence of the above argument. Experimentally, piezo-force microscopy (PFM) or in-plane piezo-force microscopy (IPPFM) allows researchers to observe the ferroelectric domains under different electric fields [210, 211], while X-ray photoemission electron microscopy (X-ray PEEM) can be used to monitor the antiferromagnetic domains simultaneously. High resolution images of both the antiferromagnetic and ferroelectric domains in (001)-oriented BiFeO$_3$ films were obtained. As mentioned previously, the spontaneous polarization of BiFeO$_3$ directs along the [111]-axis, enabling eight equivalent orientations along the four cubic diagonals. This geometry thus allows for the 180°, 109° and 71° domain switching driven by appropriate electric field, as shown in Fig. 34. Figs.35(c) and 35(d) show the PFM images of the BiFeO$_3$ film before and after the electric field poling, respectively, and 109° domain switching (regions 1 and 2) was identified in addition to the 180° and 71° domain switching (regions 3 and 4). It can be seen that the multidomain state consists of stripe regions with two different polarization directions. The PEEM images of the same regions (Figs. 35(a) and 35(b)) clearly indicate the reverse contrast in regions 1 and 2 upon the electric field poling [208]. These results demonstrate the switching of the antiferromagnetic order from the orange plane to the green plane (Fig. 34(a)) due to the 109° polarization switching. The neutron scattering on single crystal of BiFeO$_3$ also revealed that the intensity around the (1/2, -1/2, 1/2) Bragg position in the $P_{111}$ and $P_{1-11}$ domains (as shown by the lower half and upper half of the pattern in Fig. 36(a), respectively) are different, which implies that the 71° domain switching by an electric field along the [010] direction brings the rotation of the Fe spiral spin plane and then induces the flop of the antiferromagnetic sublattices [206], as shown in Fig. 36(b). These experiments unveiled the coupling between *M* and *P* at atomic level, although no global linear magnetoelectric effect exists because of no net magnetization (*<M>*=0).

These phenomena illustrated above can be understood as following. Lebeugle *et al* pointed out that a coupling energy term $E_{DM}=(P \times e_{ij}) \cdot (S_i \times S_j)$ should be included into the total energy, due to the DM interaction. This coupling energy favors the canting of $Fe^{3+}$ spins, which exactly compensates the loss of the exchange energy. Moreover, this coupling energy is zero if polarization $P$ is perpendicular to the local spin moments and maximum if it lies on the cycloid rotation plane. This picture explains reasonably the flop of antiferromagnetic domains associated with the switching of ferroelectric domains in $BiFeO_3$ thin films [206].

Another experiment on the dynamics of ferroelectric domain switching in $BiFeO_3$ films also evidenced the coupling between the ferroelectric and antiferromagnetic domains. A well known fact is that the ferroelectric domains in conventional ferroelectric films are smooth, stripe-like, and the domain width grows in proportion to the square root of the film thickness, i.e. the so-called Landau-Lifshitz-Kittel (LLK) scaling law [212]. However, qualitatively different behaviors of the ferroelectric domains from the LLK scaling in very thin $BiFeO_3$ films were observed. First, the domain walls are not straight, but irregular in shape, characterized by a roughness exponent of ~0.5-0.6 and an in-plane fractal Hausdorff dimension $H_\parallel \sim 1.4 \pm 0.1$. The average domain size appears to depart from the LLK square root dependence on the film thickness, but scales with an exponent of $0.59 \pm 0.08$ [209]. Second, the ferroelectric domains are significantly larger in size than those in other ferroelectric films of the same thickness, but closer to magnetic domains in typical magnetic materials. This implies that the ferroelectric domains are coupled with the antiferromagnetic domains in $BiFeO_3$ films. The magnons coupled with polarization (electromagnons) observed in $BiFeO_3$, to be emphasized in Sec.4, map too the coupling between the ferroelectric order and cycloidal spin order in $BiFeO_3$ [207].

In fact, dealing with the roadmaps to control the ordering state of a multiferroic system, two types of approaches are possible: phase control or domain control. In the first case, an external field is used to trigger a phase transition between two fundamentally different phases. Although this approach cannot be realized in $BiFeO_3$, the coupling between the ferroelectric domains and antiferromagnetic ones provides the second approach, i.e. external field triggers a transition between two equivalent but macroscopically distinguishable domain states. This approach together with the exchange interaction at the interfaces, makes the electric-field control of magnetism possible. An excellent example developed recently by Chu *et al* comes to $BiFeO_3$ again [213]. For a $Ta/Co_{0.9}Fe_{0.1}/BiFeO_3/SrRuO_3/SrTiO_3(001)$ heterostructure, a $10\times10\mu m^2$ region in the BFO layer was upward with a -21V biased electric voltage, as shown by the red square in Fig. 37(a). Subsequently, a $5\times5\mu m^2$ smaller area inside this region was downward poled with a +12V biased voltage, as shown by the green square. The magnetic domains in the CoFe layer exhibit two distinct regions, as we can see from the PEEM images in Fig. 37(b). These regions are the in-plane ferromagnetic domains aligned horizontally from left to right (black) and vertically from down to up (grey). The formation of the two types of domains is due to two switching events by rotation of the polarization projection on the (001) plane: the 70° in-plane switching and the 109° out-of-plane switching, which gives rise to the corresponding rotation of the antiferromagnetic order in $BiFeO_3$, as discussed in the beginning of this section [206-209]. The rotation of the antiferromagnetic order drives the reversal of magnetism of the CoFe layer via the exchange bias effect on the $BiFeO_3$-CoFe interface [213]. This approach to control the ferromagnetism by an electric field can be utilized in dynamic switching devices, i.e. back-switching the ferroelectric domains in $BiFeO_3$ and then the ferromagnetic domains in CoFe layer to the initial states by an opposite electric field. These investigations sketch a possible magnetoelectric random access memory (MeRAM) element (as shown in Fig. 38) [214]. The binary information is stored by the magnetization of the bottom ferromagnetic layer (blue), which is read by monitoring the resistance of the magnetic trilayer and written by applying a voltage across the multiferroic ferroelectric/antiferromagnetic layer (green). If the magnetization of the bottom ferromagnetic layer is coupled to the ferroelectric/antiferromagnetic layer, a reversal of the polarization $P$ in the mutliferroic layer stimulates the switching of the magnetic configuration in the trilayer from the parallel alignment to the antiparallel one, and thus the resistance from the low state $R_P$ to the high state $R_{AP}$ [214].

There are some other 3D frustrated oxides, such as $MnWO_4$ [182-190] and pyroxenes ($NaFeSi_2O_6$ and $LiCrSi_2O_6$) [191], which were found to be spiral spin order induced multiferroics. In fact, not only those oxides but also some thiospinel compounds with frustrated spin order such as $Cd(Hg)Cr_2S_4$, also exhibit multiferroicity. In addition, the magnetoelectric coupling and a colossal magnetodielectric effect were observed in these multiferroics [192-196]. For convenience to readers, we collected the main physical properties of those so far investigated multiferroics of spiral spin order and induced ferroelectricity and present them in Table II.

Although there have been extensive researches on this kind of multiferroics and several comprehensive models were developed, so far no quantitative understanding of the multiferroicity in spiral spin ordered materials has been available. For example, first principle calculation on $LiCu_2O_2$ [144] and $TbMnO_3$ [215, 216] predicts that all of these models are inadequate. Careful calculation on $TbMnO_3$ reveals that both the electronic and lattice effects have contribution to the electric polarization and the latter can be even dominant [215]. It is surprising to be disclosed that the displacements of $Mn^{3+}$ and $Tb^{3+}$ ions are generally larger than that of $O^{2-}$ ions and have significant contribution to the polarization [216], which are not possible in the framework of the current theories.

### 3.5. *Ferroelectricity in charge-ordered systems*

In parallel to the development of multiferroics of spiral (helical) spin order, another type of multiferroics, i.e. charge-ordered multiferroics, has been receiving attention too. For conventional ferroelectrics and all of those multiferroics addressed above, the ferroelectricity originates from the relative displacement between anions and cations as well as the lattice distortion associated with the second-order Jahn-Teller effect. However, an alternative mechanism — electronic ferroelectricity [217] — was proposed recently, in which the electric dipole originates from the electronic correlation rather than the covalency. This would offer an attractive possibility for novel ferroelectricity that could be controlled by the charge, spin and orbital degrees of freedom of electron.

In many narrowband metal oxides with strong electronic correlations, charge carriers may become localized at low temperature and form a periodic structure (i.e. charge-ordered state). The often cited example is magnetite $Fe_3O_4$, which undergoes a metal-insulator transition at ~125K (the Verwey transition) with a rather complex iron charge order pattern [218]. It is expected that a non-symmetric charge order may induce electric polarization. Another well studied charge-ordered materials are manganites [219]. When $LaMnO_3$ (or related compounds in which the charge of Mn ions is formally 3+) and $CaMnO_3$ (in which the Mn charge is formally 4+) are alloyed, the resulting arrangement of $Mn^{3+}$ and $Mn^{4+}$ ions can be ordered in a particular case, as shown in Fig. 39. Moreover, electrons around the atoms they occupy may have several choices among their energetically equivalent (or degenerate) electronic orbitals. This orbital degree of freedom allows for a manifold of possible electronic states to be chosen. For example, the Mn ions can occupy either of the two *d*-orbitals. However, these choices are not independent, and the charge distribution around these ions is distorted with adjacent oxygen ligands which would be dislodged once a valence electron localizes in a definite Mn *d*-orbital. Eventually, a spontaneously ordered pattern of the occupied orbitals throughout the crystal lattice (i.e. orbital ordered state) yields.

**3.5.1. *Charge frustration in LuFe₂O₄.*** The ferroelectricity associated with a charge ordered (CO) state was first demonstrated in a mixed valence oxide, $LuFe_2O_4$ [220-227]. At room temperature, $LuFe_2O_4$ has a hexagonal layered structure (space group $R\overline{3}m$, $a$=3.44Å, $c$=25.28Å) in which all Fe sites are crystallographically the same. The crystal structure consists of an alternative stacking of triangular lattices of rare-earth elements, irons and oxygens. The $Fe_2O_4$ layers and $Lu^{3+}$ ion layers stack alternatively with three $Fe_2O_4$ layers per unit cell. Each $Fe_2O_4$ layer is made up of two triangular sheets of corner-sharing $FeO_5$ trigonal bipyramids.

In $LuFe_2O_4$, an equal number of $Fe^{2+}$ and $Fe^{3+}$ ions coexists on the same site of the triangular lattice. With respect to the average $Fe^{2.5+}$ valence, $Fe^{2+}$ and $Fe^{3+}$ ions are considered to be facilitated with an excess and a deficient half electron, respectively. The Coulomb preference for pairing oppositely signed charges $Fe^{2+}$ and $Fe^{3+}$ causes the degeneracy in the lowest energy for charge configuration in the triangular lattice, and then the charge-ordered state. The charge order pattern of alternating $Fe^{2+}$:$Fe^{3+}$ layers with ratios of 2:1 and 1:2, appearing at a temperature as high as ~370 K, is shown in Fig. 40 (a). This postulated charge ordered structure allows the presence of a local electric polarization, because the centers of $Fe^{2+}$ ions and $Fe^{3+}$ ions do not coincide in the unit cell of the superstructure, as highlighted by the arrow in Fig. 40(c). The HRTEM image shown in Fig. 40(b) [222, 226] is consistent with this pattern. An electric polarization as high as ~26 μC/cm$^2$ was measured using the pyroelectric current method [220]. In response to temperature variation, significant decaying of the polarization occurs at ~250 K, the magnetic transition point, and at ~330K where the charge ordered superstructure of $Fe^{2+}$ and $Fe^{3+}$ disappears, as displayed in Fig. 40 (d) [220].

LuFe$_2$O$_4$ also exhibits remarkable magnetoelectric coupling effect. For example, remarkable response of the dielectric constant to a small magnetic field at room temperature was exampled by a change of 25% upon a field of ~1 kOe [227]. Moreover, $Fe^{2+}$ onsite crystal field excitations are sensitive to the monoclinic distortion which can be driven by temperature/magnetic field. The distortion splits further the three groups of Fe $3d$ level of $D_{3d}$ symmetry, and then a large magento-optical effects was observed [228].

Nevertheless, first principle calculation, in combination with Monte Carlo simulation, reveals another charge ordered state in connection with the Fe$_2$O$_4$ layers of LuFe$_2$O$_4$, consisting of $Fe^{2+}$ chains alternating with $Fe^{3+}$ chains in each triangular sheet. This state has almost the same stability as the charge ordered state discussed above, although it unfavors ferroelectricity and is not the ground state. The charge fluctuations associated with the inter-conversion between the two different types of charge order states could be remarkable because they are very similar in energy. In this sense, LuFe$_2$O$_4$ can be viewed as a phase-separated system in terms of the charge ordered state, consisting of two types of charge ordered domains separated by domain boundaries. The giant dielectric constant of LuFe$_2$O$_4$, as observed, may be ascribed to this kind of charge ordered state fluctuations. Given an external magnetic field, the Zeeman energy may preferentially stabilize one of the two charge ordered states because the two states most likely have different total spin moments. This explains why the dielectric response and polarization will be weakened by a magnetic field, which suppresses the charge fluctuations [221]. Experimental results also indicate that the charge fluctuations have an onset point well below the charge ordered temperature [228].

Surely, there are still enormous disputes and more investigation on the details of the proposed multiferroic origin in LuFe$_2$O$_4$ is required. For example, some results of X-ray scattering experiments revealed an incommensurate charge order with propagation vector close to (1/3, 1/3, 3/2) below 320K, which contains polar Fe/O double layers with antiferroelectric stacking [229].

### 3.5.2. *Charge/orbital order in manganites.*

Charge-ordered state is also often observed in manganites $R_xA_{1-x}MnO_3$. In addition, the charge/orbital ordered (CO-OO) state is highly favored in Ruddlesden-Popper series manganites [219]. For example, $Pr(Sr_{0.1}Ca_{0.9})_2Mn_2O_7$ is composed of bilayers of $MnO_6$ octahedra, exhibiting two CO-OO phases: high-temperature phase (CO1) and low-temperature phase (CO2). The space groups are *Amam* (with $a$=5.410, $b$=5.462, $c$=19.277Å at 405K) for the charge-disordered phase ($T>T_{CO1}$), *Pbnm* (with $a$=5.412, $b$=10.921, $c$=19.234Å at 330 K) for the CO1 phase ($T_{CO1}<T<T_{CO2}$) and *Am2m* (with $a$=10.812, $b$=5.475, $c$=19.203Å at 295 K) for the CO2 phase ($T<T_{CO2}$). The charge/orbital configurations for the three phases are shown in Fig. 41. From the synchrotron X-ray oscillation photography, it was found that with respect to the CO1 phase, the orbital stripes and zigzag chains rotate by 90° when $T$ falls down to $T_{CO2}$ [230,231]. Above $T_{CO1}$, each $MnO_6$ octahedron tilts towards the $b$-axis, as shown in Fig. 41(a). Within a single bilayer unit, pairs of tilted $MnO_6$ octahedra on the upper and lower layers line up with the shared $O^{2-}$ shifting towards the $+b$ and $-b$ directions. Such a situation causes the alternation of the Mn–O–Mn bond in the MnO plane along the $b$-axis. In the adjacent bilayer, the arrangement of the bond alternation shifts by (0, 1/2, 0). For the low-temperature phases, the structural modulation accompanied by the CO–OO is superimposed onto this structure.

For simplicity and without losing the essence of the charge polarization problem, we take into account the charge ordering in the assumed *Amam* orthorhombic lattice. For the charge order transition, as shown in Fig. 41(b) (CO1) and Fig. 41(c) (CO2), the checkerboard pattern of the CO state is superimposed onto the bond alternation pattern. Consequently, the charge polarization appears along the $b$-axis in each bilayer. In the CO1 phase, however, the CO pattern stacks along the $c$-axis with a shift by (1/2, 0, 0) with respect to the next bilayer, as shown in Fig.41(b), facilitating the inter-bilayer coupling of the polarization antiferroelectrically in nature. At $T_{CO2}$, on the other hand, the rotation of the orbital stripes is accompanied by the rearrangement of $Mn^{3+}$ and $Mn^{4+}$ ions and, thus, the CO stacking pattern, as shown in Fig. 41(c). In the CO2 phase, the charge order sequence stacks along the $c$-axis with a shift by (0, 1/2, 0) with respect to the next bilayer, coinciding with the stacking of the bond alternation. Therefore, the polarization of each bilayer along the *b*-axis is excessive, forming a charge-polarized state below $T_{CO2}$. In fact, optical SHG signals clearly demonstrate the breaking of the space-inversion symmetry. However, the direct detection of the electrical polarization by, for example, a pyroelectric current measurement, is hard to perform because of the low resistivity around $T_{CO2}$ [230].

**3.5.3. *Coexistence of site- and bond-centered charge orders.*** There has been active debate over the validity of the conventional CO picture in which the 3+ and 4+ Mn ions orbitally order in a checkerboard arrangement with the so-called CE-type antiferromagnetism (Efremov *et al.* referred this state as the site-centered order, as shown in Fig. 42(a) and Fig. 39). An alternative model of ferromagnetic Mn–Mn dimmers (the bond-ordering model of Efremov *et al,* Fig. 42(b)) was proposed for $La_{0.5}Ca_{0.5}MnO_3$ and likely identified for $La_{0.6}Ca_{0.4}MnO_3$ [232]. Recently, Efremov and his colleagues proposed an intermediate state, a kind of superposition of these two different charge-ordered patterns, and predicted the local dipole moments that add up to a macroscopic ferroelectric polarization (Fig. 42(c)) [232, 233]. In $La_{0.5}Ca_{0.5}MnO_3$, adjacent dipole moments point to opposite directions so that there is no overall electric polarization due to the moment cancellation. However, given a composition away from 50% $CaMnO_3$, the cancellation should not be complete and a net polarization could enter, as shown in the calculated phase diagram (Fig. 42(d)) [232].

So far, no direct experimental evidence with ferroelectricity in such charge-ordered manganites has been reported due to the high conductivity and possibly other unknown reasons, while indirect characterization of the ferroelectricity was reported recently [234, 235]. A so-called electric field gradient (EFG) tensor via hyperfine techniques was developed to map the whole compositional range of $Pr_{1-x}Ca_xMnO_3$ and a new phase transition occurring at a temperature between the CO point and antiferromagnetic Néel point was evidenced [234]. Although this transition can be detected in all samples with CO state, the critical temperature for the transition is suppressed upon the shift of the composition away from $x$=0.5. The principal EFG component $V_{ZZ}$ characterizing the local paraelectric susceptibility shows a sharp increase in the vicinity of this new transition due to the polar atomic vibrations.

Therefore, this new transition gives a hint of the local spontaneous polarization below the CO transition point. The refined electron diffraction microscopy data also provide indirect evidence for the electric polarization in $Pr_{0.68}Ca_{0.32}MnO_3$ [235]. The results revealed that the Zener polaron order structure is noncentrosymmetric and the relative displacements of the bound cation-anion charge pairs create permanent electric dipoles, resulting in a net permanent polarization $P_a$=4.4μC/cm$^2$. This polarization is much larger than those multiferroic manganites with spiral spin orders and thus allows for potential applications of the charge-ordered manganites. Nevertheless, the related experiments are very limited and definitely, the ferroelectricity in charge-ordered $ABO_3$ manganites seems to be a hot issue for further careful study.

Another possible and intriguing multiferroic material, benefited from the coexistence of bond-centered and site centered CO states is magnetite, $Fe_3O_4$, which undergoes a metal-insulator transition at ~125 K (the Verwey transition) with a rather complex CO pattern of $Fe^{2+}$ and $Fe^{3+}$ ions [218]. $Fe_3O_4$ crystallizes in an inverted cubic spinel structure with two distinct iron positions. The iron B sites locate inside the oxygen octahedra and contain 2/3 of the total iron ions, with equal numbers of $Fe^{2+}$ and $Fe^{3+}$ ions. These sites by themselves form a pyrochlore lattice, consisting of a network of corner-sharing tetrahedra. The iron A sites contain the other 1/3 Fe ions and are considered to be irrelevant for the CO state. The originally proposed charge order pattern consists of an alternation of $Fe^{2+}$ and $Fe^{3+}$ ions at the B-sites in the *x-y* planes and was shown to be too simple in later reports. The difficulty in determining the CO structure in $Fe_3O_4$ is related to the strong frustration of simple biparticle ordering on a pyrochlore structure and detail of the CO pattern remains to be an issue [236-239].

Alternatively, a much earlier report claimed that $Fe_3O_4$ in the insulating state below the Verwey temperature is ferroelectric and the electric polarization points to the *b*-direction [236]. The polarization leads to the formation of a ferroelectric domain structure which can be explained only by assuming a triclinic structure. Although the real microscopic origin of the ferroelectricity in magnetite remains to be unveiled, the most probable one is the coexistence of site-centered and bond-centered CO states [237-239]. In the proposed structure and charge pattern shown in Fig. 43, there exist strong fluctuations of the Fe-Fe distance (the bond length) and the site occupancy of $Fe^{2+}$ and $Fe^{3+}$ ions. The variation of the Fe-Fe bond length, in addition to the alternative occupancy of $Fe^{2+}$ and $Fe^{3+}$ ions along the <110> Fe chain on the *x*-*y* planes, results in the mixed bond- and site-centered CO chains. Such a configuration would give a non-zero contribution to the electrical polarization.

**3.5.4.** ***Charge order and magnetostriction.*** The spiral magnetic order with active antisymmetric exchange coupling and the charge ordered state are not the only possible ways towards the magnetism induced ferroelectricity. It has been postulated that the exchange striction associated with symmetric superexchange coupling plus charge-ordered state can generate ferroelectricity too. The ground magnetic order of the one-dimensional Ising spin chain with the competing nearest-neighbor ferromagnetic interaction ($J_F$) and next-nearest-neighbor antiferromagnetic interaction ($J_{AF}$) is of the up-up-down-down ($\uparrow\uparrow\downarrow\downarrow$) type if $|J_{AF}/J_F|>1/2$ [240], as shown in Fig. 44. If the magnetic ions align alternatively along the chain, the exchange striction associated with the symmetric superexchange interaction shortens the bonds between the parallel spins, while stretches those between the antiparallel spins. Ultimately, the inversion symmetry is broken and an electronic polarization yields along the chain direction [241]. $Ca_3Co_2O_6$ is a typical Ising spin chain system, and a half-doping of this compound by Mn at the Co site produces novel compound $Ca_3Co_{2-x}Mn_xO_6$, where the $Co^{2+}$ and $Mn^{4+}$ ions tend to be located in the center of oxygen cages of face-shared trigonal prisms and octahedra aligned alternatively along the *c*-axis. This is because Mn ions have a strong tendency to avoid the trigonal prismatic oxygen coordination. At *x*=1, all the Co ions are located in the trigonal prismatic sites and all the Mn ions occupy the octahedral sites, leading to the charge ordered state associated with the Ising spin chain in $Ca_3CoMnO_6$. This configuration would generate electronic ferroelectricity. In fact, a clear ferroelectric polarization was observed at 16.5K, the onset point for the magnetic order, which is signified by a broad peak in the magnetic susceptibility [242].

$Ca_3Co_{2-x}Mn_xO_6$ and the undoped compound $Ca_3Co_2O_6$ are famous for their successive metamagnetic transitions and the magnetization plateaus under magnetic field [243-247]. In agreement with the magnetization plateaus, the dielectric constant of $Ca_3Co_2O_6$ also shows plateaus [246]. For multiferroic $Ca_3Co_{2-x}Mn_xO_6$ (x~0.96), magnetization and neutron-scattering measurements revealed successive metamagnetic transitions from the $\uparrow\uparrow\downarrow\downarrow$ spin configuration under zero field to the $\uparrow\uparrow\uparrow\downarrow$ state and then the $\uparrow\uparrow\uparrow\uparrow$ state [247]. Inversion symmetry broken in the $\uparrow\uparrow\downarrow\downarrow$ state, is restored in the $\uparrow\uparrow\uparrow\downarrow$ state, resulting in disappearance of the spontaneous polarization [247].

Besides $Ca_3Co_{2-x}Mn_xO_6$, manganites $R$Mn$_2$O$_5$ as multiferroics were supposed to follow a similar mechanism for ferroelectricity generation. $R$Mn$_2$O$_5$ with $R$=Ho, Tb, Dy, Y, and Er etc, represents another kind of charge ordered manganites in addition to charge ordered ABO$_3$-type manganites [248-281]. They exhibit very complex magnetic and ferroelectric phase transitions upon temperature variation. At room temperature, TbMn$_2$O$_5$ belongs to the orthorhombic space group of *Pbam*, hosting Mn$^{3+}$ ($S$=2) ions surrounded by oxygen pyramids and Mn$^{4+}$ ($S$=3/2) ions surrounded by oxygen octahedra, as shown in Fig. 45(a). The magnetic structure of $R$Mn$_2$O$_5$ is extremely complicated and determined with multi-manifold exchange interactions, as shown in Fig. 45(b). Along the *c*-axis, the Mn spins are arranged in the five-spin loop: Mn$^{4+}$-Mn$^{3+}$-Mn$^{3+}$-Mn$^{4+}$-Mn$^{3+}$. The nearest-neighbor magnetic coupling in the loop is antiferromagnetic-type, favoring antiparallel alignment of the neighboring spins. However, because of the odd number of spins in one loop, a perfect antiparallel spin configuration cannot be possible, eventually leading to the frustrated complex magnetic structure [248].

Also, upon temperature fluctuation and external stimuli, $R$Mn$_2$O$_5$ exhibits complex magnetic transitions. From Fig. 46(a), it is seen that TbMn$_2$O$_5$ shows several magnetic and ferroelectric phase transitions accompanied by the appearance of electric polarization and dielectric anomalies along the *b*-axis. Starting from an incommensurate antiferromagnetic (ICM) ordering at $T_N$=43K with a propagation vector (~0.50, 0, 0.30), the spin configuration locks into a commensurate antiferromagnetic (CM) state at $T_{CM}$=33K with propagation vector (0.50, 0, 0.25). The dielectric constants exhibit anomalies at these magnetic transitions, too, as shown in Fig. 46(a) and (b). Spontaneous polarization arises at a temperature $T$=$T_{ferroelectric}$~38K between $T_N$ and $T_{CM}$, as shown in Fig. 46 (d). As temperature drops down to $T_{ICM}$=24K, the ICM configuration re-enters, with a sudden decrease of the polarization and a jump of the vector to (0.48, 0, 0.32). The polarization increases again with temperature decreasing down to ~10K, as shown in Fig. 46(d) [248].

Interestingly, experiments revealed that $TbMn_2O_5$ belongs to the space group *Pbam*, which meets the spatial inversion symmetry and thus would exclude ferroelectricity because of no spatial inversion breaking. While the fact that $TbMn_2O_5$ develops a spontaneous polarization is still puzzling, it was suspected that the symmetry group should be *Pb2₁m* which allows for ferroelectricity, but no direct evidence has been available [252, 253]. It was also postulated that the charge ordered state plus the commensurate magnetic order is responsible for the polarization, where the Mn spin configuration in the commensurate phase is composed of antiferromagnetic zigzag chains along the *a*-axis. Half of the $Mn^{3+}$-$Mn^{4+}$ spin pairs across the neighboring zigzags align in an approximately antiparallel manner, whereas the other half favors, more or less, the parallel alignment. The exchange striction effect drives a shift of ions (mostly $Mn^{3+}$ ions inside the pyramids) in a way that optimizes the spin-exchange energy: those ions with antiparallel spins are pulled close to each other, whereas those with parallel spins move away from each other. This leads to a distorted pattern labeled by the open black arrows in Fig. 45(b), which breaks the inversion symmetry and induces a net polarization along the *b*-axis. For the incommensurate magnetic phase, the magnetization of Mn ions in each zigzag chain is modulated along the *a*-axis and the spins in every other chain are rotated slightly toward the *b*-axis. It should be mentioned here that rare earth $Tb^{3+}$ ions also have magnetic moments and will exhibit a noncollinear magnetic order at low temperature ~10K. The net distortion associated with the Tb spins is even larger than that associated with the Mn spins. Therefore, a polarization enhancement with temperature decreasing down to ~10K was observed, as shown in Fig.46(d). A magnetic field *H* applied along the *a*-axis will force alignment of the Tb spins along the *a*-axis but leave the Mn spins nearly unchanged. This re-alignment of Tb spins makes the associated lattice distortion disappear and a rotation of the net polarization by 180°, as shown in Fig. 46(c) and (d), noting that this rotation is very quick (as shown in the inset of Fig. 46(c)) and may be utilized for memory applications [248].

Similar but slightly different multiferroic effects were observed in other $RMn_2O_5$ systems for *R*=Tm, Er, Dy, Ho, Gd, and Y [249-282]. For example, the commensurate ferroelectric phase in $TbMn_2O_5$ is replaced by an incommensurate ferroelectric phase in $ErMn_2O_5$ and $TmMn_2O_3$ [281]. The complex behavior of the electric polarization, especially the anomaly of polarization at the low-temperature commensurate-incommensurate transition, remains unclear. Researches on $YMn_2O_5$ which excludes the effects of magnetic moments at the Tb ions postulated that there are two ferroelectric phases due to the complex spin structure. The spiral spin orders ensues both in the *ac* and *bc* planes, noting that the up-up-down-down order in the *ab* plane was described above. Both types of orders can induce ferroelectric polarization, according to the KNB model and the magnetostriction mechanism, corresponding to the intermediate-temperature commensurate ferroelectric phase and low-temperature incommensurate ferroelectric phase, as shown in Fig. 47 [282]. It is possible that both the mechanisms play important roles in these complex systems.

For $RMn_2O_5$, the magnetoelectric coupling between magnetism and ferroelectricity can be even more fascinating. For example, the dielectric response of $TbMn_2O_5$ (as shown in Fig.46(b)) and $DyMn_2O_5$ to magnetic field can be very big: ~109% at 3K upon a field of ~7T [250]. This extraordinary magnetodielectric effect seems to originate from the high sensitivity of the incommensurate spin state to external perturbations. A manipulation of the magnetic structure by electric field was also observed in this type of multiferroics. For $ErMn_2O_5$, which shows its magnetic and ferroelectric transitions very similar to $TbMn_2O_5$, a static electrical field may significantly enhance the magnetic scattering intensity. The reason may be that an electric field stabilizes the ferroelectric phase, which pushes the spin configuration into the commensurate magnetic phase by modulating the direction of magnetic moment via the magnetoelectric coupling. The X-ray scattering intensity $I$ as a function of the applied electric field at ~38.5K shows the butterfly-type hysteresis, also an evidence of the manipulation/switching of magnetic structure by electric field [251].

### 3.6. *Ferroelectricity induced by E-type antiferromagnetic order*

It is well established that the time inverse symmetry imposes rather strict conditions for possible magnetic orders that can induce ferroelectricity: the magnetic structure must have enough low symmetry in order for the lattice to develop a polar axis. As a consequence, usually the spin configuration should have complicated noncollinear structures, including spiral and incommensurate ones. The noncollinear magnetic structures can be stabilized by either competing interactions (frustration) or anisotropies generated by spin-orbital coupling, which usually lead to reduced transition temperatures and weak order parameters. In turn, there is a type of so-called collinear multiferroics, which are rare so far but may be more promising since they are less prone to the obstacles mentioned above. One type of the unusual collinear multiferroics come to those with $E$-type antiferromagnetic order ($E$-phase). In perovskite manganite family $R$MnO$_3$ (space group $Pbnm$), the $E$-phase was first reported in orthorhombic HoMnO$_3$, with the spin configuration shown in Fig. 48(a), which was considered as an example to demonstrate the collinear $E$-phase induced ferroelectricity [129, 283, 284].

We first come to look at a simple model associated with the $E$-phase and understand how the ferroelectricity is generated. In the $E$-phase, the parallel Mn spins form zigzag chains in the $ab$-plane, with the chain link equal to the nearest-neighbor Mn-Mn distance. The neighboring zigzag chains along the $b$-axis are antiparallel and the $ab$-planes are stacked antiferromagnetically along the $c$-axis. The symmetric coordinates corresponding to the $E$-phase can be defined as [283]:

$$E_1 = S_1 + S_2 - S_3 - S_4 - S_5 - S_6 + S_7 + S_8, \tag{25a}$$

$$E_2 = S_1 - S_2 - S_3 + S_4 + S_5 + S_6 + S_7 - S_8, \tag{25b}$$

where $S_i$ is the spin of the $i$-th Mn atom in the magnetic unit cell, as shown in Fig. 48(a). Considering that the Mn spins in HoMnO$_3$ point along the $b$-axis, we can only consider the $b$-components of $E_{1,2}$, denoted by $E_{1,2}$. Coordinates $E_1$ and $E_2$ span an irreducible representation of the space group $Pbnm$ corresponding to $k$=(0, 1/2, 0). Taking into account of polarization $P$ as a polar vector, the Landau potential corresponding to the $E$-phase can be defined as:

$$F = a( E_1^2 + E_2^2 ) + b_1( E_1^2 + E_2^2 )^2 + b_2 E_1^2 E_2^2 c( E_1^2 - E_2^2 )P_a +$$
$$+ d( E_1^2 - E_2^2 )E_1 E_2 P_b + \frac{1}{2\chi} P^2 \qquad , \tag{26}$$

where $\chi$ is the dielectric susceptibility of the paraelectric phase and other coefficients are the phenomenological parameters of the Landau theory. Minimizing $F$ with respect to $P$ yields:

$$P_a = -c\chi( E_1^2 - E_2^2 )$$
$$P_b = -d\chi( E_1^2 - E_2^2 )E_1 E_2 , \tag{27}$$
$$P_c = 0$$

where $P_i$ is the component of $P$ along the $i$-axis. Eq.(27) shows us that the four domains in the $E$-phase space, i.e. ($\pm E_1$, $0$) and ($0$, $\pm E_2$), are all multiferroic, with polarization $P$ pointing to the $a$-axis but its sign depending on the relative balance between coordinates $E_1$ and $E_2$.

To understand the microscopic mechanism for this $E$-phase induced ferroelectricity, one has to take into account the orbitally degenerate double-exchange with $e_g$-electron per $Mn^{3+}$ ion. The Hamiltonian can be written as [283]:

$$H = -\sum_{ia\alpha\beta} C_{i,i+a} t_{\alpha\beta}^{ia} d_{ia}^{+} d_{i+a,\beta} + J_{AF} \sum_{ia} S_i \cdot S_{i+a} + \lambda \sum_i (Q_{1i}\rho_i + Q_{2i}\tau_{xi} + Q_{3i}\tau_{zi}) + \frac{1}{2} \sum_{im} \kappa_m Q_{mi}^2 ,$$

(28)

where $d_{i\gamma}$ is the annihilation operator for the $e_g$ electron on two orbitals $\gamma = \alpha(x^2-y^2)$, and $\beta(3z^2-r^2)$, $a$ is the direction of the link connecting the two nearest neighbor Mn sites, and $S_i$ the classical unit spin of $t_{2g}$ electrons of Mn sites, $C_{ij}$ the double-exchange factor arising due to the large Hund's coupling that projects out the $e_g$ electrons with spin antiparallel to $S_i$, $Q_{mi}$ represents the classical adiabatic phonon modes.

An explicit solution to this model seems challenging. Orthorhombic perovskite manganites have a $GdFeO_3$-like distorted lattice with the Mn-O-Mn bond angle ($\varphi_{ia}$) deviating from 180°. In order to include the initial structural buckling distortion in orthorhombic perovskites, the dependence of hopping parameter $t_{\alpha\beta}^{ia}$ on $\varphi_{ia}$ must be explicitly considered, and the classical adiabatic phonon modes also must be defined such that the elastic energy term is minimal for $\varphi_{ia} = \varphi_0 < 180°$, as shown in Fig. 49(a). For example, $\varphi_0$ is ~144° for $HoMnO_3$.

Monte Carlo simulation of this model Hamiltonian manifests the crucial role of the double-exchange in the formation of ferroelectric state. Because of factor $C_{i,i+a}$, electron hopping between Mn ions with opposite $t_{2g}$ spins is prohibited. The displacements of the corresponding oxygen ions perpendicular to the Mn-Mn bond (these displacements are not Jahn-Teller active) depend only on the elastic energy, favoring a small angle $\varphi_0$. On the contrary, the hopping along the ferromagnetic zigzag chains is usually allowed and the hopping energy is minimal for $\varphi_0 = 180°$. Therefore, the displacements of the oxygen ions are eventually determined by the competition between the hopping energy and elastic energy. The resultant optimized angle $\varphi$ should satisfy condition $\varphi_0 < \varphi < 180°$. Since angle $\varphi$ only depends on the bond nature (ferromagnetic or antiferromagnetic), the oxygen displacements for all zigzag chains have the same direction even though the neighboring chains have opposite spin alignment, as shown in Fig. 48(b) and Fig. 49(b). This leads to the overall coherent displacements of the O ions with respect to the Mn sublattice, i.e. ferroelectricity, as shown in right part of Fig. 49(c). Clearly, these coherent displacements, however, do not exist if the Min spin alignment is disordered as shown in the left part of Fig. 49(c) .

In fact, it was revealed by Monte Carlo simulation that the magnitude of polarization $P$ does depend on $\varphi_0$. It would be zero if $\varphi_0 = 180°$ since both the hopping energy and elastic term are optimal at $\varphi_{ia} = \varphi_0 = 180°$, as shown in Fig. 49(d). In summary, it is understood that the symmetry of zigzag spin chain in the $E$-phase orthorhombic perovskites with buckling distorted oxygen octahedra allows for the formation of a polar axis along the $a$-axis, i.e. spontaneous polarization along this direction.

The first principle calculation proved that the inequivalence of the in-plane Mn-O-Mn configurations for parallel and antiparallel spins is an efficient mechanism for driving a considerable ferroelectric polarization [284]. Moreover, in addition to the polar ionic displacement mechanism, a larger portion of the ferroelectric polarization was found to arise due to quantum-mechanical effects of electron orbital polarization [284]. Fig. 48(c) shows the charge density isosurface in the $ac$-plane in the energy region between –8eV and 0 eV (0eV is the top of the valent band) for the relaxed structure of antiferromagnetic $E$-phase ($E_I$) HoMnO$_3$ with centrosymmetric atom arrangement by first principle calculations. From the charge density distribution, in addition to the expected checkerboard-like orbital ordering, two kinds of inequivalent oxygen ions with different charge distribution are confirmed due to the energy range of hybridized Mn $e_g$ and O $p$ orbitals. This polar charge distribution would result a polarization whose direction and quality are same to the one induced by ionic displacement, and is due to symmetry breaking by the antiferromagnetic-$E$ ordering [284].

To the end of this sub-section, we look at experiments on orthorhombic HoMnO$_3$. It should be mentioned that HoMnO$_3$ is of hexagonal structure at normal ambient. However, orthorhombic HoMnO$_3$ was successfully synthesized by high pressure sintering. It does exhibit the $E$-phase below ~26K, and consequently a macroscopic polarization along the orthorhombic $a$-axis [129, 285-287], confirming the theoretical prediction. However, the rapidly enhanced polarization below 15K was argued to be related to the noncollinear spin order of $Ho^{3+}$ ions rather than the $E$-phase.

The $E$-phase and associated ferroelectricity generation were used to explain the polarization flop at the critical field $H_C$ along the $c$-axis in $TbMnO_3$, intensively coinciding with the first-order transition to a commensurate magnetic phase with propagation vector (0, 1/4, 0) [170]. Another possible case is nickelates which also exhibit $E$-phase consisting of zigzag spin chains with different direction in the $ab$-plane and different stacking mode along the $c$-axis (++--). The Landau theory analysis predicts a polarization along the $b$-axis in such $E$-phase nickelates [283].

### 3.7. *Electric field switched magnetism*

In Sec.3.4.5 we already discussed the electric field control of magnetism in spiral multiferroic materials. In spite of this, in general it is difficult to realize such a control, in particular a switching of the magnetization state. So far, no electric field switching of a magnetization between a pair of 180° equivalent states has been demonstrated. One possible reason might be that the electronic polarization appears as a second order parameter coupled to the primary order parameter (magnetization) in those multiferroics with magnetism induced ferroelectricity. It is generally believed that a realization of such a switching can be easy if the magnetization is a second order parameter coupled to polarization as the primary order parameter, i.e. the ferroelectricity-induced magnetism [288, 289], while this argument seems to be challenging in principle. Even so, it should be noted that most multiferroics addressed so far have zero or weak macroscopic magnetization because of the antiferromagnetic nature of the spin configuration, while a large magnetization will be one of the prerequisites for practical applications. These motivations make the idea of ferroelectrically induced ferromagnetism very attractive although no substantial progress along this line has been accomplished. We discuss this issue in this section.

### 3.7.1. *Symmetry consideration.* According to the discussion in Sec.3.4.2, the antisymmetric microscopic coupling between two localized magnetic moments, i.e. the DM interaction is maximized when the two magnetic moments form a 90° angle, or more accurately, when $d_{ij}$, $S_i$,

and $S_j$ form a left-handed coordinate system for determinant $|d_{ij}|>0$ in Eq.(11). However, in these compounds under consideration, the Heisenberg-type interaction $E_{ij}^H = J_{ij} \cdot (S_i \cdot S_j)$ is usually much stronger than the DM interaction. For $J_{ij}=J_{ji}$, the Heisenberg interaction favors an angle of either 0 or 180° between $S_i$ and $S_j$, therefore, the presence of the DM interaction can only lead to a small canting of these interactive moments, corresponding to a weak macroscopic magnetization, i.e., weak ferromagnetism.

Now we discuss the DM factor $d_{ij}$. If the midpoint between the interactive moments is an inversion center, $d_{ij}$ is identically zero. For conventional ferroelectrics of interest today, a small polar structural distortion away from a centro-symmetric paraelectric structure exists. If the midpoint between two neighboring magnetic ions in the paraelectric structure is an inversion center, this symmetry will be broken by the ferroelectric distortion, which actually "switches on" the DM interaction (a nonzero $d_{ij}$) between the two ions, i.e. switches on a nonzero magnetization. This criterion was coined as the structural-chemical criterion, and the material specific parameter $D$ defined in Eq.(11) can be identified by polarization $P$. In summary, a ferroelectric distortion can generate a weak magnetization when the phenomenological invariant

$$E_{DML} = P \cdot (M \times L),\tag{29}$$

is allowed in the free energy of an antiferromagnetic-paraelectric phase. Consequently, at the ferroelectric transition point, once polarization $P$ becomes nonzero, the system gains an energy $E_{DML}$ by simultaneously generating a nonzero $M$. Moreover, if it is possible to reverse polarization $P$ using an electric field without varying the direction of vector $L$ (in Eq.(29)), magnetization $M$ will certainly reverse in order to minimize the total free energy. Therefore, the DM interaction (i.e. invariant term Eq.(11)) allows a possibility for electric field induced switching of magnetization. Note that some other symmetry operation associated with the ferroelectric transition or electric field induced sequences would result in $d_{ij}$=0, which unfortunately will prevent such a macroscopic magnetization from appearance [263]. In the following two subsections, we illustrate some examples to demonstrate such an effect.

### 3.7.2. Electric polarization induced antiferromagnetism in BaNiF$_4$. The first example is

BaNiF$_4$, which was theoretically predicted to exhibit ferroelectrically induced magnetic order [289]. BaNiF$_4$ is a representative of the isostructural family of barium fluorides of chemical formula Ba$M$F$_4$ with $M$=Mn, Fe, Co, Ni, Zn, or Mg, etc. They crystallize in the base-centered orthorhombic structure with space group $Cmc2_1$. The magnetic unit cell is doubled in comparison with the chemical unit cell, and contains four magnetic Ni ions arranged in sheets perpendicular to the $b$-axis. The cations within each sheet form a puckered rectangular grid, with the magnetic moments of neighboring cations aligned antiferromagnetically but all parallel to the $b$-axis. The coupling between different sheets is weak.

Nevertheless, first principle calculation including the spin-orbit coupling predicts that the collinear spin configuration with all spins aligned along the $b$-axis is unstable and the moments assume a noncollinear configuration where all spins are slightly tilted toward the $\pm c$-axis. Although this issue remains to be further clarified experimentally, the weak spin canting can be explained by the DM interaction. The magnetic space group of BaNiF$_4$ does not allow the occurrence of weak ferromagnetism, nevertheless a nonzero DM factor between the magnetic nearest neighbors along the $c$-axis, $d_c$, is available, in spite of such DM interaction along the $a$-axis vanishes. The canting due to the DM interaction generates a weak antiferromagnetic order parameter $L_c=s_1+s_2-s_3-s_4$ in addition to experimentally observed (primary) antiferromagnetic order parameter $L_{ab}=s_1-s_2-s_3+s_4$. Following the theory of the DM interaction described above, the DM coupling between order parameters $L_{ab}$ and $L_c$ on the macroscopic level can be written as:

$$E_{macro}^{DM}=D\cdot(L_{ab}\times L_c),  \tag{30}$$

Surely, an inclusion of this DM term in the free energy allows a control of $L_c$ by electric field. In fact, computation shows that no canting of the magnetic moments in the nonpolar structure is possible and the resultant magnetic order corresponds to a collinear structure, as shown in Fig. 50(a). For a polar distorting structure, the magnetic order becomes nonlinear, as shown in Fig. 50(b). When polarization $P$ is reversed in the calculation, clear orientation dependence of $L_c$ on $P$ is obtained, if order parameter $L_{ab}$, is fixed, as shown in Fig. 50(c). This does indicate a reversal of $L_c$ upon a reversal of $P$ driven by electric field [289].

### 3.7.3. *Electric polarization induced weak ferromagnetism in FeTiO$_3$.* The second example

comes to FeTiO$_3$. Before discussing this system, we look at BiFeO$_3$ first, which was described carefully in the earlier sections, because BiFeO$_3$ is a starting example for designing multiferroics with ferroelectrically induced weak ferromagnetism.

In paraelectric BiFeO$_3$ with space group R3c, Bi ions occupy the A sites with the Wyckoff position 2$a$ of local site symmetry *32* (as shown in Fig. 51(a)) whereas magnetic Fe ions occupy the B-sites with the Wyckoff position 2$b$ of inversion symmetry. The Fe spins order ferromagnetically within the antiferromagnetically coupled (111) planes of magnetic easy axis perpendicular to the [111] direction. Although in the paraelectric phase of BiFeO$_3$, the symmetry operator I transforms each Fe sublattice onto itself, i.e. I$L$=I($S_1$-$S_2$)=$L$, as shown in Fig. 51(b), in this case, the invariant $E_{PLM}$ is forbidden by the symmetry (the paraelectric point group is *2′/m′* or *2/m* for which weak ferromagnetism is allowed). In other words, the midpoints between the magnetic sites are not the inversion centers, as shown in Fig. 51(b). First principle calculation reveals that for BiFeO$_3$, the sign of vector coefficient $D$ defined in Eq.(27) is independent of the polar distortion, instead, it is determined by a rotational (non-polar) distortion of the oxygen octahedral network [290].

The situation would be entirely different if we place the magnetic ions on the A sites which are ordered similarly so that the magnetic criterion is still satisfied, as shown in Fig. 51 (c). This corresponds to the A-site magnetism, and one has I$L$=-$L$, i.e. the midpoints between the magnetic sites are the inversion centers. Placing a ferroelectrically active ion such as Ti$^{4+}$ on the B site would then satisfy the structural-chemical criterion. Although the magnetic point group in the paraelectric phase becomes *2/m′ (2′/m)* in which weak ferromagnetism is also forbidden, a ferroelectric distortion by design, via term $E_{DLM}$, would favor a lower symmetry *m′ (m)*, thus allowing the weak ferromagnetism, as shown in Fig. 51(c). The high pressure metastable phase of FeTiO$_3$ and MnTiO$_3$ [290-293] with space group R3c meets the criterions above, and provides the possibility of realizing the electric field switching of magnetization.

In fact, first principle calculation along this line is quite optimistic and direct. For FeTiO$_3$, a paraelectric phonon of symmetry type $A_{2u}$ can be polarized along [111] direction in the $R\overline{3}c \rightarrow R3c$ transition. One highly unstable mode in the *A2u* phonons consisting of displacements of the Fe ions and Ti ions against oxygen was also predicted, which is similar to other R3c ferroelectrics such as $BiFeO_3$ and $LiNdO_3$. A spontaneous polarization as big as 80-100 $\mu C/cm^2$, together with a ferroelectric transition point as high as 1500~2000K, was estimated. More exciting is the chirality change of the S-O-S bonds (as shown in Fig. 52) associated with the variation of polarization $P$ in orientation, was revealed in the calculation [288].

Although $BaNiF_4$ and $FeTiO_3$ were predicted to be multiferroics with ferroelectrically induced magnetism, so far no experimental evidence has been available due to the challenge of sample synthesis. High quality samples and experimental verification of these predictions are urgently needed, so that a substantial step towards practical control and switching of magnetism by electric field can be made.

### 3.8. *Other approaches*

Before ending this long section, we make some remarks on other strategies of integrating the two functions: ferroelectricity and magnetism, into one single phase compound. The unveiled physics may shed light on design and synthesis of novel multiferroics.

### 3.8.1. *Ferroelectricity in DyFeO₃.*
It has been known that in orthorhombic $DyAlO_3$ there exists a large linear magnetoelectric component [294, 295]. However, the Al ions are diamagnetic at the ground state, and then these materials do not show spontaneous **P** and magnetization **M**. Substitution of a magnetic ion at B site may produce the multiferroic state. Recently, researchers found the magnetic field induced ferroelectricity in orthorhombic $DyFeO_3$ [296]. The magnetic structures of $DyFeO_3$ are shown in Fig.53. Below $T_r$~37K, the Fe spins align antiferromagnetically in configuration $A_xG_yC_z$ where the G-type and A-type components of Fe spins direct toward the *b*-axis and *a*-axis respectively, while the C-type one along the *c*-axis. Upon further cooling, magnetization $M$ shows another anomaly at $T_N^{Dy}$ ~4 K, corresponding to the antiferromagnetic ordering of Dy moments in the $G_xA_y$ configuration. Moreover, below $T_r$, A magnetic field $H>H_r^{Fe}$ along the *c*-axis causes the configuration change to $G_xA_yF_z$ so as to produce a weak ferromagnetic component along the *c*-axis. Under $H$=30kOe along the *c*-axis, a large $P$  only along the *c*-axis below $T_N^{Dy}$ was observed, as shown in Fig. 54(a). Moreover, with increasing $H$, polarization $P$ as normal linear magnetoelectric component increases monotonically from zero, but shows an anomaly at $H=H_r^{Fe}\sim24$kOe. The extrapolated value of $P$ from the data within the region of $H> H_r^{Fe}$ backforward to $H=0$ is nonzero, as shown by the dashed line in Fig. 54(b). This demonstrates the existence of spontaneous polarization $P$. In fact, the multiferroic state can be further confirmed by the P-E hysteresis, as shown in Fig. 54(c) and (d) [296].

The orientation relationship between $P$ and $M$ ($P||M$) in DyFeO$_3$, which is different from those in spiral magnets such as DyMnO$_3$, together with the disappearance of $P$ at $T_N^{Dy}$ and anomaly of $P$ at $H_r^{Fe}$, suggests that the mechanism for ferroelectricity in DyFeO$_3$ is different from the inverse DM interaction and depends on both the magnetic structures of Dy and Fe ions. It is postulated that the exchange striction between those adjacent Fe$^{3+}$ layer and Dy$^{3+}$ layer with the interlayer antiferromagnetic interaction (see Fig. 53(c) and (d)) results in the multiferroic phase. The ferromagnetic sheets formed by Fe and Dy ions stack along the $c$-axis. For the A$_y$ component along the $b$-axis, the spins on the Fe layer become parallel to the moments on one of the nearest neighbor Dy layers and antiparallel to those moments on another nearest neighbor layer. As a result, the Dy layers should displace cooperatively toward the Fe layers with antiparallel spins, via the exchange striction. Then the polarization along the $c$-axis appears [296].

### 3.8.2. Ferroelectricity induced by A-site disorder.

Perovskite lattice instabilities are often described by the tolerance factor $t = (r_A + r_O)/\sqrt{2}(r_B + r_O)$, where $r_A$、$r_B$ and $r_O$ are the A-site, B-site and O ionic radii, respectively. Conventional ferroelectric materials such as BaTiO$_3$ often have $t>1$, indicating that the B site ion is too small for ideal cubic structure. Assisted by the covalent hybridization with O ions, B-site ions deviate from the center positions and then cause ferroelectric polarization. The ferroelectrics with lone pair mechanism, as stated in Sec.3.2, have $t<1$ and the ferroelectricity is from off-centering of the A-site ions. Without Pb or Bi, perovskite structures of $t<1$ generally have tilted BO$_6$ octahedra instead of the A-site off-centering. Unfortunately, the magnetic perovskite materials often have $t<1$ and tilted BO$_6$ octahedra because those ions with $d$ electrons are generally larger than $d^0$ ions, and then not ferroelectrically active.

However, first principle calculations suggest that the octahedral tilting is prevented in $KNbO_3$-$LiNbO_3$ alloys with the average tolerance factor significantly smaller than one, because K ions and smaller Li ions distribute randomly in the lattice, which is coined as A-site disorder. The ferroelectricity appears to originate from the large off-centering of Li ions, contributing significantly to the difference between the tetragonal and rhombohedral ferroelectric states and yielding a tetragonal ground state even without strain coupling [297].

Based on above discussion, it is predicted that $(La,Lu)MnNiO_6$ with $t$ <1 exhibits polar-type lattice distortion [298]. This polar behavior arises from the frustration of the octahedral tilting instabilites due to the mixture of A-site cations of different sizes and the fact that the coherence length for the A-site off-centering is shorter than that for the tilting instabilities [298]. On the other hand, $Mn^{3+}$ and $Ni^{3+}$ ions can occupy the B-sites in an order form, resulting in the double perovskite structure. The superexchange interaction between $Mn^{3+}$ and $Ni^{3+}$ is ferromagnetic [86, 87]. Due to these mechanisms, $(La,Lu)MnNiO_6$ may exhibit large ferroelectric polarization and ferromagnetism simultaneously. However, again it is difficult to synthesize $(La,Lu)MnNiO_6$ because of the phase separation and competing phases, which often occur for perovskite oxide materials with mismatching A-site species. Therefore, so far no experimental evidence with this A-site disorder induced multiferroicity has been available.

### 3.8.3. *Possible ferroelectricity in graphene*. Besides those approaches in transitional metal oxides substantially addressed above, some other approaches may be also utilized to synthesize novel multiferroic materials. For example, an electronic phase with coexisting magnetic and ferroelectric orders in graphene ribbons with zigzag edges is predicted [299-303]. The physics lies in that the coherence of the Bardeen-Cooper-Schrieffer (BCS) wave function for electron-hole pairs in the edge bands, available in each spin channel, is related to the spin-resolved electric polarization [299]. Although the total polarization may vanish due to the internal phase locking of the BCS state, strong magnetoelectric effects are expectable. By placing the graphene between two ferromagnetic dielectric materials, theoretical analysis predictsthat the magnetic interaction at the interface affects the graphene band structure and leads to an effective exchange bias between the magnetic layers, which is

strongly dependent of the electronic properties (particularly of the position of the electrochemical potential, i.e. the Fermi level) of the graphene layer. Therefore, an external electric field (the gate bias) can modulate the exchange bias [304].

**3.8.4. *Interfacial effects in multilayered structure*.** Interfacial effects, perhaps different from the macroscopic mechanical transfer process, can be exemplified in multiferroic superlattices, and then significant magnetoelectric effects can be expected. For example, first principle calculation predicts that, in the ferroelectric/ferromagnetic multilayers such as $Fe/BaTiO_3$ structure, the bond fluctuation on the ferroelectric/ferromagnetic interface will modulate the interfacial magnetization upon the polarization reversal due to the interface bonding sensitive to the atomic displacements on the interface [305, 306]. Similar effects are predicted by first principle calculation in $Fe_3O_4/BaTiO_3$ oxide-superlattice, too [307]. The effects of the charge imbalance and strain as well as oxygen vacancies on the interfaces of superlattice, may play important roles [308]. Moreover, first-principle calculation claims that even for the Fe(001), Ni (001) and Co (0001) films, an external electric field can induce remarkable changes of the surface magnetization and surface magnetocystalline anisotropy, originating from spin-dependent screening of electric field at the metal surface, as shown in Fig. 55 [309]. However, these effects still need experimental demonstration. Another approach to multiferroics is the so called tricolor multilayered oxides structures. Tricolor multilayered structure (i.e. ABCABC…) without ferroelectric layer, where at least one layer or one interfacial layer should be ferromagnetic, such as $LaAlO_3/La_{0.6}Sr_{0.4}MnO_3/SrTiO_3$ structure, exhibits multiferroicity on the ferromagnetic interface. The details of the tricolor multilayered structure can refer to recent literature [310-314].

## 4. Elementary excitation in multiferroics: electromagnon

For condensed matters, it is well established that any spontaneous breaking of symmetry will induce novel elemental excitation [315]. For conventional ferroelectrics, a displacive structural phase transition is associated with one of the transverse optical (TO) phonons softening with its frequency, corresponding to the square root of the inverse order parameter susceptibility $\chi(0)$, i.e. $\omega^2 \propto 1/\chi(0)$ [5]. Here, a soft polar phonon directly couples to the divergent dielectric susceptibility and broken spatial inversion symmetry. Spin waves (i.e. magnons) are the characteristic excitations of the magnetic structure. It is expected that the simultaneous breaking of the spatial reversion symmetry and time inversion symmetry and, thus, the strong coupling between the magnetic and lattice degrees of freedom can lead to complex excitations. In this setting, the character associated with the soft mode is less obvious since the multiferroic order does not arise from pure structural degrees of freedom but from their complex interplay with magnetism. Thus, the collective excitation directly reflecting the inverse DM mechanism is the rotation mode of the spiral plane that is driven by electric field, and a consequence fundamentally different from ferroelectric and spin excitation exists: electro-active magnons, or electromagnons (i.e. the spin waves that can be excited by a.c. electric field). This kind of elemental excitations due to the magneto-dielectric interaction was theoretically predicted more than thirty years ago [316], but no other experimental observation have been made until very recently [317-336].

### 4.1. *Theoretical consideration*

We first outline the theoretical framework of electromagnons, developed recently [317]. From the KNB theory, the spin supercurrent in noncollinear magnets, $j_s \propto S_i \times S_j$, leads to the electric polarization defined by $P \propto e_{ij} \times j_s$, with $e_{ij}$ the unite vector connecting two sites $i$ and $j$. An effective Hamiltonian describing the coupling between spins and atomic displacement $u_i$ may take the following form:

$$H = H_1 + H_2 + H_3 + H_4$$
$$H_1 = -\sum_{m,n} J(R_m - R_n) S_m \cdot S_n ,$$
$$H_2 = -\lambda \sum_m (u_m \times e_z) \cdot (S_m \times S_{m+1}) , \qquad (31)$$
$$H_3 = \sum_m (\frac{\kappa}{2} u_m^2 + \frac{1}{2m} P_m^2),$$
$$H_4 = \sum_m D(S_m^y)^2 .$$

where $H_1$ denotes the Heisenberg interaction with $R_i$ and $S_i$ the coordinates and spin moment of site $i$; the spin-lattice interaction $H_2$ stems from the relativistic spin-orbit interaction and corresponds to the DM interaction once the static displacement $<u_m>$ is nonzero and the inversion symmetry is broken, $u_m$ is regarded as the lowest frequency representative coordinate relevant to polarization $P$, i.e. the TO phonons, $P = e^* u_m$ with a Born charge $e^*$; in term $H_3$, $\kappa$ and $m$ are the spring constant and effective mass of $u_m$; term $H_4$ deals with the easy-plane spin anisotropy with anisotropic factor $D$.

This Hamiltonian allows a helical spin ordering with decreasing temperature, corresponding to the softening and condensation of the spin bosons. The phonon mode $u_x$ does not show any frequency softening, but the spontaneous polarization is realized through the hybridization of $u_x$ with the spin bosons. One may assume that the spins are on the easy plane, i.e., $S_n^z = S\,cos(QR_n + \phi)$, $S_n^x = S\,sin(QR_n + \phi)$, $S_n^y = 0$, where $Q$ is the spiral wave number and $\phi$ is the phase angle. Also, the equations of motion for spins and displacements can be derived out from the Hamiltonian. Considering the lowest temperature region with spin order and converting the lattice into a rotating local coordinate system ($\xi,\ \eta,\ \zeta$ ) and momentum space ($q$), one has the equations of motion:

$$\dot{S}_q^{\eta} = -A(q)S_q^{\xi}$$
$$\dot{u}_q = p_q / m$$
$$\dot{S}_q^{\xi} = B(q)S_q^{\eta} - i\lambda S^2\left[\frac{e^{iQa}-e^{-iQa}}{2i}e^{i\phi}u_{q-Q} + \binom{Q \to -Q}{\phi \to -\phi}\right],$$
$$\dot{p}_q = -\kappa u_q + i\lambda S\left[\frac{e^{iQa}-e^{i(q-Q)a}}{2i}e^{i\phi}S_{q-Q}^{\eta} + \binom{Q \to -Q}{\phi \to -\phi}\right]$$

$$(32)$$

with

$$A(q) = 2S\left[\frac{2J(Q)-J(Q+q)-J(Q-q)}{2} + \frac{2\lambda^2 S^2}{\kappa}sin^2(\frac{qa}{2})\times sin^2(Qa)\right],$$
$$B(q) = 2S\left[J(Q)-J(q)+\frac{\lambda^2 S^2}{\kappa}sin^2(Qa) + K\right]$$

$$(33)$$

From the equations of motion, one can evidently see the coupling between the spin wave modes and electric polarization. Here, $S^{\eta}$ and $S^{\xi}$ are the canonical variables and form a harmonic oscillator at each $q$ in the rotate frame. The spin wave at $q$ is coupled with the phonon $u$ at $q \pm Q$, or $u_q$ is coupled to $S^{\eta}$ at $q \pm Q$. The uniform lattice displacement $u_o^{\,y}$ is coupled to $e^{-i\phi}S_{Q}^{\eta}-e^{i\phi}S_{-Q}^{\eta}$, which corresponds to the rotation of both the spin plane and the direction of polarization along the $z$-axis. This mode is the Goldstone mode with frequency $\omega=0$ if $K=0$. The spin wave mode at $q=0$ corresponds to the sliding mode, i.e., spiral phason.

The dynamic dielectric function can be obtained by the retarded Green function and it has the poles at $\omega_{\pm}$, given by

$$\omega_{\pm}^2 = \frac{1}{2}\left(\omega_0^2 + \omega_p^2 \pm \sqrt{(\omega_0^2 + \omega_p^2)^2 - 4A(Q)K\omega_0^2}\right)$$

$$\omega_0 = \sqrt{\kappa/m}$$

$$\omega_p = \sqrt{A(Q)B(Q)}$$

(34)

where $\omega_0$ is the frequency for the original phonon and $\omega_p$ is the frequency of the mode $e^{-i\phi}S_Q^\eta - e^{i\phi}S_{-Q}^\eta$. In the limit of $\lambda << \omega_0^2$, one can see from the above equation that there are two modes contributing to the dielectric function. One is the phonon mode with frequency $\omega_+ \approx \omega_0$ which is high and doesn't show any softening. The dielectric function is most likely contributed from the other mode, i.e. the $z$-axis rotation mode (spin wave mode) at $\omega_- \approx \sqrt{A(Q)K}$, which is hybridized with the polarization mode $u_o{}^y$.

The theoretical study on the elementary excitation based on the symmetry analysis and the Landau theory also gives similar results [318]. The above discussion combined with a realistic estimation of materials parameters allows one to calculate the frequency of the collective mode, i.e., the electromagnon. As well established already, ferroelectricity induced by the spin order is usually observable in incommensurate spiral/helical spin ordered systems. This special spin order can be suppressed by external magnetic field, thus the corresponding electromagnon can be wiped out. In consequence, a significant response of the reflection spectrum, ranging from *d.c.* up to terahertz frequency range, to an external magnetic field, can be expected. Along this line, dielectric spectroscopy under an external magnetic field can be a roadmap to reveal electromagnons in multiferroics. In fact, preliminary experiments to disclose this electromagnon excitation, reported recently, were quite successful and good consistency between experimental observation and theoretical prediction is evidenced, to be given below.

## 4.2. *Electromagnons in spiral spin ordered (Tb/Gd)MnO₃*

For multiferroics, the frequency dependence of dielectric response usually shows a broad relaxation-like excitation. The characteristic frequency for $GdMnO_3$ is $\nu_0 = 23(\pm 3)$ cm$^{-1}$ and for TbMnO$_3$ it is $v_0$=20($\pm$3) cm$^{-1}$, as shown in Fig. 56. The dielectric response of this excitation increases with decreasing temperature and becomes saturated once the low-temperature spiral magnetic phase enters. Upon a magnetic field, both the imaginary and real parts of the dielectric constant will be suppressed. In particular, the excitation will be suppressed when the *a.c.* component of the electric field *e* is rotated from *e*||*a* to *e*||*b*, given a constant magnetic field. However, it remains unchanged when the a.c. component of magnetic field *h* is rotated from *h*||*c* to *h*||*b*, as shown in the inset of Fig. 56(c) [319]. The significant sensitivity of the excitation to the a.c. electric field demonstrates the strong coupling between magnetic and lattice degrees of freedom, reflecting the close correlation of spin structure and electric polarization and thus providing the possible evidence for electromagnons in multiferroics [319].

These experimental results are also quantitatively consistent with theoretical predictions. For TbMnO$_3$, from the spin wave dispersion data observed in the neutron scattering and electron spin resonance (ESR) spectroscopy, the exchange coupling $J_1$ was estimated to be 8$SJ_1$=2.4 meV and the spin-lattice coupling was $\lambda$~1.0 meV/Å. The Born charge was assumed to be 16*e* where *e* is the bare unit charge. Then the evaluated frequency of the collective mode in TbMnO$_3$ is $\omega_-$ ~ 10 cm$^{-1}$, which is of the same order with experimental data (~20 cm$^{-1}$).

Inelastic neutron scattering (INS) is the most powerful technique to disclose the magnetic excitation in spin systems [320]. The INS dispersion relations for spin-wave excitations in TbMnO$_3$ along the *a*-axis and *c*-axis of the *Pbnm* lattice in paraelectric sinusoidal phase and in ferroelectric spiral phase, respectively, were presented in Fig. 57(a) and (b). Clearly, the three low-lying magnons are revealed, as shown in Fig. 57(c), in which the lowest energy one is the sliding mode of spiral. The other two modes at 1.1meV and 2.5meV correspond to the rotations of the spiral rotation plane, as shown in Fig. 57(d). The latter two modes are coupled with the electric polarization and the outcome is in perfect agreement with the infrared spectroscopy result. This is a hybridized phonon-magnon excitation (i.e. electromagnon.)

It should be mentioned that some other methods such as the far-infrared spectroscopy were recently utilized for probing electromagnons in spiral multiferroics, such as in Eu$_{1-x}$Y$_x$MnO$_3$ [321, 322] and GdMnO$_3$ [323], which also demonstrate the existence of the spin(magnon)-lattice(phonon) coupling and electromagnons in perovskite *R*MnO$_3$. It is worthy of noting that the elementary excitation in $R$MnO$_3$ remains ambiguous although the experimental results are also quantitatively consistent with theoretical predictions. According to the inverse DM mechanism, the $k=0$ magnon mode responding to the rotation mode of the spiral plane should be active for the $E$-vector perpendicular to the spin spiral plane. Then, the polarization selection rule for the electromagnon, i.e. the absorption band in Eu$_{1-x}$Y$_x$MnO$_3$ with $P||a$, should be $E^\omega||c$. However, the absorption band with $E^\omega||a$ was observed in Eu$_{1-x}$Y$_x$MnO$_3$. This discrepancy appeals for further research. For example, the wide range optical spectra on Eu$_{1-x}$Y$_x$MnO$_3$ revealed that the possible candidate of origin for this absorption band is the two-magnon excitation driven by electric field [324].

### 4.3. *Electromagnons in charge frustrated $RMn_2O_5$*

Additional evidence on electromagnons comes from the far-infrared transmission spectra for YMn$_2$O$_5$ and TbMn$_2$O$_5$. TbMn$_2$O$_5$ (YMn$_2$O$_5$) favors the paramagnetic/paraelectric state at $T>41(45)$K, the commensurate magnetic order and ferroelectric order at $24(20)$K$<T<38(41)$K, and the incommensurate magnetic order and ferroelectric order below $24(20)$K. The far-infrared transmission data revealed a clear electromagnon excitation feature below the lowest phonon centered at ~97cm$^{-1}$ and the strongest absorption near 10cm$^{-1}$: 7.9cm$^{-1}$ for YMn$_2$O$_5$ and 9.6cm$^{-1}$ for TbMn$_2$O$_5$, as shown in Fig. 58 [325].

### 4.4. *Spin-phonon coupling in hexagonal $YMnO_3$*

Although the ferroelectric ordering and magnetic ordering in hexagonal $R$MnO$_3$ is not concomitant, there exists a strong interplay between the two order parameters, as discussed in Sec.3.4. It is reasonable to postulate that the spin-phonon coupling in hexagonal $R$MnO$_3$ is strong. Looking at such a coupling in YMnO$_3$, characterized by thermal conductivity, it was observed that the thermal conductivity exhibits an isotropic suppression in the cooperative paramagnetic state, followed by a sudden increase upon the magnetic ordering. This unprecedented behavior without any associated structural distortion is probably the consequence of a strong dynamic coupling between the acoustic phonons and low-energy spin fluctuations in geometrically frustrated magnets [326]. Some other experiments, such as thermal expansion [327], Raman scattering [328], and ultrasonic measurement [329] also revealed the existence of a giant spin-lattice (phonon) coupling.

Also, such a coupling can be probed by inelastic neutron scattering which plays an important role in studying the elementary excitation in YMnO$_3$ [330]. Fig. 59(a) shows the magnetic structure of YMnO$_3$ and the dashed lines in Fig. 59(b) plots the magnon dispersion of three modes along the $a^*$-direction (as shown in Fig. 59(a)) (the O$_T$O$_P$ tilting direction involved in the ferroelectric distortion) measured by neutron scattering (symbols). The dispersions of the transverse phonon mode mainly polarized along the $c$-axis in the ferroelectric phase with propagation along the $a^*$-axis, obtained at 200K (triangles) and 18K (circles), are shown in Fig. 59(b), together with the optical phonon mode (squares) and three magnon modes (dashed line). It is evident that a gap in the phonon dispersion opens at $q_0 \sim 0.185$ and a crossing of the 200K phonon dispersion with the magnon mode 2 arises at $q_{across} \sim 0.3$. Moreover, the gap opens mainly below $T_N$, indicating its coupling with the magnetic subsystem. Fig. 59(c) shows the nuclear dynamical structure factor revealing the phonon-like component of the hybrid excitation and a jump from the lower to the upper mode is observed, providing a natural interpretation of the experimentally observed gap. These data reveal a strong coupling between spins and phonons and possible electromagnons, i.e. the hybridization between the two types of elementary excitations, in hexagonal manganites [330].

### 4.5. Cycloidal electromagnons in BiFeO$_3$

BiFeO$_3$ is similar to YMnO$_3$ in the sense that the ferroelectricity and magnetism originate from different ions. However, as illustrated in Sec.3.4.4, the ferroelectricity in BiFeO$_3$ is closely related to the cycloidal antiferromagnetic order, implying probably an intimate relationship between the electric polarization and spin wave excitations (magnons), i.e. the electromagnons [207, 331-333]. BiFeO$_3$ exhibits a G-type antiferromagnetic order which is subjected to a long-range modulation associated with a cycloidal spiral of a periodicity of $\sim$62nm. The spiral propagates along the $[10\bar{1}]$ direction with the spin rotation within the $(\bar{1}2\bar{1})$ plane, as shown in Fig. 8(c) and (d) [207].

Recently, low-energy Raman scattering spectroscopy was used to unveil the magnon spectra of BiFeO$_3$ [207]. Although no phonons below 50 cm$^{-1}$ are expected, several peaks in the Raman spectra were observed. The two configurations with parallel polarization and crossed polarization on the (010) plane produced spectra with two distinct sets of peaks, as shown in Fig. 60(a). The two sets of peaks, respectively corresponding to two species of spin wave excitations lying in (cyclone modes) and out of (extra-cyclone modes) the cycloidal plane, exhibit distinctive dispersive energy curves that depend on their coupling to the electric polarization, as shown in Fig. 60(b). The antiferromagnetic magnon zone folding, induced by the periodicity of the cycloidal spin order, leads to a very simple expression for the energy level structure of the cyclone mode. This cyclone mode remains gapless, as expected from the antiferromagnetic ordering, but a gap is expected for the extra-cyclone mode due to the pinning of the cycloidal plane by the polarization. The experimental results do fit this picture and an extra-cyclone mode with gap was unambiguously assigned, demonstrating the cycloidal electromagnons, as shown in Fig. 60(b).

The elementary excitations in multiferroics will significantly affect the physical properties, which reveals new possibility for applications. For example, the magnetic sublattice precession is coherently excited by picoseconds thermal modification of the exchange energy during detecting the magnetic resonance mode in multiferroic Ba$_{0.6}$Sr$_{1.4}$Zn$_2$Fe$_{12}$O$_{22}$ using the time-domain pump-probe reflectance spectroscopy. This excitation induces the modulation of the material's dielectric tensor and then a dynamic magnetoelectric effect [334].

Besides those examples cited above, more experiments did reveal the strong spin-phonon (lattice) coupling in other multiferroics, such as the two dimensional triangular CuFeO$_2$ [335]. These experiments unveiled the existence of electromagnons in a broad category of materials. However, a comprehensive understanding of their origins, conceptual pictures, and dynamics, seems far from sufficient. One key point is whether the origin of electromagnon excitations is the DM exchange interaction. Some works pointed out that the electromagnon excitation in multiferroic orthorhombic $R$MnO$_3$ should result from the Heisenberg coupling between spins despite the fact that the polarization arises from the much weaker DM exchange interaction [336].

## 5. Ferrotoroidic systems

In practice, ferromagnetism, ferroelectricity and ferroelasticity are widely utilized in modern technology. The three functions are always called fundamental ferroicity. One common character for these functions is the domain structure associated with the spontaneous magnetization, polarization and elastic transform, respectively. These domains are the key units for data memory. For example, ferromagnetic domains are memory units in computer hard disks and ferroelectric domains are found in ferroelectric random access memories. Recently, the fourth ferroicity —ferrotoroidicity— was proposed as being one of the fundamental ferroicity, and consequently the fourth kind of ferroics —ferrotoroidics— was addressed [337-348].

### 5.1. *Ferrotoroidic order*

As well known, a magnetic toroidic moment is generated by a vortex of a magnetic moment, such as atomic spins or orbital currents, which can be represented by a time-odd polar (or "axiopolar") vector $T = \frac{1}{2} \sum_i r_i \times S_i$ where $r_i$ and $S_i$ are the $i$-th magnetic moment and its positional vector, respectively [337-339]. This toroidic vector $\boldsymbol{T}$ changes its sign upon both time inversion and space inversion operations and is generally associated with a "circular" or "ring-like" arrangement of spins. The concept of magnetic torodic moment can be sketched by a ring-shape torus with an even number of current windings which exhibit a magnetic toroidic moment $\boldsymbol{T}$ (the green arrow in Fig. 61(a)) perpendicular to the ring plane. In a magnetic toroidic system, it is possible to induce a magnetization $M$ by an electric field $E$ and a polarization $P$ by a magnetic filed $H$, which is one of the reasons why much attention has been paid to magnetic toroidic systems. For example, in the system shown in Fig. 61(a), a magnetic field along the ring plane drives a congregation of the current loops in one direction and, eventually, an electric polarization along this direction appears, as shown in Fig. 61(b).

A system in which the toroidic moments are aligned spontaneously in a cooperative way is coined as a ferrotoroidic system. The macroscopic vector $\boldsymbol{T}$ of this system can also be used as the order parameter for various d.c./optical magnetoelectric phenomena, which describe the genuinely electronic couplings between an electric field and a magnetic field. For details, the toroidic moment $\boldsymbol{T}$ describes the coupling between polarization $P$ and magnetization $M$ and one can easily derive out $\boldsymbol{T} \propto P \times M$ for multiferroics of ferroelectric and ferromagnetic orders. However, it should be mentioned here that a nonzero macroscopic $\boldsymbol{T}$ does not necessarily require the co-existence of $P$ and $M$. For example, $GaFeO_3$ is a prototypical ferrotoroidic system, as shown in Fig. 62(a) and (b) [340-343]. It is pyroelectric in nature with the built-in electronic polarization along the $b$-axis in the orthorhombic cell, and its spontaneous magnetization stems from the ferromagnetic arrangement of Fe spins. However, the displacements of two Fe-ions sites are opposite, as if it was antiferroelectric. In this case, a macroscopic toroidal moment is present but its magnitude is larger than $P \times M$. This is also one of the reasons why antiferromagnetics or antiferroelectrics are categorized into the components of mutlferroics. On the other hand, the difference between ferrotorodics and multiferroics is disputed since the definition for each of the two types of ferroics remains unclear. For example, typical ferrotoroidic $GaFeO_3$ has been regarded as typical multiferroic or magnetoelectric material [341].

In fact, any physical system can be characterized by its behavior upon spatial and temporal reversals. Ferromagnetics and ferroelectrics correspond to the systems whose order parameters change their sign upon the temporal and spatial reversal, respectively. For a ferroelastic system, no such change occurs under both the two reversals, as shown in Fig. 63. It is apparent that the three fundamental ferroic orders correspond to three of the four parity-group representations and the residual one should be assigned as the ferrotoroidic order which changes sign under both the two reversals. This is the reason to coin ferrotoroidics as the fourth type of fundamental ferroics and the relationship between ferrotoroidics and multiferroics can be highlighted. The multiferroics are spatial- and time-asymmetric because of the coexistence of two order parameters: one violating the spatial reversal symmetry and the other breaking the temporal reversal symmetry.

It is well known that ferroelasticity is always related to ferroelectricity, and similarly ferrotorodicity is intrinsically linked to antiferromagnetism because of its vortex nature. Fig. 64 shows four simple and typical antiferromagnetic systems, where (a) and (b) have equal and opposite toroidal moments and the antiferromagnetic arrangement in (c) also has a toroidal moment, while the arrangement in (d) does not.

Besides GaFeO$_3$, LiCoPO$_4$ and LiNiPO$_4$ also are the typical ferrotoroidics [344-348]. LiCoPO$_4$ has a olivine crystal structure with *mmm* symmetry in a paramagnetic state. The Co$^{2+}$ ions are located at coordinate like (1/4+$\varepsilon$, 1/4, -$\delta$) where $\varepsilon$ and $\delta$ are small displacements allowed by the *mmm* symmetry, as shown in Fig. 65(a). At 21.8K, the Co$^{2+}$ ions order in a compensated antiferromagnetic configuration with spins along the *y*-axis while the symmetry changes to *mmm'*. Moreover, recent neutron scattering data revealed a rotation of the spins by $\varphi$=4.6° away from the *y*-axis and a reduction of symmetry down to 2′ with *x* as the twofold axis. This magnetic order is not the helical type and all magnetic moments order antiparallely with $\mathbf{S_n}$||(0, cos$\varphi$, sin$\varphi$), contributing to the weak magnetism along the *y*-axis.

LiCoPO$_4$ exhibits a ferrotoroidical order in the *x-z* plane, as shown in Fig. 65(b). The spin part of the toroidal moment is described by $T = \frac{1}{2}\sum_n r_n \times S_n$ with $r_n$ the radius vector and $S_n$ the spin of the *n*-th magnetic ions, taking the center of the unit cell as the origin. Note that only the components of $S_n$ that are oriented perpendicularly to $r_n$ contribute to $\mathbf{T}$, as shown by the green arrows in Fig. 65(b). Clearly, the contribution of the spins in ions 1 and 3 are contrary to the contribution of the spins in ions 2 and 4. However, the clockwise contribution from ions 1 and 3 is larger than the anticlockwise contribution from ions 2 and 4 because $r_{1,3}$>$r_{2,4}$, leading to a residual toroidal moment $T_y$ perpendicular to the *x-y* plane. Any sign reversal of either $S_x$ or $\varphi$ will result in the reversal of order parameters of antiferromagnetism and ferrotoroidicity (±*l* and ±$\mathbf{T}$). In fact, recent experiments using resonant X-ray scattering demonstrated the existence of ferrotoroidical moment in this system, noting that LiNiPO$_4$ is very similar to LiCoPO$_4$, although the spins in LiNiPO$_4$ are aligned along the *z*-axis.

## 5.2. *Magnetoelectric effect in ferrotoroidic systems*

The toroidic moment $\mathbf{T}$ (i.e. the coupling between *P* and *M*) can cause some interesting optical magnetoelectric effects. One of them originates from the polarization component induced by optical magnetic field as an analog of the magnetoelectric coupling in optical frequency. The normal Farady (or Kerr) rotation, as shown in Fig. 62(c), stems from the dichroism or birefringence with respect to the right-hand or left-hand circularly polarized light. The optical magnetoelectric effect refers to the dichroism/birefringence with respect to the light propagation vector, irrespective of the light polarization, as shown in Fig. 62(c). Another important feature of the optical magnetoelectric effect is the second-order nonlinear optical activity. Due to the presence of the toroidal moment $T$, the second harmonic (SH) light with polarization in parallel to $T$ can be generated (Fig. 62(c)) in addition to the ordinary SH light polarized along the $P$ direction. Eventually, the incident light polarized along the $T$ direction can generate the SH components polarized along the $P$ and $T$ directions, respectively. This $T$-induced SH component may reverse its phase upon the magnetization reversal. Consequently, the polarization of the SH light can rotate depending on the magnetization direction or equivalently of the toroidic moment direction. This nonlinear Kerr rotation can be used to sensitively probe the toroidic moment or the breaking of the inversion symmetry.

Both $LiCoPO_4$ and $LiNiPO_4$ exhibit very large magnetoelectric coupling and the low-temperature symmetry of the magnetic ground state allows the existence of a linear magnetoelectric effect [344, 345]. For example, in $LiNiPO_4$, the magnetoelectric tensor $\alpha$ has two nonzero components $\alpha_{xz}$ and $\alpha_{zx}$ [345] (correspondingly for $LiCoPO_4$ subscript $z$ should be replaced by $y$). Fig. 66 shows the electric polarization along the $z$-axis under an external magnetic field along the $x$-axis below the magnetic transition point ~20K. It is evident that a relatively large magnetic field along the $x$-axis can induce a large polarization along the $z$-axis. More exciting is that the relationship between the polarization along the $z$-axis and magnetic field along the $x$-axis exhibits a butterfly loop around the magnetic transition point and this loop disappears at a lower temperature. It is well known that the butterfly loop always corresponds to the appearance of spontaneous moments, as in ferromagnetics and ferroelectrics, and this phenomenon demonstrates the existence of the macroscopic and spontaneous toroidical moments.

## 5.3. *Observation of ferrotoroidic domains*

Domain structure and wall apply to ferrotoroidics too although the spin order is essentially antiferromagnetic. A ferrotoroidical system can exhibit a ferrotoroidical domain which is independent of an antiferromagnetic domain because of the different symmetries in these systems. Take LiCoPO$_4$ as an example again. The antiferromagnetic ordering reduces the symmetry from *mmm* to *mmm′* and the number of symmetry operations from sixteen to eight, corresponding to two antiferromagnetic domains (±*l*). The spin rotation around *x* reduces the symmetry to 2′ and number of symmetry operations to two, corresponding to two ferrotoroidical (FTO) domains (±*T*) [348].

The second harmonic generation (SHG) appears to be a powerful tool in detecting domain structure in ferrotoroidics. For the first order approximation, sign reversal of order parameter *O* will induce the reversal of SHG susceptibility $\chi(O)$. This means a 180° phase shift of the SHG light from opposite domains, which allows one to identify the domain structure. Given the fact that different ferroic orders correspond to different symmetries and then $\chi(O)$, it is possible to image different domains coexisting by polarization analysis. This approach was demonstrated recently in LiCoPO$_4$ using the SHG technique, where the ferrotoroidical domains were successfully imaged, providing direct evidence for ferrotoroidicity as a kind of fundamental ferroicity. Fig. 67(a) shows the $\chi_{zzz}$ image obtained at 2.25eV for a nearly single antiferromagnetic domain in LiCoPO$_4$ (100) single crystal, where the single antiferromagnetic domain with a single antiferromagnetic domain wall at the lower left, shown by the dark line (black patch in the center of the sample is damage), is mapped.

The images using SHG light from $\chi_{yyz}$ and $\chi_{zyy}$ components exhibit completely different patterns. Fig. 67(b) gives the images using SHG light from $\chi_{yyz}+\chi_{zyy}$. Extra patterns with bright or dark areas are observed in the single antiferromagnetic domain region, indicating the existence of other ferroic domain structures except the antiferromagnetic domain. Moreover, a rotation of the detected SHG polarization around *x* by 90° (i.e. transform from $\chi_{yyz}+\chi_{zyy}$ to $\chi_{yyz}-\chi_{zyy}$) leads to a reversal of the brightness of all regions (as shown in Fig. 67(c)). This is because the rotation changes the sign of the $\chi_{zyy}$-contribution which inverts the interference and, thus, the contrast between the $\chi_{zyy}$- and $\chi_{yyz}$-contributions. This reversal is possible only if the SHG contributions responsible for the interference stem from independent sources like the antiferromagnetic and ferrotoroidical domains. The extra domain structure was regarded as the ferrotoroidical domain, as sketched in Fig. 67(d). It is noted that there are three kinds of domains in this sample. With respect to the largest domain (antiferromagnetic,$+l$; FTO,$+\boldsymbol{T}$: labeled as "++" in figure), the red domains have ($+l$, $-\boldsymbol{T}$) and the blue domain has ($-l$, $-\boldsymbol{T}$) [348].

Although the existence of ferrotoroidic order was demonstrated by experimental identification of ferrotoroidical domains and other relevant evidences, there do exist several important and confusing questions on ferrotoroidicity. One of them is the exact microscopic definition of the ferrotoroidic moment, as done for ferroelectric polarization and magnetization. While it was claimed that the "toroidization" represents the toroidal moment per unit cell volume, the periodic boundary condition in the bulk periodic case leads to a multivaluedness of the toroidizaiton and only the toroidizaiton differences are observable quantities [349]. Based on the concept of Berry phase, it was presented that a geometric characterization of the ferrotoroidic moment, in terms of a set of Abelian Berry phases, provides a computational method to measure the ferrotoroidic moment [350]. So far, no well accepted exact definition of the ferrotoroidic moment has been proposed.

# 6. Potential applications

Multiferroics, or ferrotoroidics, simultaneously exhibit ferroelectricity and magnetism and provide alternative ways to encode and store data using both electric polarization and magnetization. Even more exciting is the mutual control between the electric polarization and magnetization due to the strong magnetoelectric coupling between them in multiferroics. Consequently, huge potential applications in sensor industry, spintronics and so on, are stimulated and then expected.

## 6.1. *Magnetic field sensors using multiferroics*

The easiest, and most direct, application of multiferroics is to utilize the sensitivity of electric polarization (voltage) to an external magnetic field, for developing a magnetic field sensor, as shown in Fig. 68(a). And a prototype read head using multiferroic materials is shown in Fig. 68(b). Even more attractive is the reversed process of order parameter (i.e. the control of magnetization by external electric field or electric polarization.) For example, Multiferroics can provide a novel means for modulating the phase and amplitude of millimeter wavelength signals passing through a fin-line waveguide. The fin-line is a rectangular waveguide loaded with a slab of dielectric material at the center of the waveguide. Conventional means for magnetic parameters control implies cumbersome electromagnets. Magnetoelectric materials provide the possibility of tuning magnetic parameters by voltage. Applying a voltage across the slab results in a shift in the absorption line for the multiferroic material thus allowing to modulate phase and amplitude of the propagating wave with the electric field.

Unfortunately, magnetization switching by electric field/polarization seems to be very difficult or insignificant. On the other hand, almost all of present multiferroic materials are antiferromagnetic and exhibit a small macroscopic magnetic moment. So it is challenging to detect the tiny influence of an external electric field on magnetization and the change of electric polarization directly.

Given the fact that a ferromagnetic layer can be pinned by its antiferromagnetic neighbor, and most multiferroics are antiferromagnetic, it is possible to utilize this pinning effect to monitor magnetization switching by electric field/polarization [20]. To do so, a soft ferromagnetic layer can be deposited on an antiferromagnetic multiferroic film, as shown in Fig. 69. Utilizing the magnetoelectric coupling of the multiferroic film, one applies an external electric field to modulate the magnetization of the multiferroic film, and eventually switch the magnetization of the soft ferromagnetic layer due to the magnetic pinning. By this way the magnetoelectric process can be realized as a read out operation of information. Following this roadmap, NiFe alloy film deposited on (0001) epitaxial $YMnO_3$ film was reported and the magnetic pinning and exchange bias in this structure was confirmed [351, 352].

## 6.2. *Electric field control of exchange bias by multiferroics*

Utilizing multiferroics to control the transport behaviors of spin-valve structures represents a promising direction towards the potential applications of multiferroicity. We first briefly present the physics of exchange bias effect associated with spin valve structure which is simplified as a bilayer structure consisting of a ferromagnetic layer in contact with an antiferromagnetic layer, and then discuss how to couple multiferroics into this structure.

There are two general manifestations of exchange interactions that have been observed on the interface between the ferromagnetic layer and an antiferromagnetic one. The first is an exchange bias of the magnetic hysteresis as a consequence of pinned uncompensated spins on the interface, which is the practical interest of conventional antiferromagnetic layer in spin-valve structures [353]. The exchange bias manifests itself by a shift of the hysteresis along the magnetic field axis for the ferromagnetic layer. The second is an enhancement of the coercivity of the ferromagnetic layer as a consequence of enhanced spin viscosity or spin drag effect.

Within a simple model on the exchange bias effect, the exchange field $H_E$ depends on the interface coupling $J_{eb}=J_{ex}S_FS_{AF}/a^2$, where $J_{ex}$ is the exchange parameter, $S_F$ and $S_{AF}$ are the moments of the interfacial spins in the ferromagnetic/antiferromagnetic layers, respectively, $a$ is the unit cell parameter of the antiferromagnetic layer. $H_E$ also depends on magnetization $M$ and thickness $t_F$ of the ferromagnetic layer, the anisotropy factor $K_{AF}$ and thickness $t_{AF}$ of the antiferromagnetic layer. These dependences can be formulated as [353]:

$$H_E = \frac{-J_{eb}}{\mu_0 M t_F}\sqrt{1-\frac{J_{eb}^2}{4K_{AF}^2 t_{AF}^2}} = H^\infty\sqrt{1-\frac{1}{4\Re^2}} \ , \tag{35}$$

where $H^\infty$ is the effective field and $\Re=K_{AF}t_{AF}/J_{eb}$ is the normalized factor. However, this model has a long standing discrepancy with experimental observation, while the random field model proposed by Malozemoff [354] and the concept of multidomain structure with the antiferromagnetic layer give relatively better consistent with experiments. In the case of $\Re\gg1$, $H_E$ is then given by [353,354]:

$$H_E = H^\infty = -\frac{\xi J_{ex}S_{AF}S_F}{\mu_0 M t_F aL} , \tag{36}$$

where $L$ is the domain size of the antiferromagnetic layer and $\xi$ the pre-factor depending on the domain shape and average number $z$ of the frustrated interaction paths for each uncompensated interfacial spins.

If we replace the antiferromagnetic layer with a multiferroic layer, such as $BiFeO_3$ which is of ferroelectric-order and antiferromagnetic-order and the antiferromagnetic domains are crossly coupled with ferroelectric domains, as discussed already in Sec.3.4.5, a multiferroic spin valve structure is developed. As discussed in Sec.3.4.5, an external electric field will drive motion and/or switching of the ferroelectric domains, and thus modulate the coupled antiferromagnetic domains. In this case, the exchange bias effect can be controlled by means of electric field instead of magnetic field in conventional spintronics. This approach thus allows a possibility to modulate/switch the magnetization of the ferromagnetic layer in the spin valve structure. As shown below, recent experiments demonstrated the applicability of this approach.

### 6.2.1. *Exchange bias in CoFeB/BiFeO₃ spin valve structure.*

**6.2.1.** *Exchange bias in CoFeB/BiFeO₃ spin valve structure.* The related experiments focused on the exchange bias effect for a ferromagnetic CoFeB layer at 300K deposited on an adjacent antiferromagnetic $BiFeO_3$ film [355, 356]. Fig. 70(a) shows the hysteresis loops of different CoFeB/$BiFeO_3$ structures and significant exchange bias was observed. Microscopically, the X-ray photoelectron emission microscopy and piezoresponce force microscopy were utilized to map the antiferromagnetic domains and ferroelectric domains of $BiFeO_3$. A linear variation of the exchange field with the inverse antiferromagnetic domain size was evaluated, excellently consistent with theoretical prediction (Eq.(36)), as shown in Fig. 70(b). Simultaneously, a fitting of the experimental data gives $\xi$=3.2 which hints the existence of uncompensated spins on the ferromagnetic/antiferromagnetic interfaces.

Regarding the magnetic moment on the interface, polarized neutron reflectivity (PNR) investigation revealed that an interface layer of $2.0\pm0.5$ nm in thickness carries a magnetic moment of $1.0\pm0.5\mu_B/f.u.$. However, within the framework of the Malozemoff model, the interfacial moment due to the pinned uncompensated spins is $m_s^{pin}=2S_{AF}/aL\sim0.32\mu_B/nm^2$, only 1% of the measured moment by polarized neutron reflectivity. This indicates that majority of the uncompensated spins on the interface are non-pinned and they can rotate with the spins in the CoFeB layer, resulting in a significantly enhanced coercivity. This means that the coercivity and magnetization of the ferromagnetic layer in the spin valve structure can be manipulated by controlling the number or density of non-pinned uncompensated spins on the interface, while the latter can be modulating the effective interfacial anisotropy or antiferromagnetic domain size of the $BiFeO_3$ layer. As pointed out above, the antiferromagnetic domain size of the $BiFeO_3$ layer depends on its polarization or electric field applied it on it [356]. That is the strategy to the electric field modulated exchange bias in CoFeB/BiFeO$_3$ spin valve structure.

**6.2.2. *Exchange bias in Py/YMnO$_3$ spin valve structure.*** Beside the CoFeB/BiFeO$_3$ spin valve structure reviewed above, a similar experiment on a $Cr_2O_3$/ferromagnetic alloy bilayer structure, in which $Cr_2O_3$ is a magnetoelectric compound rather than a multiferroic, was also reported [357]. However the detected signal was very tiny, while a significant effect observed in Pt/YMnO$_3$/Py, as shown in the inset of Fig. 71(b), was recently reported. In this structure, YMnO$_3$ is the pinning layer and Py is the soft ferromagnetic alloy [358]. Fig. 71(a) plots the magnetic hysteresis (*M-H* loops) measured under different electric fields at *T*=2K. The loop shift from the origin point indicates an exchange-bias field of ~60 Oe under zero electric field ($V_e$=0), noting that the magnetization and exchange-bias field depend on temperature. Upon an electric field applying across the YMnO$_3$ layer, the shift of the *M-H* loop gradually disappears, indicating suppression of the exchange-bias field and coercivity. At $V_e$=1.2V, the loop becomes asymmetric and narrow. Moreover, the electric field induced magnetization reversal was also realized in this structure, which is evident by the decrease of magnetization with an increasing electric field from zero until $V_e$=0.4V, at which the magnetization changes its sign (i.e. switching,), as shown Fig. 71(b). Unfortunately, this process is irreversible and no back-switching of the magnetization to the initial *M*>0 state upon decreasing of the electric field from the maximum value was observed.

The transport behavior of the Pt/YMnO$_3$/Py structure modulated by an external electric field is shown in Fig.72, where the anisotropic magnetoresistance (AMR) at 5K under various electric fields are presented with *R* the resistivity and $\theta_a$ the angle between measuring magnetic field $H_a$ and electric current $\boldsymbol{J}$ ($\theta_a$=0 corresponds to $\boldsymbol{J}\|H_a$). The increasing electric field $V_e$ results in an additional $R(\theta_a)$ minimum at ~270° because the electric field mimics the effect of the increasing temperature/magnetic field, and then reduces the uniaxial exchange-bias based energy barrier [358]. These results reveal a genuine electric-field effect on the exchange bias in $YMnO_3$/Py heterostructure and may be utilized in spintronics.

### 6.3. *Multiferroic/semiconductor structures as spin filters*

Multiferroic/semiconductor heterostructures are attractive due to some novel effects. In fact, much effort has been directed toward synthesizing and characterizing $YMnO_3$ thin films as potential gate dielectrics for semiconductor devices [359-361]. The most widely studied system is a $YMnO_3$/GaN heterostructure because $YMnO_3$ and GaN both have hexagonal symmetry [359]. So far, however, less attention has been paid to the role of the heterostructure interface. First principle calculation predicts different band offsets at the interface between antiferromagnetically ordered $YMnO_3$ and GaN for the spin-up and spin-down states. This behavior is due to the interface-induced spin splitting of the valence band. The energy barrier depends on the relative orientation of the electric polarization with respect to the polarization direction of the GaN substrate, suggesting an opportunity to create a magnetic tunnel junction in this heterostructure [362, 363].

### 6.4. *Four logical states realized in tunneling junction using multiferroics*

Ferroelectric random-access memories (FeRAMs) represent one of the typical device for ferroelectric applications in recent years, favored by 5 ns access speed and 64Mbytes memory density. The disadvantage of ferroelectric random-access memories is the destructive read and reset operation. By comparison, magnetic random access memories (MRAMs) have been lagging far behind ferroelectric random-access memories, mainly because of the slow and high-power read/write operation. Multiferroics offer a possibility to combine the advantages of ferroelectric random-access memories and magnetic random access memories in order to compete with electrically erasable programmable read-only memories (EEPROMs). Recently, Fert and his group developed a novel magnetic tunneling junction (MTJ) in which multiferroic $La_{0.1}Bi_{0.9}MnO_3$ (LBMO) was used as the insulating barrier, and ferromagnetic half-metal $La_{2/3}Sr_{1/3}MnO_3$ (LSMO) and Au were used as the bottom and top electrodes, respectively [364,365]. The structure and energy level of this new magnetic tunneling junction are sketched in Fig.73. The ferroelectricity and ferromagnetism of the as-prepared ultra thin $La_{0.1}Bi_{0.9}MnO_3$ film down to 2nm in thickness were identified. This magnetic tunneling junction exhibits normal tunneling magnetoresistance (TMR) effect (i.e. the resistance is low when the magnetization of bottom electrode $La_{2/3}Sr_{1/3}MnO_3$ is aligned with that of $La_{0.1}Bi_{0.9}MnO_3$, and higher when their magnetizations are antiparallel), as shown in Fig. 74.

In addition to the normal tunneling magnetoresistance, one may expect a modulation of resistance by the ferroelectricity of the $La_{0.1}Bi_{0.9}MnO_3$ film (i.e. the electroresistance effect.) The bias-voltage dependence on the current for two different bias sweep directions (as shown by the arrow in Fig. 75(a)) exhibit significant hysteresis (i.e. the tunneling current is smaller when the voltage is swept from +2V to -2V.) The electric field has huge effect on the tunneling magnetoresistance value, which is evident by the high resistance at a +2V voltage than rather at a -2V voltage. Consequently, it is possible to obtain four different resistance states at a low bias voltage in this tunneling magnetoresistance structure by combining the tunneling magnetoresistance and electroresistance effect, as shown in Fig. 75(d). This prototype device allows for an encoding of quaternary information by both ferromagnetic and ferroelectric order parameters, and a non-destructive reading by the resistance measurement [364]. This paves the way for novel reconfigurable logic spintronics architectures and an electrically controlled readout in quantum computing schemes using the spin filter effects [365].

## 6.5. *Negative index materials*

One other application, among many, is associated with negative index materials (NIM). Materials that simultaneously display negative permittivity and permeability, often referred to as negative index materials, have been presently receiving special attention because the interaction of such materials with electromagnetic radiations can be described by a negative index of refraction [366]. To date, experimental realization of negative index has only been gained in metamaterials composed of high frequency electrical and magnetic resonant reactive circuits that interact in the microwave band [366]. A lot of effort has also been directed to the far infrared band. Using an ideal model in which both ferromagnetic and ferroelectric resonances are available, a negative index of refraction in the THz region using a finite difference method in time-domain (FDTD) was predicted [367]. These results favor the capability of the mechanical phase in a multiferroic material to control the phase between the electric field $E$ and magnetic field $H$, and thus manipulates the direction of power propagation that identifies multiferroics as a possible source for a negative index of refraction.

## 7. Conclusion and open questions

In summary, because of the promising application potentials of magnetoelectric coupling and mutual control between two or more fundamental ferroic order parameters in data memories/storages and their significance in condensed matter and materials sciences, multiferroic and ferrotoroidical materials have attracted a large effort from physicists and material scientists. Several breakthroughs and milestones have been accomplished due to this upsurge in interest. We conclude this state-of-the-art review with a highlight of some important challenges that remain unresolved. Comprehensive approaches to them are needed in order to accelerate this active and exciting field of multiferroicity:

a) For $BiFeO_3$, one of the rare room-temperature multiferroics, the relationship between the spontaneous polarization and incommensurate cycloid spin order needs further study. Is the "lone pair" mechanism sufficient to account for the polarization in $BiFeO_3$? Can the room-temperature multiferroicity of $BiFeO_3$ provide some clues to search for novel room-temperature multiferroics? What is the physical mechanism for the strong coupling between the ferroelectric polarization and incommensurate spin order? These problems will shed light on discovery of novel room-temperature multiferroics and their practical applications.

b) The mechanism of ferroelectricity in hexagonal manganites remains unclear. For hexagonal manganites, disputes on the relatively large ferroelectric polarization are active, and the polarizations originating respectively from the electronic orbitals and lattice distortion need more clarification. How closely is the ferroelectricity in $YMnO_3$ linked with the frustrated triangular spin lattice? Moreover, the mechanism for electric field control of the magnetic phase in $HoMnO_3$ and the nature of the $Ho^{3+}$ magnetism remain confusing.

c) Although several microscopic models have been proposed to explain the ferroelectricty in spiral spin ordered multiferroics, they are far from sufficient to illustrate all of those abundant phenomena observed experimentally, in particular in the quantitative sense. The ferroelectricity in the $e_g$ systems like $LiCu_2O_2$ is still a controversial issue, and the multiferroicity associated with either the easy-plane type or easy-axis type $120^o$ spiral spin order in triangular lattices is not fully understood even in a qualitative sense.

d) Special and continuous attention has to be paid to mechanisms with which the ferroelectricity and ferromagnetism can be effectively integrated, in particular for charge-ordered multiferroics. So far no quantitative theory on the ferroelectricity in $LuFe_2O_4$ is available, and the predicted ferroelectricity in $Pr_{1-x}Ca_xMnO_3$ and $Pr(Sr_{0.1}Ca_{0.9})_2Mn_2O_7$ is waiting for direct experimental evidence. For $RMn_2O_5$, a full understanding of the ferroelectricity origin seems to be extremely challenging.

e) Ferroelectricity in antiferromagnetic $E$-phase and weak ferromagnetism induced by ferroelectricity remain to be theoretical concepts and no reliable experimental evidence is available. The antiferromagnetic $E$-phase induced ferroelectricity in orthorhombic $TbMnO_3$ or $YMnO_3$ remains unclear and needs further clarification. High quality samples of noncentrosymmtric $MnTiO_3$ and $FeTiO_3$ have not yet been available even by high pressure synthesis. Appealing for high quality materials for experimental and theoretical investigations is appreciated.

f) Complex elementary excitations in multiferroic materials have yet to be explored. New elemental excitations — electromagnons — are expected and have been confirmed by preliminary experiments. However, a comprehensive understanding of their origins, conceptual pictures and dynamics, is still lacking. So far no practical prediction of these element excitations, in terms of their potential applications, has been given.

g) Although quite a number of multiferroics have been synthesized and characterized, almost all of them exhibit either small/net spontaneous magnetization or electric polarization. The observed electric polarization in multiferroic manganites is nearly two orders of magnitude less than typical ferroelectrics, which is too small to be practically applicable. The magnetic order state in multiferroic and ferrotoroidical systems usually is antiferromagnetic-type. Moreover, the temperature for the coexistence of ferroelectricity and magnetism and, thus, the mutual control between them, remains very low, although recent work revealed that CuO seems to be a multiferroic with the ferroelectric Curie point as high as 220K. These issues essentially hinder multiferroics from practical applications at room temperature.

h) Basically, the magnetoelectric coupling and mutual control between ferroelectricity (polarization) and magnetism (magnetization) for most multiferroics remain weak. Although the mutual control has been identified in some multiferroic systems, few of them show the reversal of polarization upon a magnetic field reversal, which is very useful in practical applications. Moreover, the inverse process (i.e. the magnetization switching driven by electric field/polarization) seems to be difficult either. The major challenge is to search for novel materials and mechanisms to realize the effective mutual control between these ferroic order parameters.

i) Owing to the advanced techniques for materials synthesis and fabrication, the objects of modern condensed matter physics and material sciences have been extended to artificial structures, such as nanoscale quantum dots/wires/wells and superlattices, etc. The domain/interface engineering has been in rapid development. Novel multiferroics stemming from new mechanisms for the magnetoelectric coupling/mutual control between these ferroic order parameters can be fabricated upon artificial design. The physics and novel giant effect associated with these new artificial structures given the coexistence of two or more ferroic orders, can be very promising in future investigations.

j) Our understanding of ferrotoroidical systems is still quite preliminary. Up until now, there

has been no unified and clear definition of the macroscopic toroidal moment in ferrotoroidical systems. The relationship between ferrotoroidicity and multiferroicity remains unclear and should be clarified in future.

k) Practical applications of multiferroic and ferrotoroidical materials seem to be challenging, although some possible prototype devices, in storages, sensors, spintronics and other fields, have been proposed. Not only the mutual control between the ferroic order parameters but also some additional effects (e.g., the control of exchange bias by electric field) deserve extensive exploration in future.

**Acknowledgement**

The invaluable support from Professors N. B. Ming and D. Y. Xing in Nanjing University is gratefully acknowledged. We appreciate the stimulating discussions with Dr. C. W. Nan, Dr. X. G. Li, and Dr. X. M. Chen. This work is supported by the National Natural Science Foundation of China (50832002, 50601013), the National Key Projects for Basic Researches of China (2006CB921802, 2009CB623303, 2009CB929501), and the 111 Project of MOE of China (B07026).

**Figure Captions**

Fig.1. Sketches of ferroelectricity and ferromagnetism integration as well as the mutual control between them in multiferroics. Favored multiferroics would offer not only excellent ferroelectric polarization and ferromagnetic magnetization (polarization-electric field hysteresis and magnetization-magnetic field hysteresis) but also high quality polarization-magnetic field hysteresis and magnetization-electric field hysteresis (Reprinted with permission from Ref.[14], Elsevier, Copyright (2007)).

Fig.2. Relationship between ferroelectricity (polarization $P$ and electric field $E$), magnetism (magnetization $M$ and magnetic field $H$), and ferroelasticity (strain $\varepsilon$ and stress $\sigma$): their coupling and mutual control in solid or condensed matters represent the cores of multiferroicity (Reprinted with permission from Ref.[16], AAAS, Copyright (2005)).

Fig.3. Lattice structures of high temperature paraelectric phase (left) and low temperature ferroelectric phase (right) of perovskite $BaTiO_3$. In the ferroelectric phase, the B-site Ti ions shift from the centro-symmetric positions, generating a net polarization and thus ferroelectricity.

Fig.4. (a) Orbital configuration of O-TM-O chain unit (TM is the transitional metal ion) in perovskite $ABO_3$ cell and (b) the corresponding energy levels. The B-site TM ions with $d^0$ configuration tend to move toward one of the neighboring oxygen anions to form a covalent bond.

Fig.5. (a) Ferroelectric $P$-$E$ loops of $Pb(Fe_{0.5}Nb_{0.5})O_3$ thin films (Reprinted with permission from Ref.[45], American Institute of Physics, Copyright (2007)), (b) ferromagnetic $M$-$H$ loop of $Pb(Fe_{0.5}Nb_{0.5})O_3$ single crystal at $T$=3 K, (c) dielectric constant as a function of temperature for $Pb(Fe_{0.5}Nb_{0.5})O_3$ single crystal, measured at a frequency of $10^4$Hz, (d) roughly linear

behavior between dielectric variation $\delta\varepsilon$ and squared magnetization $M^2$ between $T$=130 K and 143 K (see text for details) (Reprinted with permission from [47], http://link.aps.org/doi/10.1103/PhysRevB.70.132101, American Physical Society, Copyright (2004)).

Fig.6. A summary of experimental results on BiMnO$_3$. (a) X-ray $\theta$-$2\theta$ diffraction spectra at various temperatures; (b)~(d) lattice parameters, thermal analysis YG and DTA, and resistivity as a function of temperature, respectively; (e) Magnetic $M$-$H$ hysteresis and (f) hysteresis of magnetodielectric effect $\Delta\varepsilon(\mu_0H)/\varepsilon(0)$ against magnetic field at various temperatures (Reprinted with permission from Ref.[46], http://link.aps.org/doi/10.1103/PhysRevB.67.180401, American Physical Society, Copyright (2003)).

Fig.7. (a) Valence electron localization functions projected onto the Bi-O and Mn-O planes for cubic BiMnO$_3$ (left column) and cubic LaMnO$_3$ (right column). (b) Valence electron localization functions for monoclinic BiMnO$_3$. The blue end of the scale bar corresponds to no electron localization while the white end to a complete localization (Reprinted with permission from Ref.[51], American Chemical Society, Copyright (2001)).

Fig.8. (a) Lattice structure of BiFeO$_3$: Bi ion shifting along the [111] direction and the distorted FeO$_6$ octahedra surrounding the [111] axis. Polarization $P$ points along the [111] direction, indicated by the arrow. (b) Measured $P$-$E$ loop for BiFeO$_3$ single crystal (Reprinted with permission from Ref.[63], American Institute of Physics, Copyright (2007)). (c) and (d) Spin configuration of BiFeO$_3$. The spiral spin propagation wave vector $q$ is along the [10$\bar{1}$] direction and the polarization is along the [111] direction. These two directions define the ($\bar{1}$2$\bar{1}$) cycloidal plane on which the spin rotation proceeds, as shown by the shaded region in (c) and (d) (Reprinted with permission from Ref.[207], http://link.aps.org/doi/10.1103/PhysRevLett.101.037601, American Physical Society,

Fig.9. Measured magnetic *M-H* hysteresis loops of BiFeO$_3$ nanoparticles with different sizes at *T*=300 K. Open circles denote the bulk sample. Solid circles open up triangles, open rectangles, and solid down-triangles denote the samples with grain sizes of 4 nm, 15 nm, 25 nm, and 40 nm, respectively. The inset shows the saturated magnetization $M_s$ (open circles) and the difference ($\Delta M$, solid circles) between $M_s$ of the nanoparticles and the bulk samples. (Reprinted with permission from Ref.[82], American Institute of Physics, Copyright (2007)).

Fig.10. (a) Lattice structure of ferroelectric YMnO$_3$, with the arrows indicating the direction of ion shift from the centrosymmetry positions (Reprinted with permission from Ref.[106], Macmillan Publishers Ltd: Nature Materials, Copyright (2004)). (b) Electronic configuration of Mn ions in the MnO$_5$ pyramid of YMnO$_3$.

Fig.11. Crystal structures of (a) paraelectric phase and (b) ferroelectric phase of YMnO$_3$. The spheres and pyramids represent Y ions and MnO$_5$ pyramids respectively. The arrows indicate the direction of ion shift from the centrosymmetry positions, and the numbers are the bond length. (Reprinted with permission from Ref.[106], Macmillan Publishers Ltd: Nature Materials, Copyright (2004)).

Fig.12. Coupled magnetic and ferroelectric domain structures observed in YMnO$_3$. YMnO$_3$ has four types of 180° domains denoted by (+*P*, +*l*), (+*P*, -*l*), (-*P*, -*l*) and (-*P*, +*l*), respectively, where ±*P* and ±*l* are the independent components of the ferroelectric and AFM order parameters (Reprinted with permission from Ref.[113], Macmillan Publishers Ltd: Nature, Copyright (2002)).

Fig.13. (a) Spin configurations and lattice symmetry of HoMnO$_3$ in different temperature ranges with and without electric field. The red arrows represent the Ho spins and the yellow arrows for the Mn spins (See text for details and reprinted with permission from Ref.[119],

Macmillan Publishers Ltd: Nature, Copyright (2004)). (b) Dielectric constant as a function of temperature for HoMnO$_3$, indicating three anomalies. (c) Dielectric constant as a function of temperature for HoMnO$_3$ under different magnetic fields. (Reprinted with permission from Ref.[116], http://link.aps.org/doi/10.1103/PhysRevLett.92.087204, American Physical Society, Copyright (2004)).

Fig.14. (a) Sinusoidal (upper) and spiral (lower) spin order for a one-dimensional spin chain with competing exchange interactions. (b) Geometric spin frustration in a two-dimensional triangular lattice. The DM interaction and as-generated polarization (indicated by open read arrows) in La$_2$CuO$_4$ and $R$MnO$_3$ are illustrated in (c) and (d) (Reprinted with permission from Ref.[19], Macmillan Publishers Ltd: Nature Materials, Copyright (2007)).

Fig.15. A schematic illustration of the competing interactions involved in the Hamiltonian proposed for multiferroic manganites by Sergienko and Dagotto [135]. The middle part explains the double exchange and super-exchange interactions among the Mn 3d orbitals. The lower part shows the phonon modes of oxygen ions, which are coupled to the $t_{2g}$ electrons of Mn ions by the DM interaction. The upper part figures the modes of the Jahn-Teller distortion.

Fig.16. Monte Carlo simulation of multiferroicity based on the Sergienko-Dagotto model Hamiltonian. (a) Simultaneous ferroelectric and magnetic transitions characterized by polarization $P$ and AFM structural factor $S(\pi/2, \pi/2)$. (b) Spin configuration of the spiral ordered state and oxygen ion displacement (ferroelectric polarization). The arrows indicate the direction of the Mn spins and the filled circles represent the oxygen ions (Reprinted with permission from Ref.[135], http://link.aps.org/doi/10.1103/PhysRevB.73.094434, American Physical Society, Copyright (2006)).

Fig.17. (a) Crystal structure of LiCu$_2$O$_2$ and (b) its spin ordered configuration with multifold exchange interactions. The blue lines indicate the quais-1D spin ladders consisting of Cu ions.

The red spheres represent the oxygen ions and gray spheres denote the Li ions. Spiral arrangements of the Cu spin ladders and corresponding polarization under zero magnetic field (c) and 9.0 Tesla (d) applied along the *b*-axis (Reprinted with permission from Ref.[142], http://link.aps.org/doi/10.1103/PhysRevLett.98.057601, American Physical Society, Copyright (2007)).

Fig.18. Measured physical properties of $LiCu_2O_2$ as a function of temperature: (a) magnetic susceptibility along the *b*-axis and its temperature derivative, (b) dielectric constant along the *c*-axis, (c) polarization along the *c*-axis and that along the *a*-axis (d) under various magnetic fields as numbered (Tesla) (Reprinted with permission from Ref.[142], http://link.aps.org/doi/10.1103/PhysRevLett.98.057601, American Physical Society, Copyright (2007)).

Fig.19. (a) Lattice structure and three spin arrangements in Ni sublattice for $Ni_3V_2O_8$. The LTI (low temperature insulator) phase exhibits a spin spiral structure which can induce ferroelectric polarization *P* along the *b*-axis, while the spins in the HTI (high temperature insulator) and CAF (canted antiferromagnetic) phases are collinear. (b) Phase diagram of magnetic field against temperature for $Ni_3V_2O_8$ under magnetic field along the *a*-axis and *c*-axis respectively. (c) Polarization along the *b*-axis as a function of temperature and magnetic field applied along the *a*-axis and *c*-axis, respectively (Reprinted with permission from Ref.[151], http://link.aps.org/doi/10.1103/PhysRevLett.95.087205, American Physical Society, Copyright (2005)).

Fig.20. Crystal lattice structure (a) and $120^{o}$ spin ordered state (b) in triangular-lattice $RbFe(MoO_4)_2$. In (a) the spin interactions are denoted by $J_1'$, $J'$, $J$, and $J_2'$ (Reprinted with permission from Ref.[153], http://link.aps.org/doi/10.1103/PhysRevLett.98.267205, American Physical Society, Copyright (2007)).

Fig.21. Crystal lattice structures of $CuCrO_2$ with delafossite structure (a) and $(Li/Na)CrO_2$ with ordered rock salt structure (b) (Reprinted with permission from Ref.[155], http://link.aps.org/doi/10.1103/PhysRevLett.101.067204, American Physical Society, Copyright (2008)).

Fig.22. (a) Symmetry elements in $CuCrO_2$ with space group $R\bar{3}m$: two-fold rotation axis $2$, reflection mirror $m$, and three-fold rotation axis along the $c$-axis with inversion center. (b) Symmetry elements (left) and a schematic figure (right) of the $120^o$ spin ordered structure with (110) spiral plane (Reprinted with permission from Ref.[155], http://link.aps.org/doi/10.1103/PhysRevLett.101.067204, American Physical Society, Copyright (2008)).

Fig.23. A summary of experimental results on $CuFeO_2$: (a) phase diagram of magnetic field against temperature (the inset in the top right corner is the crystal structure), (b) ac magnetic susceptibility (the inset is the dimensional dilation), and (c) dielectric constant measured in parallel and perpendicular to the $c$-axis as a function of temperature, respectively, (d) polarization perpendicular to the $c$-axis as a function of magnetic field at several temperatures (from upper to bottom: $T$=2 K, 7 K, 9 K, 10 K, 11 K) (Reprinted with permission from Ref.[156], http://link.aps.org/doi/10.1103/PhysRevB.75.100403, American Physical Society, Copyright (2007)).

Fig.24. (a) Spin configuration of the incommensurate collinear sinusoidal spin ordered state at $T$=35 K (upper) and the spiral spin ordered state at 15 K (middle: the $bc$-plane, lower: the three-dimensional) in $TbMnO_3$. The measured magnetization and specific heat, modulation wave-number, dielectric constant and polarization along the $a$-axis, $b$-axis, and $c$-axis, respectively, are shown in (b), (c), (d), and (e) (Reprinted with permission from Ref.[162], Macmillan Publishers Ltd: Nature, Copyright (2003)).

Fig.25. Dielectric constants (a) and (b) and polarizations (c) and (d) along the *c*-axis and *a*-axis as functions of temperature under different magnetic fields for TbMnO$_3$ (Reprinted with permission from Ref.[162], Macmillan Publishers Ltd: Nature, Copyright (2003)).

Fig.26. (a) Various multiferroic domain walls conceivable in DyMnO$_3$. (b) Calculated domain wall structure between the $P||+c$ and $P||+a$ domains. Blue and red arrows represent the Mn spins and local polarizations, respectively. The color gradation represents the angle of local polarization relative to the *a*-axis (Reprinted with permission from Ref.[181], http://link.aps.org/doi/10.1103/PhysRevLett.102.057604, American Physical Society, Copyright (2009)).

Fig.27. Magnetization (a)~(c) and magnetic field induced changes of polarization along the *c*-axis (d)~(f) for TbMnO$_3$, as a function of external magnetic field along the *a*-axis, *b*-axis, and *c*-axis, respectively, at various temperatures. The inset of (a) shows a magnified view of the high field region (Reprinted with permission from Ref.[169], http://link.aps.org/doi/10.1103/PhysRevB.71.224425, American Physical Society, Copyright (2005)).

Fig.28. Phase diagram of temperature against Mn-O-Mn bonding angle $\phi$ (corresponding to different rare earth ionic radius) for manganites *R*MnO$_3$. The inset shows the wave numbers of spiral spin order for these manganites (Reprinted with permission from Ref.[163], http://link.aps.org/doi/10.1103/PhysRevLett.92.257201, American Physical Society, Copyright (2004)).

Fig.29. Physical properties of CoCr$_2$O$_4$: (a) spin configuration and polarization of the conical spin ordered state, and (b) crystal lattice, electronic and spin structures. The measured hysteresis loops of magnetization and polarization against external magnetic field at two temperatures are shown in (c) and (d). (e) Switching (reversal) of polarization induced by time-dependent magnetic field. The upper part of (e) illustrates the spiral spin and polarization

structures        (Reprinted        with        permission        from        Ref.[197],
http://link.aps.org/doi/10.1103/PhysRevLett.96.207204, American Physical Society,
Copyright (2006)).

Fig.30. (a) Temperature ($T$)-dependence of electric polarization ($\boldsymbol{P}$) along the $[\bar{1}\,10]$
direction and magnetization $\boldsymbol{M}$ along the [001] direction in $CoCr_2O_4$ below 30K. $\boldsymbol{P}$ suddenly
switches sign when cooling across 14 K without changing signs of $\boldsymbol{M}$ and $\boldsymbol{Q}$. (b) and (c)
$H$-dependence of $\boldsymbol{M}$ and $\boldsymbol{P}$ at 20 K and 10 K, respectively (Reprinted with permission from
Ref.[201], http://link.aps.org/doi/10.1103/PhysRevLett.102.067601, American Physical
Society, Copyright (2009))

Fig.31. (a) Relationship between the principal superexchange interaction $J$ and the Cu-O-Cu
bond angle $\varphi$ in low-dimensional cuprates. (b) Schematic drawing of the commensurate
collinear (AF1) and incommensurate noncollinear (AF2) antiferromagnetic spin orders in
CuO. (c) Measured polarization as a function of temperature in CuO (Reprinted with
permission from Ref.[202], Macmillan Publishers Ltd: Nature Materials, Copyright (2008)).

Fig.32. Measured polarization of $Ba_2Mg_2Fe_{12}O_{22}$ as a function of the rotation angle $\phi$ of
magnetic field with respect to the [001] direction and the rotation angle $\theta$ of magnetic field
with respect to the [120] direction. The magnetic field rotates horizontally (A-D) and
vertically (E-H) in the shaded planes shown in A and D (Reprinted with permission from
Ref.[205], AAAS, Copyright (2008)).

Fig.33. Reversal of electric polarization $P_c$ and spiral spin order induced by external electric
field along the $c$-axis in $TbMnO_3$. (a) Scattering intensities and (b) spin configurations of the
spin order states with different polarization $P_c$. See details in text (Reprinted with permission
from Ref.[180], http://link.aps.org/doi/10.1103/PhysRevLett.98.147204, American Physical
Society, Copyright (2007)).

Fig.34. Four equivalent electric polarization directions of $BiFeO_3$ crystal. The numbers in each figure indicate the reversal angles relative to the polarization along the [111] direction. The shaded planes represent the AFM plane perpendicular to the spiral spin planes (Reprinted with permission from Ref.[208], Macmillan Publishers Ltd: Nat. Mater., Copyright (2006)).

Fig.35. PEEM images of $BiFeO_3$ film before (a) and after (b) electric poling as indicated by arrows; and IPPFM images of the same area before (c) and after (d) the electric poling, noting the 109°-ferroelectric domain switching (regions 1 and 2) and 180°- and 71°-domain switching (regions 3 and 4). (e) PFM image of the same area with polarization labeled (Reprinted with permission from Ref.[208], Macmillan Publishers Ltd: Nat. Mater., Copyright (2006)).

Fig.36. (a) Neutron scattering intensity in the adjacent $P_{111}$ (lower half) and $P_{1-11}$ (upper half) domains in $BiFeO_3$ single crystal. (b) Schematic drawing of the planes of spin rotations and cycloids $k_1$ vector for the two polarization domains with the domain wall (light gray plane) (Reprinted with permission from Ref.[206], http://link.aps.org/doi/10.1103/PhysRevLett.100.227602, American Physical Society, Copyright (2008)).

Fig.37. (a) In-plane PFM image showing the ferroelectric domain structure of $BiFeO_3$ with a large (10μm, red-line square) and small (5 μm, green-line square) electrically switched region. (b) Corresponding XMCD-PEEM image for the CoFe film grown on the electrically written $BiFeO_3$ film. (c) Schematic drawings of the two adjacent domains in [001]-oriented BFO. (Reprinted with permission from Ref.[213], Macmillan Publishers Ltd: Nature Materials, Copyright (2008))

Fig.38. A possible multiferroic random access memory element using antiferromagnetic multiferroic materials (Reprinted with permission from Ref.[214], Macmillan Publishers Ltd: Nature Materials, Copyright (2008)).

Fig.39. Charge/orbital ordered structure of $Pr_{0.5}Ca_{0.5}MnO_3$ at low temperature (Reprinted with permission from Ref.[219], AAAS, Copyright (2000)).

Fig.40. Atomic configuration of charge order state of $LuFe_2O_4$ on the *ab* plane (a) and in three-dimensional space (c). The red arrow in (c) indicates the direction of polarization. (b) Transmission electron diffraction pattern of $LuFe_2O_4$ along the $[1\,\overline{1}\,0]$ direction at *T*=20 K. ((a) and (b) were reprinted with permission from Ref.[222], http://link.aps.org/doi/10.1103/PhysRevLett.98.247602, American Physical Society, Copyright (2007), and (c) was reprinted with permission from Ref.[19], Macmillan Publishers Ltd: Nature Materials, Copyright (2007)). (d) Electric polarization of $LuFe_2O_4$ as a function of temperature under two different cooling field modes (Reprinted with permission from Ref.[220], Macmillan Publishers Ltd: Nature, Copyright (2005)).

Fig.41. Synchrotron X-ray oscillation photographs (upper row), and schematic charge/orbital ordered configurations (middle row) as well as lattice structures (lower row) of $Pr(Sr_{0.1}Ca_{0.9})_2Mn_2O_7$ at three different temperatures (Reprinted with permission from Ref.[230], Macmillan Publishers Ltd: Nature Materials, Copyright (2006)).

Fig.42. Site-centered (a) and bond-centered (b) charge/orbital ordered phases as well as the superposition of the two ordered phases (c) for mixed-valence manganites. The green circles represent the Mn ions, the blue circles for the rare earth ions, and the red circles for the oxygen ions. The arrow indicates the direction of polarization *P*. (d) Predicted phase diagram of $Pr_{1-x}Ca_xMnO_3$. Abbreviations FM, C, CE, and A represent the ferromagnetic, C-type, CE-type and A-type antiferromagnetic phases, respectively. The yellow region is predicted to exhibit ferroelectricity (Reprinted with permission from Ref.[232], Macmillan Publishers Ltd: Nature Materials, Copyright (2004))

Fig.43. Structure and polarization of charge-ordered $Fe_3O_4$. In the *xy* chains the $Fe^{2+}$ and $Fe^{3+}$

ions (filled and open circles) align alternatively, and simultaneously there is an alternation of short and long Fe-Fe bonds (the black arrows indicate the direction of Fe ion shift and red arrows indicate the direction of polarization).

Fig.44. (a) One dimensional chain with alternating charges (charge-ordered state) and up-up-down-down spin structure. (b) Magnetostriction effect, which shortens the ferromagnetic bonds and generates a ferroelectric polarization.

Fig.45. (a) Crystal structures of TbMn$_2$O$_5$ on the *ab* plane (left) and *a(b)-c* plane (right). Five types of magnetic exchange interactions are denoted by $J_1$, $J_2$, $J_3$, $J_4$, and $J_5$, respectively. (b) Spin (solid arrows) configuration and crystal distortion (electric polarization, open arrows) of TbMn$_2$O$_5$ (Reprinted with permission from Ref.[249], http://link.aps.org/doi/10.1103/PhysRevB.71.214402, American Physical Society, Copyright (2005)).

Fig.46. (a) Temperature dependence of magnetic susceptibilities and dielectric constants along the *a*-axis*, b*-axis*,* and *c*-axis, respectively, as well as specific heat for TbMn$_2$O$_5$. (b) Dielectric constant along the *b*-axis as a direction of magnetic field along the *a*-axis. (c) and (d) Polarization along the *b*-axis as a function of magnetic field along the *a*-axis (Reprinted with permission from Ref.[248], Macmillan Publishers Ltd: Nature, Copyright (2004)).

Fig.47. Magnetic structure of the low-temperature incommensurate ferroelectric phase of YMn$_2$O$_5$. (a), (b) and (c) represent the *ab*, *ac, bc* planes, respectively. (d) Polarization induced by the magnetic striction mechanism in the *ab* plane (Reprinted with permission from Ref. [282], http://link.aps.org/doi/10.1103/PhysRevB.78.245115, American Physical Society, Copyright (2008)).

Fig.48. (a) Spin structures of two AFM *E*-phases in perovskite HoMnO$_3$. The arrows on the Mn atoms (blue spheres) denote the directions of their spins, and the direction of polarization

is signed by the black arrows. The red spheres denote the O atoms. (Reprinted with permission from Ref.[283], http://link.aps.org/doi/10.1103/PhysRevLett.97.227204, American Physical Society, Copyright (2006)). (b) In-plane ferroelectric configuration of AFM $E$-phase ($E_I$) HoMnO$_3$. The small red spheres denote the O atoms. The bigger spheres represent the Mn atoms, and the regions shaded by blue and pink color denote the AFM coupled spin zigzag chains. The green and yellow arrows represent the directions of ionic displacement of Mn (left) and O (right) respectively, and the resultant polarization is denoted by the thick arrow at the bottom. (c) The $ac$-plane charge density isosurface plot in the energy region between −8eV and 0 eV (0eV is the top of the valent band) for the relaxed structure of AFM $E$-phase ($E_I$) HoMnO$_3$ by first principle calculations. (Reprinted with permission from Ref.[284], http://link.aps.org/doi/10.1103/PhysRevLett.99.227201, American Physical Society, Copyright (2007)).

Fig.49. Monte Carlo simulation of the AFM $E$-phase induced polarization. (a) Starting configuration of a Mn-O-Mn bond. (b) A Monte Carlo snapshot of the $E$-phase at $T$=0.001. The arrows on the Mn ions denote their spin and the ferromagnetic zigzag chains are shown by solid red lines. (c) Local arrangement of the Mn-O-Mn bonds with (left) disordered Mn spins and (right) opposite Mn spin chains. The arrows indicate the oxygen displacements, open and cross circles denotes the direction of Mn spins. (d) Dependence of polarization on the starting Mn-O-Mn angle $\phi_0$ (Reprinted with permission from Ref.[283], http://link.aps.org/doi/10.1103/PhysRevLett.97.227204, American Physical Society, Copyright (2006)).

Fig.50. (a) Collinear magnetic structure of BaNiF$_4$ extracted from experimental observations. (b) Canting spin ordered structure, i.e., weak magnetic order obtained from first-principle calculations including the spin-orbit coupling. (c) Reversal of polarization in (b) leads to a reversal of the canted magnetic moments and thus to a reversal of vector $L_c$ (Reprinted with permission from Ref.[289], http://link.aps.org/doi/10.1103/PhysRevB.74.020401, American Physical Society, Copyright (2006)).

Fig.51. (a) Crystal structure and symmetry elements of paraelectric $BiFeO_3$ with space group $R\overline{3}c$. (b) Spin structure and symmetry elements of $BiFeO_3$. (c) Spin structure and symmetry elements of $FeTiO_3$ (Reprinted with permission from Ref.[288], http://link.aps.org/doi/10.1103/PhysRevLett.100.167203, American Physical Society, Copyright (2008)).

Fig.52. Chiral nature of the $S_1$-O-$S_2$ bonds of $FeTiO_3$ in the ferroelectric phase with polarization $P$ up (left), polarization $P$ down (right), and in the paraelectric phase (middle) (Reprinted with permission from Ref.[288], http://link.aps.org/doi/10.1103/PhysRevLett.100.167203, American Physical Society, Copyright (2008)).

Fig.53. Magnetic structures of $DyFeO_3$ below $T_N^{Fe}$ under magnetic field (along the $c$-axis) $H<H_r^{Fe}$ ((a) and (b)) and $H>H_r^{Fe}$ ((c) and (d)), respectively. In (b) and (d), the magnetic structures of Dy ions are different from (a) and (c), and then a reversed polarization appears in (d). (Reprinted with permission from Ref.[296], http://link.aps.org/doi/10.1103/PhysRevLett.101.097205, American Physical Society, Copyright (2008))

Fig.54. (a) Temperature dependence of polarization of $DyFeO_3$ along the $a$-axis, $b$-axis, $c$-axis under magnetic field of 30 kOe ($>H_r^{Fe}$) along the $c$-axis. Dotted line shows the polarization along the $c$-axis under a magnetic field of 500 Oe ($<H_r^{Fe}$). (b) Magnetic field (along the $c$-axis) dependence of the residual polarization obtained by $P$-$E$ loops (filled circles) and the displacement current measurement (solid line) at $T$=3K. The dashed line is the extrapolated polarization curve in the regions of $H>H_r^{Fe}$ towards $H$=0. (c) and (d) Magnetic field (along the $c$-axis) dependence of the $P$-$E$ loops measured under $H||c$ and $E||c$ configurations with different frequency by a Sawyer-Tower bridge (Reprinted with permission from Ref.[296], http://link.aps.org/doi/10.1103/PhysRevLett.101.097205, American Physical Society, Copyright (2008))

Fig.55. Difference between the spin densities for a 15 monolayer-thick Fe film with and without external electric field (*E*) of 1V/Å, i.e. *Δσ=σ(E)- σ(0)*. (Reprinted with permission from Ref.[309], http://link.aps.org/doi/10.1103/PhysRevLett.101.137201, American Physical Society, Copyright (2008))

Fig.56. Dielectric spectra of $GdMnO_3$ and $TbMnO_3$ at different temperatures under various combinations of electric and magnetic fields (Reprinted with permission from Ref.[319], Macmillan Publishers Ltd: Nature Physics, Copyright (2006)).

Fig.57. Dispersion relations of spin wave excitations in paraelectric (a) and ferroelectric (b) phases of $TbMnO_3$, respectively. The dash lines are the dispersion relations of $LaMnO_3$ for comparison. (c) Spectra of element excitations in paraelectric and ferroelectric phases of $TbMnO_3$. (d) Three magnons in the ferroelectric spiral spin order phase of $TbMnO_3$ (Reprinted                         with                         permission                         from                         Ref.[320], http://link.aps.org/doi/10.1103/PhysRevLett.98.137206, American Physical Society, Copyright (2007)).

Fig.58. Far-infrared optical transmission spectra for $YMn_2O_5$ (a), and for $TbMn_2O_5$ (b) at different temperatures under various combinations of electric and magnetic fields (Reprinted with permission from Ref.[325], http://link.aps.org/doi/10.1103/PhysRevLett.98.027202, American Physical Society, Copyright (2007)).

Fig.59. (a) Schematic of the spin order and atomic positions in $YMnO_3$. The squares represent oxygen ions, and the circles indicate Mn ions. (b) Dispersion of phonons and magnons in $YMnO_3$. The dashed lines indicate the measured magnon dispersions along the $a^*$-axis in (a). Triangles and circles represent the phonon dispersions obtained at *T*=200K and 18K respectively. The squares indicate the optical phonon mode. The gap in the phonon dispersion

opens at $q_0$~0.185, and the crossing of the 200K phonon dispersion with the magnon mode arises at $q_{cross}$~0.3. (c) Nuclear dynamical structure factor calculated as a function of wave vector along ($q$, 0,6) and energy. A jump from the lower mode to the upper mode, which results in experimentally observed gap, occurs (Reprinted with permission from Ref.[330], http://link.aps.org/doi/10.1103/PhysRevLett.99.266604, American Physical Society, Copyright (2007)).

Fig.60. (a) Raman spectra of spin excitations in $BiFeO_3$. The equally and non-equally spaced modes at low frequencies correspond to the $\Psi$ and $\phi$ cycloidal modes selected out using parallel (∥) and crossed (⊥) polarizations. The inset shows the superposition of these two kinds of modes on another sample. (b) $\Psi$ (circles) and $\phi$ (squares) cycloidal mode frequency as a function of the mode index $n$, respectively (Reprinted with permission from Ref.[207], http://link.aps.org/doi/10.1103/PhysRevLett.101.037601, American Physical Society, Copyright (2008)).

Fig.61. Magnetic toroidic moment in a simple system: (a) a ring-shape torus with an even number of current windings exhibits a toroidic moment **T** (the green arrow), (b) a magnetic field along the ring plane induces the congregating of the current loops in one direction and eventually an electric polarization along this direction (Reprinted with permission from Ref.[348], Macmillan Publishers Ltd: Nature, Copyright (2007)).

Fig.62. Crystal lattice structure (a) and schematic magnetic toroidic moments (b) of $GaFeO_3$, (c) four kinds of magnetic-optical effects (Reprinted with permission from Ref.[14], Elsevier, Copyright (2007))

Fig.63. Relations between the ferroic orders and the space-/time- reversal (Reprinted with permission from Ref.[348], Macmillan Publishers Ltd: Nature, Copyright (2007))

Fig.64. Possible antiferromagnetic spin orders: (a) and (b) have equal and opposite toroidal

moments, and the antiferromagnetic arrangement in (c) also has a toroidal moment, while the arrangement shown in (d) has not.

Fig.65. Arrangements of spins of $Co^{2+}$ ions on the $yz$ plane (a) and the $xz$ plane (b) for the ground state of LiCoPO$_4$. The solid and open circles represent the Co ions at $x\sim3/4$ and $x\sim1/4$ positions respectively. The gray arrows are the spins of Co ions (Reprinted with permission from Ref.[348], Macmillan Publishers Ltd: Nature, Copyright (2007)).

Fig.66. Relations between electric polarization along the z-direction and magnetic field along the $x$-direction at different temperatures adjacent to the magnetic transition point of LiNiPO$_4$ (Reprinted with permission from Ref.[346], http://link.aps.org/doi/10.1103/PhysRevB.62.12247, American Physical Society, Copyright (2000)).

Fig.67. Images of a single antiferromagnetic domain in LiCoPO$_4$ (100) single crystal, obtained using SHG light at 10K. (a), (b) and (c) are the images by SHG light from $\chi_{zzz}$、 $\chi_{yyz}+\chi_{zyy}$ and $\chi_{yyz}-\chi_{zyy}$ at 2.25eV. (d) Three kinds of domains in this sample, and their relations to the largest domain (AFM,+$l$; FTO,+$T$: shown as "++" in the figure), the red domains have (+$l$, -$T$) and the blue domain has (-$l$, -$T$). The black patch in the centre of sample in all figures is damage defect (Reprinted with permission from Ref.[348], Macmillan Publishers Ltd: Nature, Copyright (2007)).

Fig.68. (a) Multiferroic materials as probe of magnetic field. The middle layer (the white layer) is multiferroic, and the upper and lower layers (gray layers) are ferromagnetic metals. An external magnetic field will induce the electric polarization perpendicular to the magnetic field direction, and then a voltage. (b) The read-head device using the probe in (a). The blue layer is the magnetic media (magnetic disk) and the black arrows in it indicate two opposite bits.

Fig.69. Schematic of a soft ferromagnetic layer deposited on multiferroic antiferromagnetic film. The external electric field induces a variation in magnetization of the antiferromagnetic multiferroic film, and eventually results in the reversal of magnetization in the soft ferromagnetic layer due to the magnetic pinning effects (Reprinted with permission from Ref.[20], Macmillan Publishers Ltd: Nature Materials, Copyright (2007)).

Fig.70. Exchange bias in CoFe/BiFeO$_3$ system. (a) Magnetic field dependence of magnetization of CoFeB/BiFeO$_3$/SrTiO$_3$(001) multilayer (upper left), CoFeB/BiFeO$_3$/SrTiO$_3$(111) (upper right), CoFeB/BiFeO$_3$/La$_{0.7}$Sr$_{0.3}$MnO$_3$/SrTiO$_3$(001) (lower left), and CoFeB/BiFeO$_3$/La$_{0.7}$Sr$_{0.3}$MnO$_3$/SrTiO$_3$(111) (lower right), respectively. (b) Dependence of the exchange field on the inverse of the domain size for BiFeO$_3$ films. L$_{FE}$ and L$_{AF}$ represent the sizes of the FE and AF domains. (c) Thickness dependence of exchange field for CoFeB/BiFeO$_3$ grown on SrTiO$_3$ (001) (Reprinted with permission from Ref.[355], http://link.aps.org/doi/10.1103/PhysRevLett.100.017204, American Physical Society, Copyright (2008)).

Fig.71. Measured *M-H* loops (a) and magnetization (b) of exchange biased Pt/YMnO$_3$/Py structure under different external electric field $V_e$. The inset in (a) shows the relation between magnetization and temperature, and the insert in (b) is the multilayered structure (Reprinted with permission from Ref.[358], http://link.aps.org/doi/10.1103/PhysRevLett.97.227201, American Physical Society, Copyright (2006)).

Fig.72. Anisotropic magnetoresistance (AMR) of Pt/YMnO$_3$/Py structure measured at *T*=5K under different electric fields, where $\theta_a$ is the angle between magnetic field $H_a$ and electric current *J* ($\theta_a$=0° corresponds to *J*||$H_a$) (Reprinted with permission from Ref.[358], http://link.aps.org/doi/10.1103/PhysRevLett.97.227201, American Physical Society, Copyright (2006)).

Fig.73. Structure and energy landscape of a new magnetic tunneling junction (MTJ) in which multiferroic $La_{0.1}Bi_{0.9}MnO_3$ (LBMO) was used as insulating barrier and ferromagnetic half metal $La_{2/3}Sr_{1/3}MnO_3$ (LSMO) and Au were used as the bottom and top electrodes respectively (Reprinted with permission from Ref.[364], Macmillan Publishers Ltd: Nature Materials, Copyright (2007)).

Fig.74. Measured tunneling magnetoresistance effects in $La_{2/3}Sr_{1/3}MnO_3$/ $La_{0.1}Bi_{0.9}MnO_3$/Au multilayer structure. The red curve represents the resistivity and the black one represents the magnetoresistance ratio (Reprinted with permission from Ref.[364], Macmillan Publishers Ltd: Nature Materials, Copyright (2007)).

Fig.75. (a) Influence of external electric field on the tunneling current in $La_{2/3}Sr_{1/3}MnO_3$/ $La_{0.1}Bi_{0.9}MnO_3$/Au junctions. The arrows denote the sequence for electric field application. (b) Dependence of the tunneling electroresistance effect (TER) and tunneling magnetoresistance (TMR) on external electric field $V_{dc}$. (c) Measured TMR upon an electric bias of +2V and -2V respectively. (d) Four states of resistance in the junction (Reprinted with permission from Ref.[364], Macmillan Publishers Ltd: Nature Materials, Copyright (2007)).

**Table I. A list of multiferroics excluding those multiferroics induced by spiral spin order (listed in Table II)**

| Compound | Crystal Structure (Space group) | Magnetic ions | Mechanism for multiferroics | Ferroelectric polarization | Ferroelectric transition temperature | Magnetic transition temperature | References |
|---|---|---|---|---|---|---|---|
| $RFe_3(BO_3)_4$ (R=Gd,Tb,et al) | $R32$ | $R^{3+}$, $Fe^{3+}$ | ferroelectric-active $BO_3$ group | $P_a \sim 9$ $\mu C/cm^2$ (under 40 kOe magnetic field) | $\sim 38$ K | $\sim 37$ K | [37,38] |
| $Pb(B_{1/2}B'_{1/2})O_3$ (B=Fe,Mn,Ni,Co; B'=Nb,W,Ta) | $Pm3m$ | B' | B ions induced ferroelectricity, B' ions induced magnetism | $\sim 65$ $\mu C/cm^2$ | $\sim 385$ K | $\sim 143$ K | [42-45,47] |
| $BiFeO_3$ | $R3c$ | $Fe^{3+}$ | Lone pair at A-site | $P_{[001]} \sim 75 \mu C/cm^2$ | $\sim 1103$ K | $\sim 643$K | [58-84] |
| $BiMnO_3$ | $C2$ | $Mn^{3+}$ | Lone pair at A-site | $\sim 20$ $\mu C/cm^2$ | $\sim 800$ K | $\sim 100$ K | [51-57] |
| $Bi(Fe_{0.5}Cr_{0.5})O_3$ | - * | $Cr^{3+}$ | Lone pair at A-site | $\sim 60$ $\mu C/cm^2$ | - * | - * | [90,91] |
| $(Y,Yb)MnO_3$ | Hexagonal $P6_3cm$ | $Mn^{3+}$ | Geometric ferroelectricity | $\sim 6$ $\mu C/cm^2$ | $\sim 950$K | $\sim 77$ K | [102-106] |
| $HoMnO_3$ | Hexagonal $P6_3cm$ | $Mn^{3+}$ | Geometric ferroelectricity | $\sim 5.6$ $\mu C/cm^2$ | $\sim 875$ K | $\sim 76$ K for $Mn^{3+}$ $\sim 5$ K for $Ho^{3+}$ | [116-119] |
| $InMnO_3$ | Hexagonal $P6_3cm$ | $Mn^{3+}$ | Geometric ferroelectricity | $\sim 2$ $\mu C/cm^2$ | $\sim 500$ K | $\sim 50$ K | [123,124] |
| $YCrO_3$ | Monoclinic $P21$ | $Cr^{3+}$ | Geometric ferroelectricity (?) | $\sim 2$ $\mu C/cm^2$ | $\sim 475$ K | $\sim 140$ K | [125] |
| Orthorhombic Y(Ho)MnO_3 | Orthorhombic | $Mn^{3+}$ | $E$-type antiferromagnetism | $\sim 100$ $\mu C/m^2$ | $\sim 28$ K | $\sim 28$ K | [129,130] |
| $Pr_{1-x}Ca_xMnO_3$ | $Pnma$ | $Mn^{3+}$, $Mn^{4+}$ | Site and bond centered charge-order | $\sim 4.4$ $\mu C/cm^2$ [&] | $\sim 230$ K | $\sim 230$ K for charge ordered state | [232-235] |
| $Pr(Sr_{0.1}Ca_{0.9})_2Mn_2O_7$ | *Am2m* | $Mn^{3+}$, $Mn^{4+}$ | Charge/orbital order | - * | - * | $T_{COI} \sim 370K$ $T_{CO2} \sim 315K$ | [230,231] |
| $LuFe_2O_4$ | $R\bar{3}m$ | $Fe^{2+}$, $Fe^{3+}$ | Charge frustration | $\sim 26~\mu C/cm^2$ | $\sim 330$ K | $\sim 330K$ for charge ordered state | [220-227] |
| $Ca_3Co_{2-x}Mn_xO_7$ | *R3c* | $Co^{2+}$, $Mn^{4+}$ | Charge ordered state plus magnetostriction | $\sim 90~\mu C/m^2$ | $\sim 16.5$ K | $\sim 16$ K | [242] |
| $RMn_2O_5$ (R=Y, Tb, Dy, et al) | *Pbam* | $Mn^{3+}$, $Mn^{4+}$ | Charge ordered state plus magnetostriction | $\sim 40~\mu C/cm^2$ | $\sim 38$ K | $T_N$=43 K $T_{CM}$=33 K $T_{ICM}$=24 K | [248-281] |
| $(Fe,Mn)TiO_3$ | *R3c* (high pressure phase) | $Fe^{3+}$, $Mn^{3+}$ | Polarization induced weak ferromagnetism | - * | - * | - * | [290] |
| $DyFeO_3$ | *Pbnm* | $Fe^{3+}$, $Dy^{3+}$ | Magnetostriction between adjacent antiferromagnetic Dy and Fe ions | $\sim 0.4\mu C/cm^2$ (under 90kOe magnetic field) | $\sim 3.5$ K | $T_N^{Dy} \sim 3.5$ K $T_N^{Fe} \sim 645$ K | [296] |

**\* No expreiemtal data available.**

**& Assumed from the image and data of the refined electron diffraction microscopy.**

**Table II. A list of multiferroics with spiral spin order induced ferroelectricity**

| Compound | Crystal structure | Magnetic ions | Spiral spin wave vector $q$ | Ferroelectric temperature (K) | Spontaneous polarization ($\mu C/m^2$) | References |
|---|---|---|---|---|---|---|
| $LiCu_2O_2$ | Orthorhombic ($Pnma$) | $Cu^{2+}$ | (0.5, 0.174, 0) | <23 | $P_c$=4 | [130] |
| $LiCuVO_4$ | Orthorhombic ($Pnma$) | $Cu^{2+}$ | (0, 0.53, 0) | <3 | $P_a$=20 | [134,135] |
| $Ni_3V_2O_8$ | Orthorhombic ($mmm$) | $Ni^{2+}$ | (0.28, 0, 0) | 3.9~6.3 | $P_b$=100 | [136] |
| $RbFe(MoO_4)_2$ | Triangular ($P\bar{3}m1$) | $Fe^{3+}$ | (1/3, 1/3, 0.458) | <3.8 | $P_c$=5.5 | [140] |
| $CuCrO_2$, $AgCrO_2$ | Delafossite ($R\bar{3}m$) | $Cr^{3+}$ | (1/3, 1/3, 0) | <24 | 30 [b] | [142] |
| $NaCrO_2$, $LiCrO_2$ | Ordered sock salt ($R\bar{3}m$) | $Cr^{3+}$ | (1/3, 1/3, 0) and (-2/3, 1/3, 1/2) | <60 | Antiferroelectricity | [142] |
| $CuFeO_2$ | Delafossite ($R\bar{3}m$) | $Fe^{3+}$ | ($b$, $b$, 0) $b$=0.2-0.25 | <11 | $P$=300 ($\perp$c) (H=6-13T) [a] | [143] |
| $Cu(Fe,Al/Ga)O_2$ Al/Ga=0.02 | Delafossite ($R\bar{3}m$) | $Fe^{3+}$ | ? | <7 | $P_{[110]}$=50 | [144-146] |
| $RMnO_3$ (R=Tb,Dy) | Orthorhombic ($Pbnm$) | $Mn^{3+}$ | (0, $k$, 1) $k$=0.2-0.39 | <28 | $P_c$=500 | [147-165] |
| $CoCr_2O_4$ | Cubic spinel ($m3m$) | $Cr^{3+}$ | $(b, b, 0)$ $B=0.63$ | <26 | $P_c=2$ | [181] |
| $AMSi_2O_6$ ($A$=Na,Li;$M$=Fe,Cr) | Monoclinic ($C2/c$) | $Fe^{3+}$ $Cr^{3+}$ | ? | <6 | $P_b=14$ | [174] |
| $MnWO_4$ | Monoclinic ($Pc/2$) | $Mn^{2+}$ | (-0.21, 0.5, 0.46) | 7~12.5 | $P_b=55$ | [166] |
| $CuO$ | Monoclinic ($C2/c$) | $Cu^{2+}$ | (0.506, 0, -0.843) | 213~230 | $P_b=150$ | [185] |
| $(Ba,Sr)_2Zn_2Fe_{12}O_{22}$ | Rhomboheral Y-type hexaferrite | $Fe^{3+}$ | $(0, 0, 3d)$ $0<d<1/2$ | <325 | 150 (H=1T) [a] | [183] |
| $Ba_2Mg_2Fe_{12}O_{22}$ | Rhomboheral Y-type hexaferrite | $Fe^{3+}$ | //[001] | <195 | $P_{[120]}=80$ (H=0.06-4T) [a] | [184] |
| $ZnCr_2Se_4$ | Cubic spinel | $Cr^{3+}$ | $(b, 0, 0)$ | <20 | -[a] | [179] |
| $Cr_2BeO_4$ | Orthorhombic | $Cr^{3+}$ | $(0, 0, b)$ | <28 | 3 [b] | [180] |

[a] External magnetic field is needed to induce the spiral spin order and then the ferroelectricity.
[b] Polycrystalline samples.

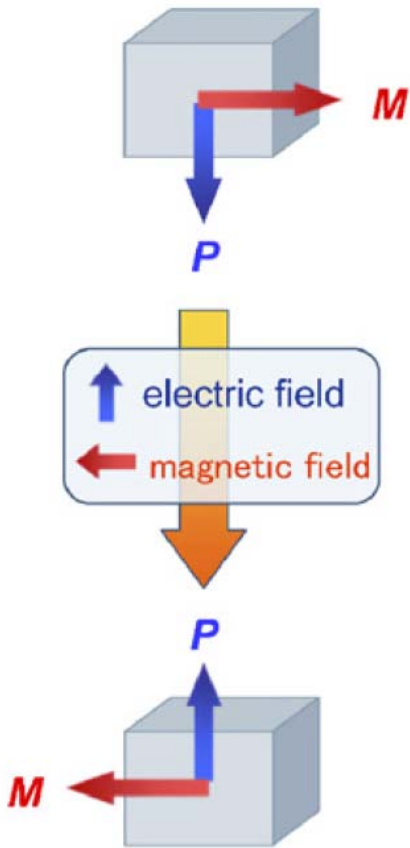

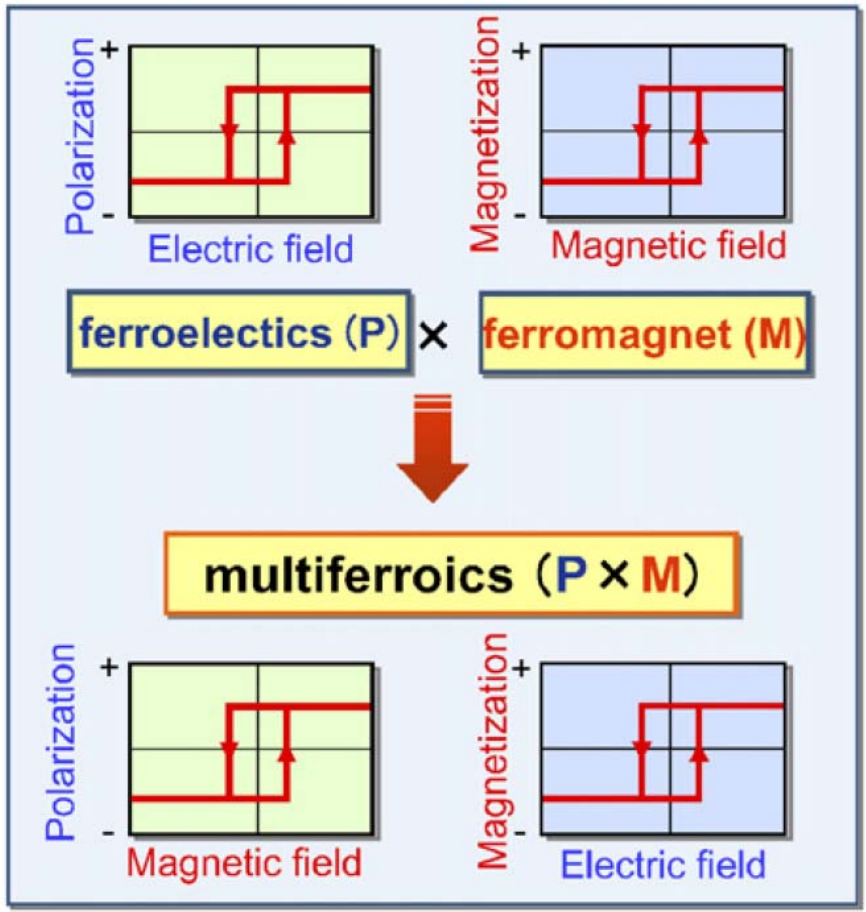

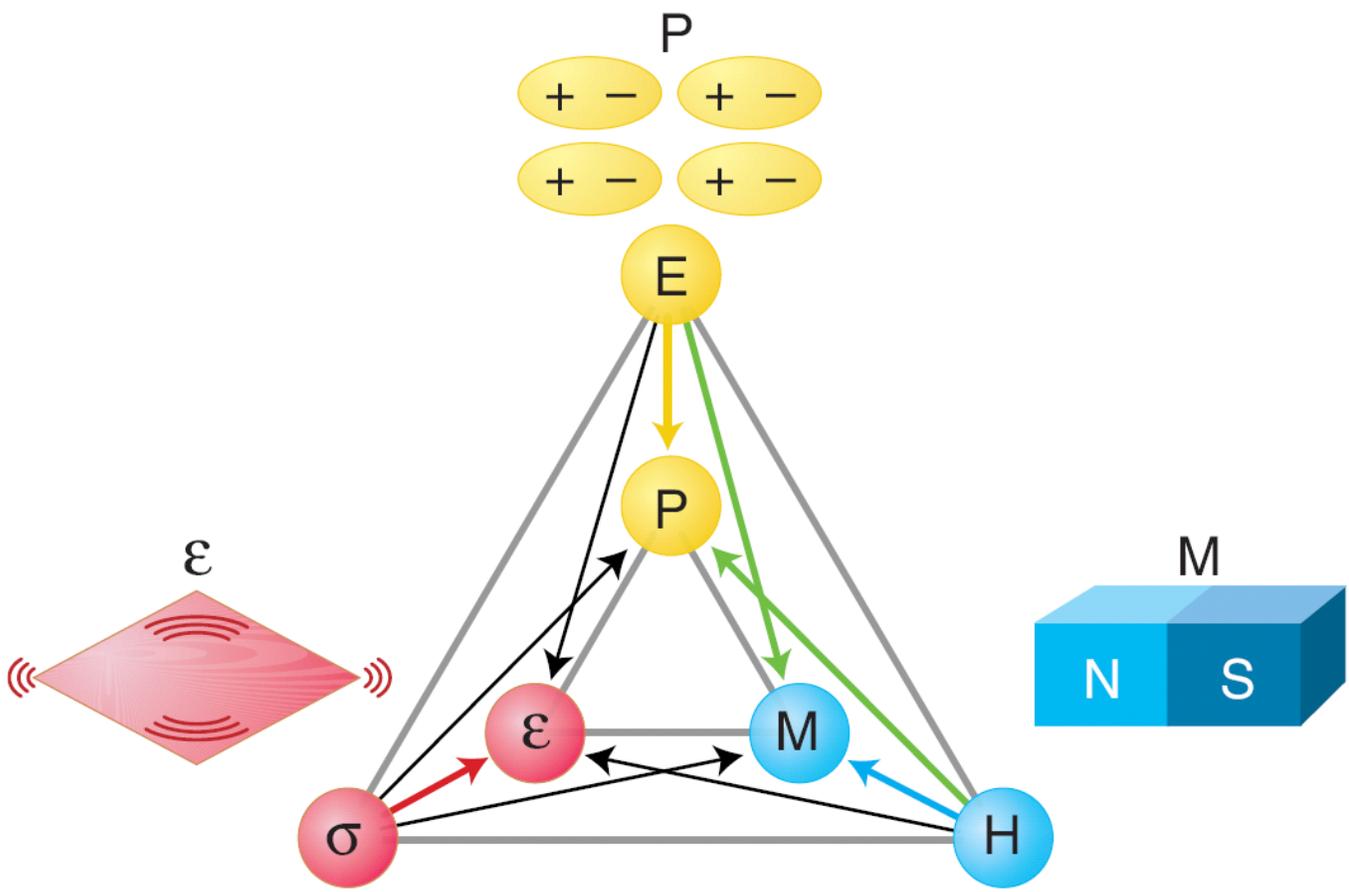

<u>Cubic Perovskite</u>

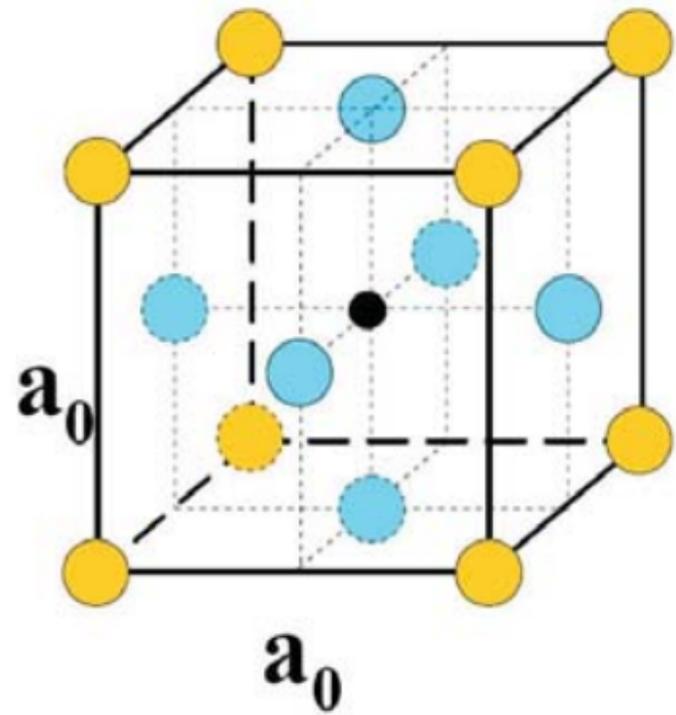

$a_0$

$a_0$

When T<Tc

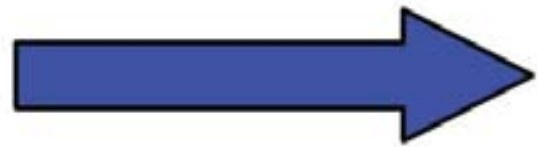

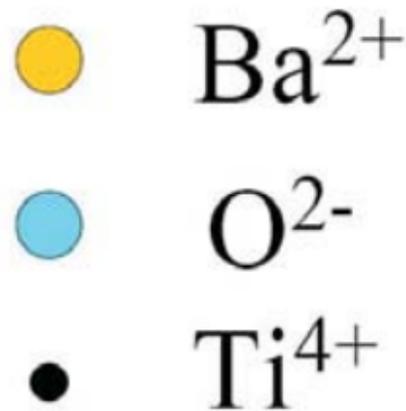

○ $Ba^{2+}$

○ $O^{2-}$

• $Ti^{4+}$

<u>Tetragonal</u>

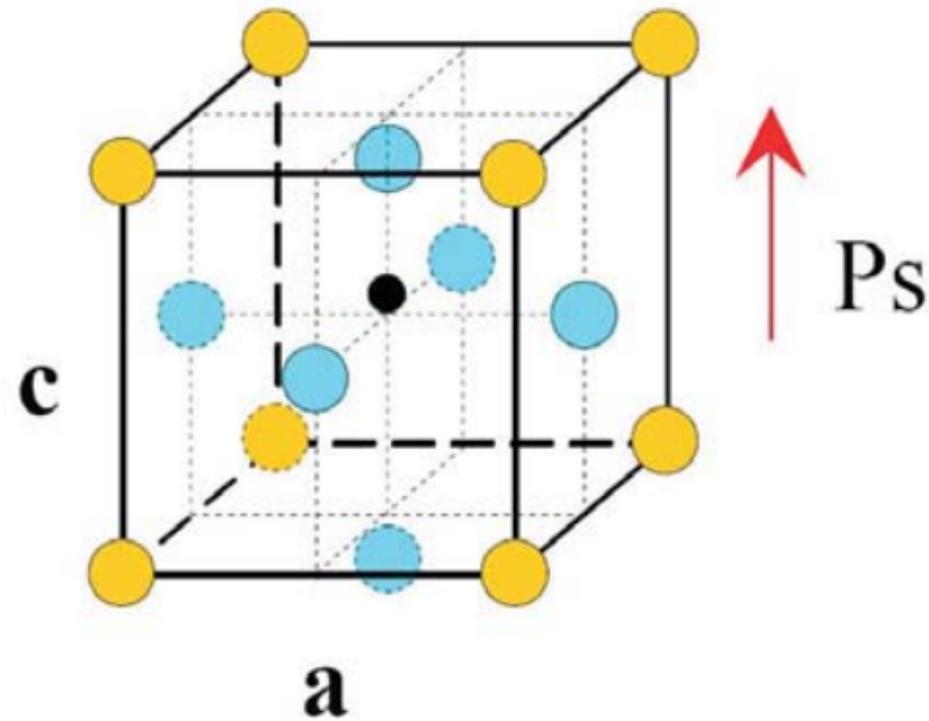

Ps

c

a

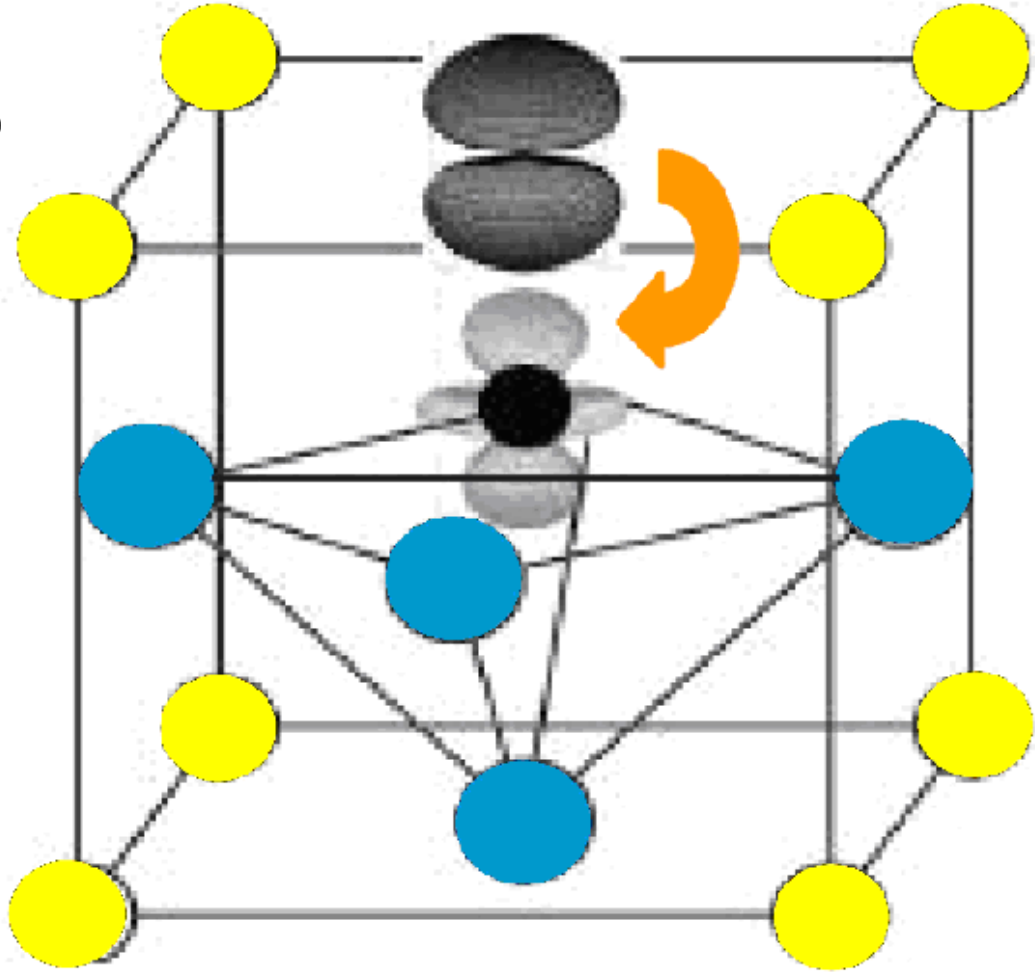

(a)

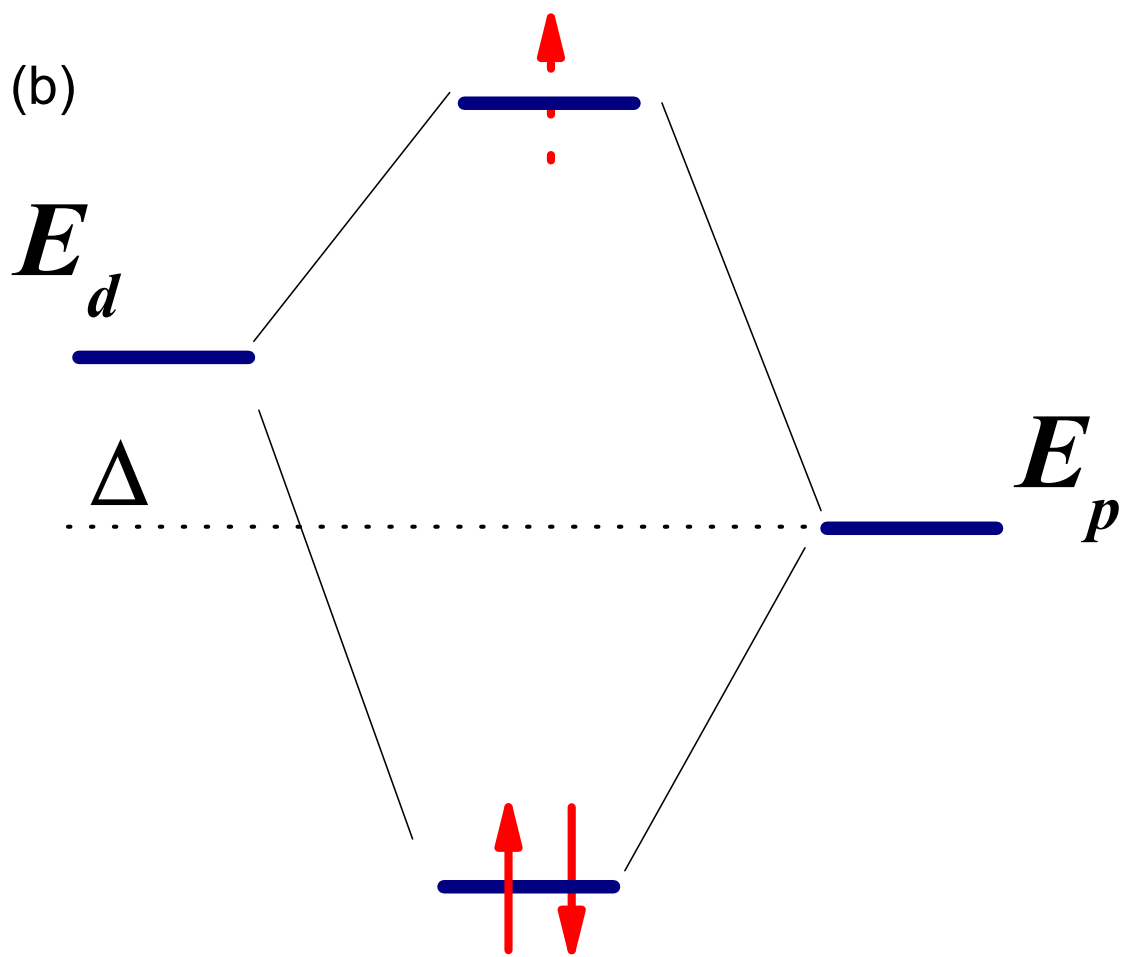

(b)

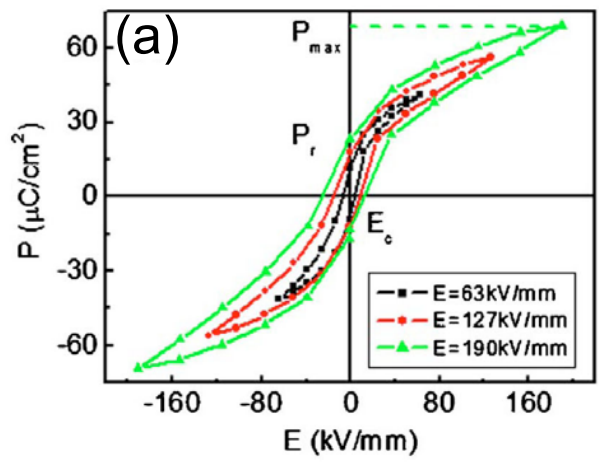

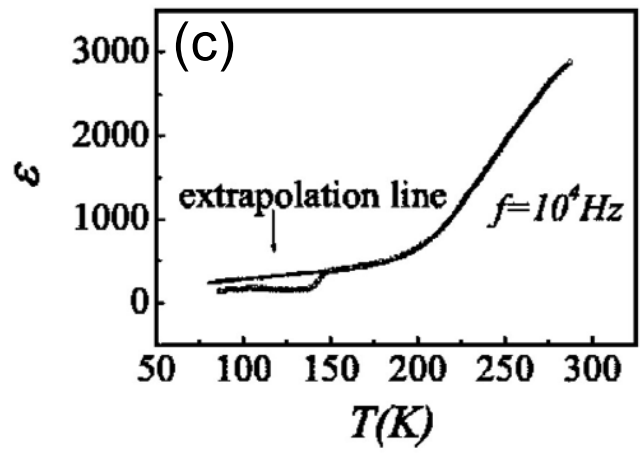

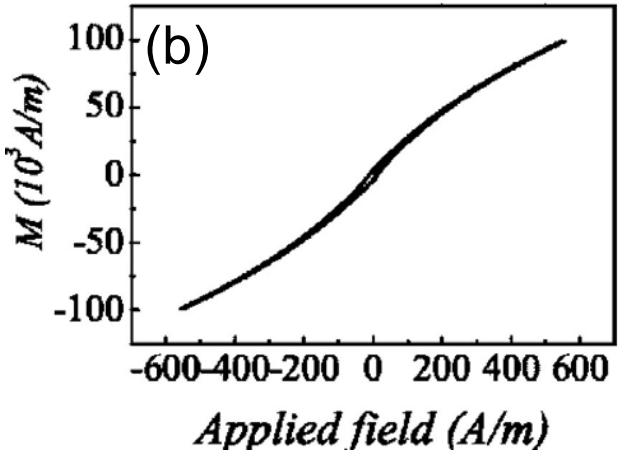

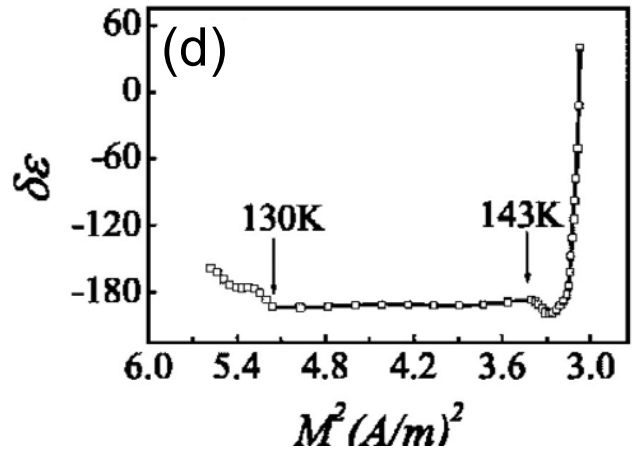

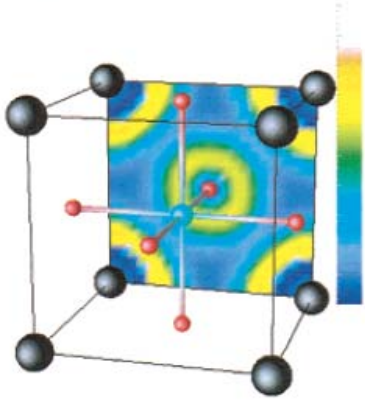
Cubic BiMnO₃, Bi-O plane

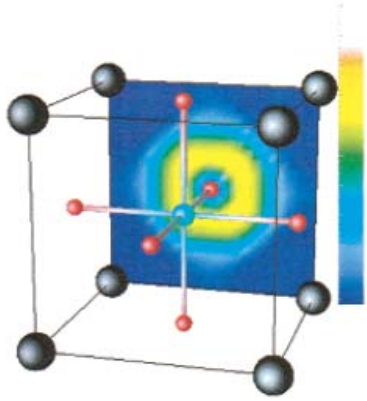
Cubic LaMnO₃, La-O plane

**(a)**

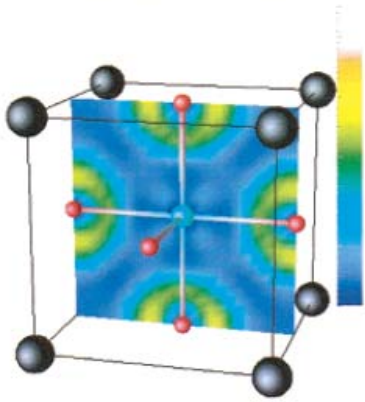
Cubic BiMnO₃, Mn-O plane

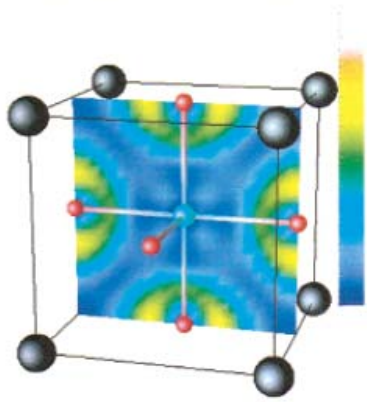
Cubic LaMnO₃, Mn-O plane

**(b)**

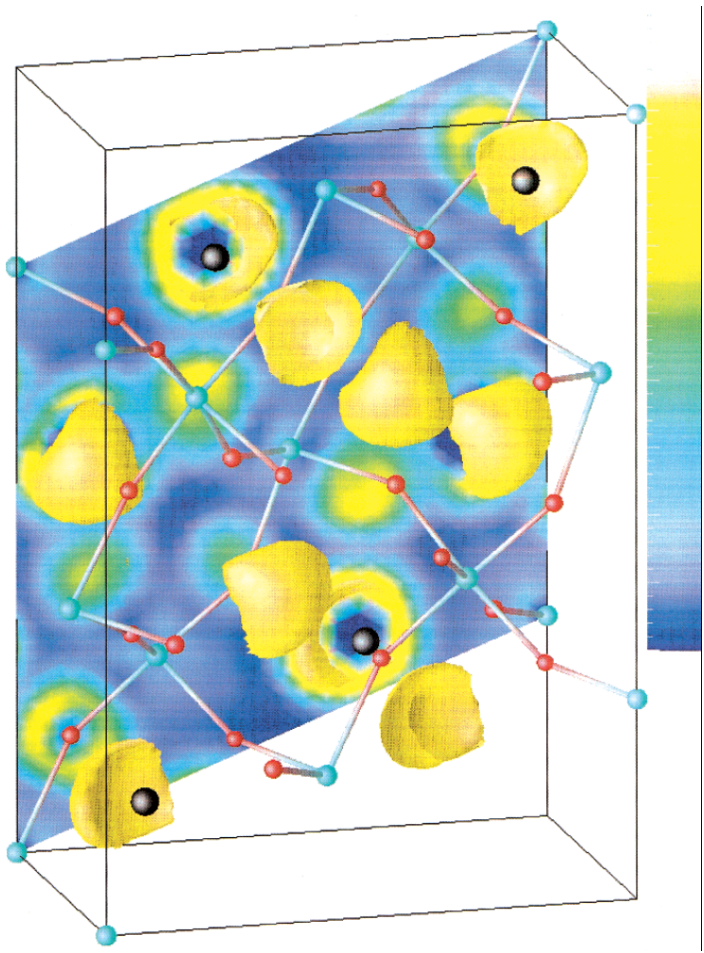

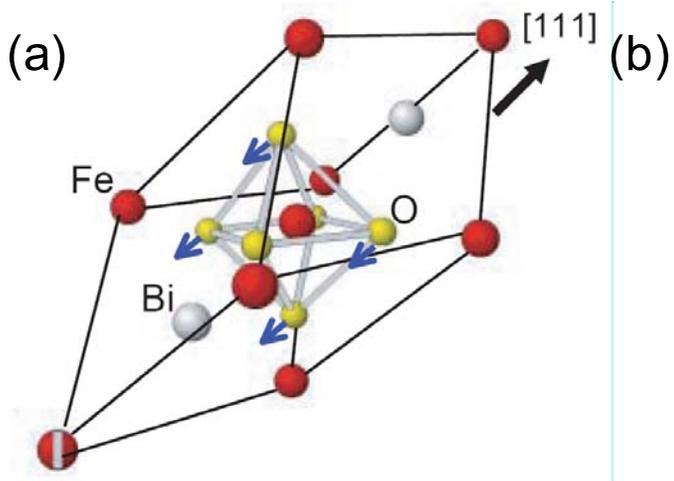

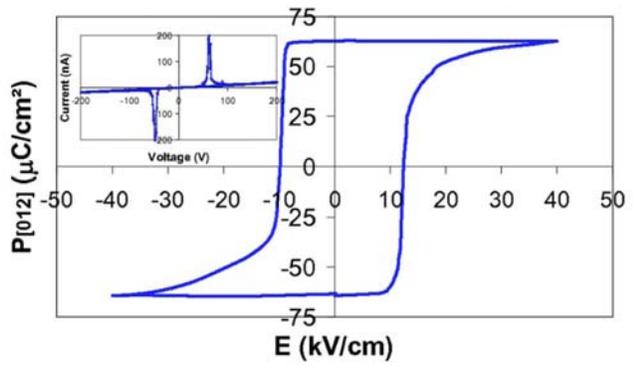

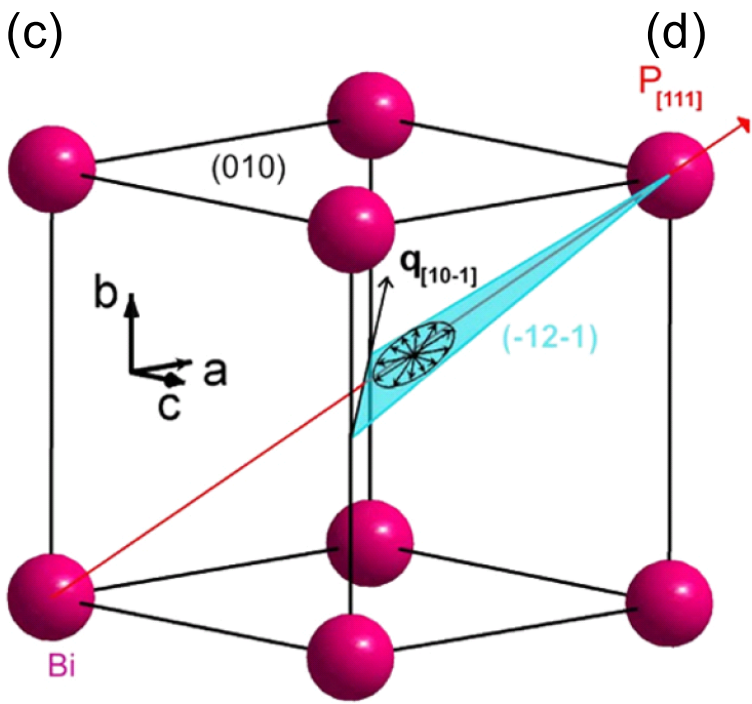

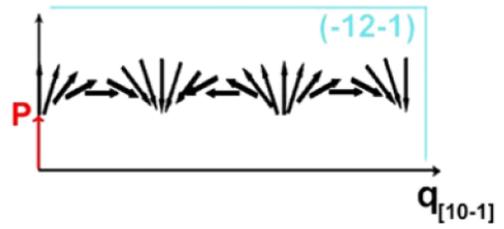

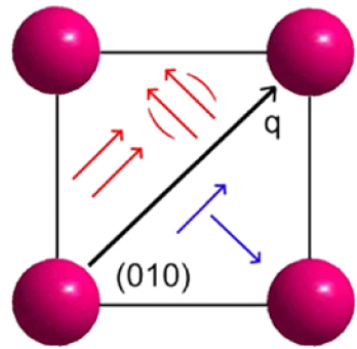

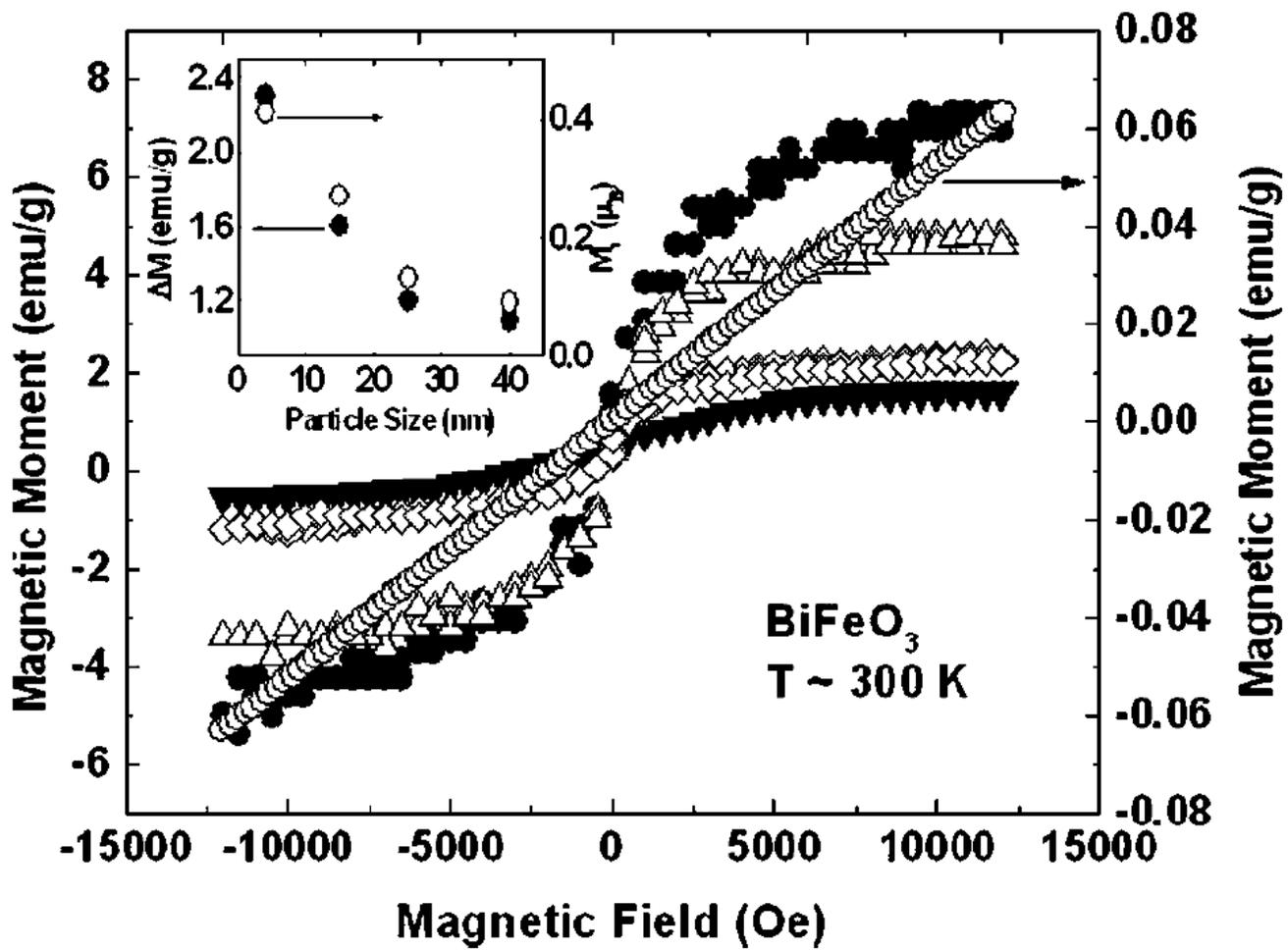

(a)

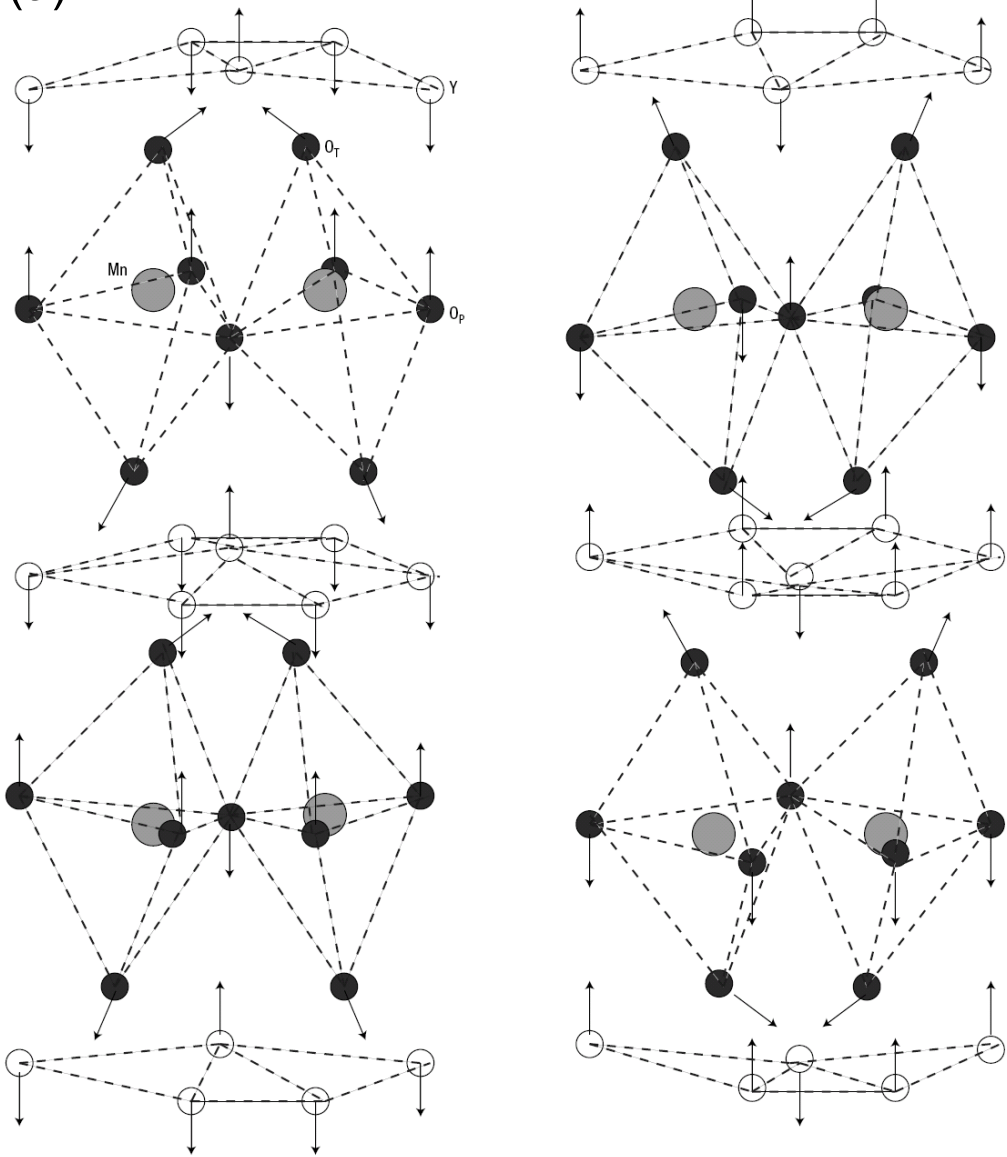

(b)

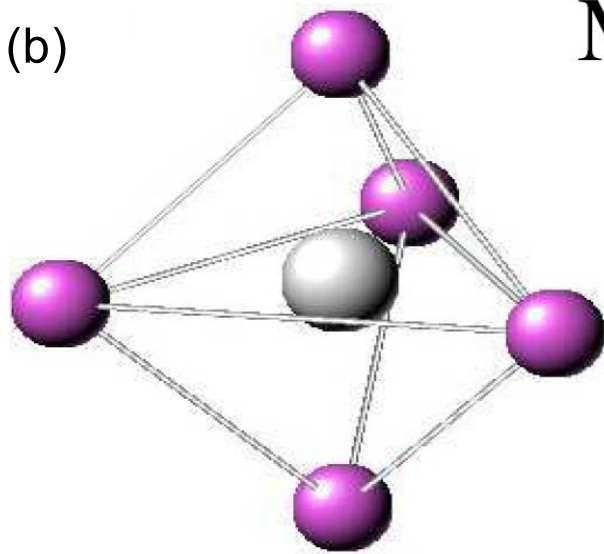

$MnO_5(D_{3h})$

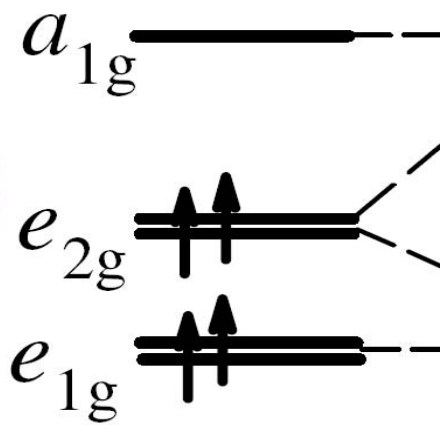

$a_{1g}$

$e_{2g}$

$e_{1g}$

**a**     Paraelectric

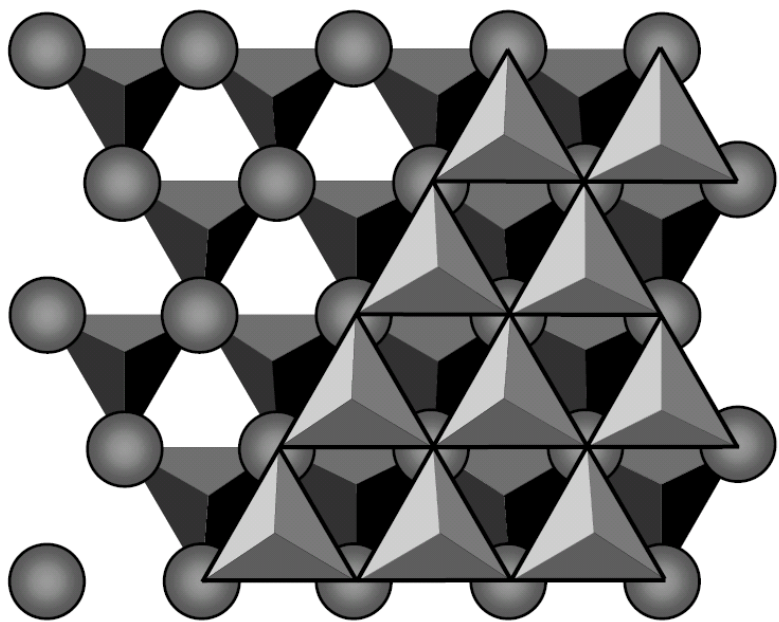

**b**     Ferroelectric

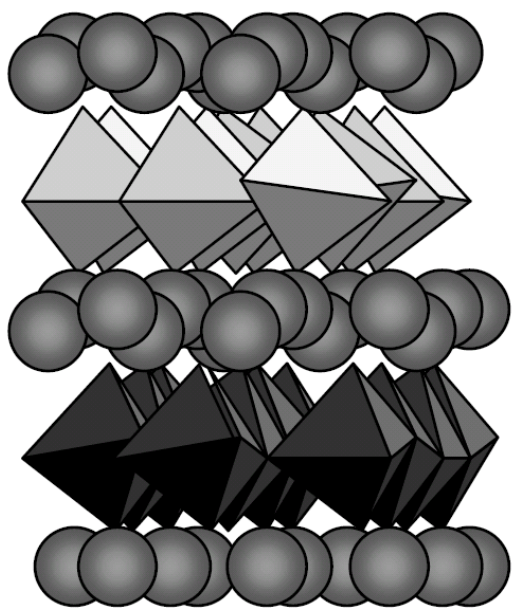

Centrosymmetric

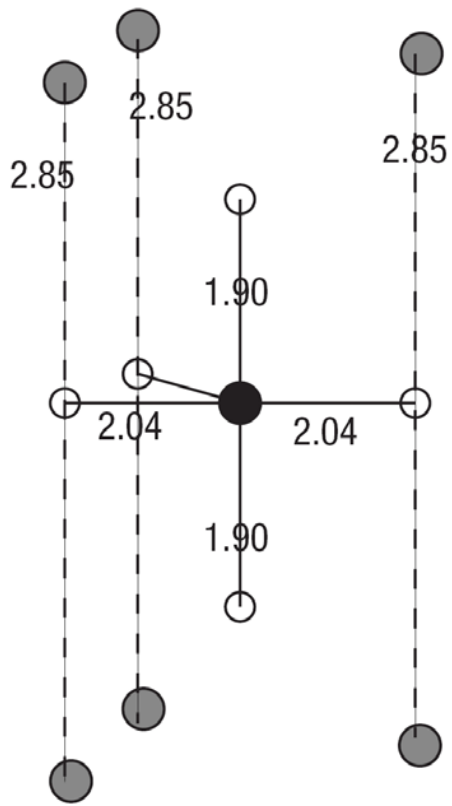

Ferroelectric

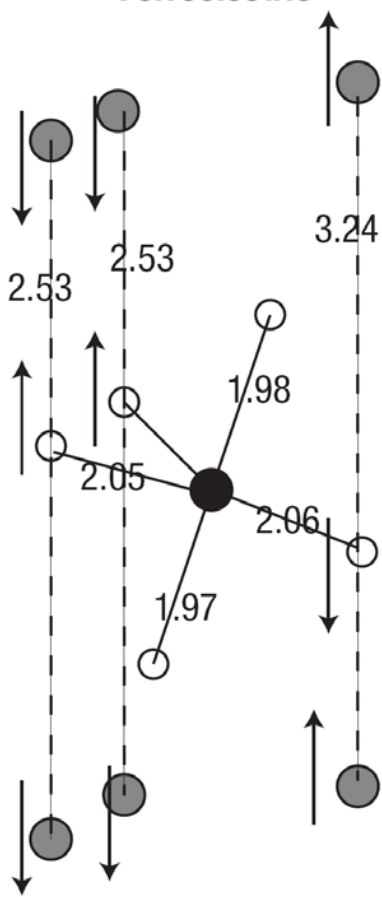

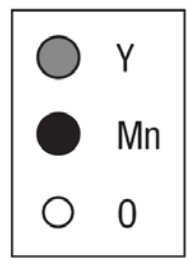

Y

Mn

O

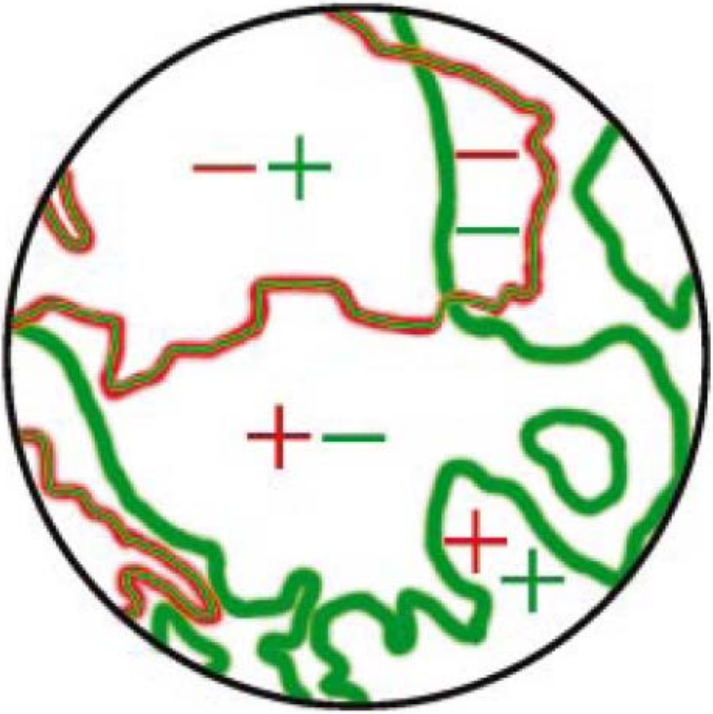

0.5 mm

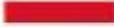 FEL

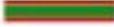 FEL
& AFM

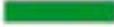 AFM

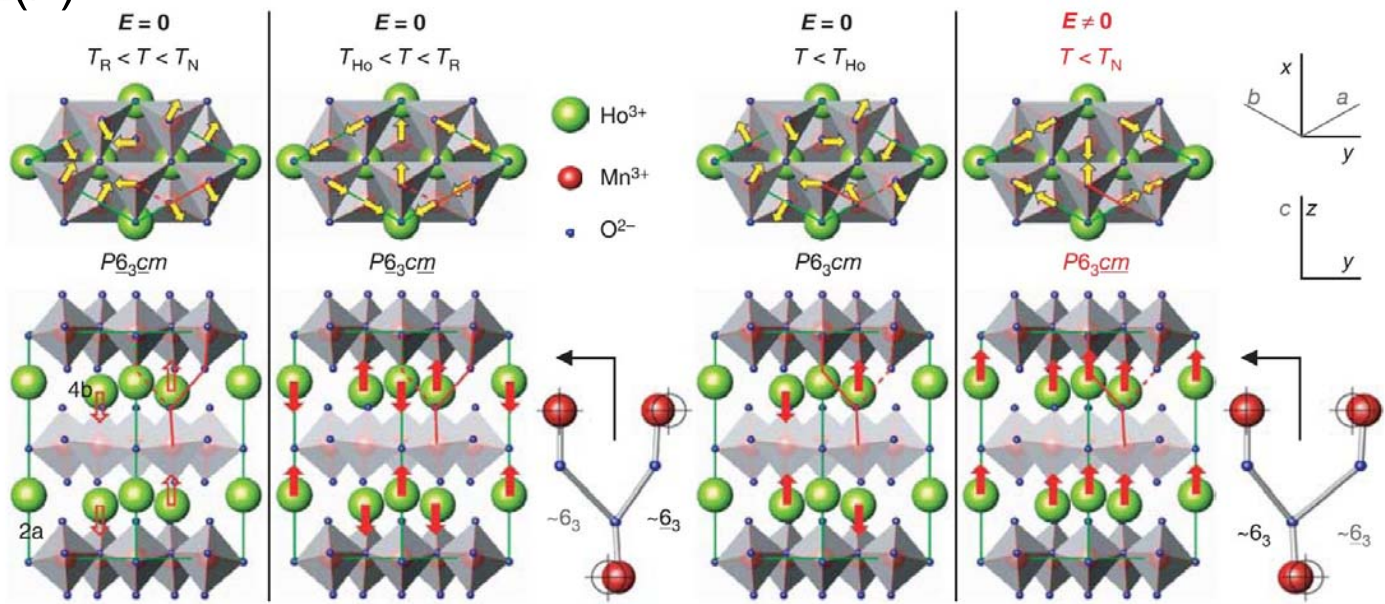

(a)

| E = 0 | E = 0 | E = 0 | E ≠ 0 |
|---|---|---|---|
| $T_R < T < T_N$ | $T_{Ho} < T < T_R$ | $T < T_{Ho}$ | $T < T_N$ |
| $P6_3cm$ | $P6_3c\underline{m}$ | $P6_3cm$ | $P6_3c\underline{m}$ |

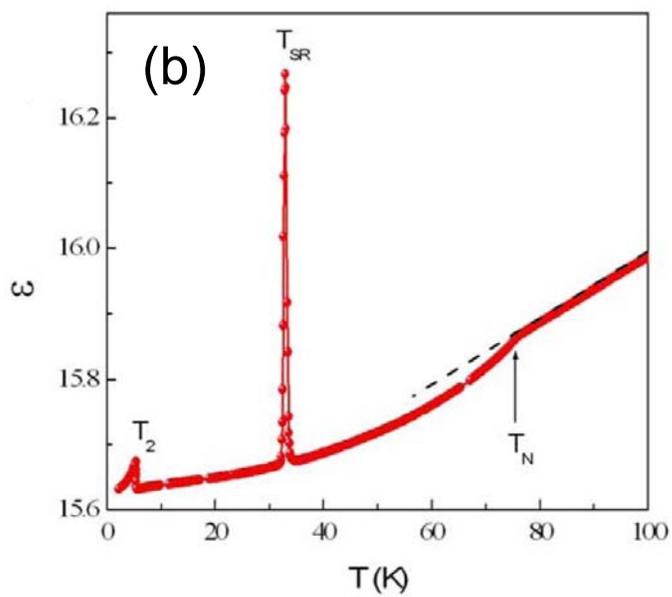

(b)

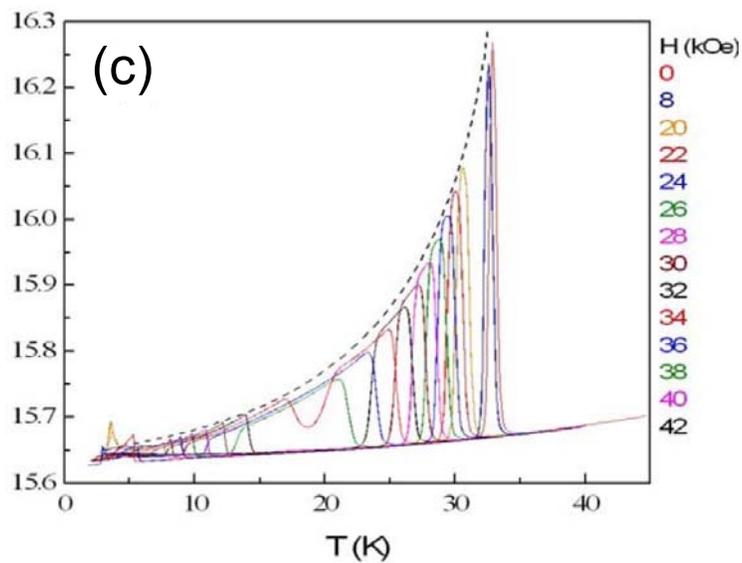

(c)

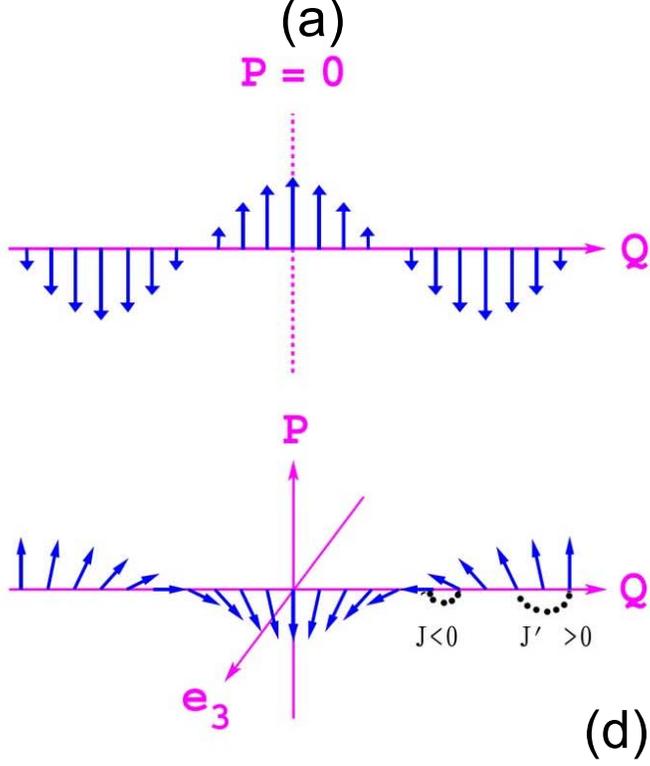

(a)

P = 0

Q

P

$e_3$

J < 0    J' > 0

(b)

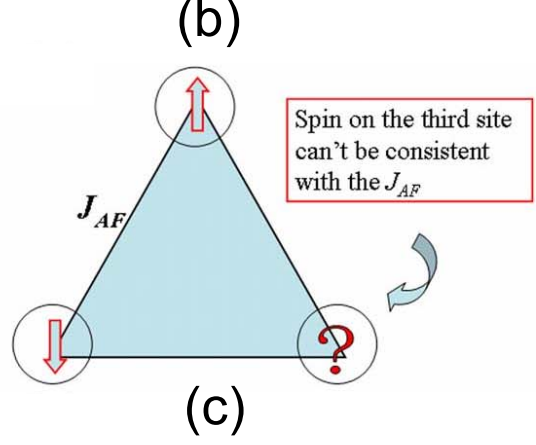

$J_{AF}$

Spin on the third site can't be consistent with the $J_{AF}$

?

(c)

Effects of Dsyaloshinskii–Moriya interaction

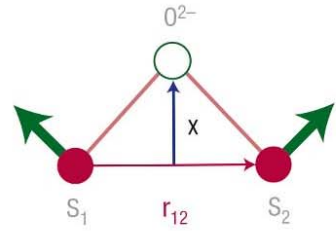

$O^{2-}$

x

$S_1$    $r_{12}$    $S_2$

(d)

Weak ferromagnetism (LaCu$_2$O$_4$)

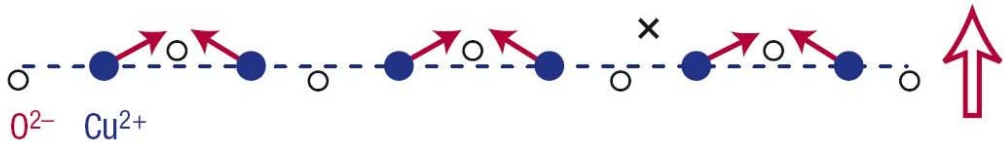

$O^{2-}$   $Cu^{2+}$

×

Weak ferroelectricity (RMnO$_3$)

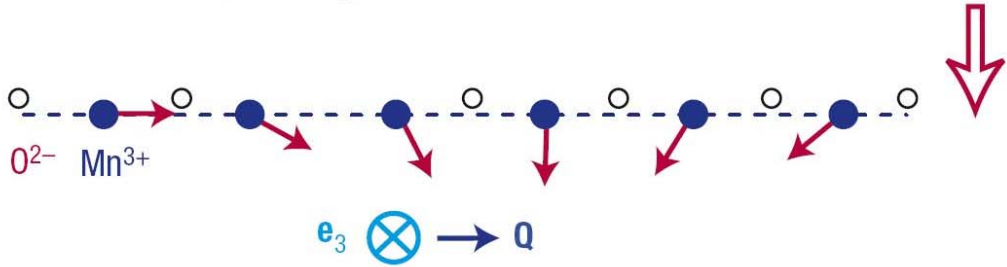

$O^{2-}$   $Mn^{3+}$

$e_3$  $\otimes$  $\longrightarrow$ Q

# Schematic figure of the model Hamiltonian

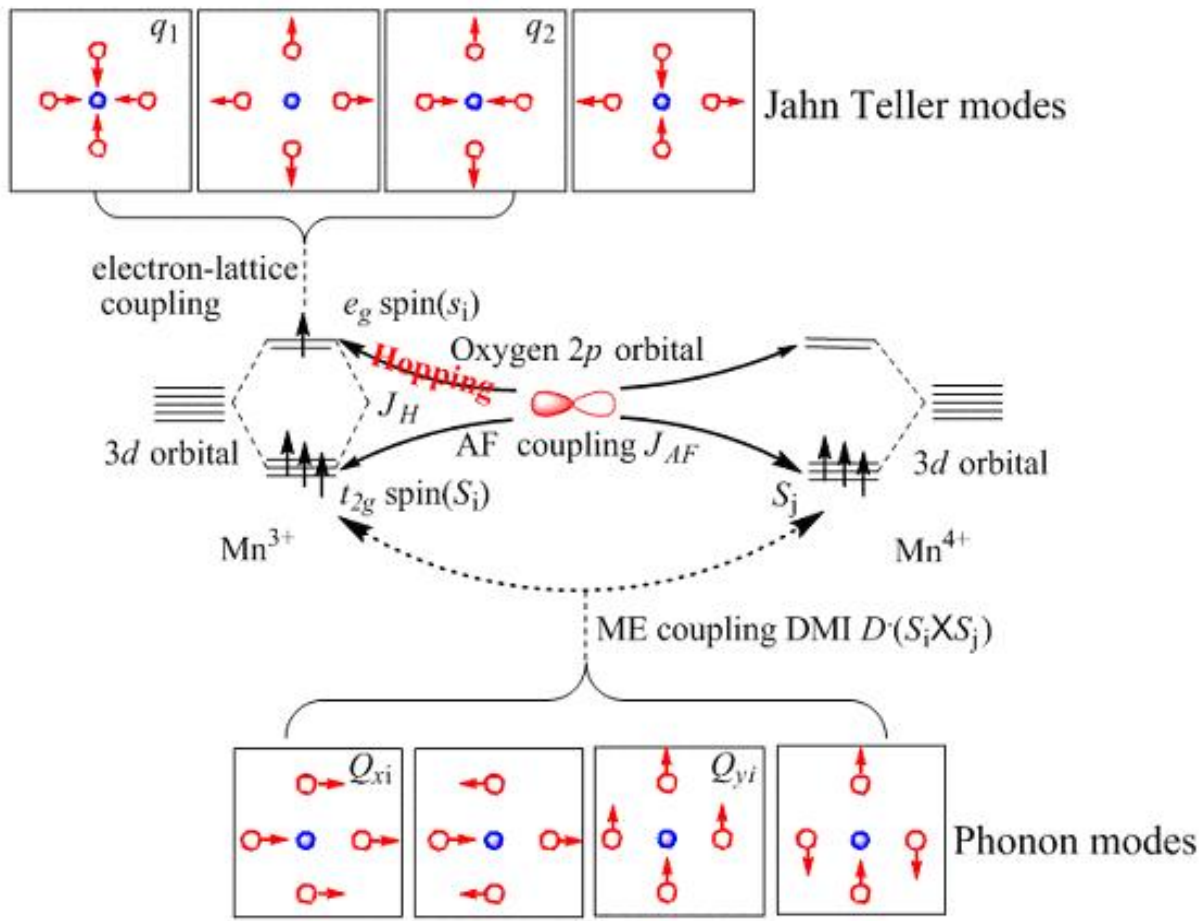

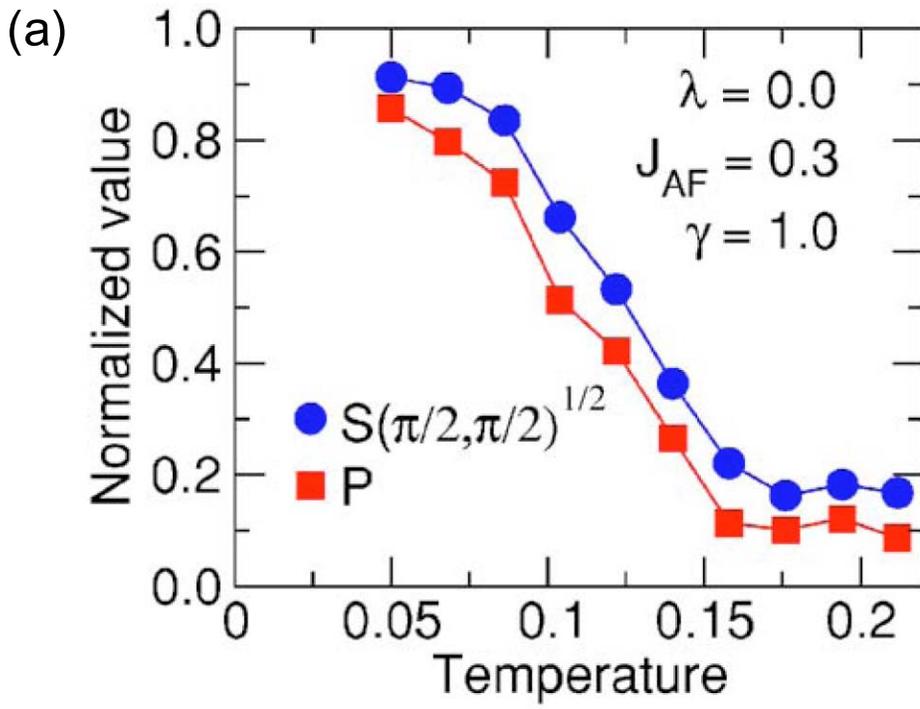

(a)

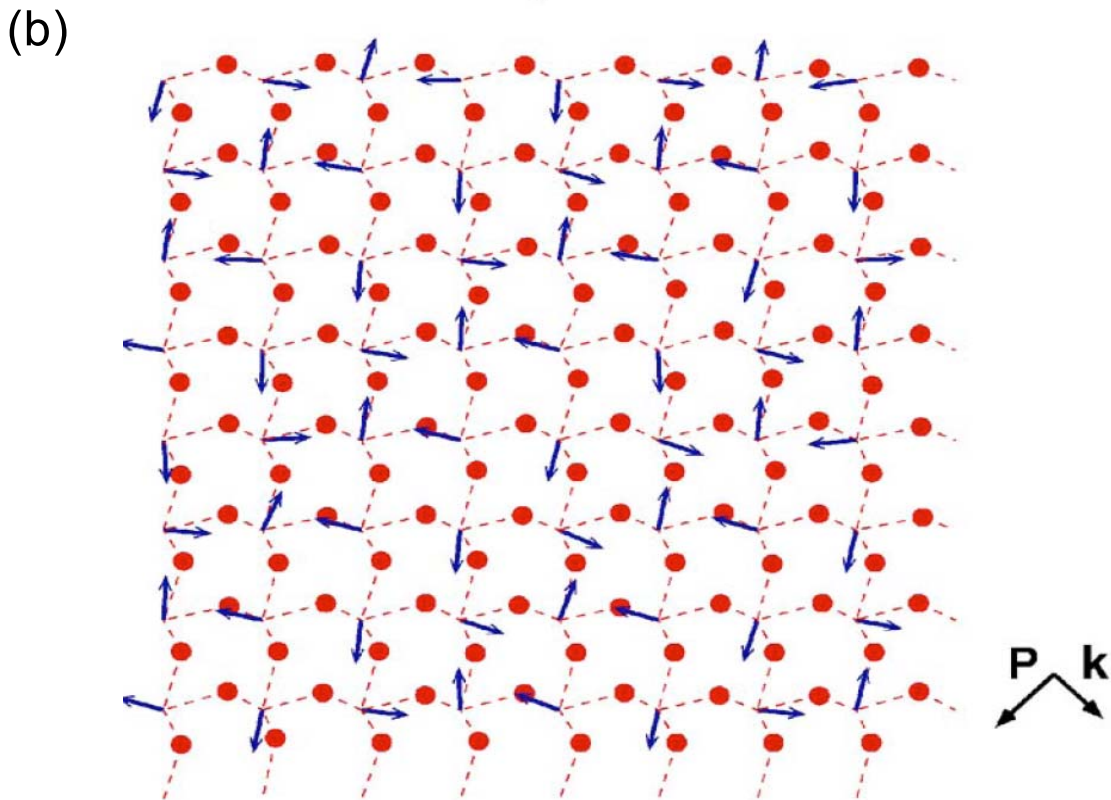

(b)

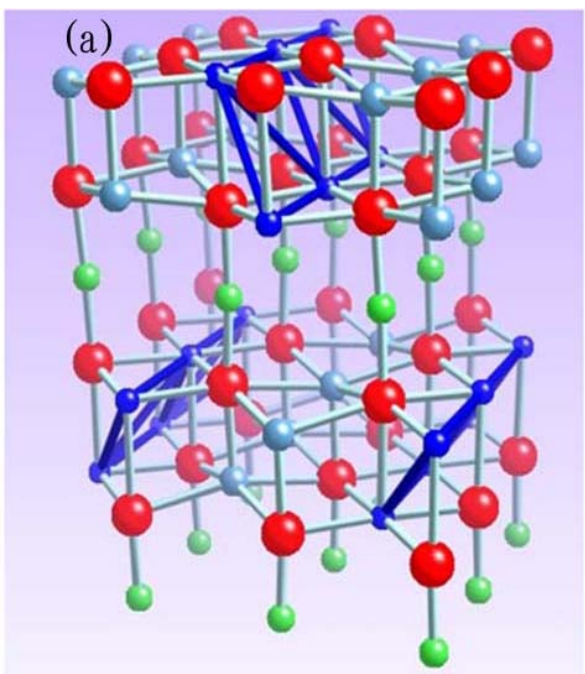

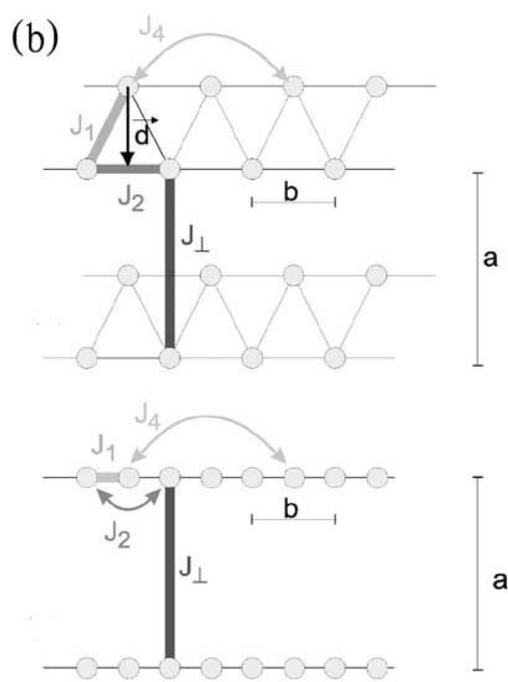

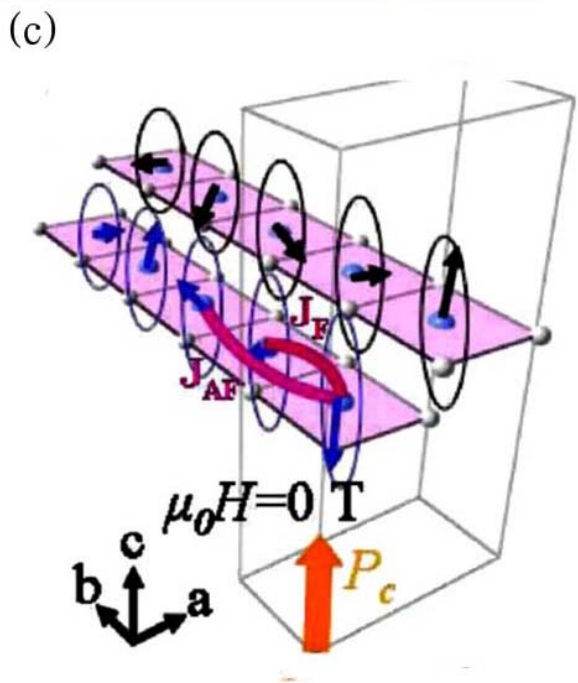

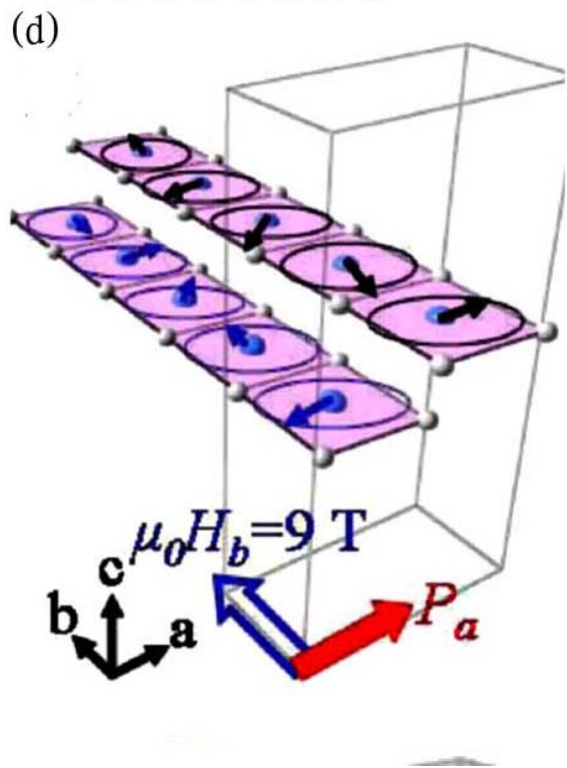

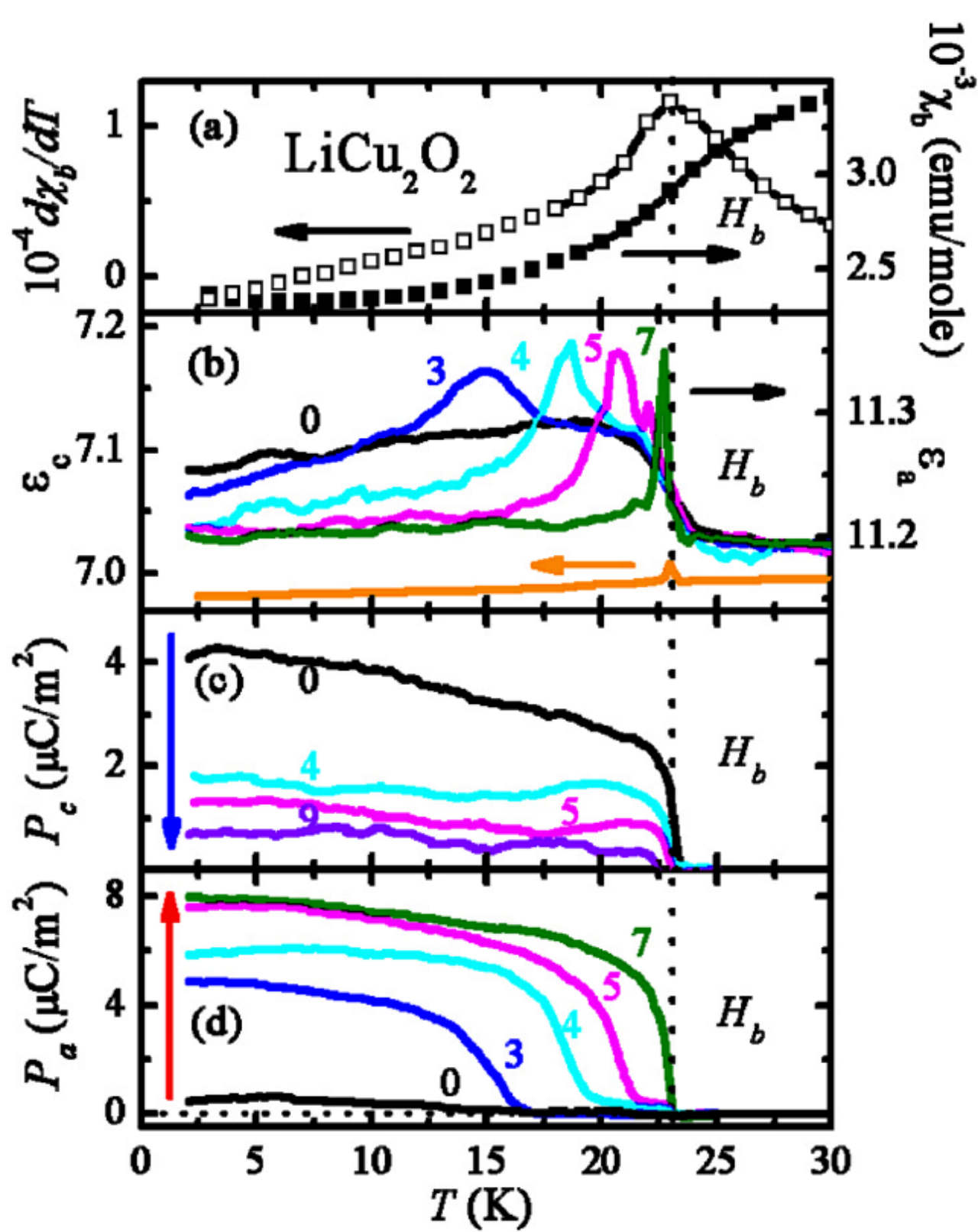

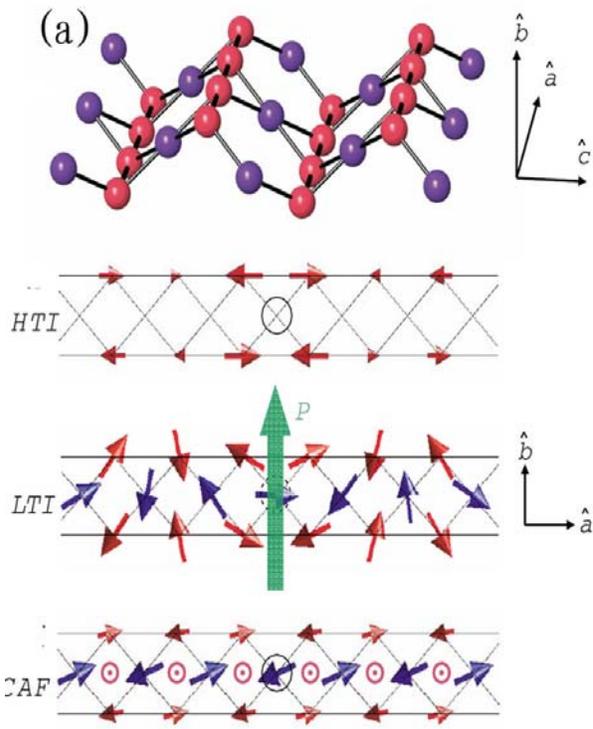

HTI

LTI

CAF

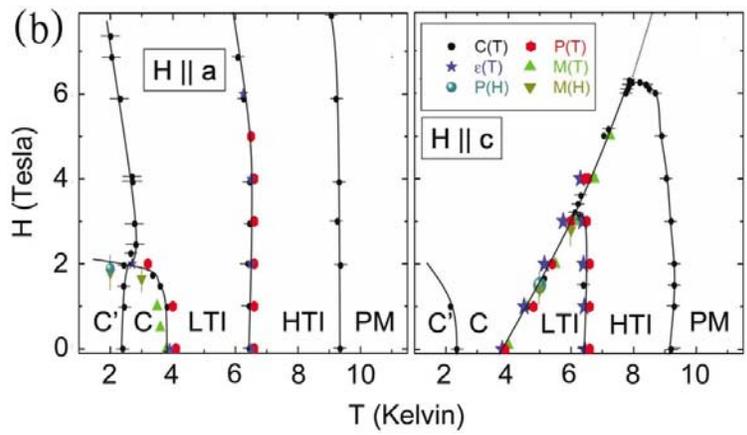

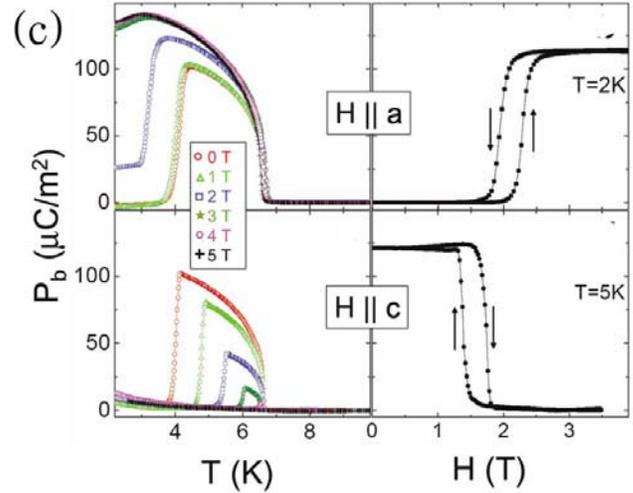

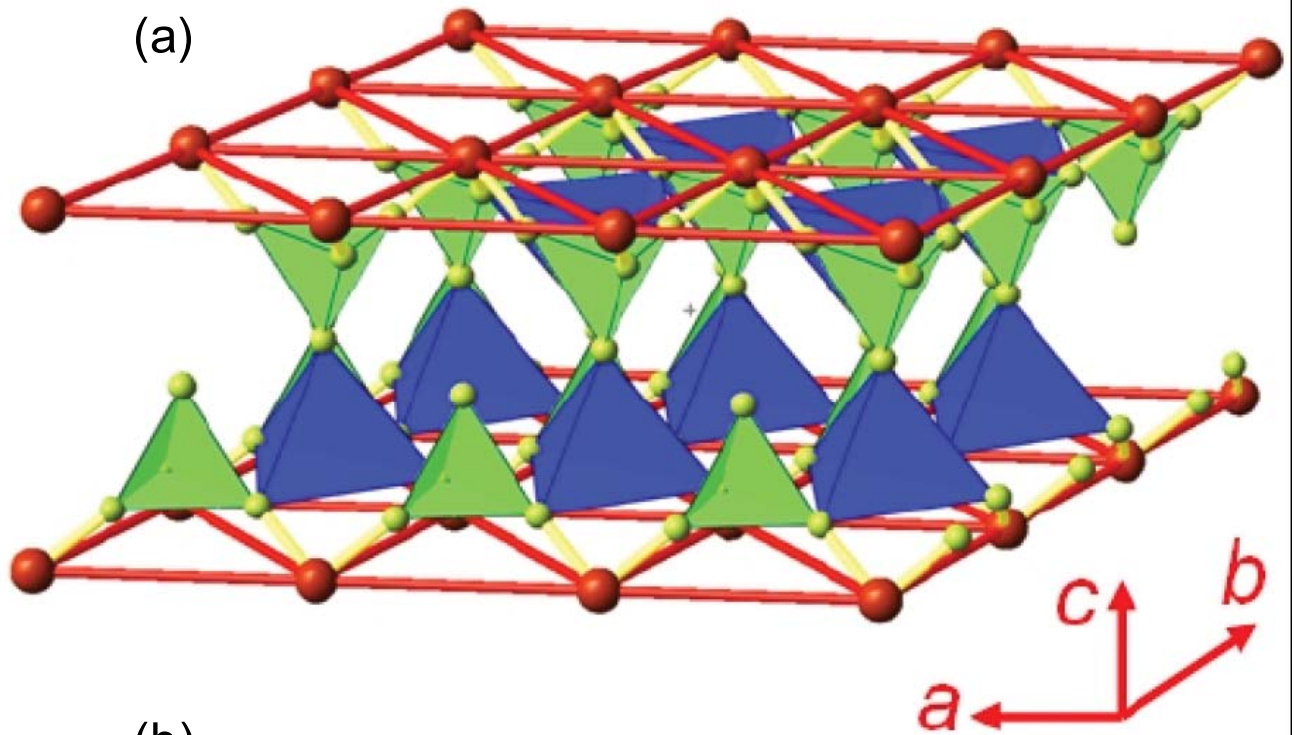

(a)

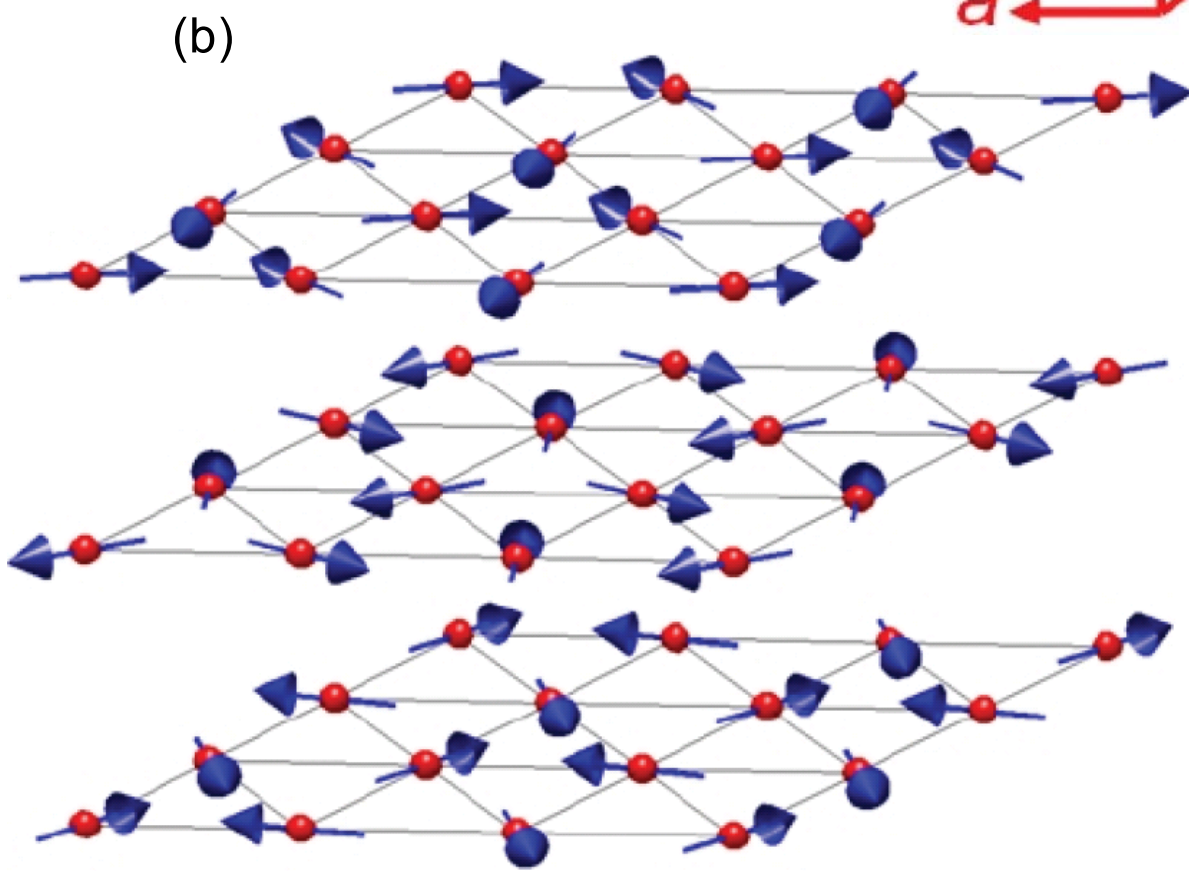

(b)

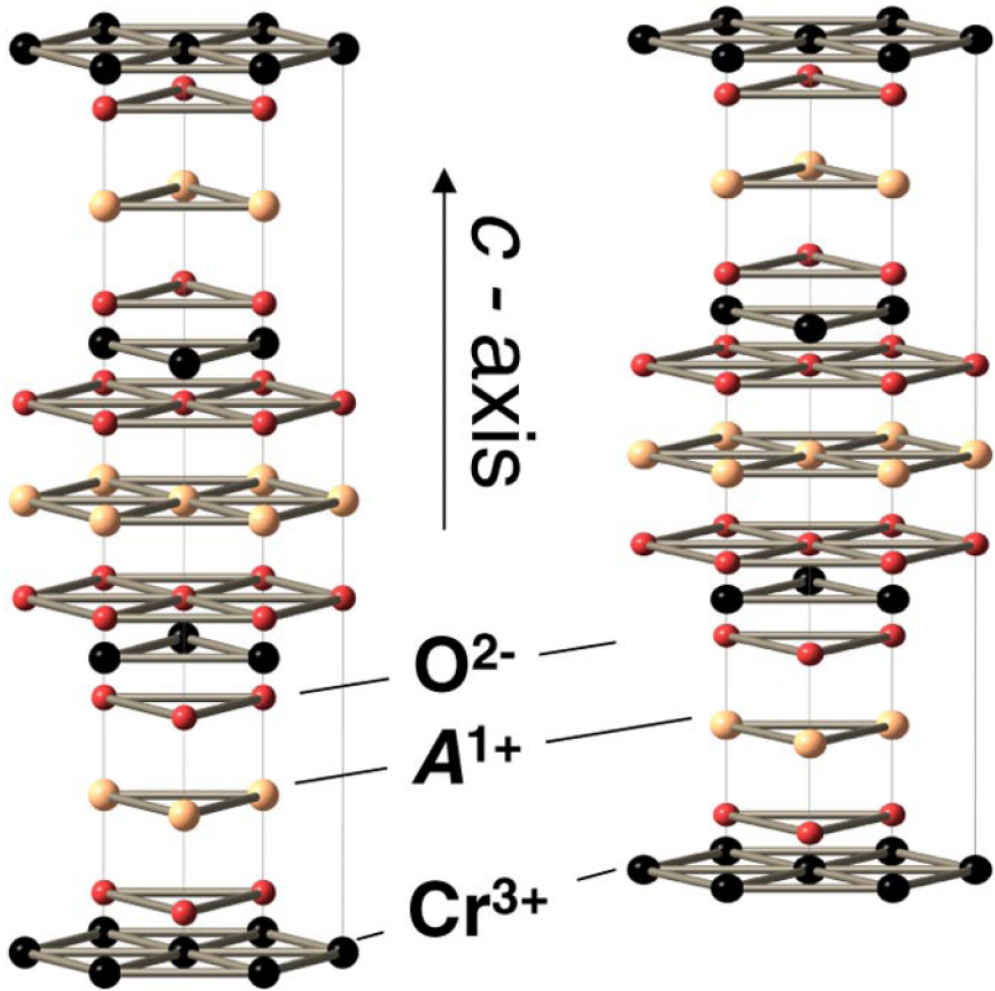

$c$ - axis

$O^{2-}$

$A^{1+}$

$Cr^{3+}$

(a) Delafossite    (b) Ordered Rock Salt

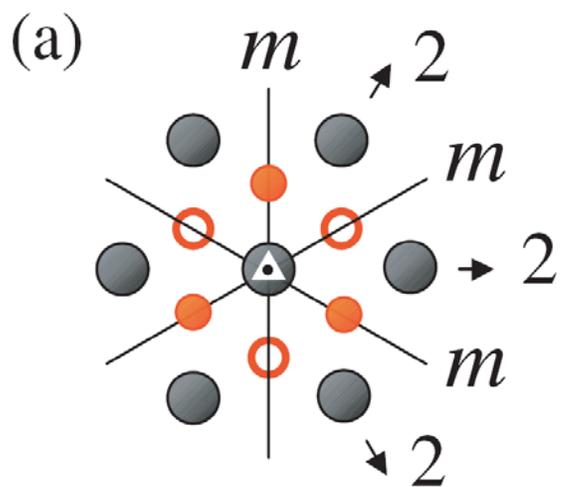

(a)

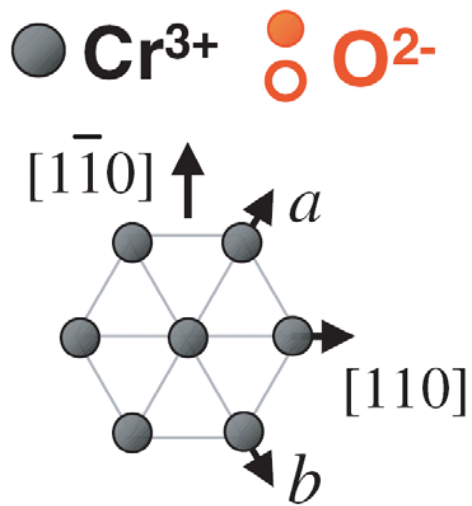

Cr³⁺ O²⁻

[1̄10] ↑  a

[110]

b

(b)

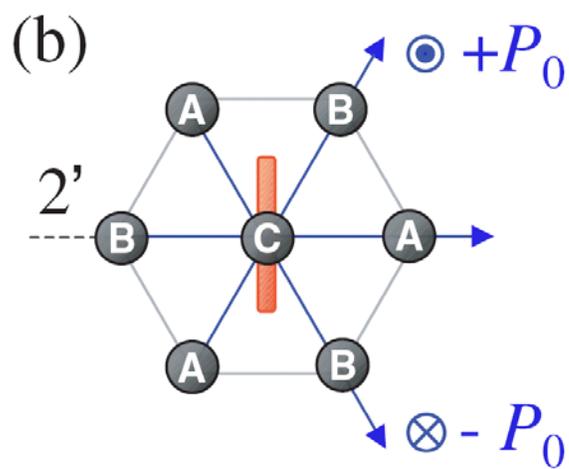

2'   +P₀   ⊙

⊗ −P₀

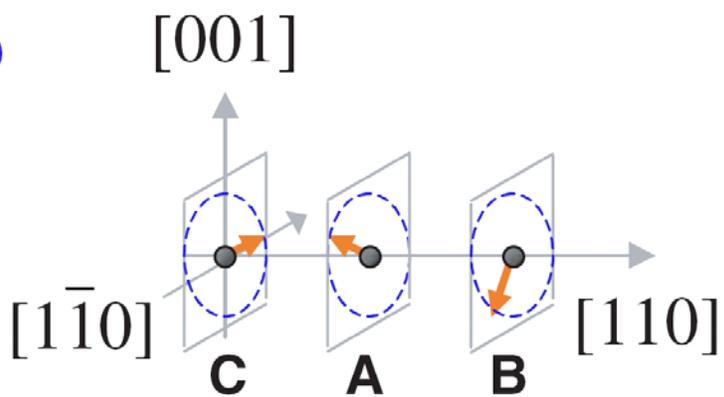

[001]

[1̄10]

[110]

C    A    B

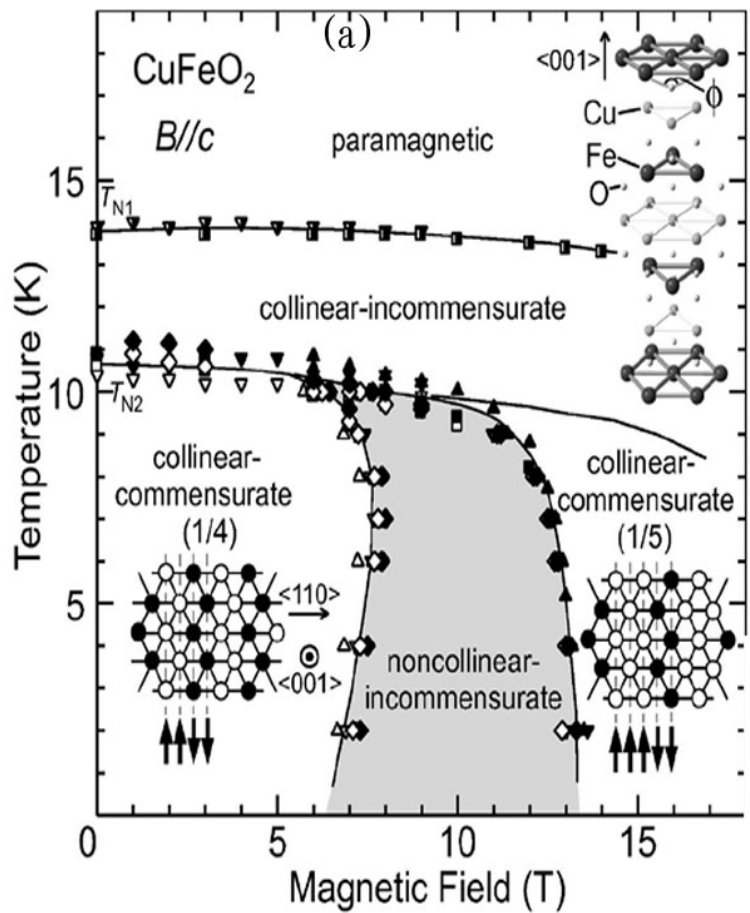
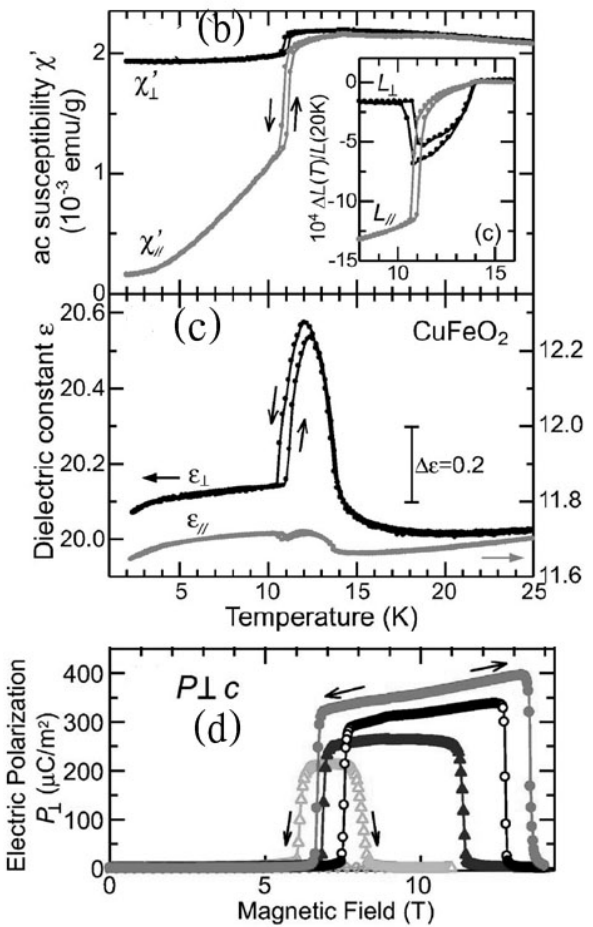

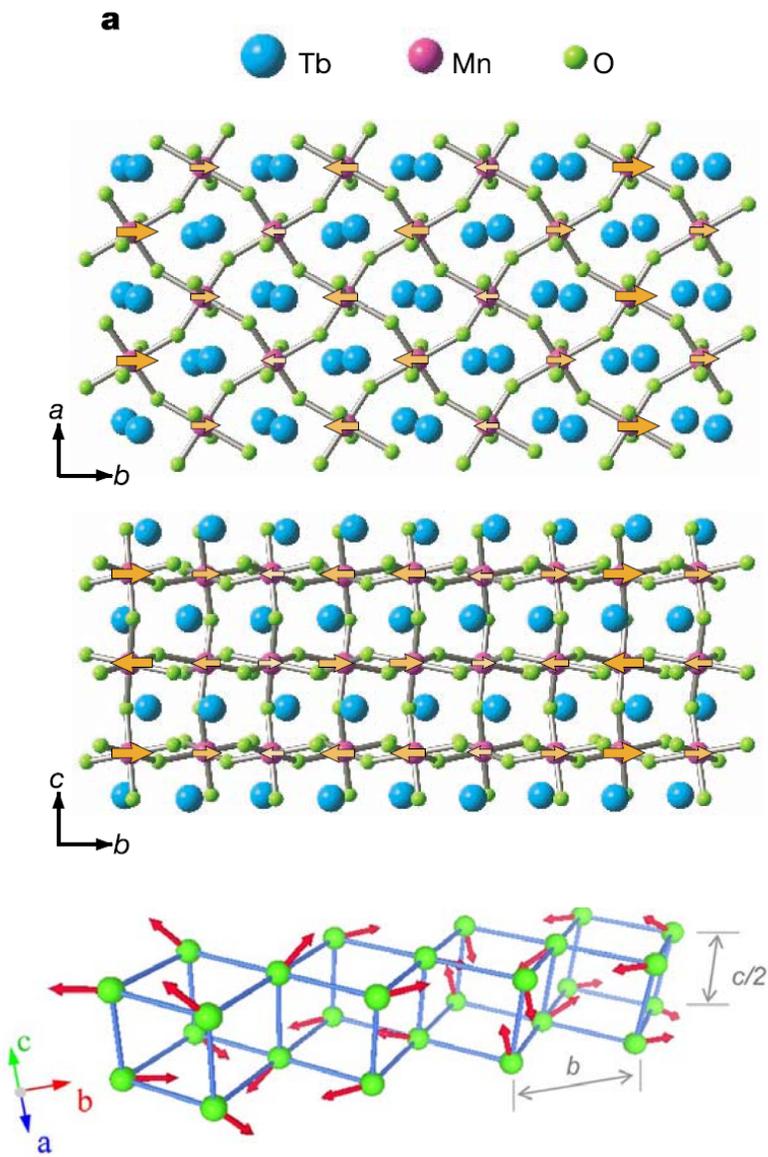

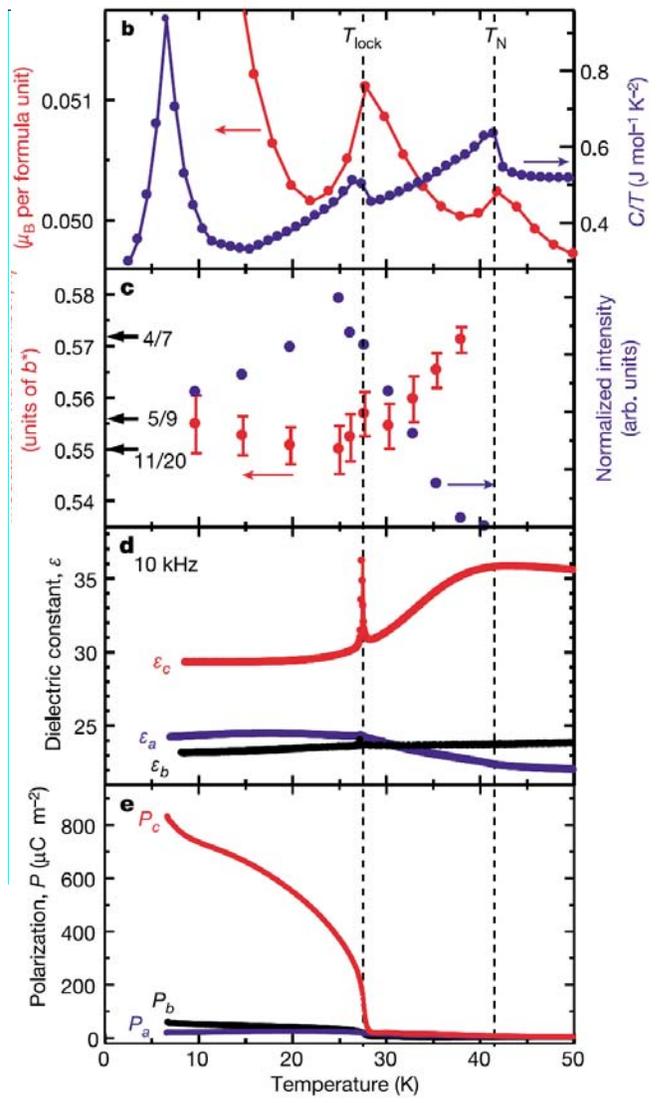

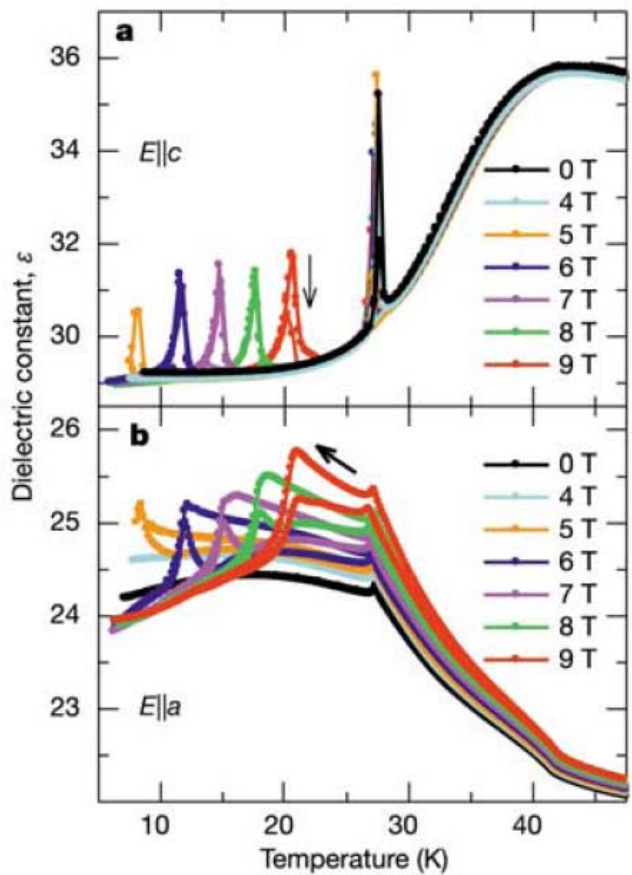
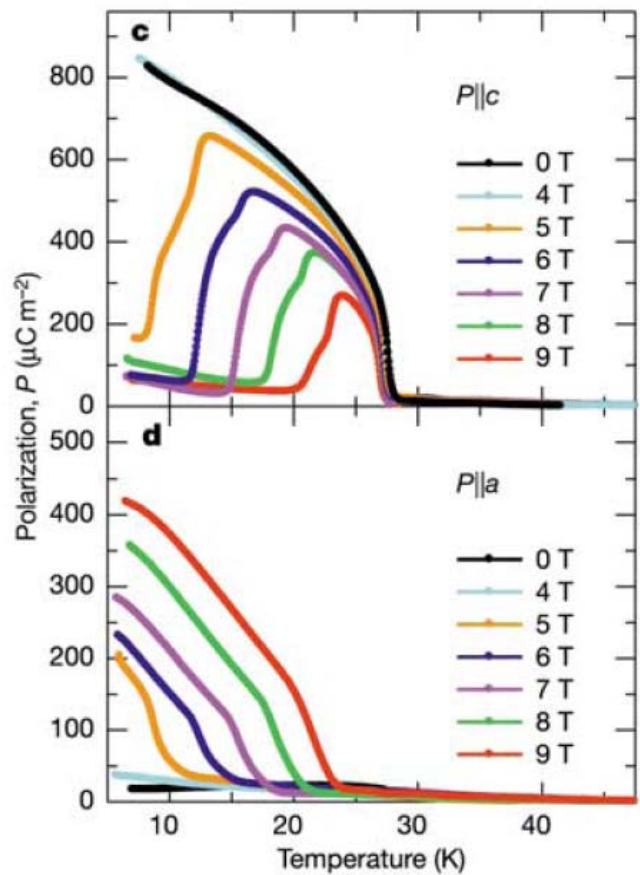

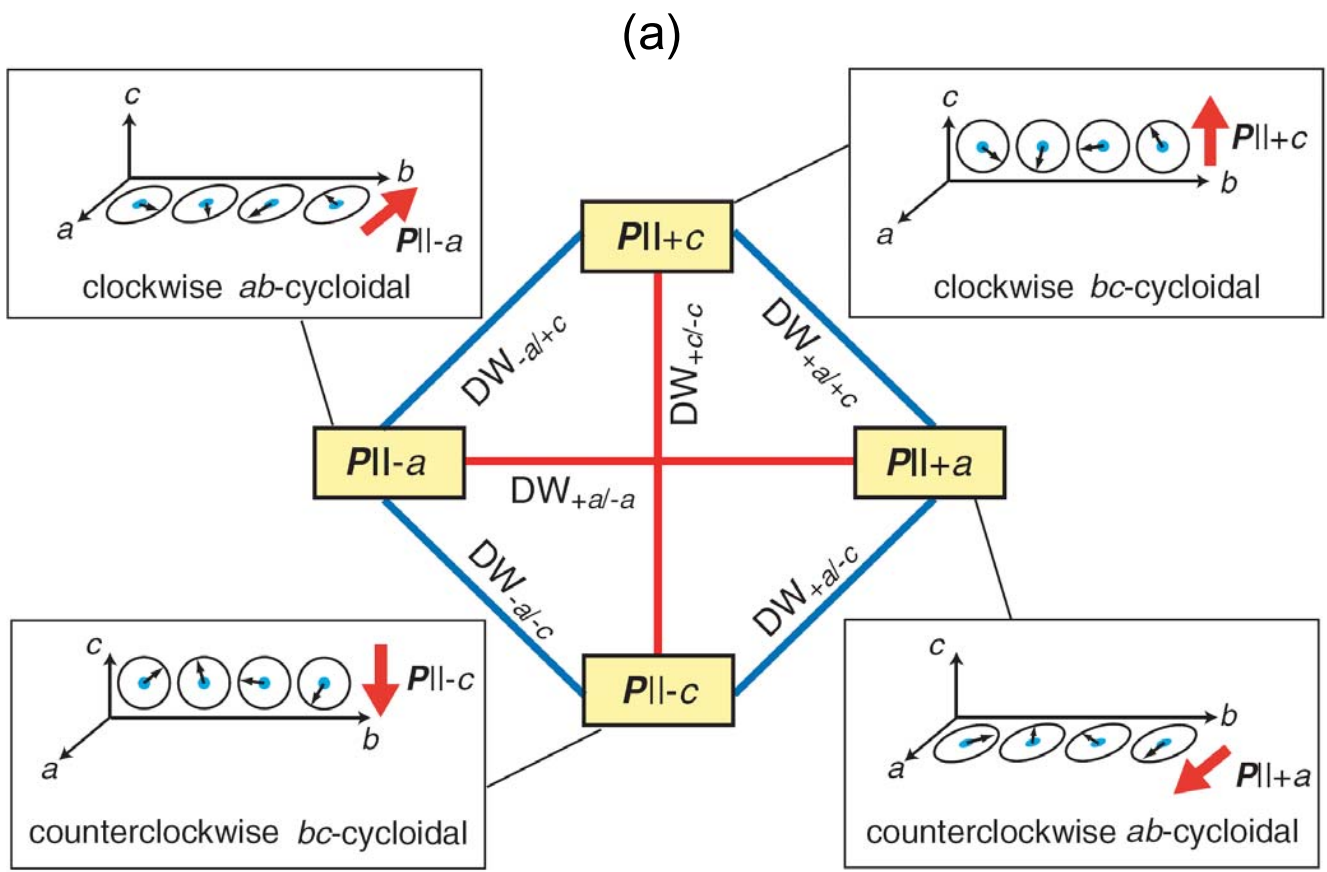

(a)

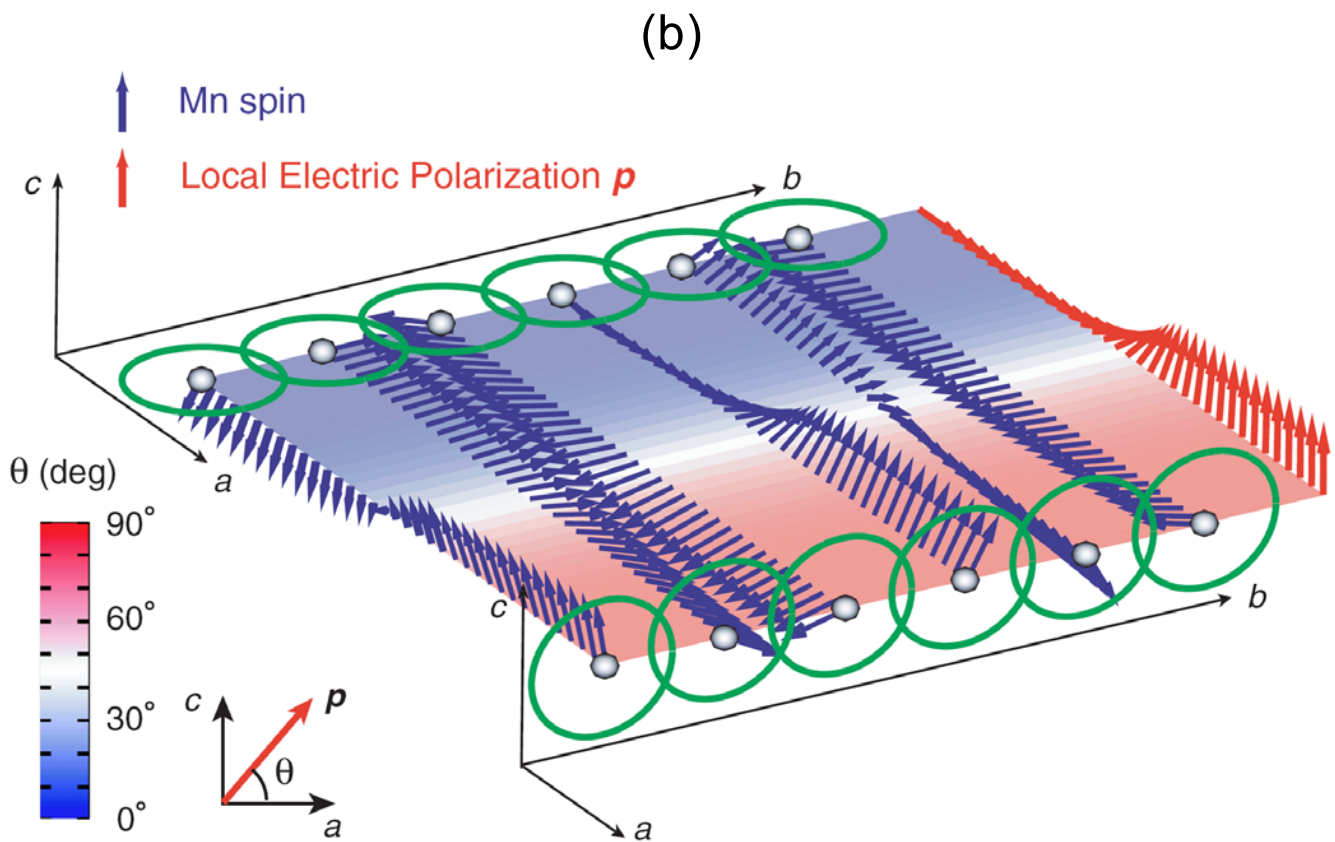

(b)

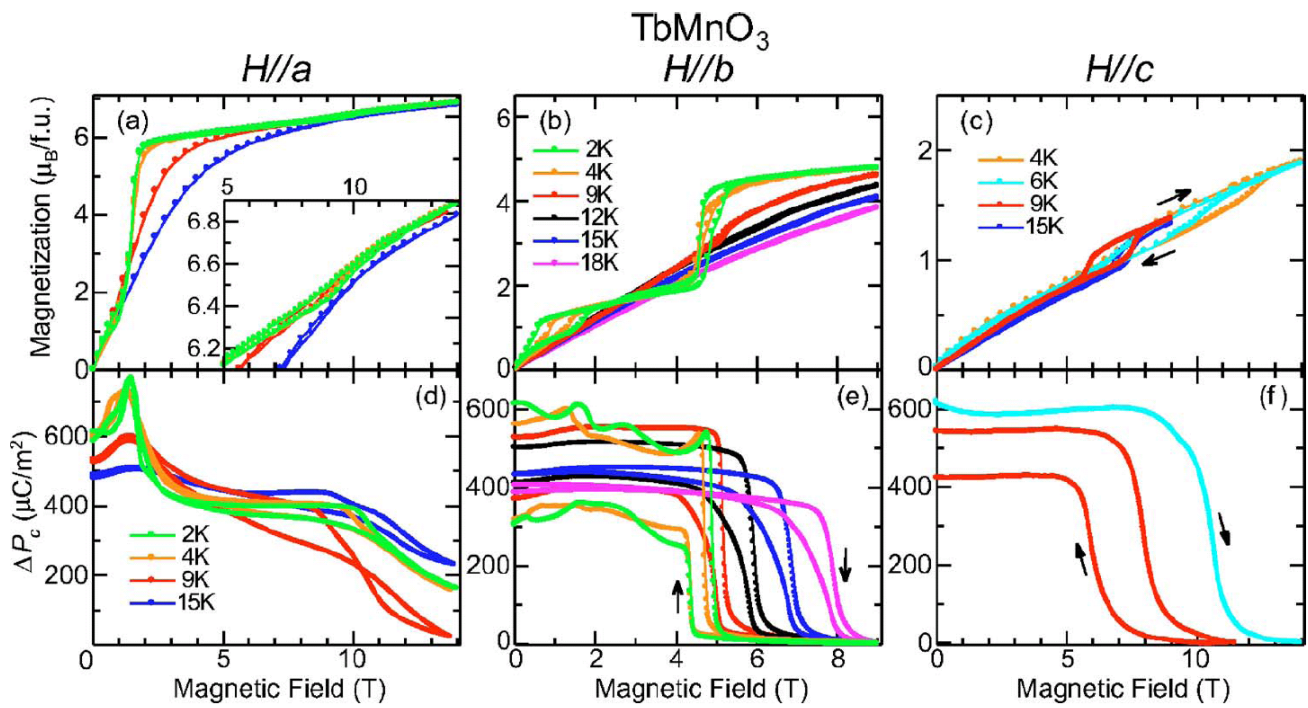

TbMnO$_3$

$H//a$

(a)

(d)

$H//b$

(b) 2K 4K 9K 12K 15K 18K

(e)

$H//c$

(c) 4K 6K 9K 15K

(f)

Magnetization ($\mu_B$/f.u.)

$\Delta P_c$ ($\mu$C/m$^2$)

Magnetic Field (T)

2K 4K 9K 15K

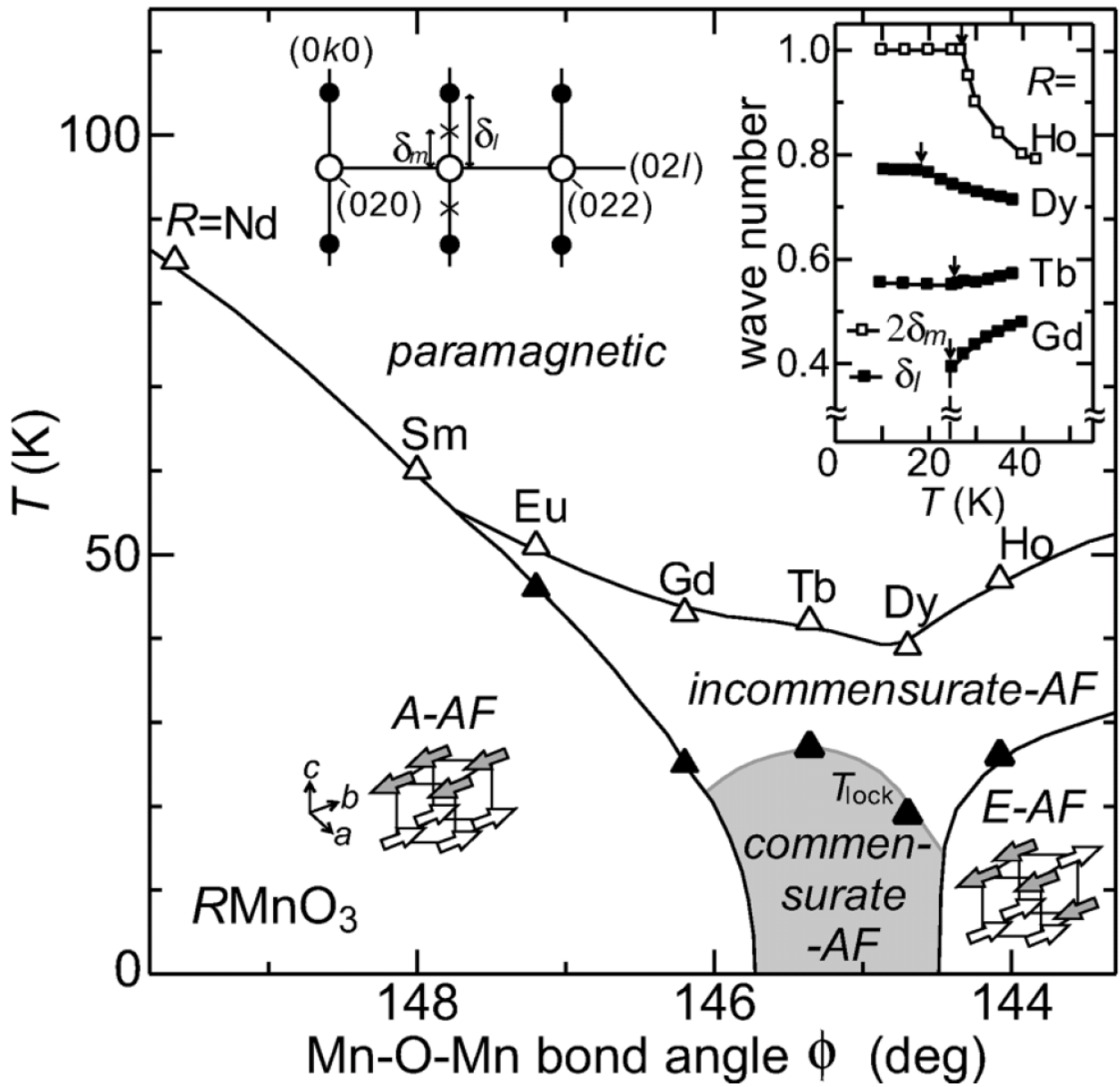

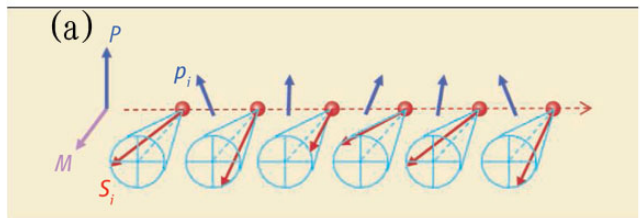

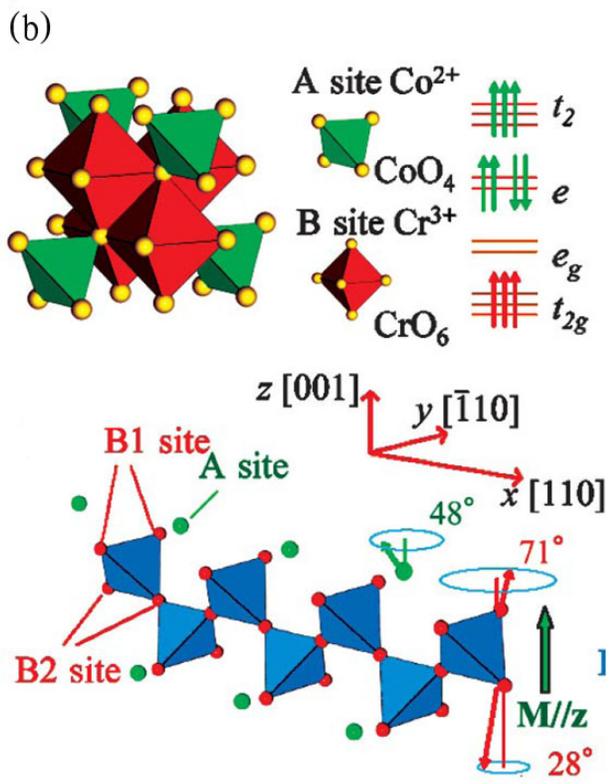

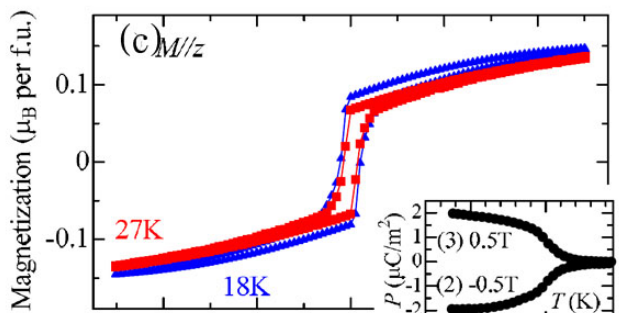

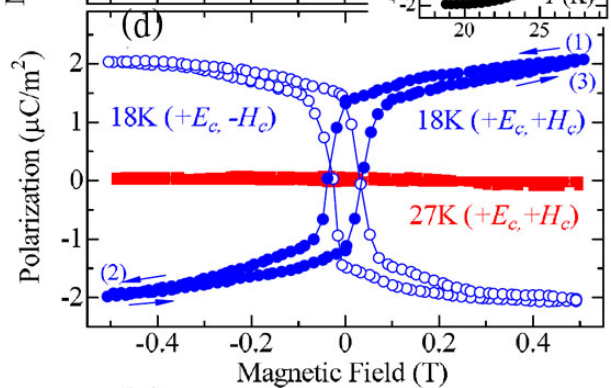

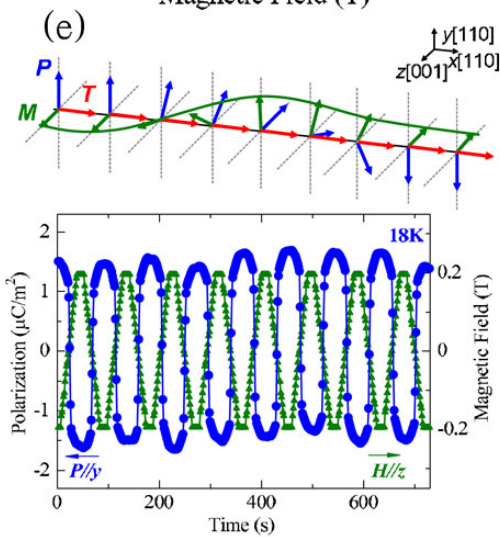

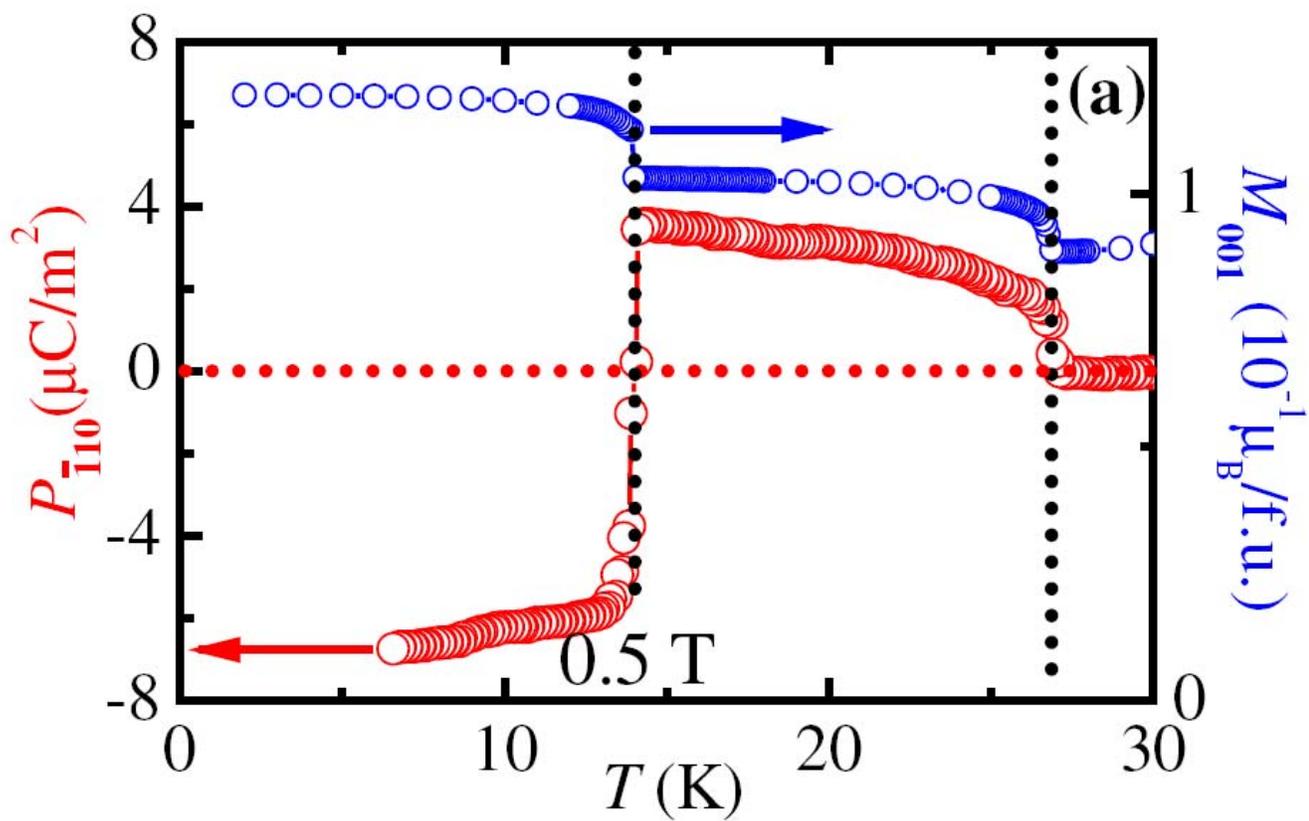

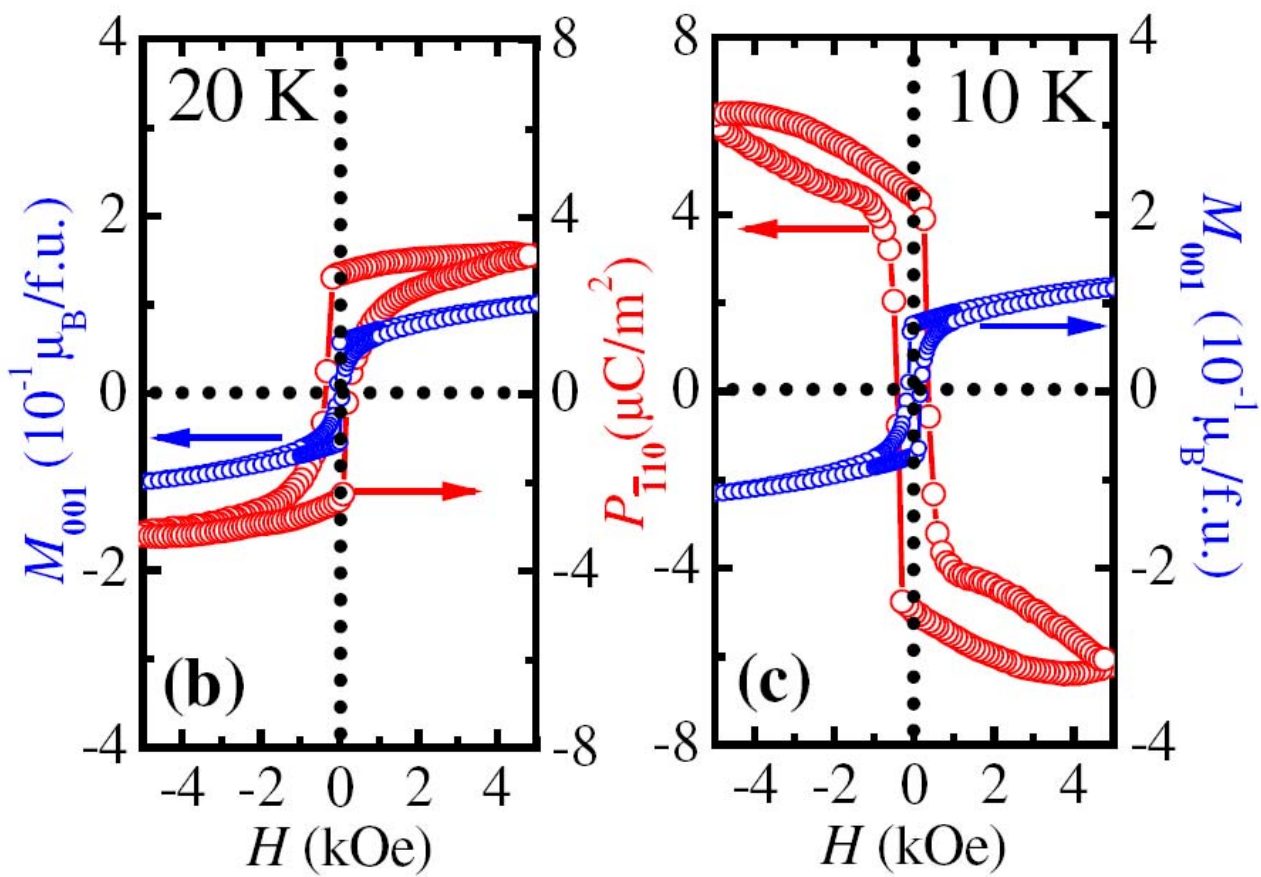

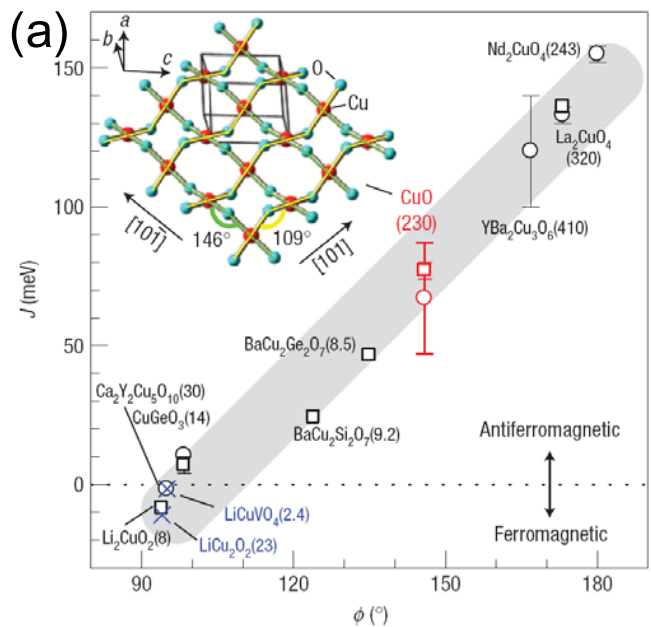

(a)

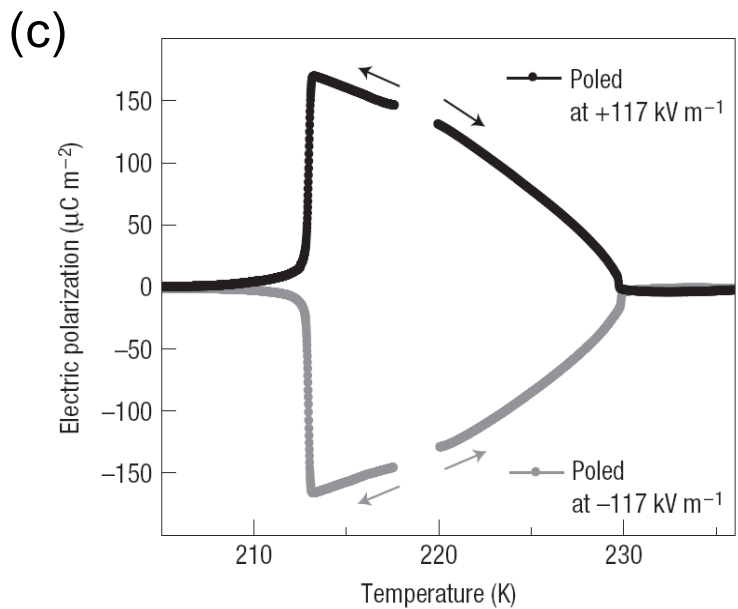

(c)

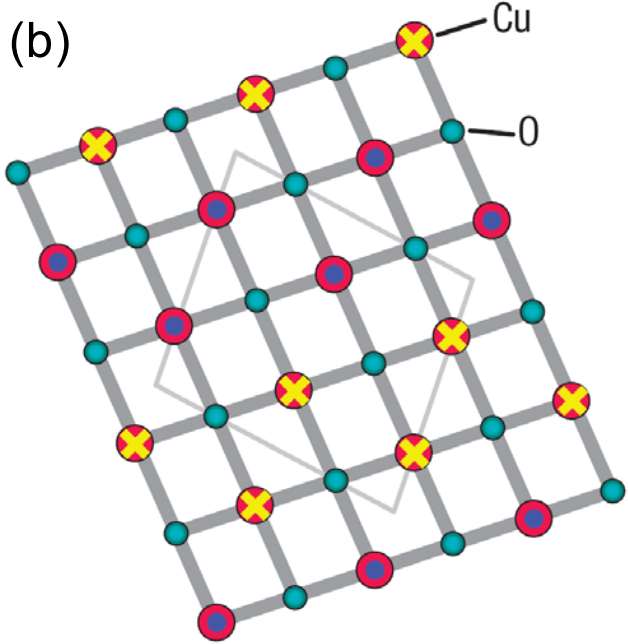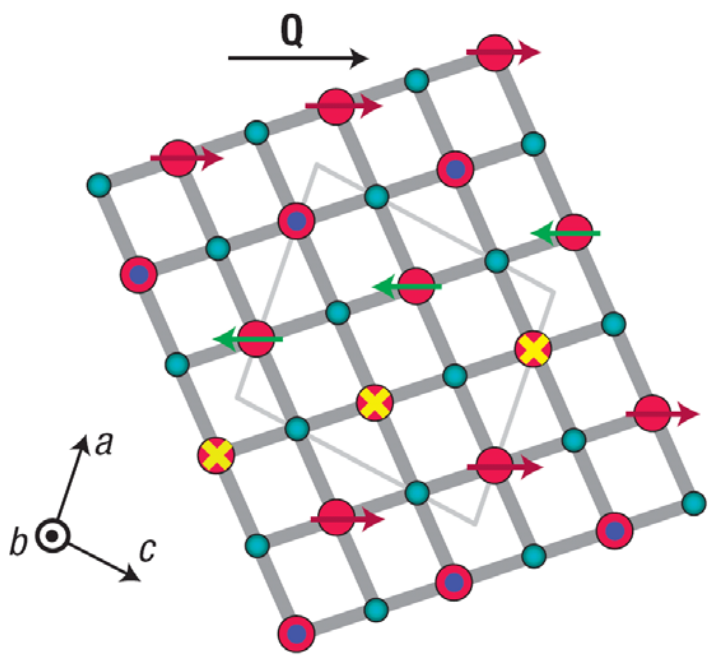

(b)

AF1 (commensurate collinear)

AF2 (incommensurate spiral)

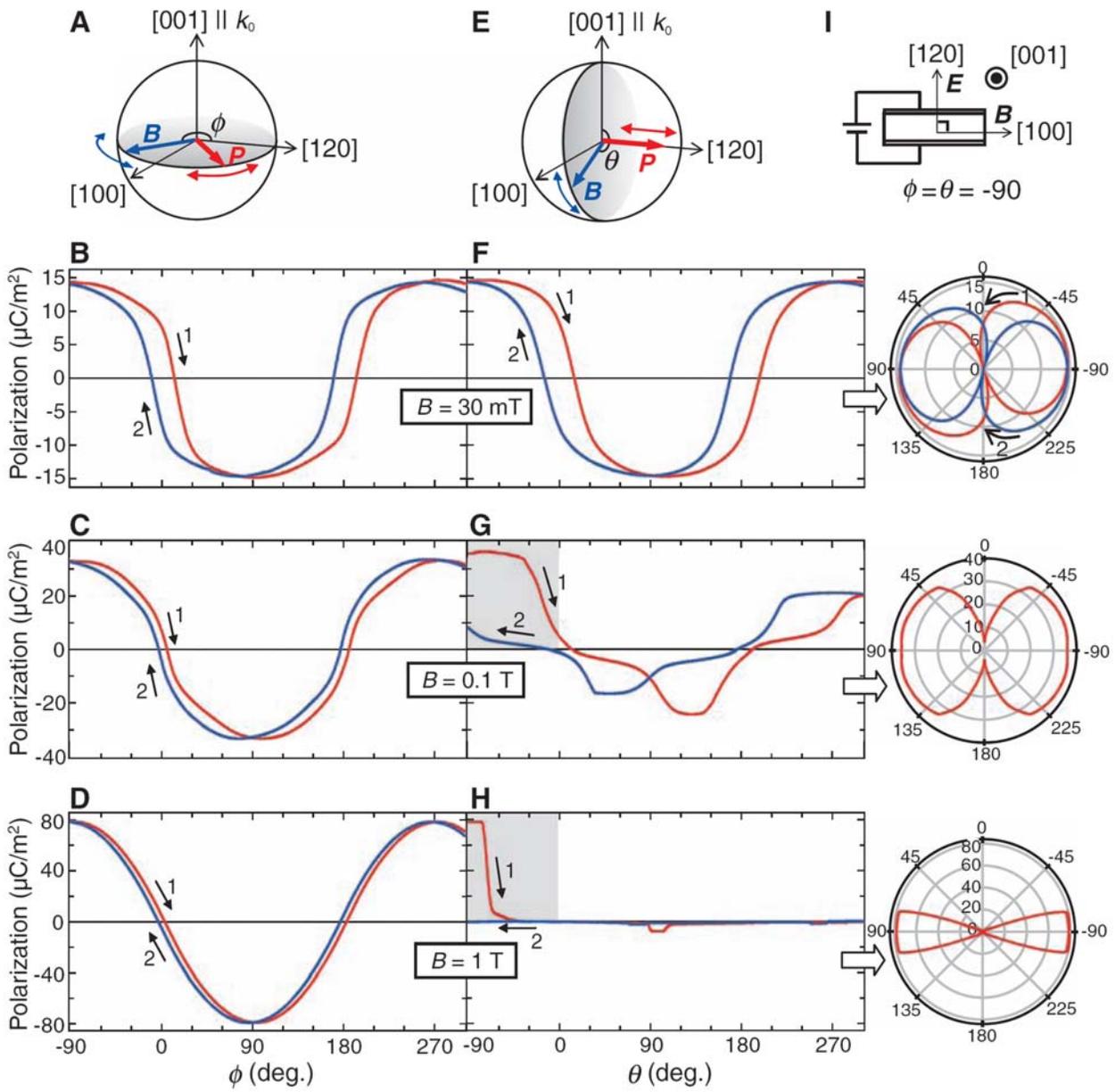

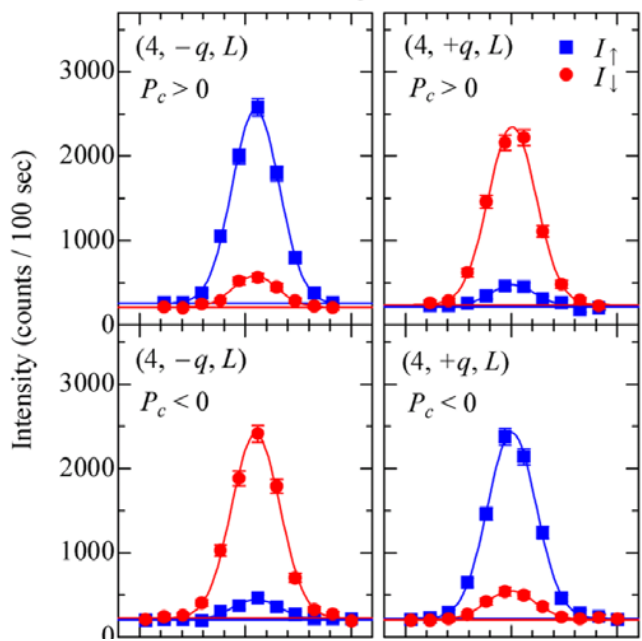

(a) TbMnO₃   T = 9 K

(b)

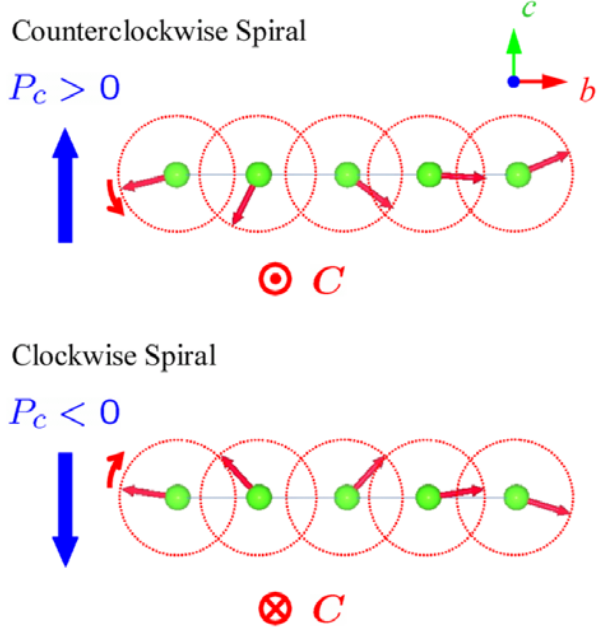

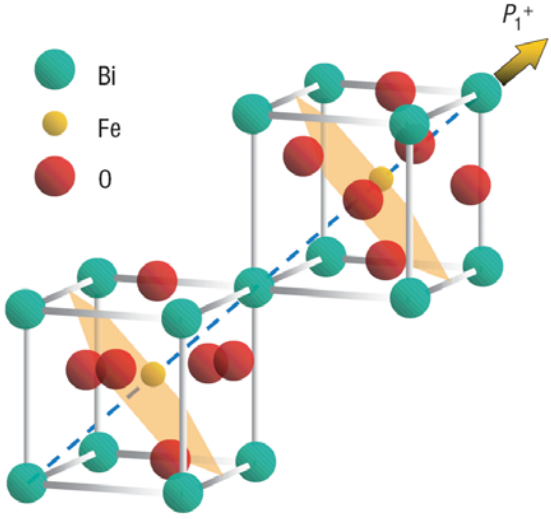

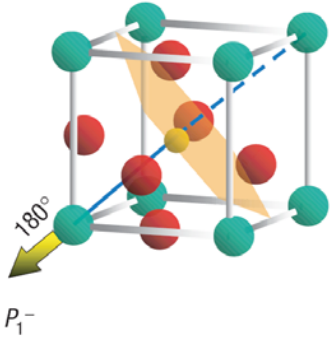
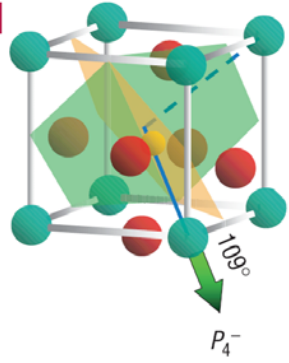
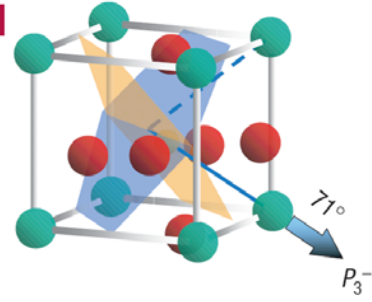

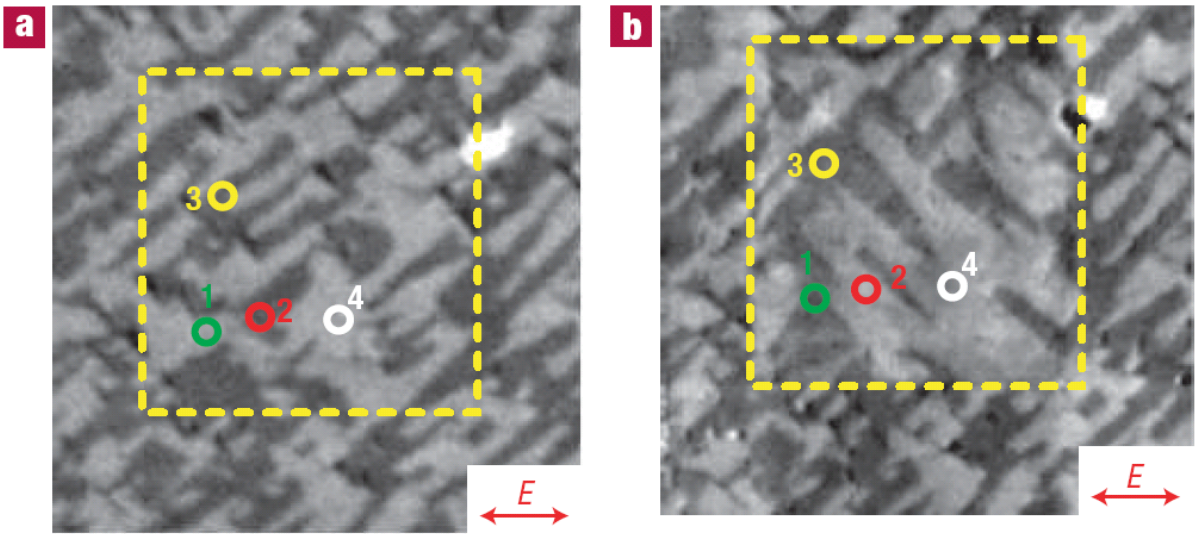

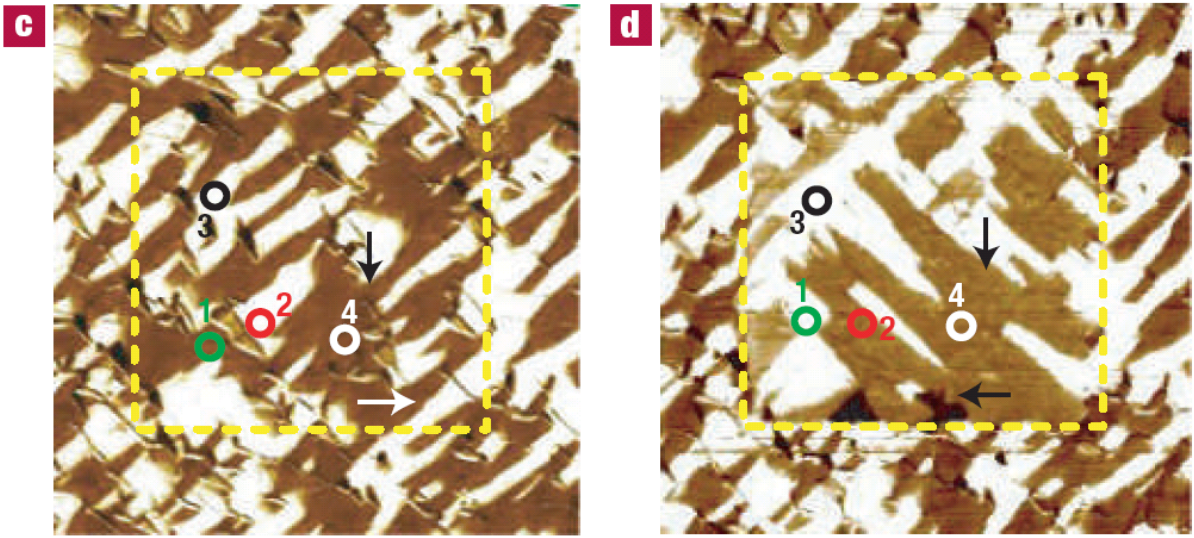

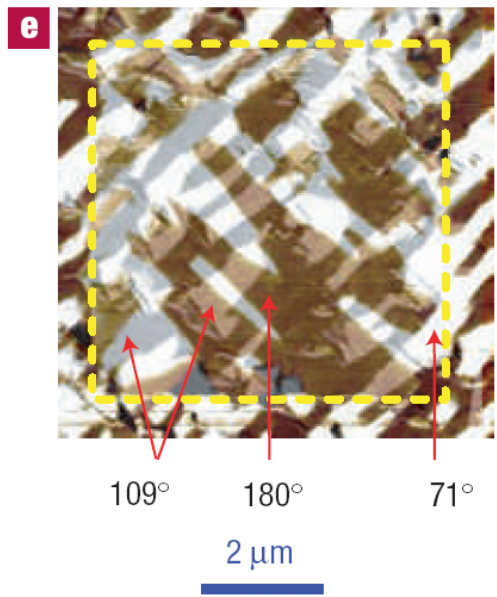

109°    180°    71°

2 μm

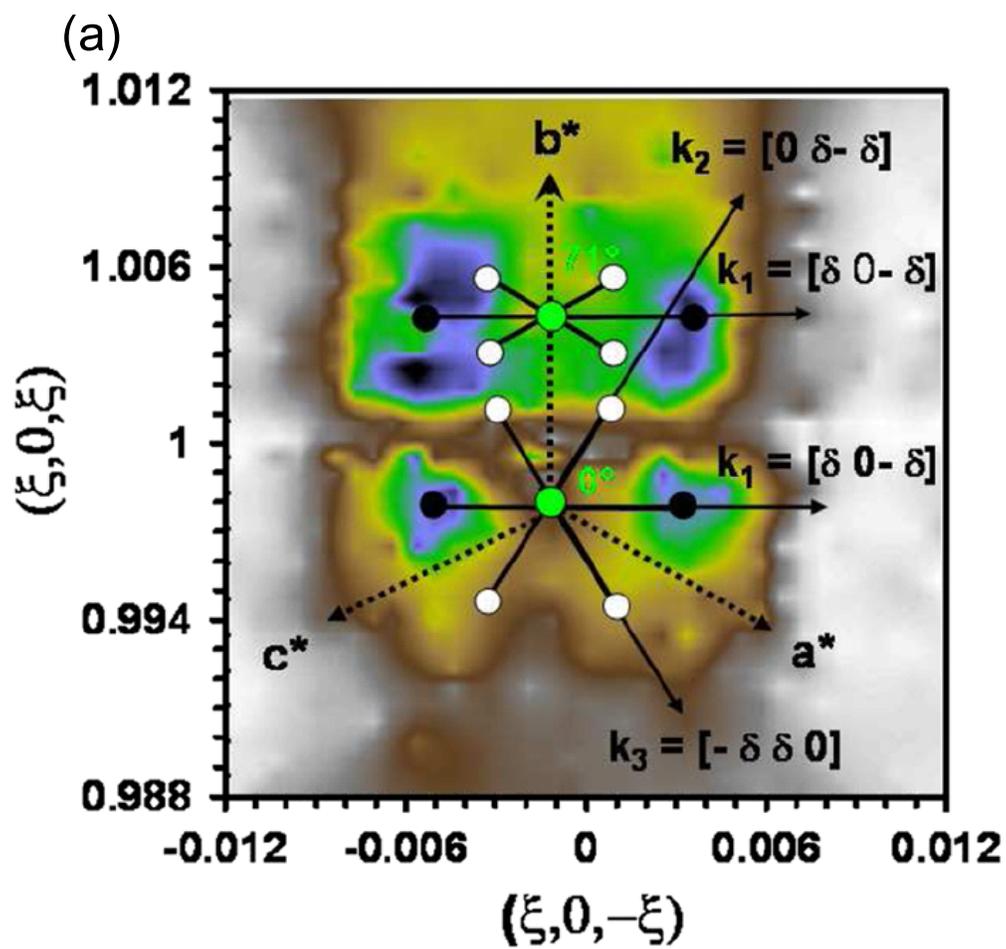

(a)

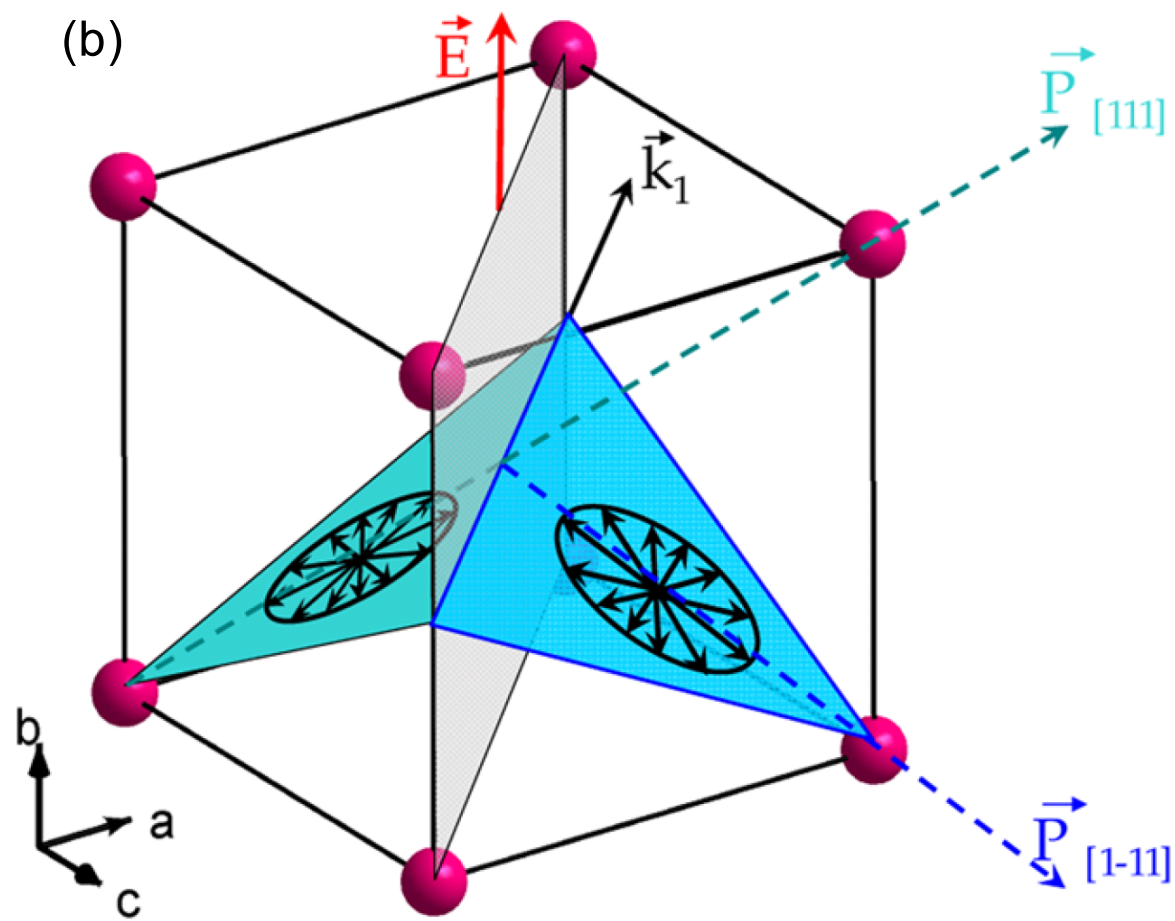

(b)

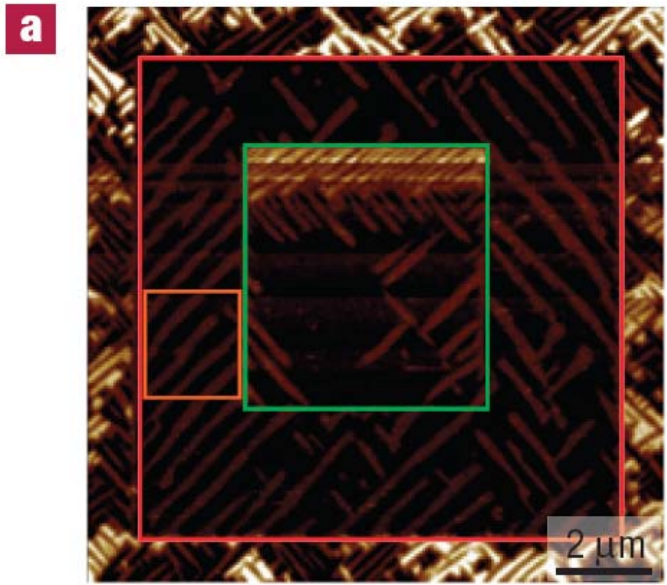

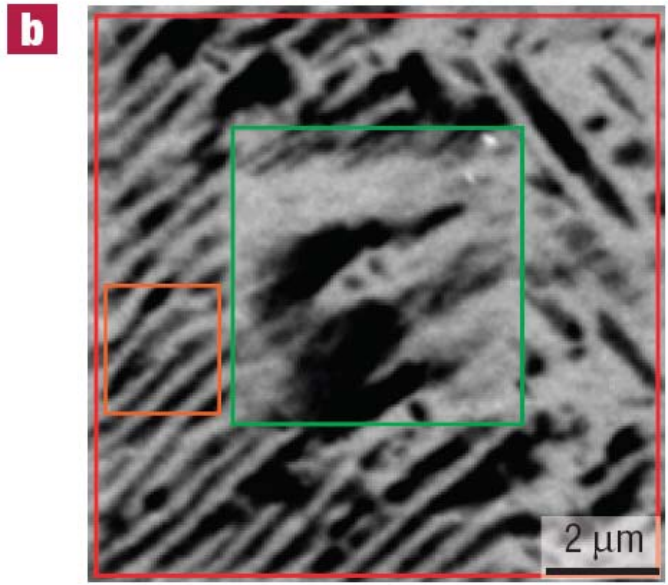

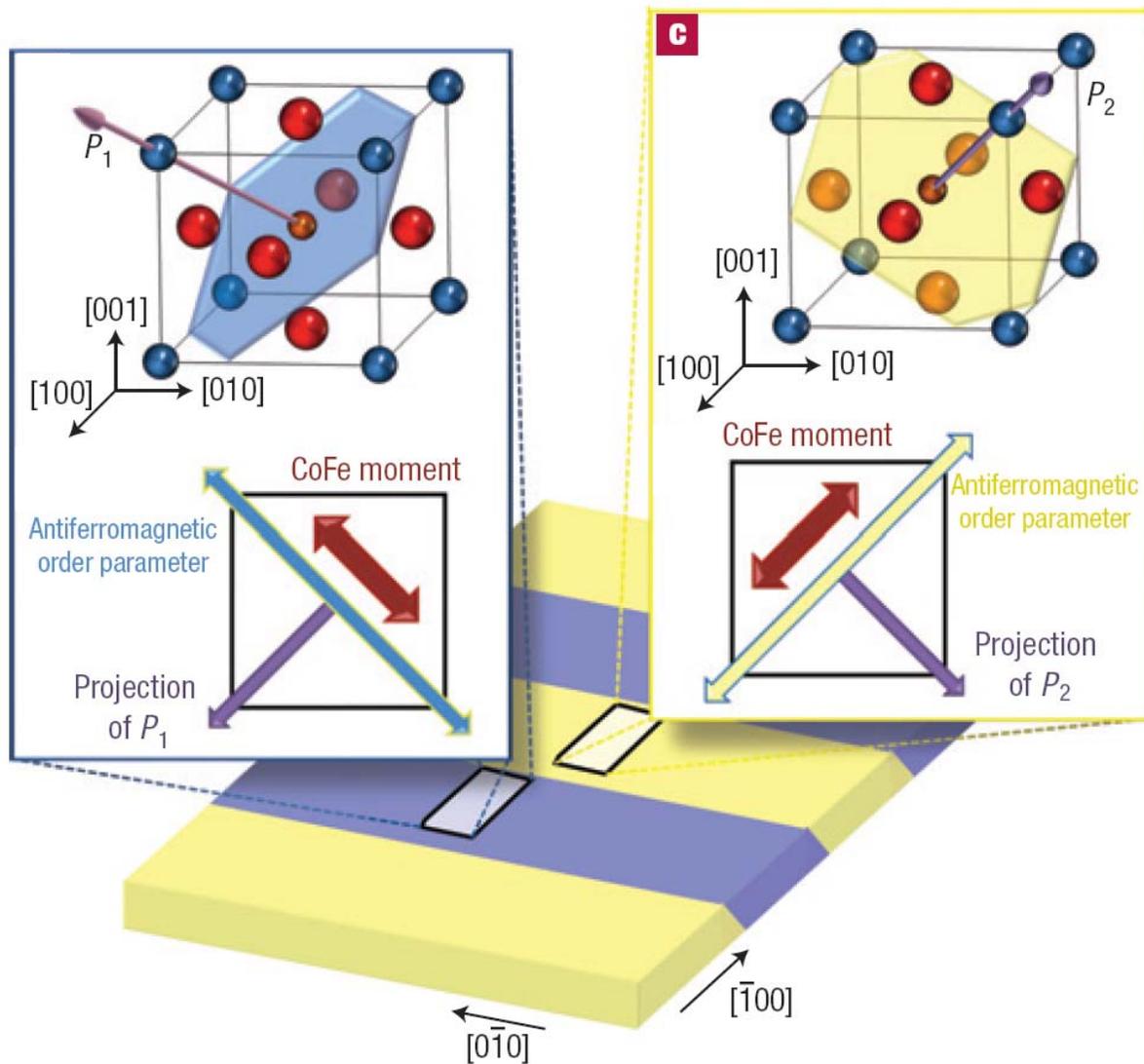

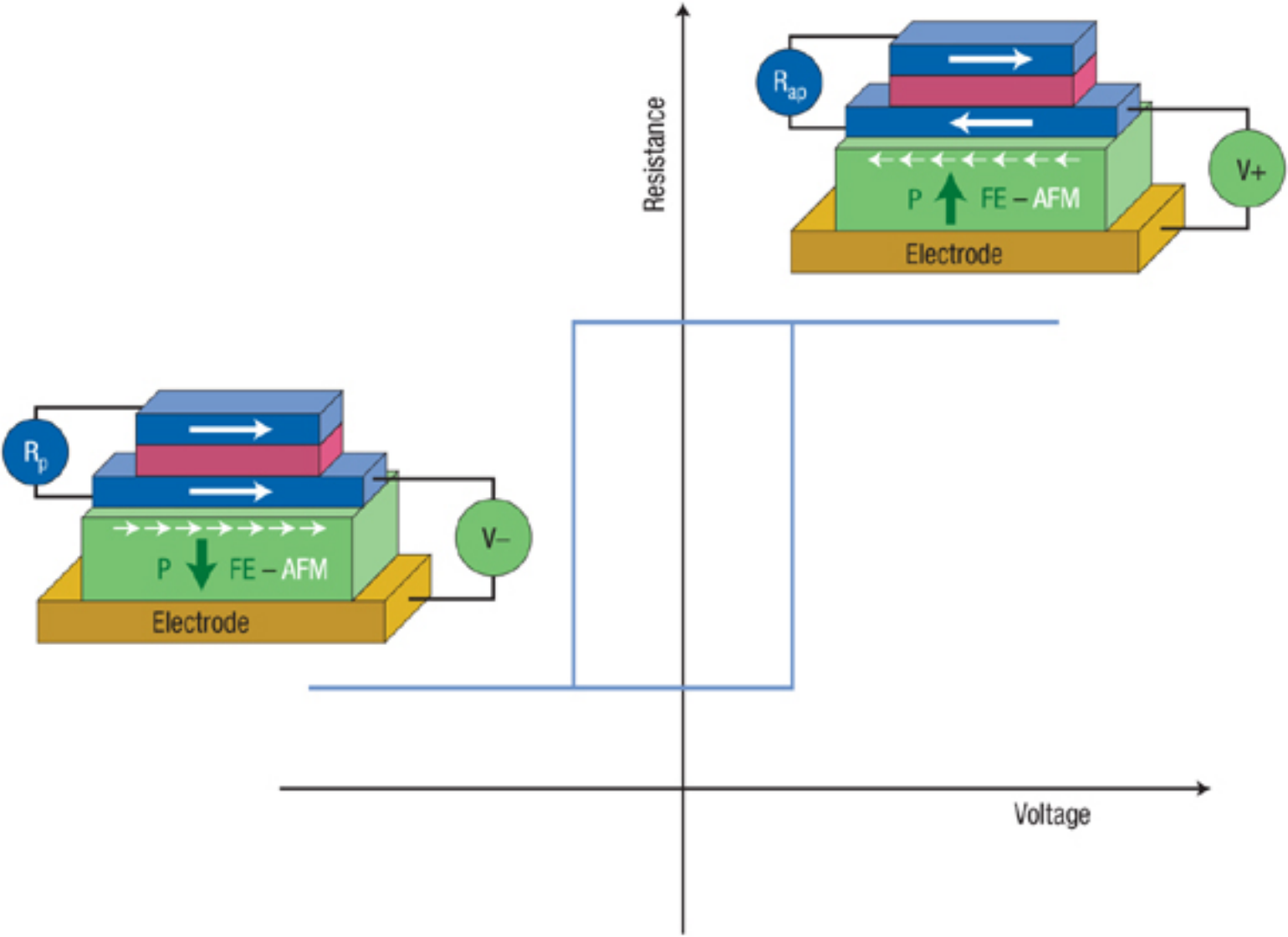

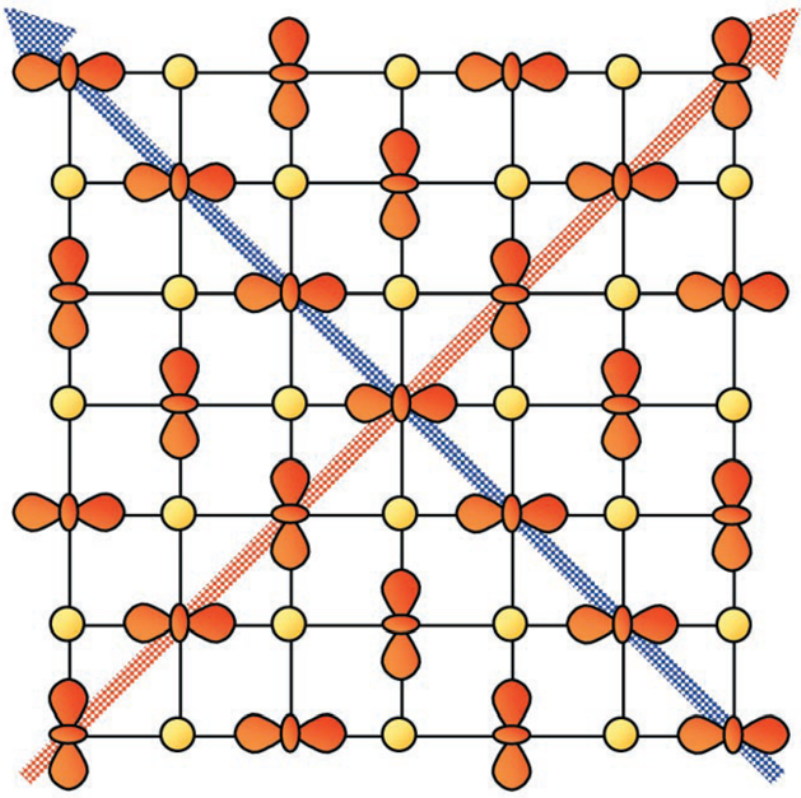

Mn³⁺

Mn⁴⁺

$a$

$b$

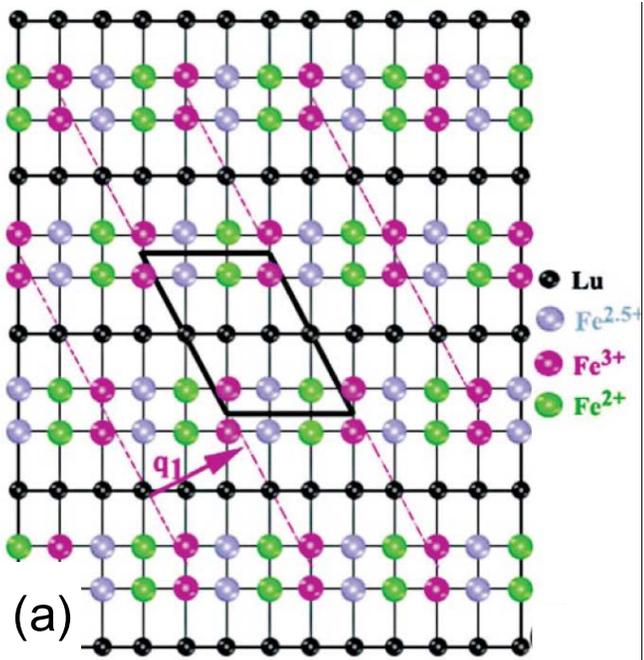

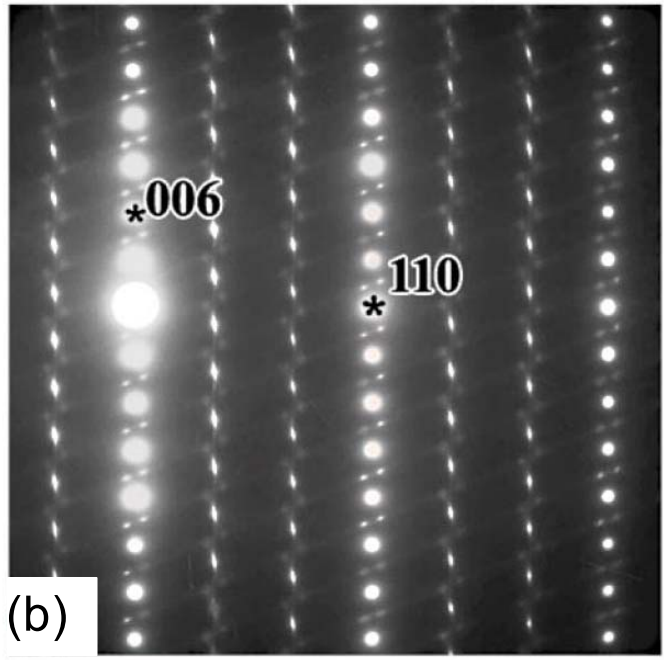

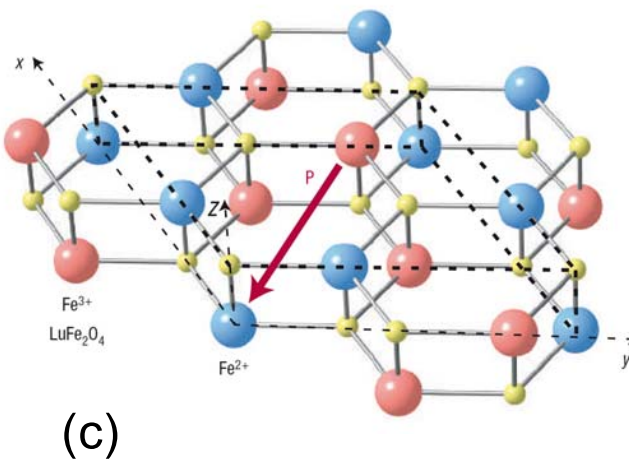

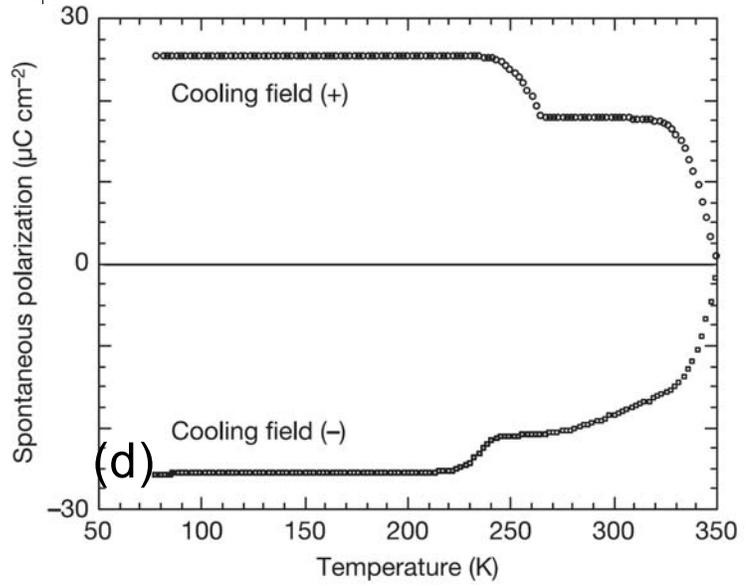

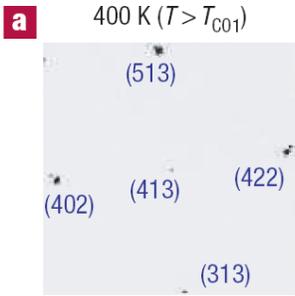

**a** 400 K ($T > T_{CO1}$)

(513)

(402)    (413)    (422)

(313)

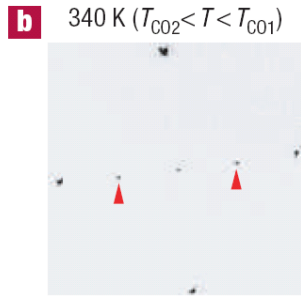

**b** 340 K ($T_{CO2} < T < T_{CO1}$)

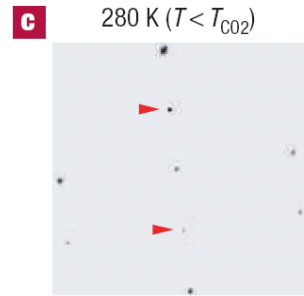

**c** 280 K ($T < T_{CO2}$)

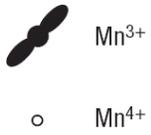

⬥ Mn³⁺

○ Mn⁴⁺

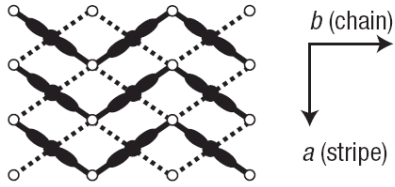

$b$ (chain)

$a$ (stripe)

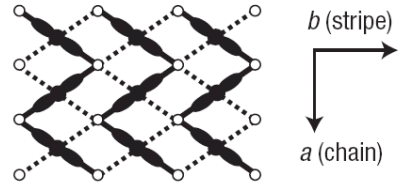

$b$ (stripe)

$a$ (chain)

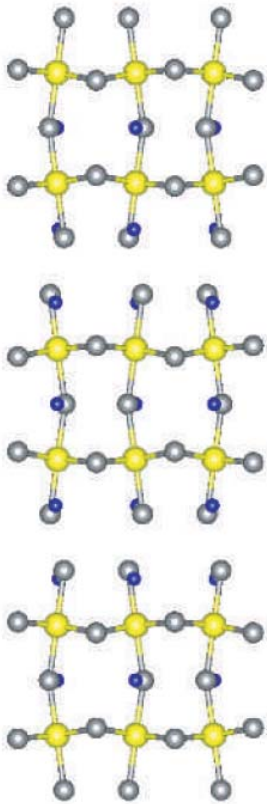

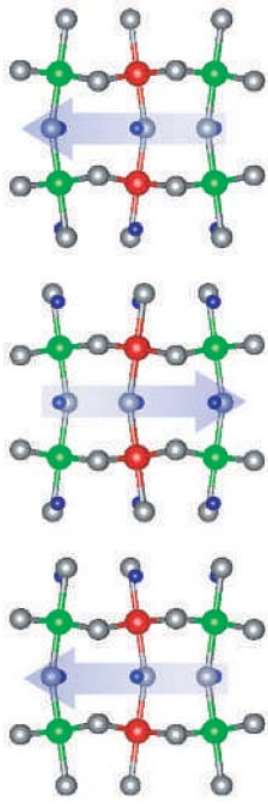

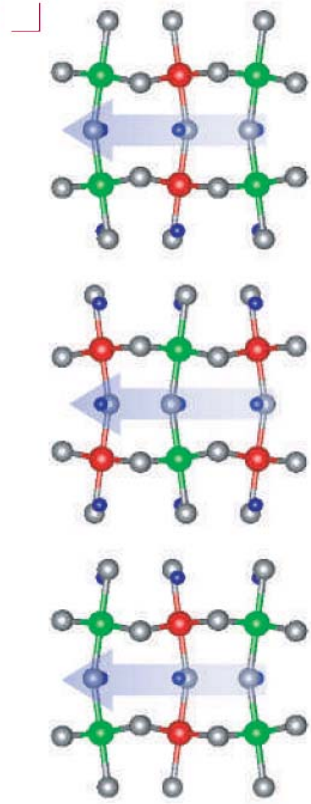

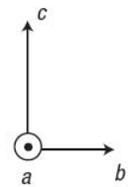

● Pr, Sr, Ca

● Mn³·⁵⁺

● Mn³⁺

● Mn⁴⁺

● O

$c$

$b$

$a$

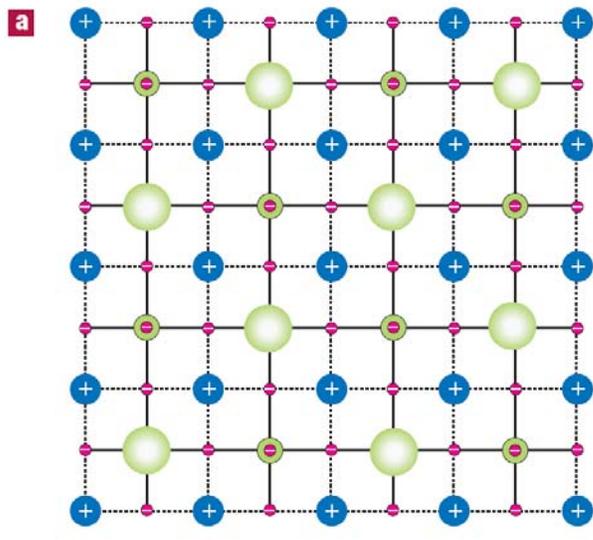

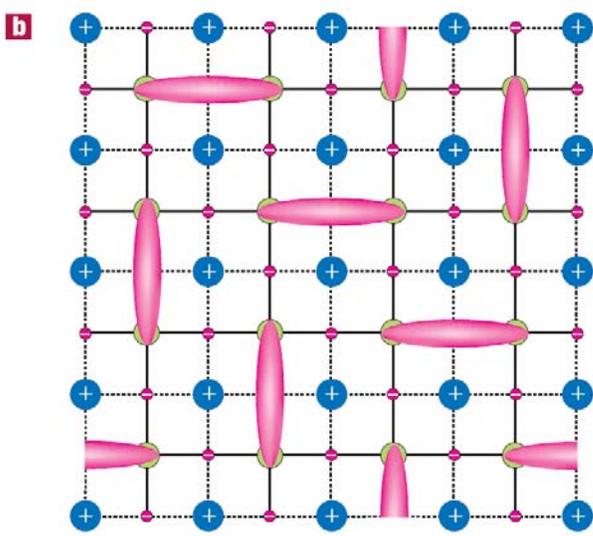

+

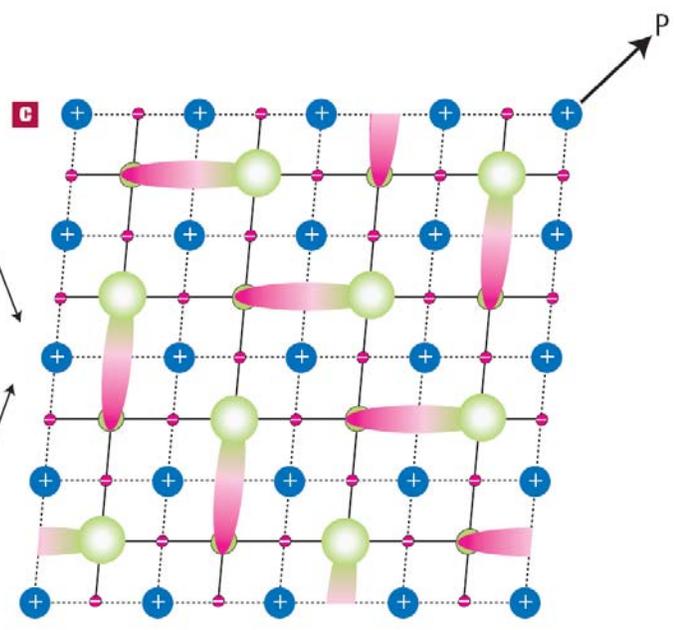

P

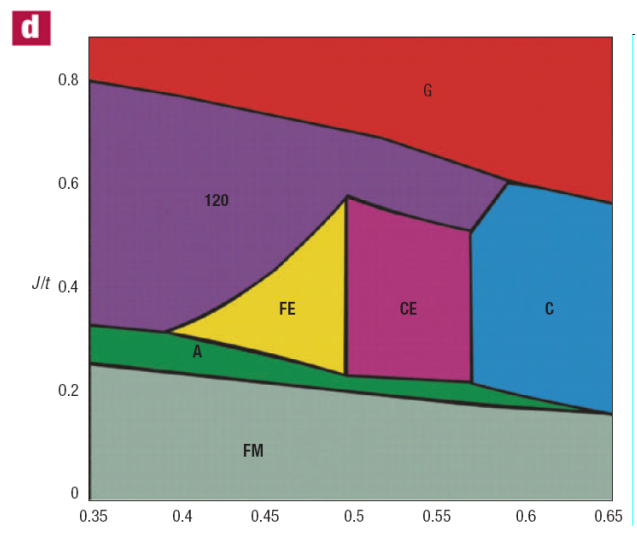

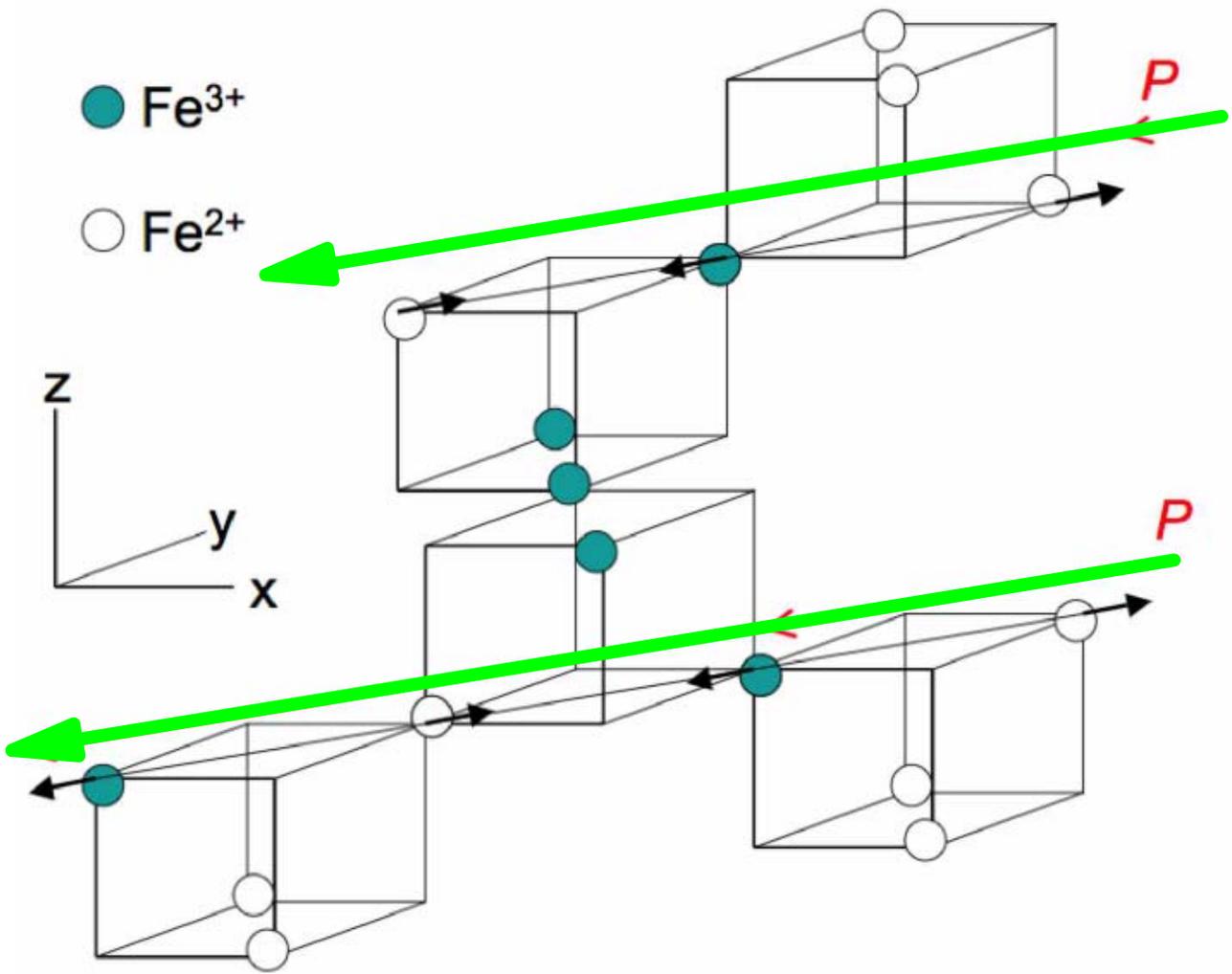

**(a)**

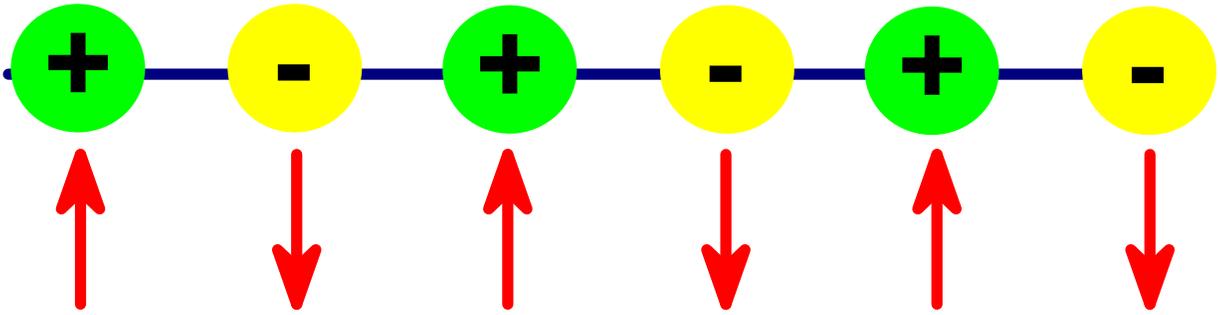

**(b)**

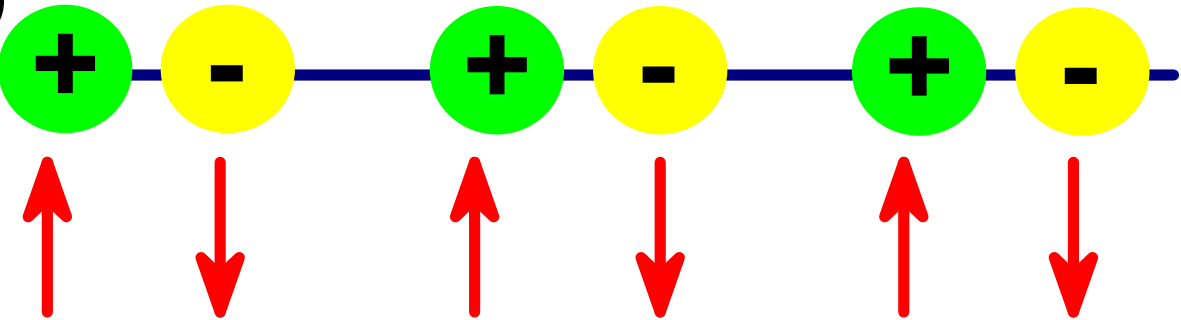

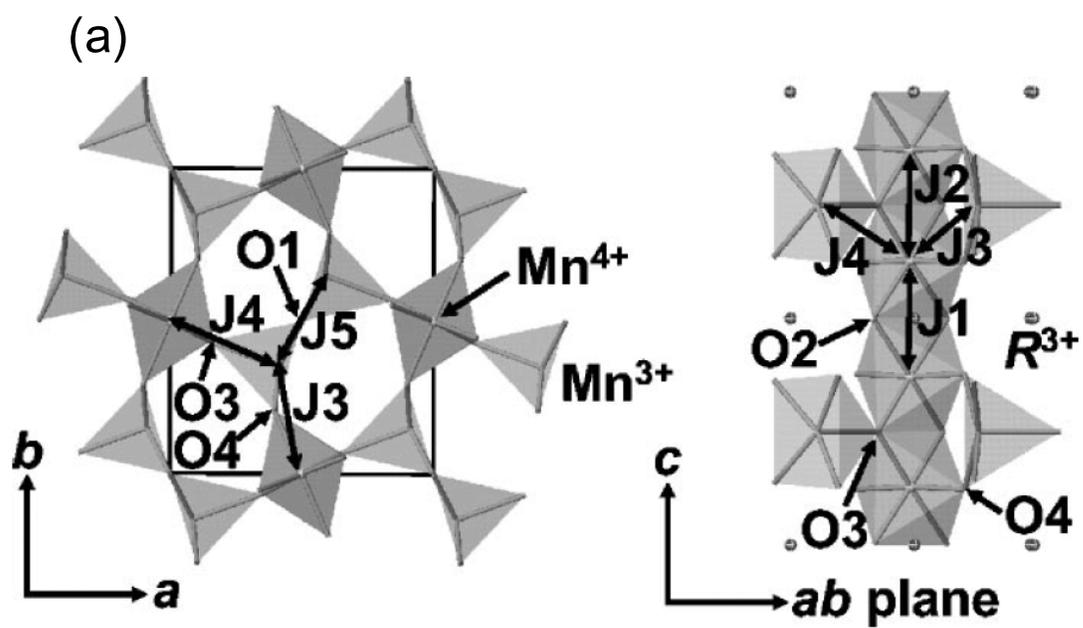

(a)

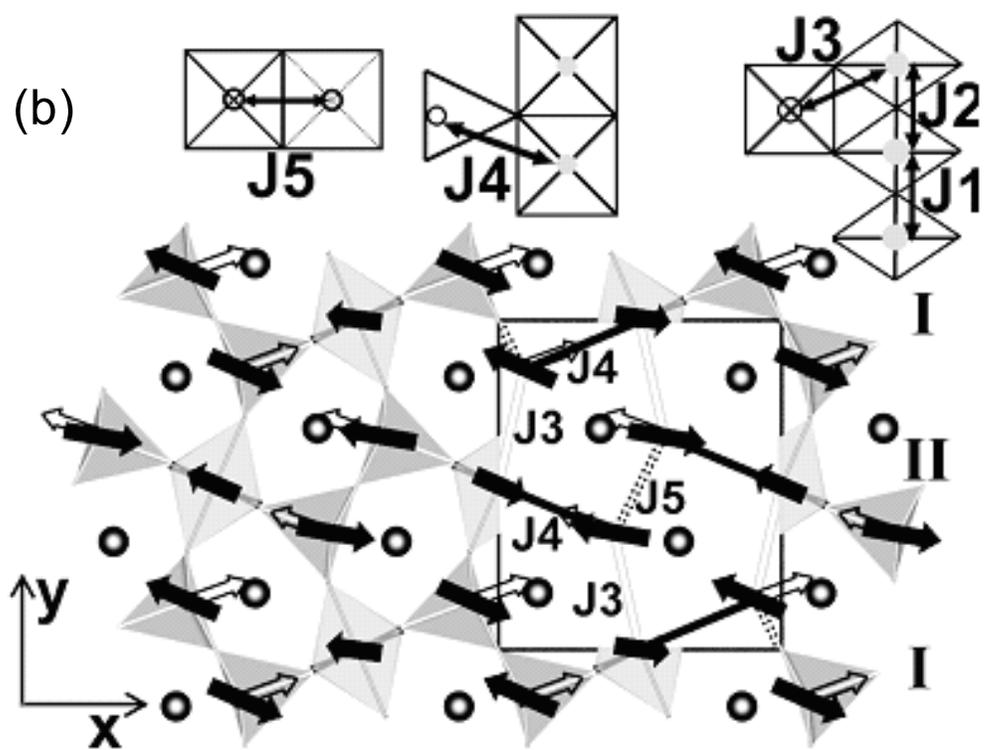

(b)

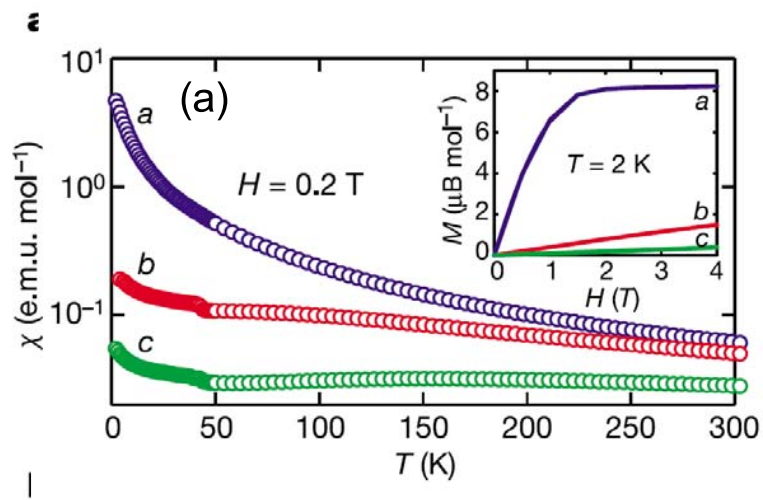

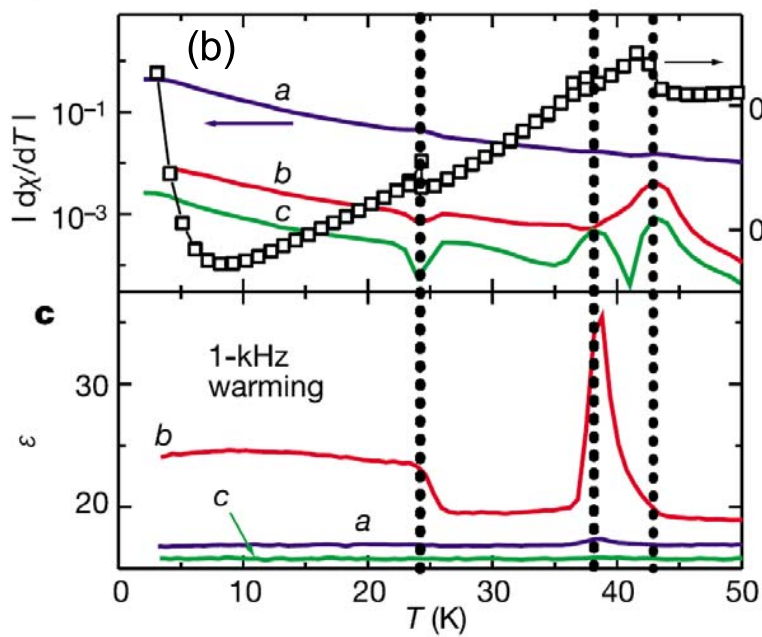

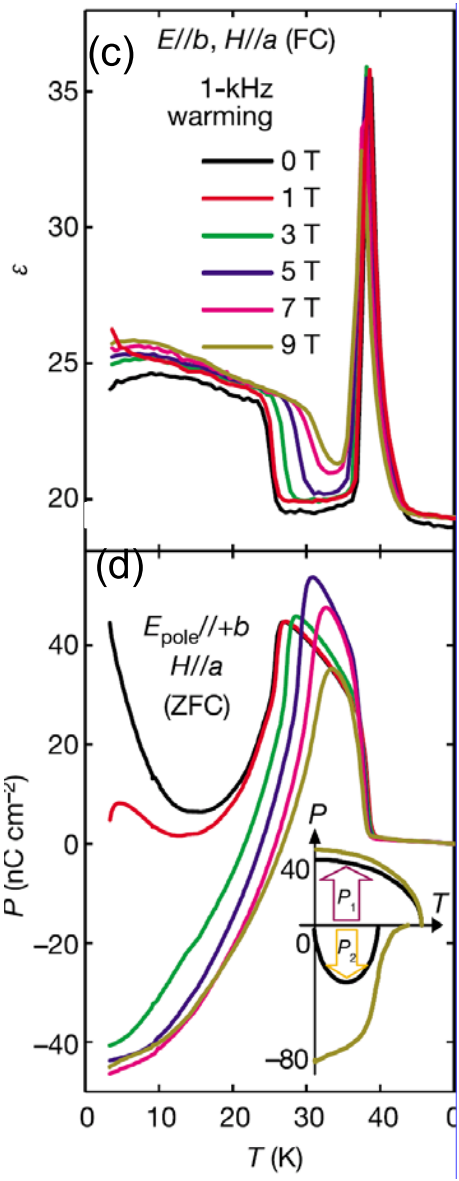

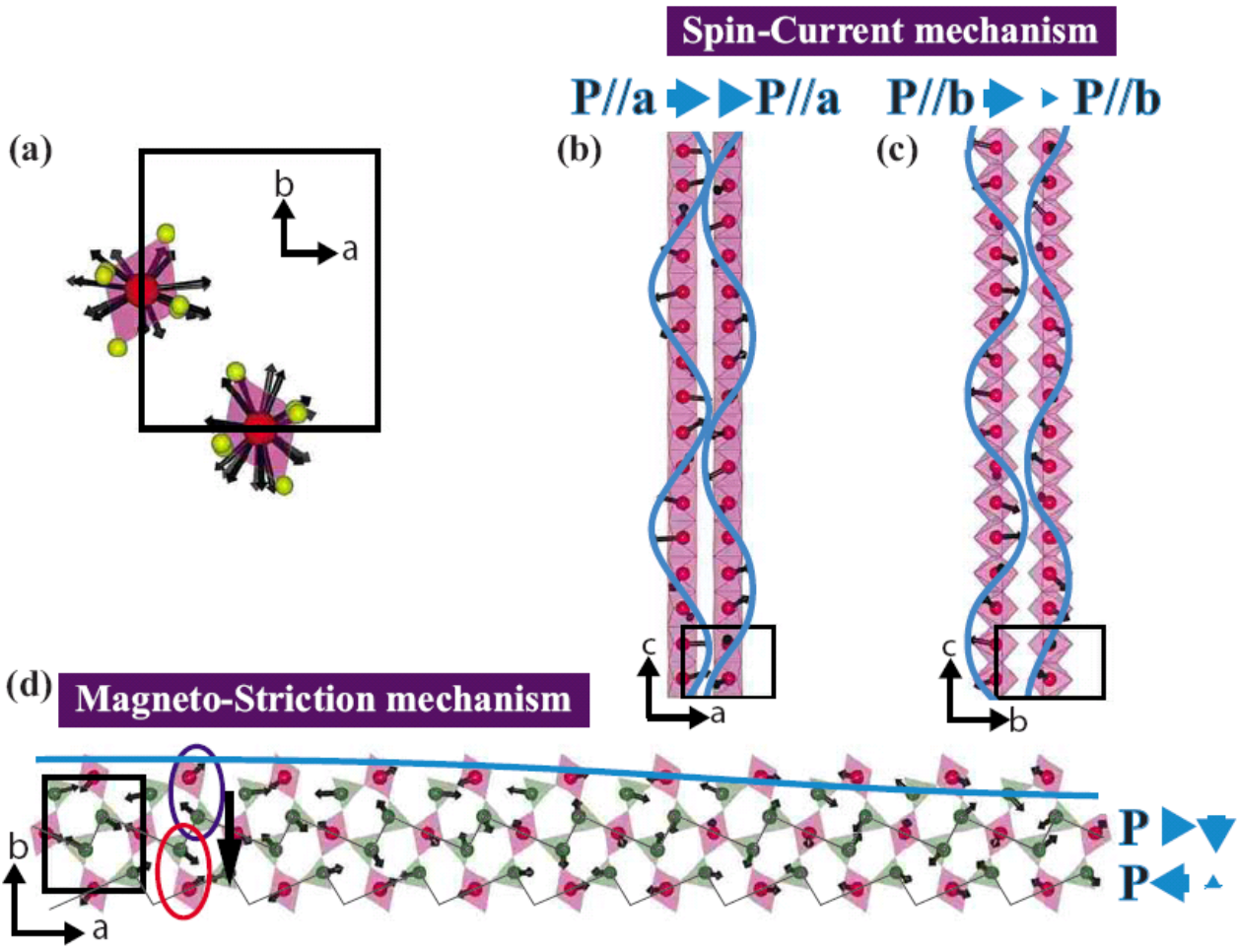

**Spin-Current mechanism**

P//a ▶▶ P//a    P//b ▶ ▶ P//b

(a)

(b)

(c)

**(d) Magneto-Striction mechanism**

P ▶▼

P ◀▲

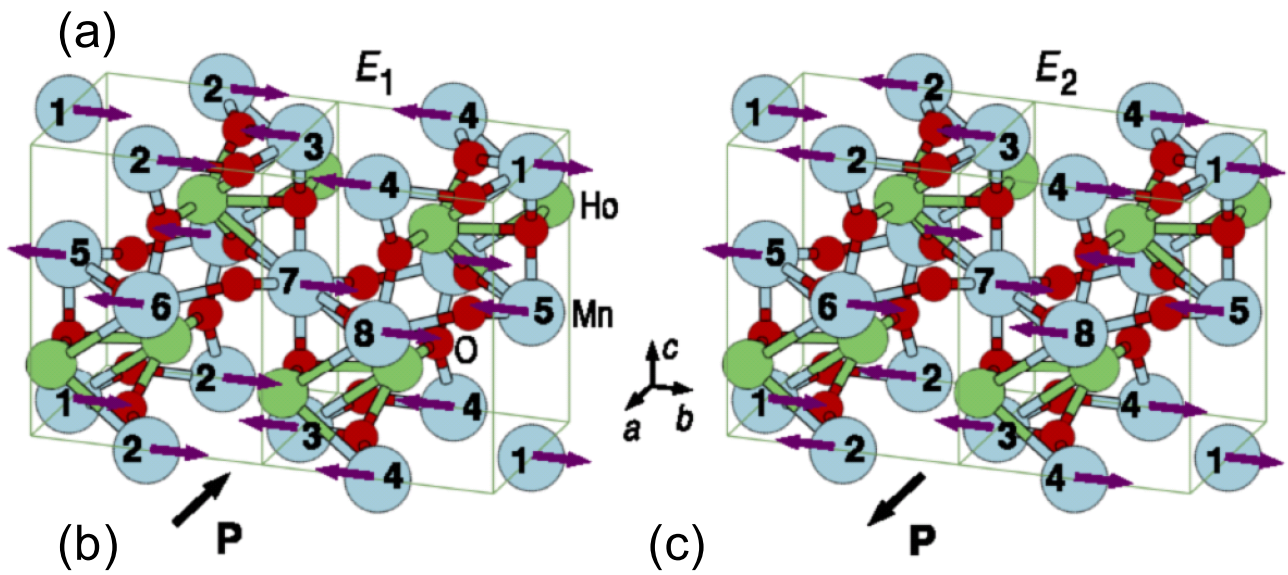

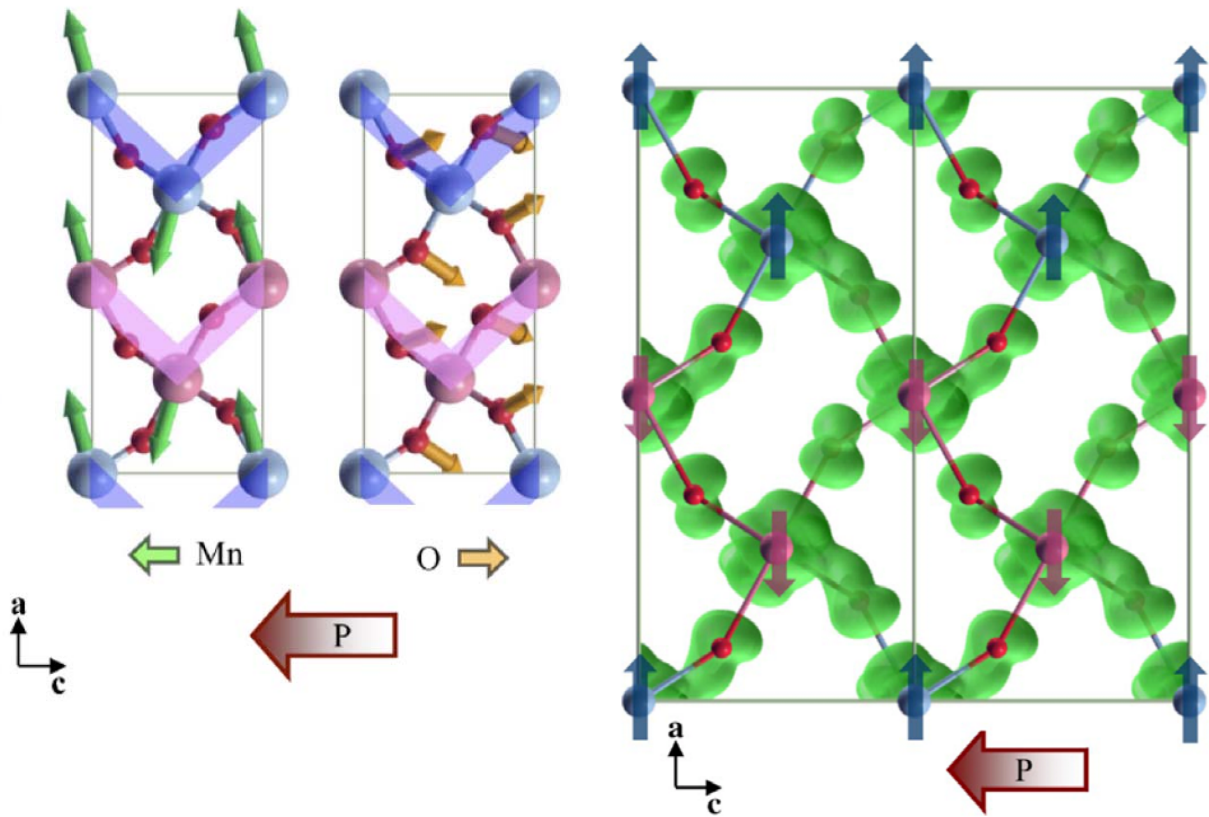

(a)

(b)

(c)

(d)

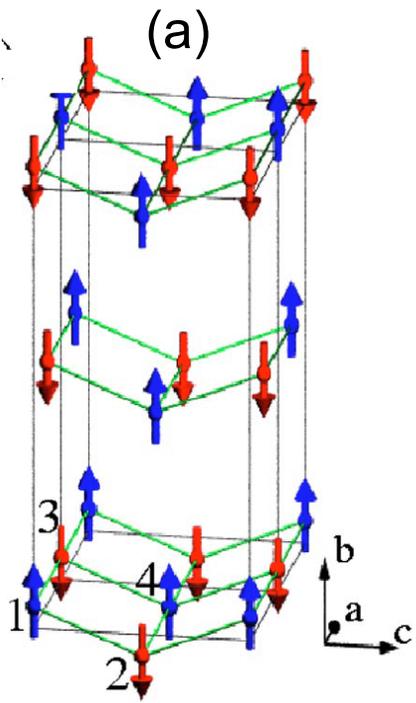
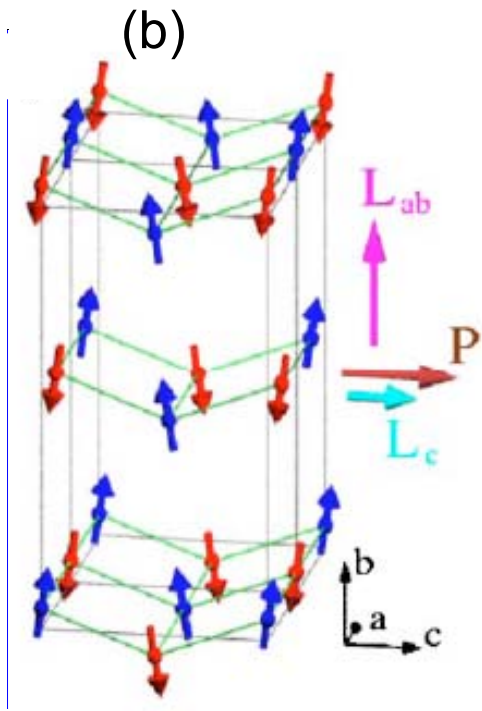
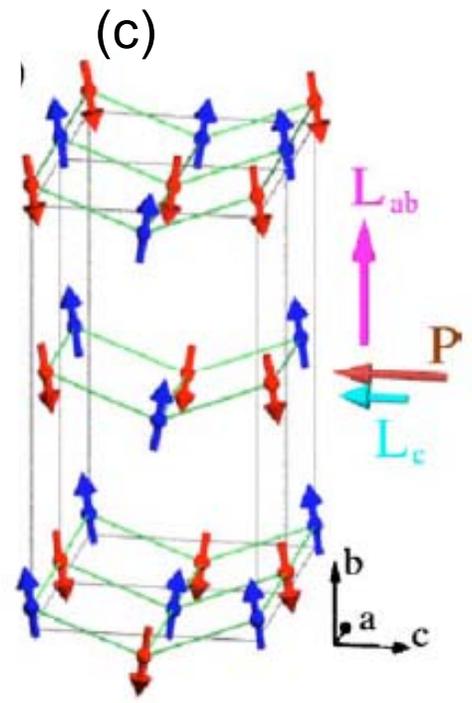

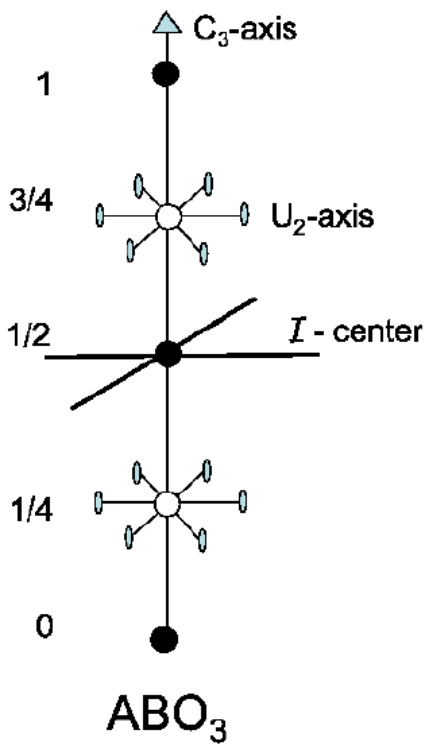

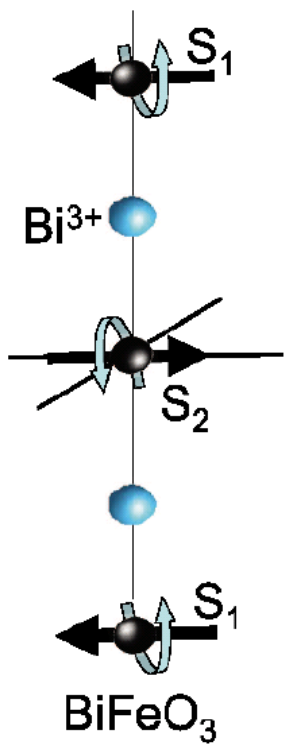

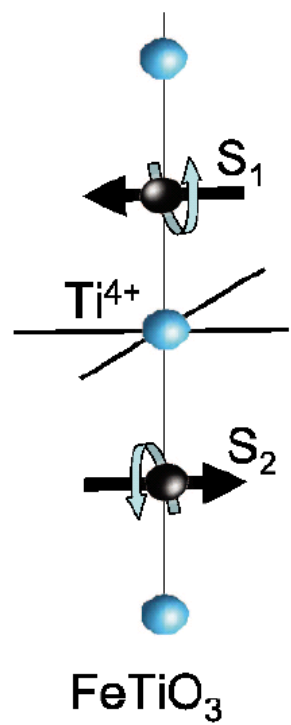

3/4     $U_2$-axis

1/2     $I$ - center

1/4

0

**ABO₃**

○ A-site: (1/4,1/4,1/4) (3/4,3/4,3/4)

● B-site: (0,0,0) (1/2,1/2,1/2)

(a)

$C_3$-axis

$S_1$

$Bi^{3+}$

$S_2$

$S_1$

**BiFeO₃**

$L = S_1 - S_2$

$I\,L = I\,(S_1 - S_2)$
$= (S_1 - S_2)$
$= +L$

(b)

$S_1$

$Ti^{4+}$

$S_2$

**FeTiO₃**

$L = S_1 - S_2$

$I\,L = I\,(S_1 - S_2)$
$= (S_2 - S_1)$
$= -L$

(c)

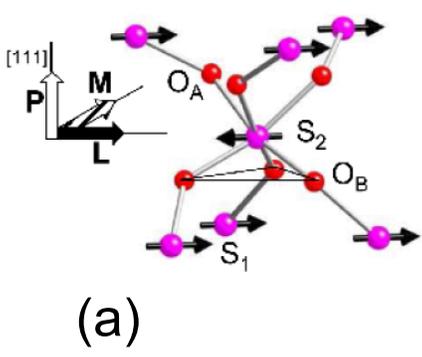 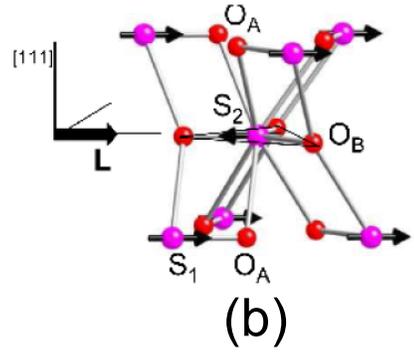 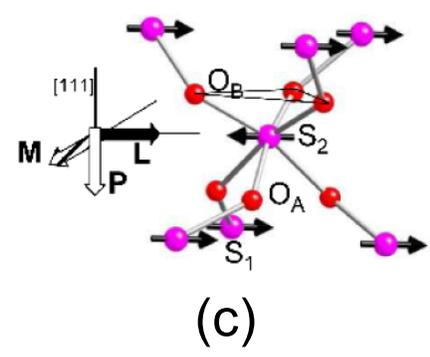

(a)                    (b)                    (c)

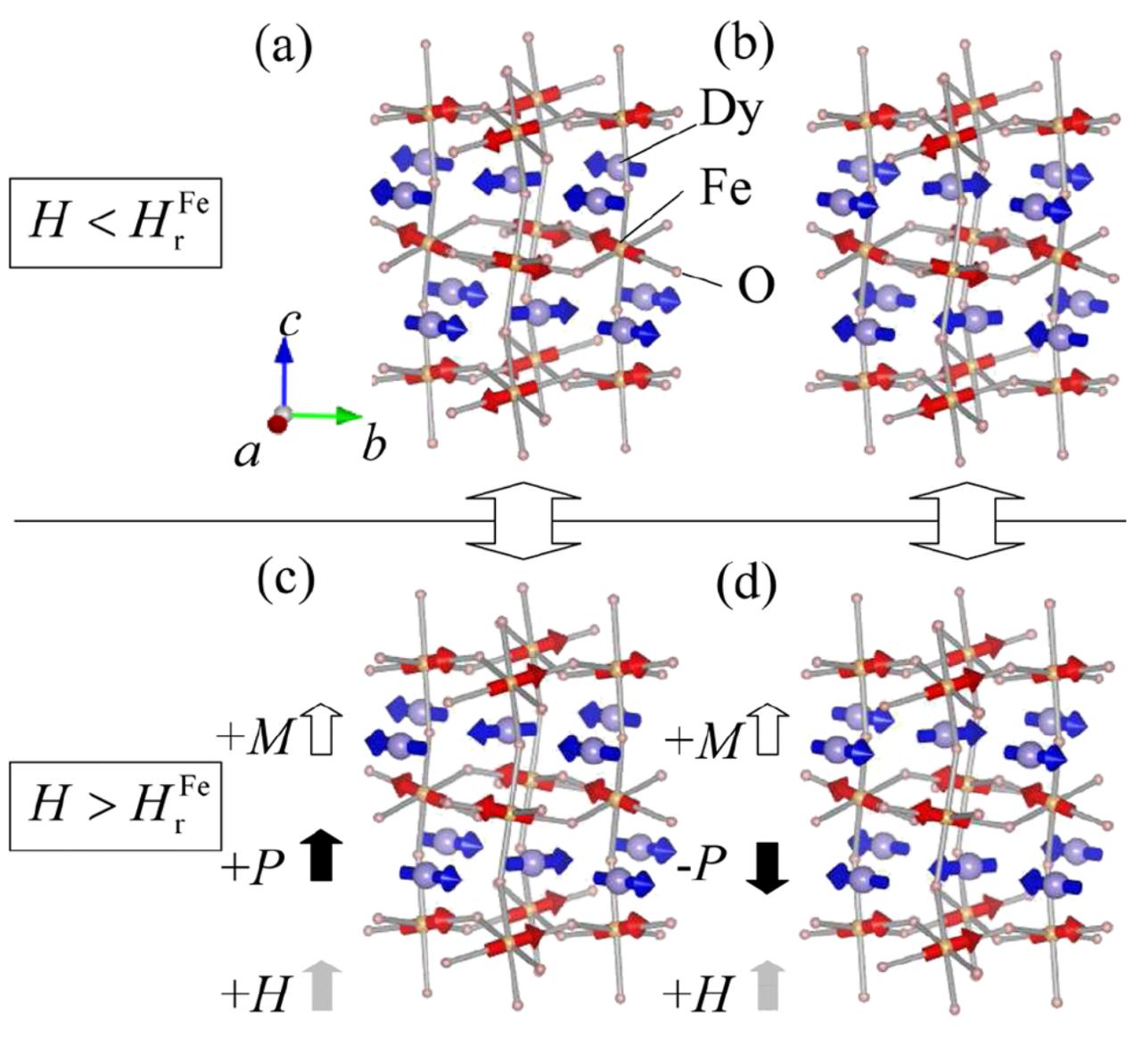

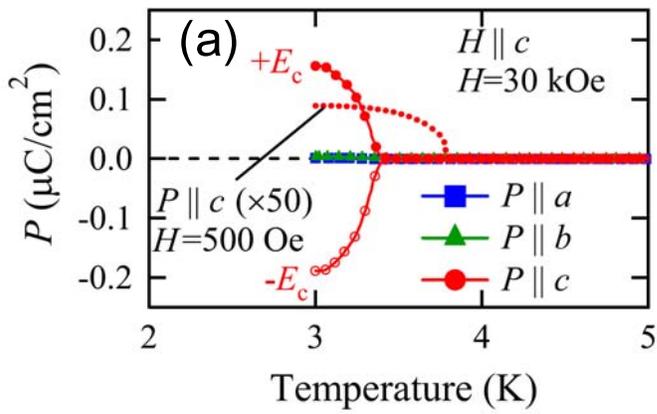

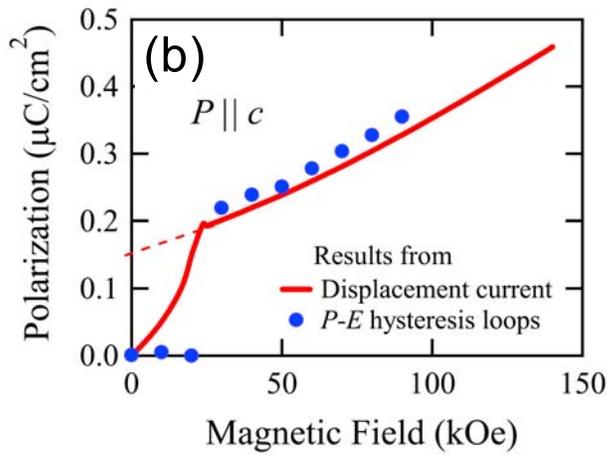

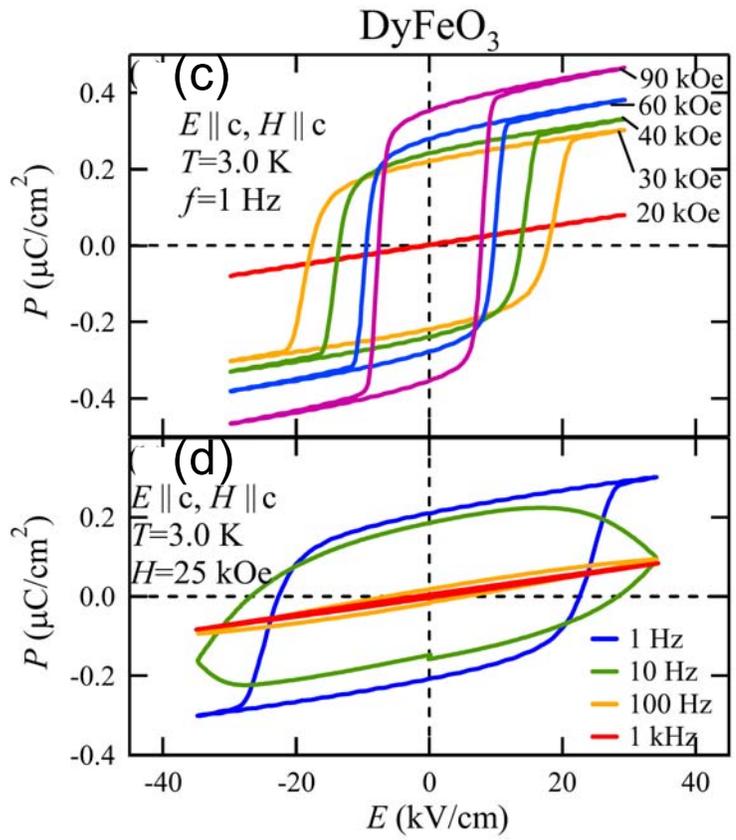

DyFeO$_3$

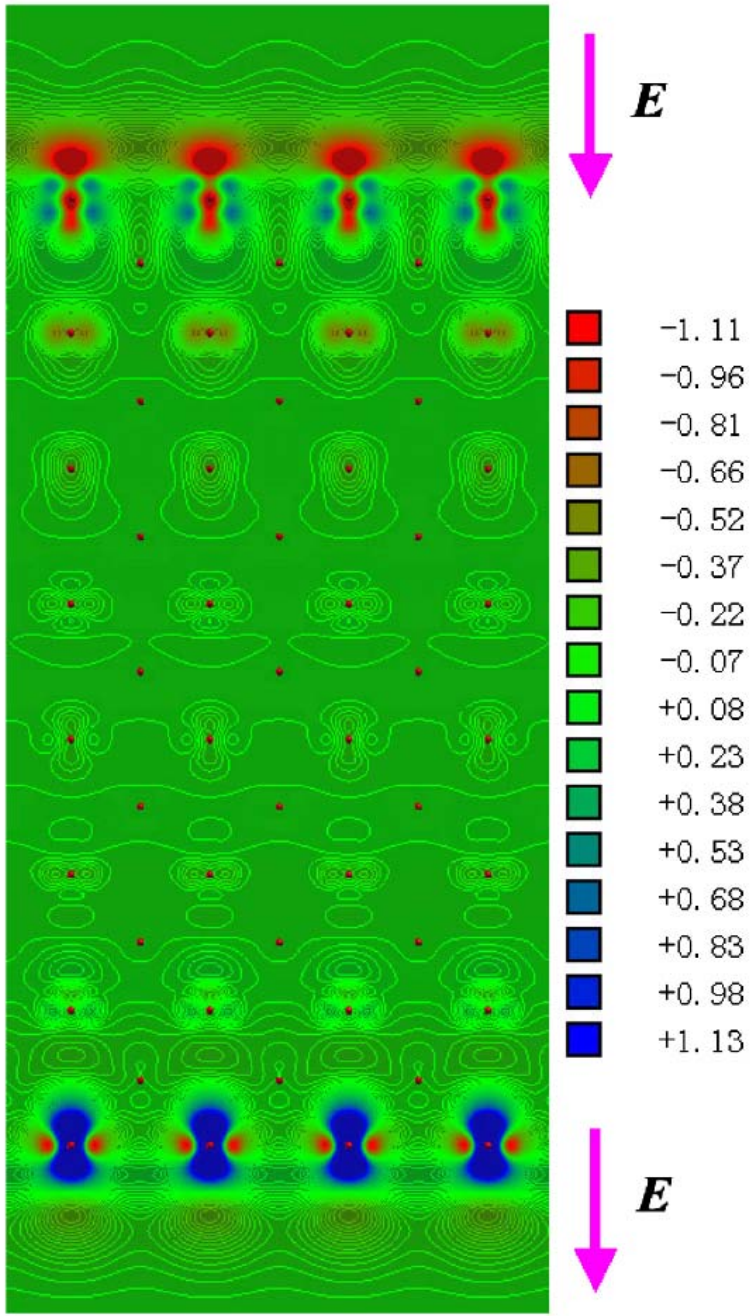

| | |
|---|---|
| ■ | -1.11 |
| ■ | -0.96 |
| ■ | -0.81 |
| ■ | -0.66 |
| ■ | -0.52 |
| ■ | -0.37 |
| ■ | -0.22 |
| ■ | -0.07 |
| ■ | +0.08 |
| ■ | +0.23 |
| ■ | +0.38 |
| ■ | +0.53 |
| ■ | +0.68 |
| ■ | +0.83 |
| ■ | +0.98 |
| ■ | +1.13 |

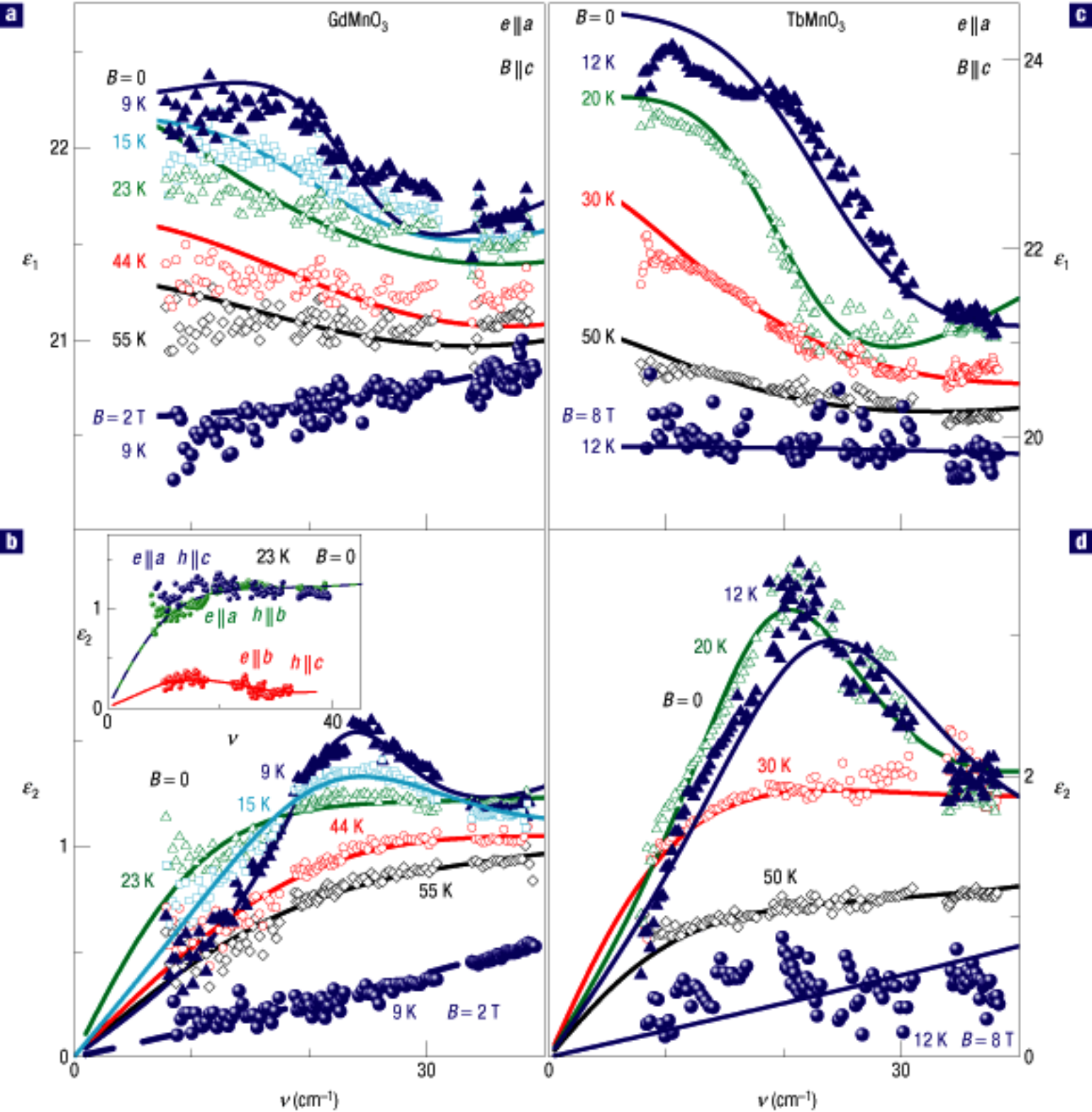

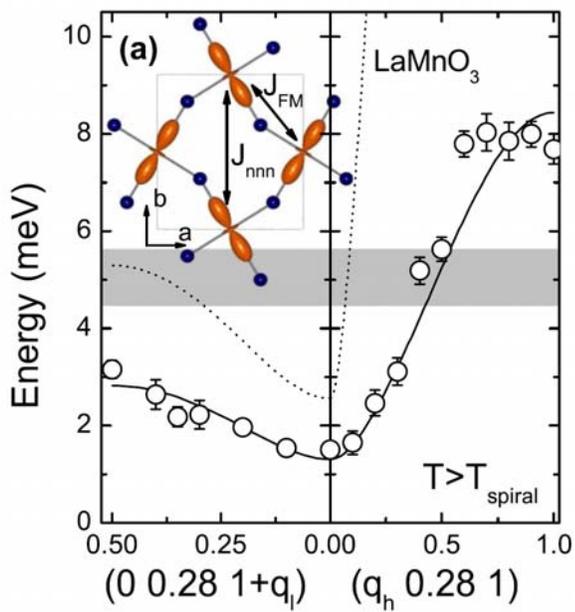
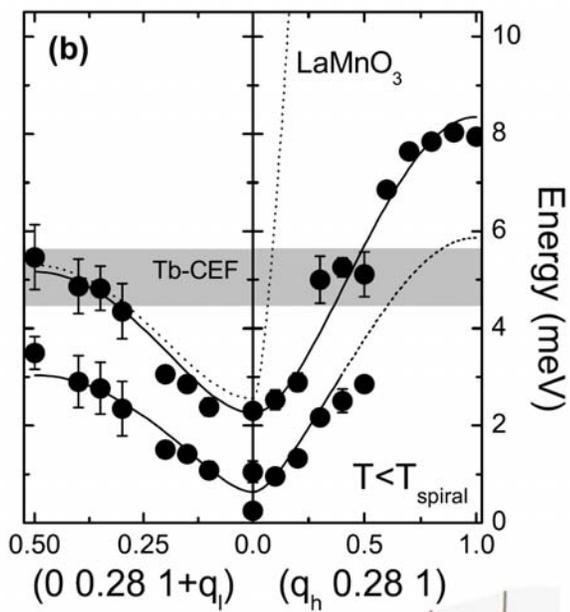
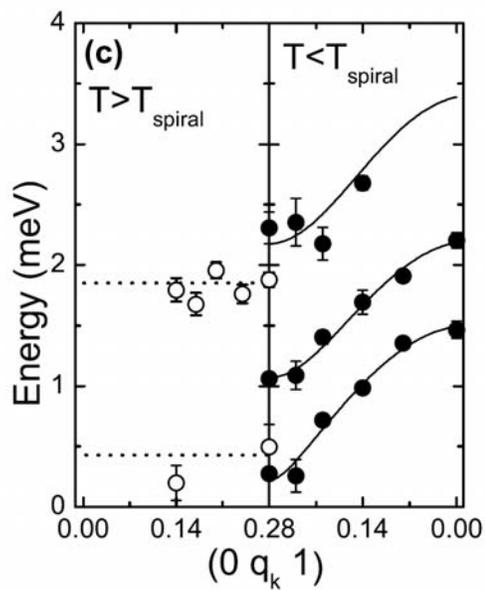
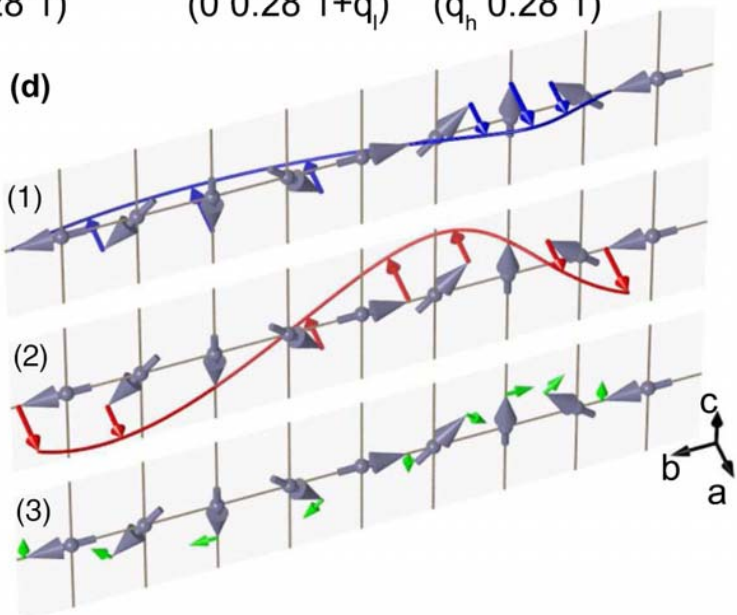

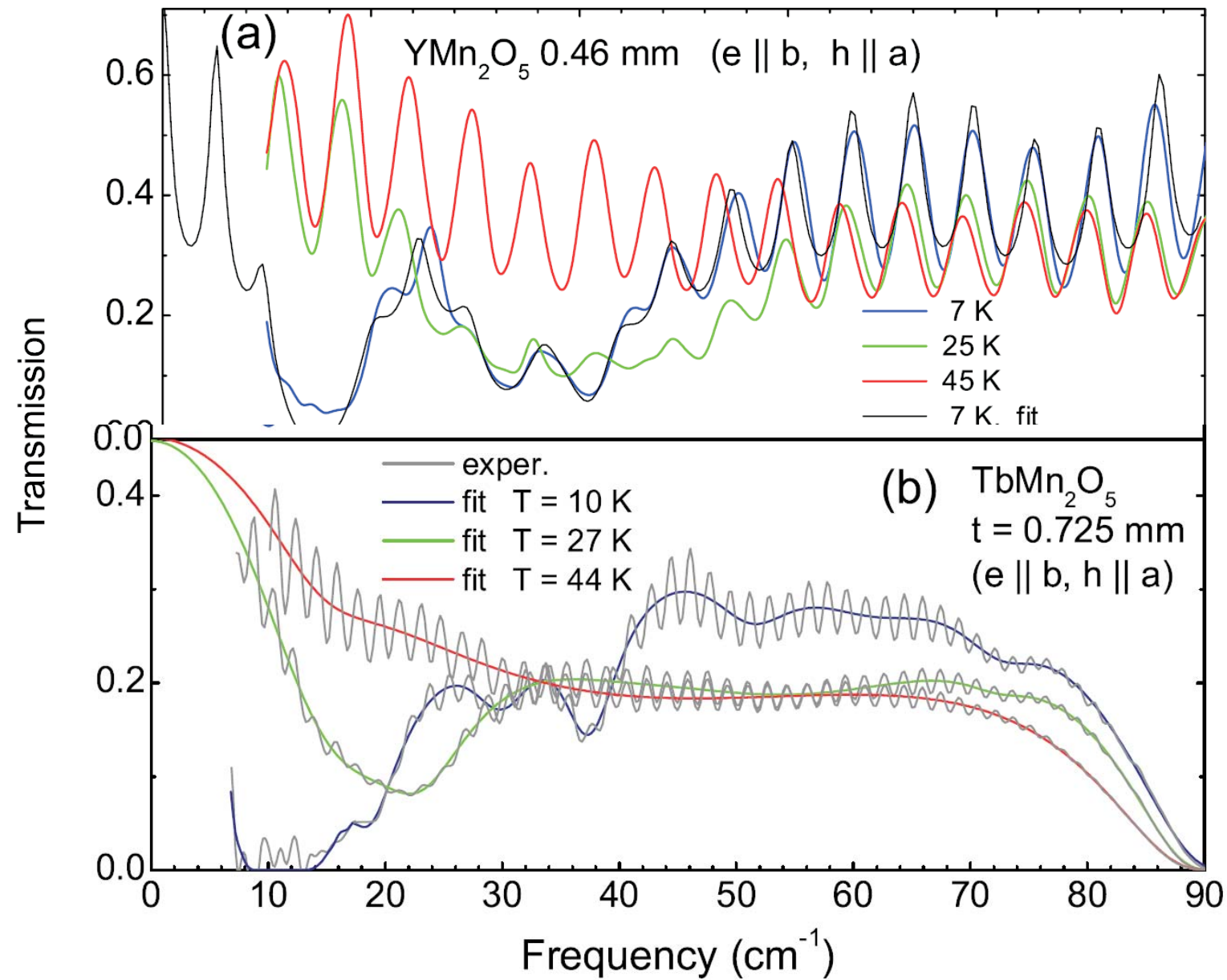

**(a)** YMn$_2$O$_5$ 0.46 mm (e ∥ b, h ∥ a)

- 7 K
- 25 K
- 45 K
- 7 K. fit

**(b)** TbMn$_2$O$_5$ t = 0.725 mm (e ∥ b, h ∥ a)

- exper.
- fit   T = 10 K
- fit   T = 27 K
- fit   T = 44 K

Transmission

Frequency (cm$^{-1}$)

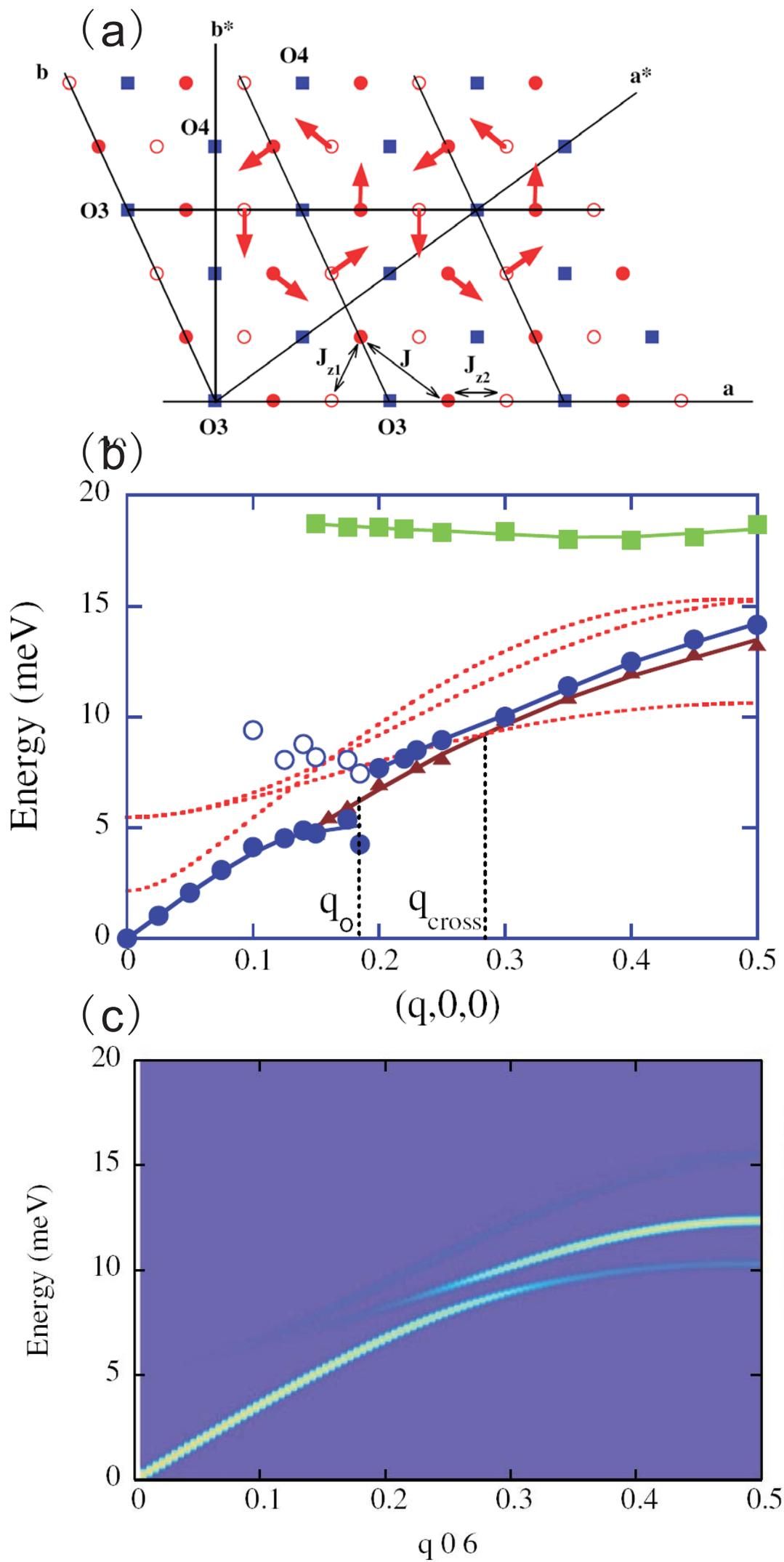

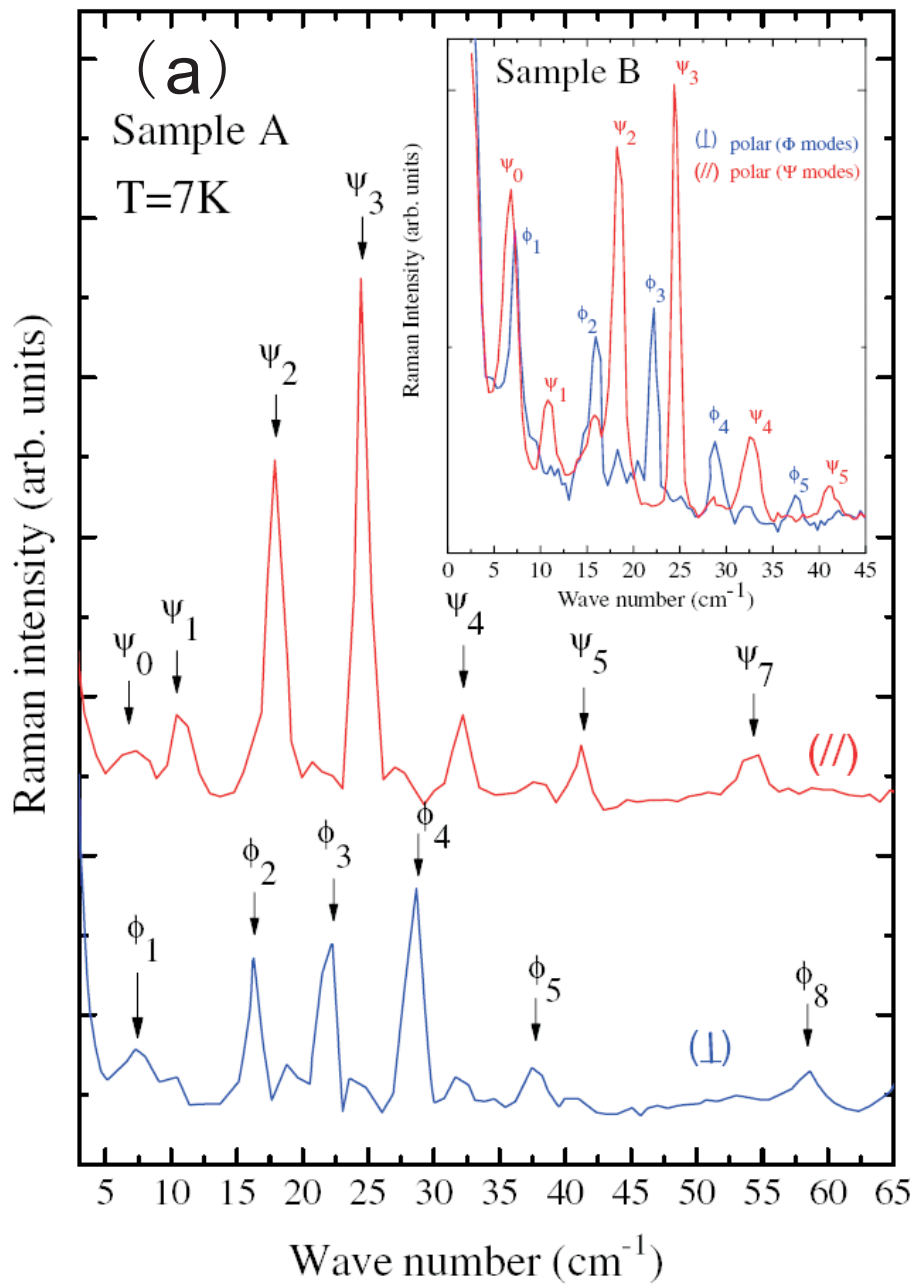

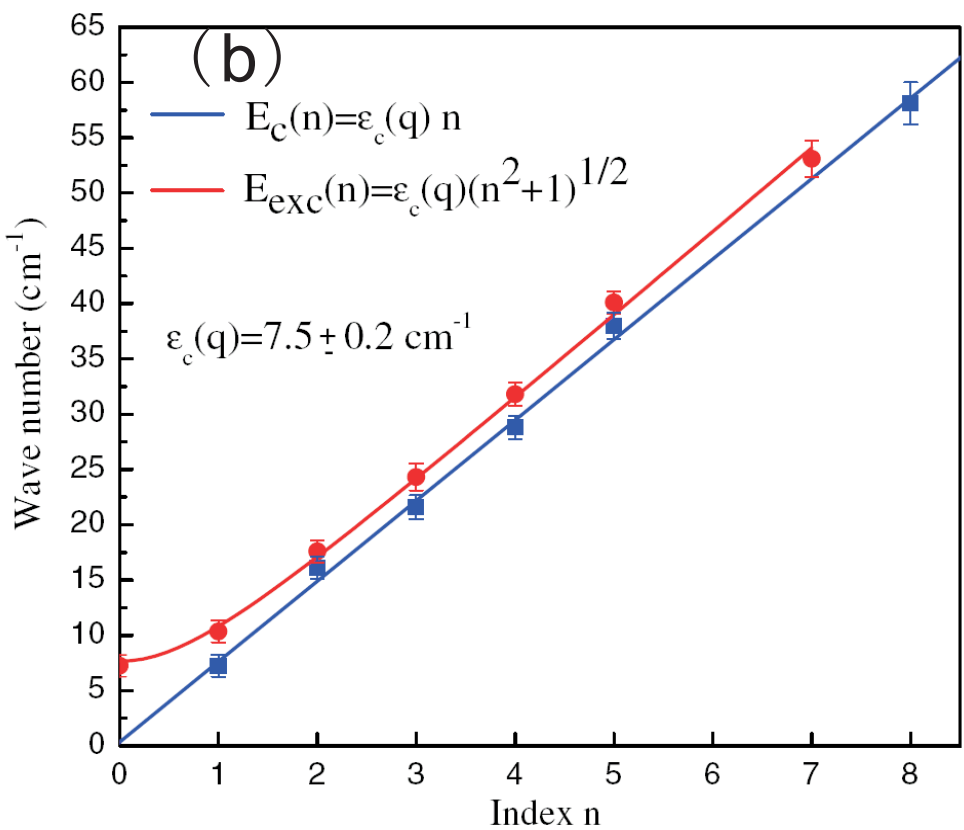

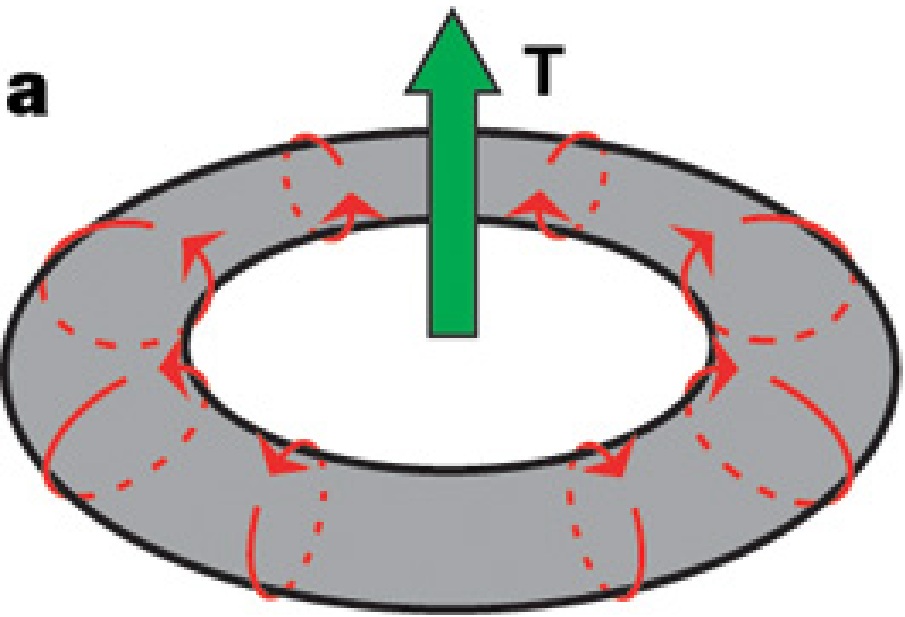

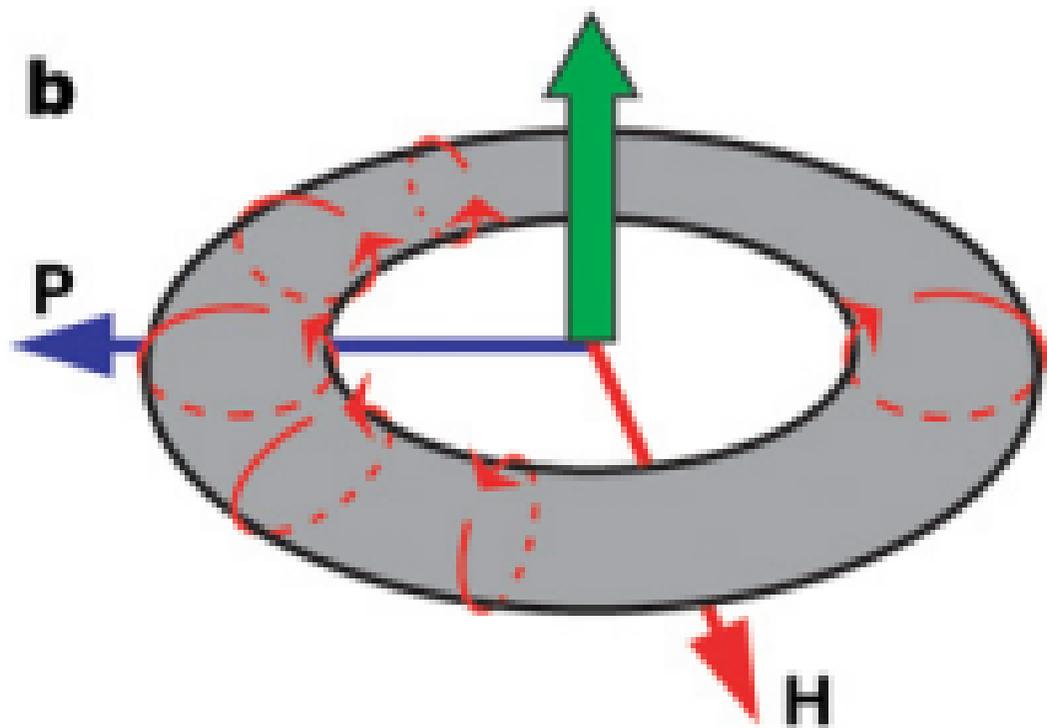

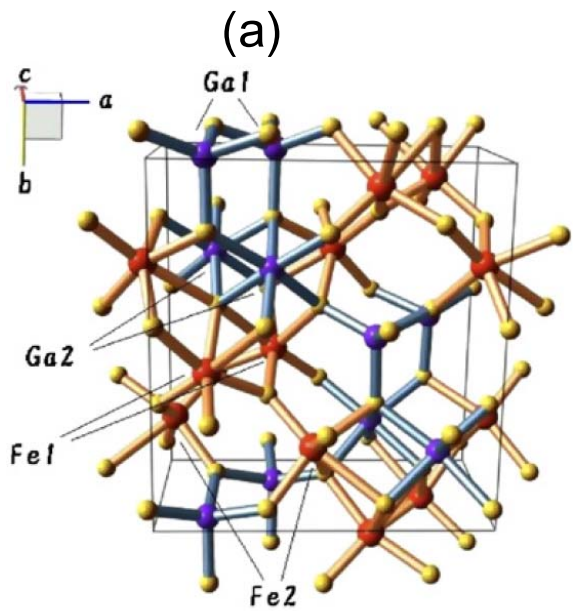

(a)

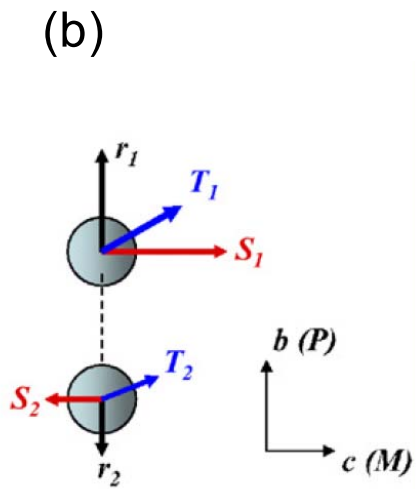

(b)

(c)

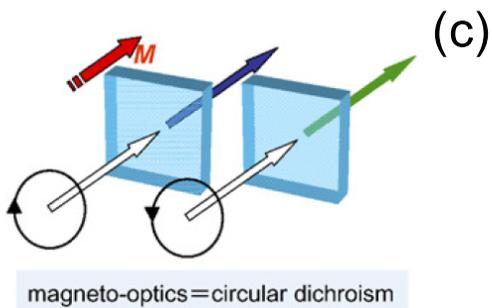

magneto-optics＝circular dichroism

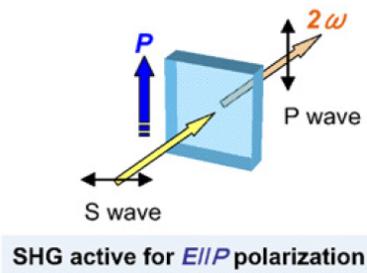

**SHG active for *E∥P* polarization**

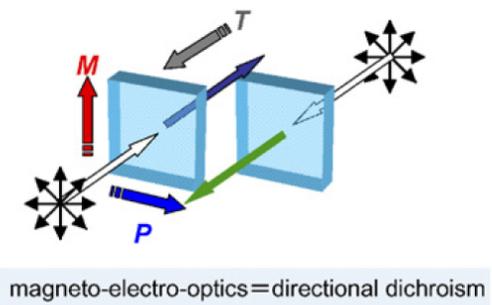

magneto-electro-optics＝directional dichroism

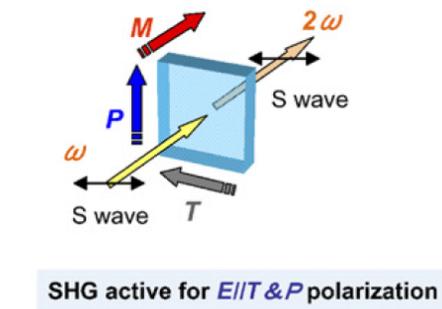

**SHG active for *E∥T＆P* polarization**

|  | Space | |
| --- | --- | --- |
| Time | Invariant | Change |
| Invariant | Ferroelastic 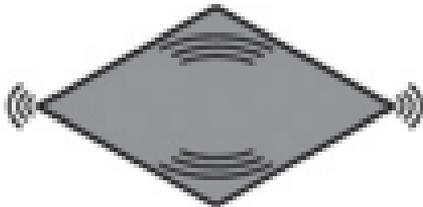 | Ferroelectric 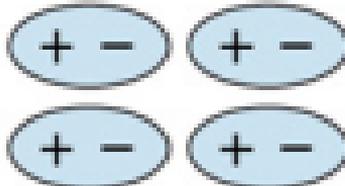 |
| Change | Ferromagnetic 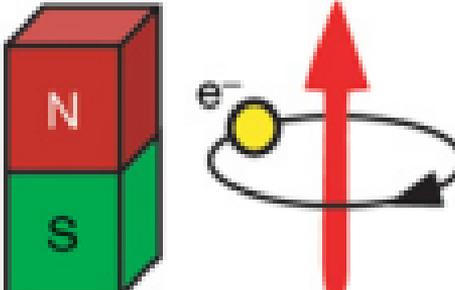 | Ferrotoroidic 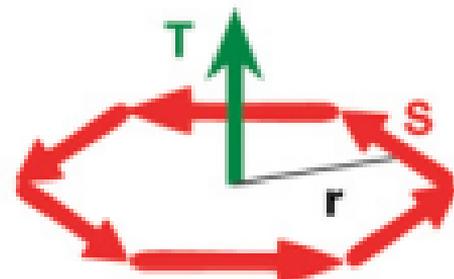 |

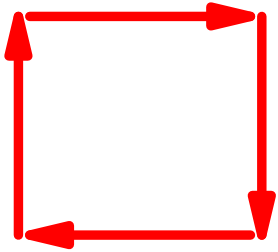

(a)

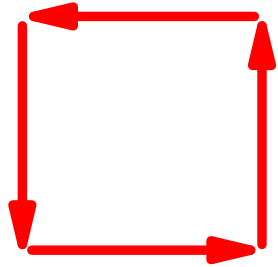

(b)

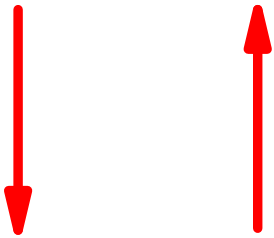

(c)

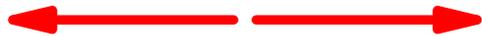

(d)

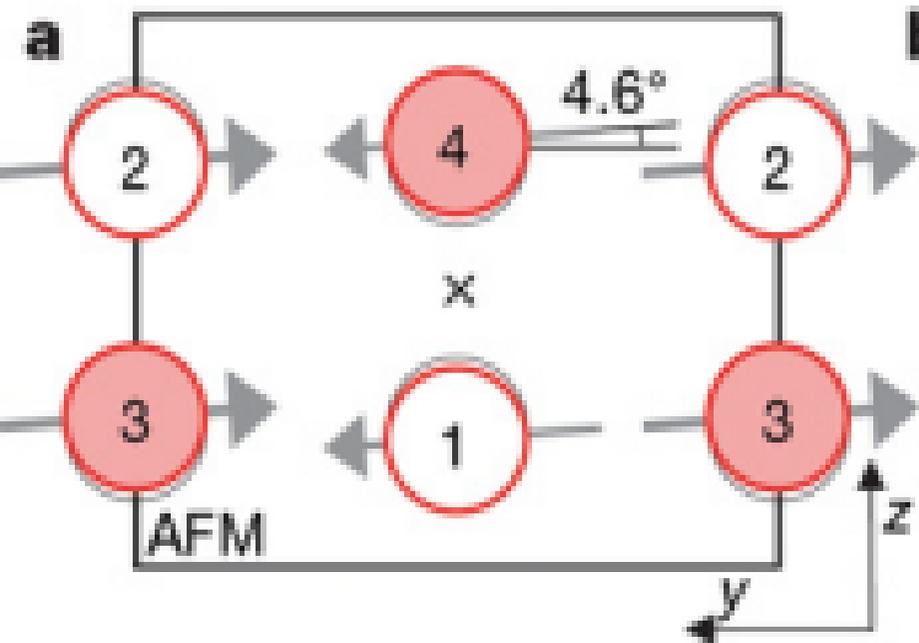
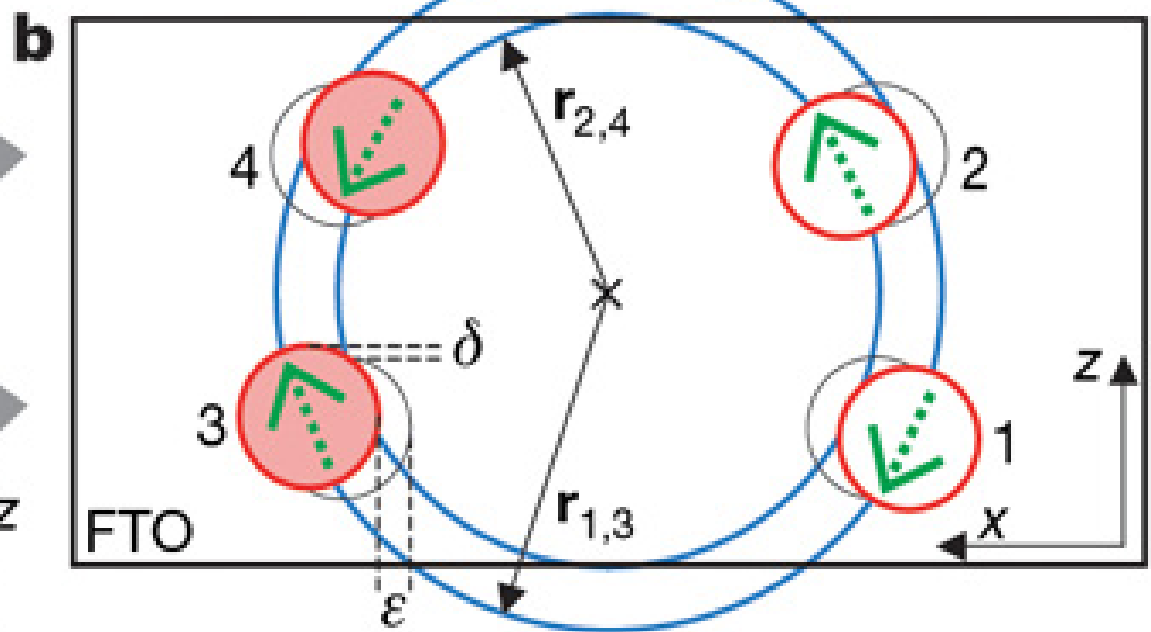

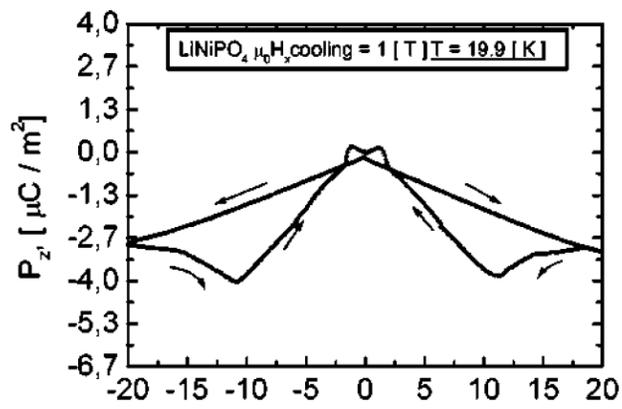

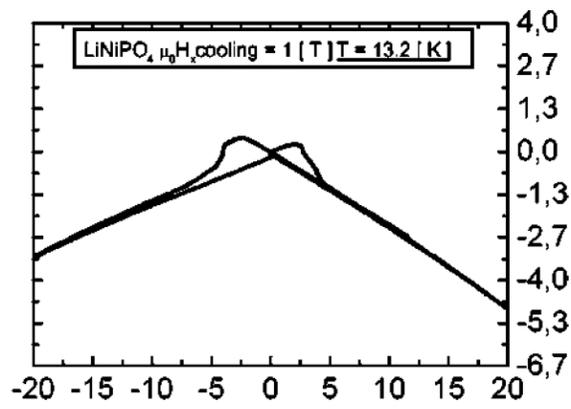

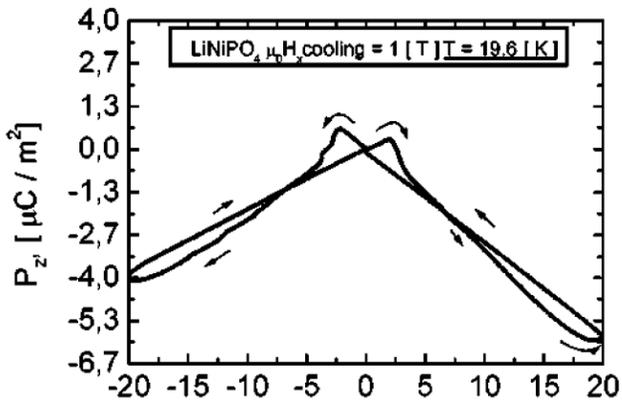

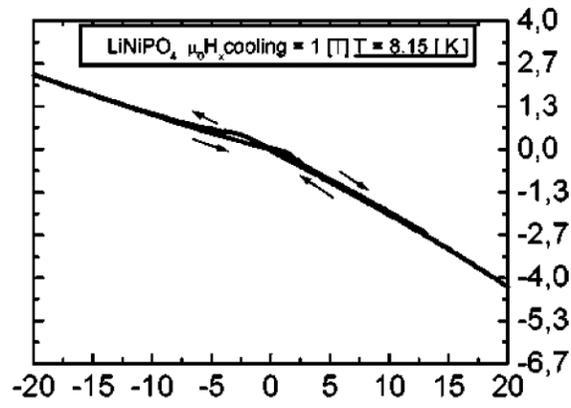

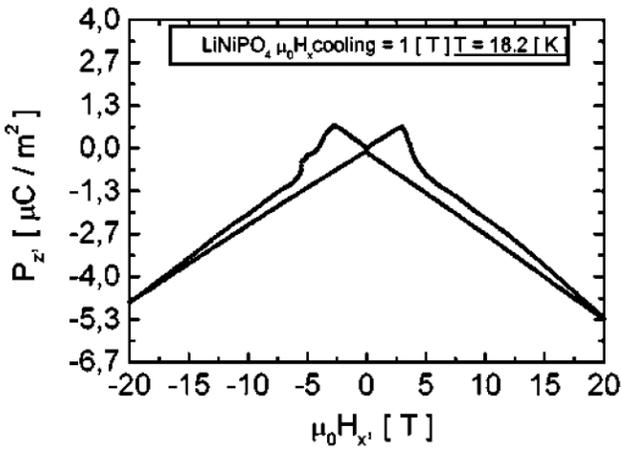

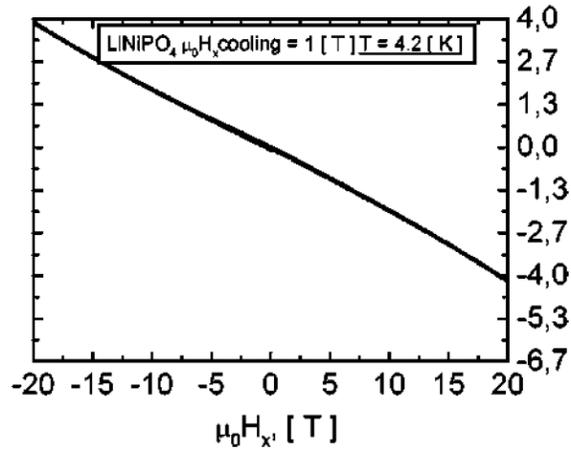

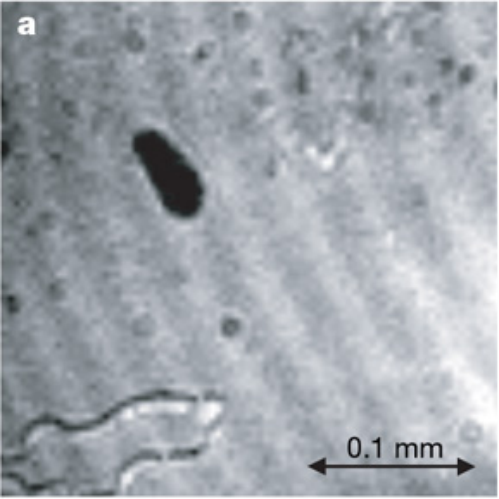
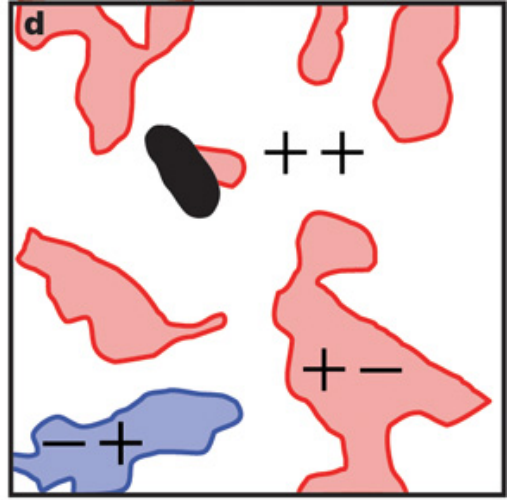
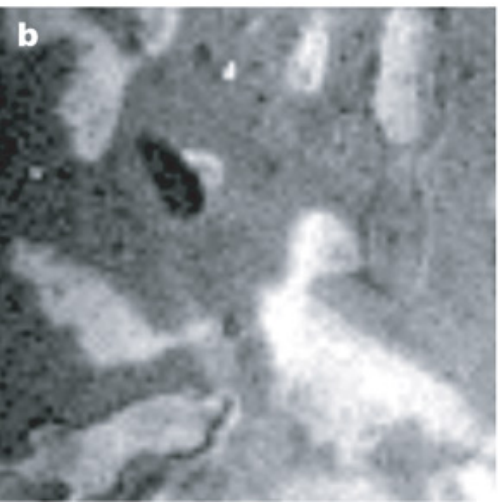
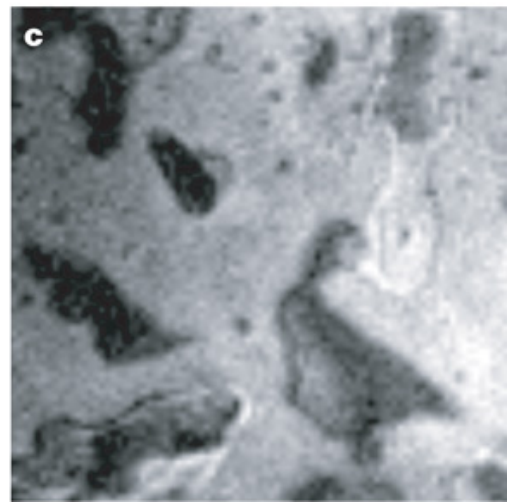

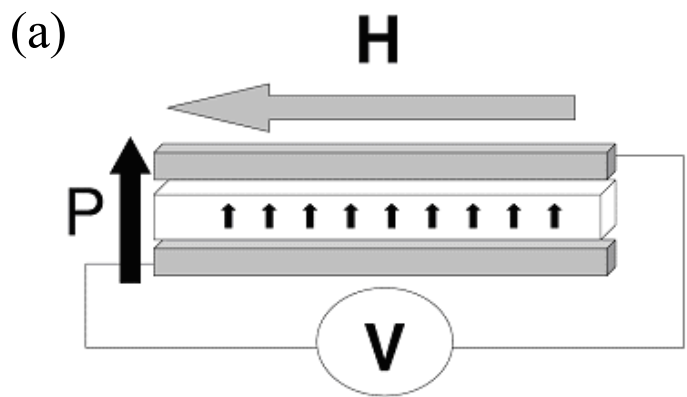

(a)

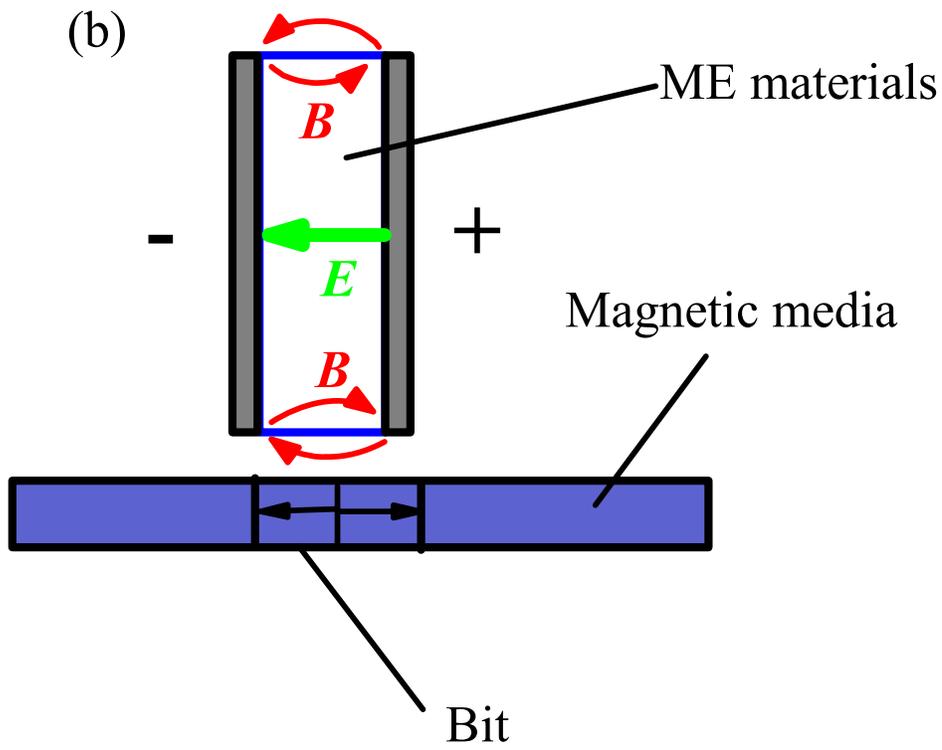

(b)

ME materials

Magnetic media

Bit

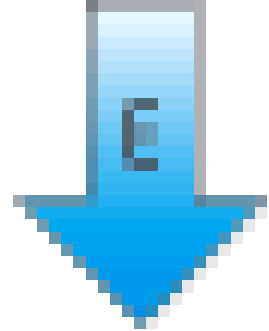
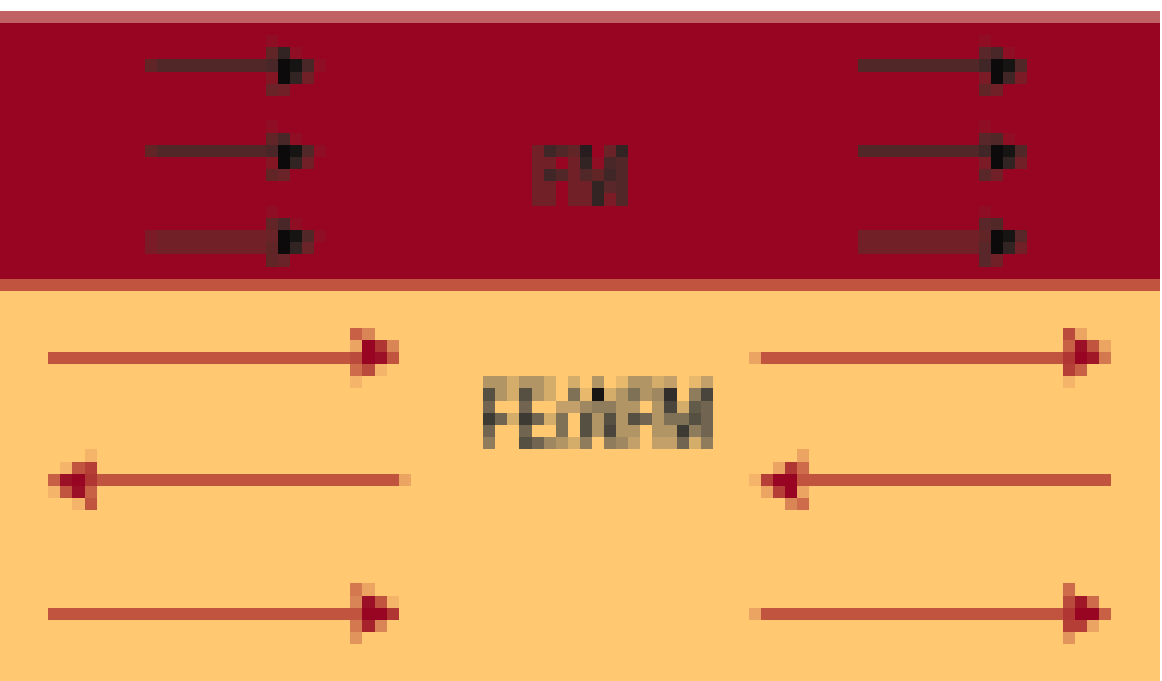
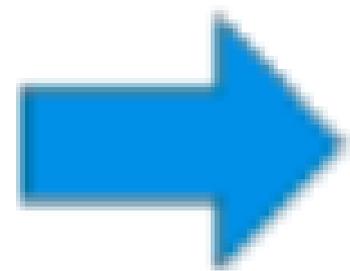
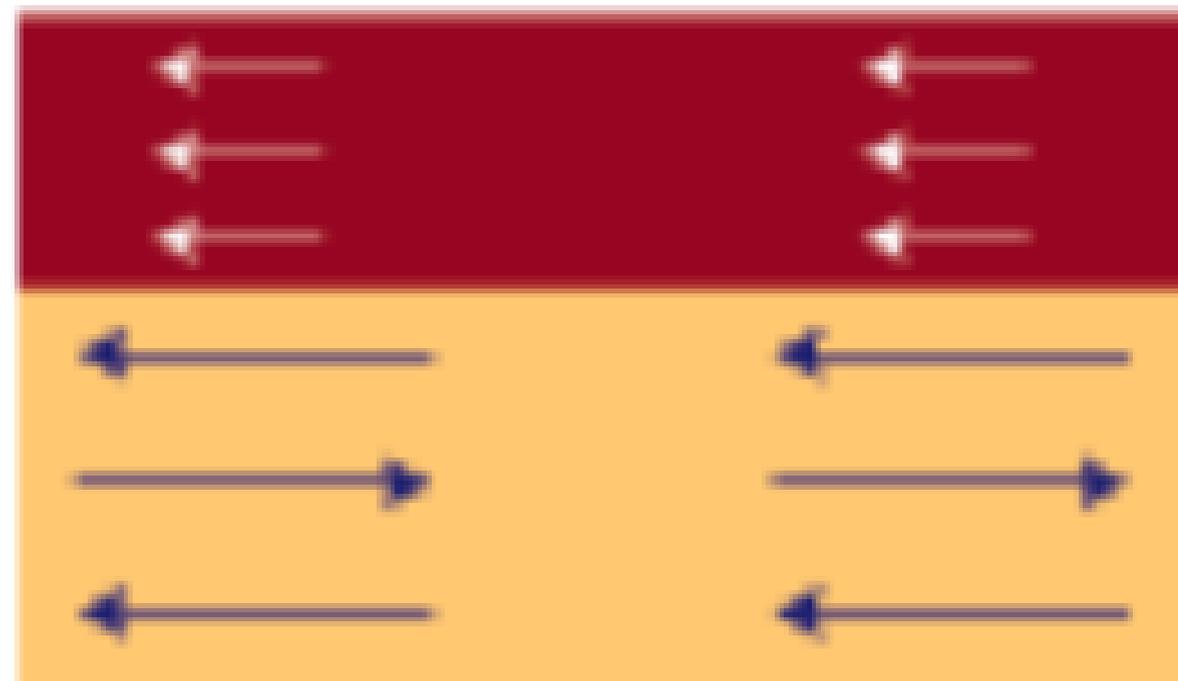

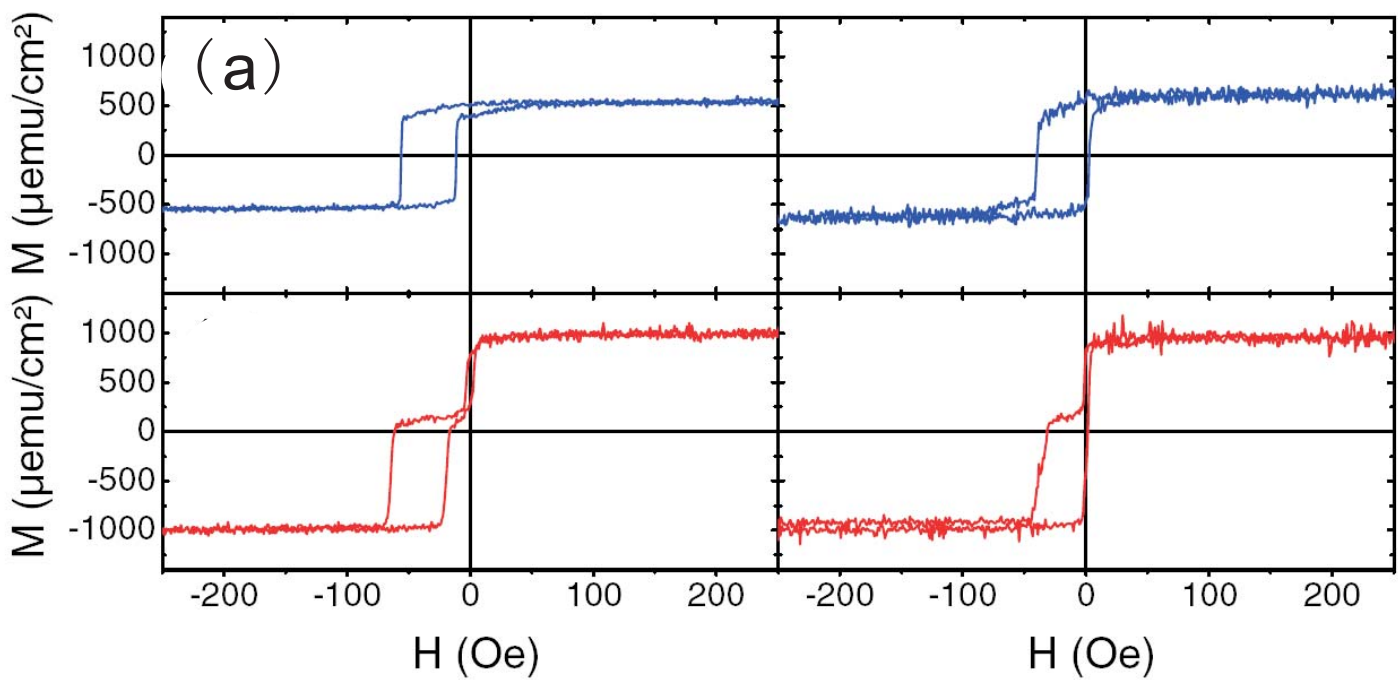

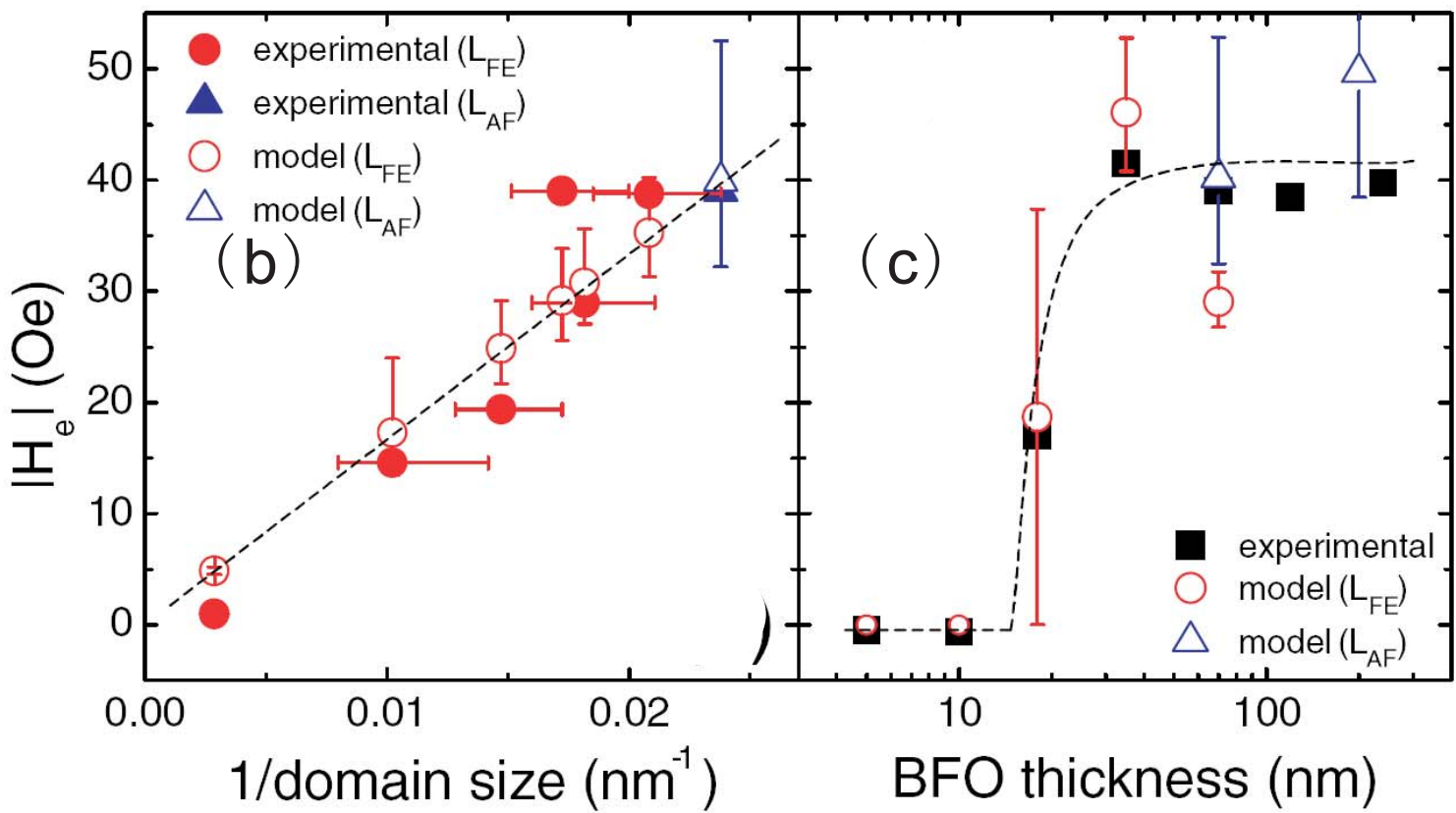

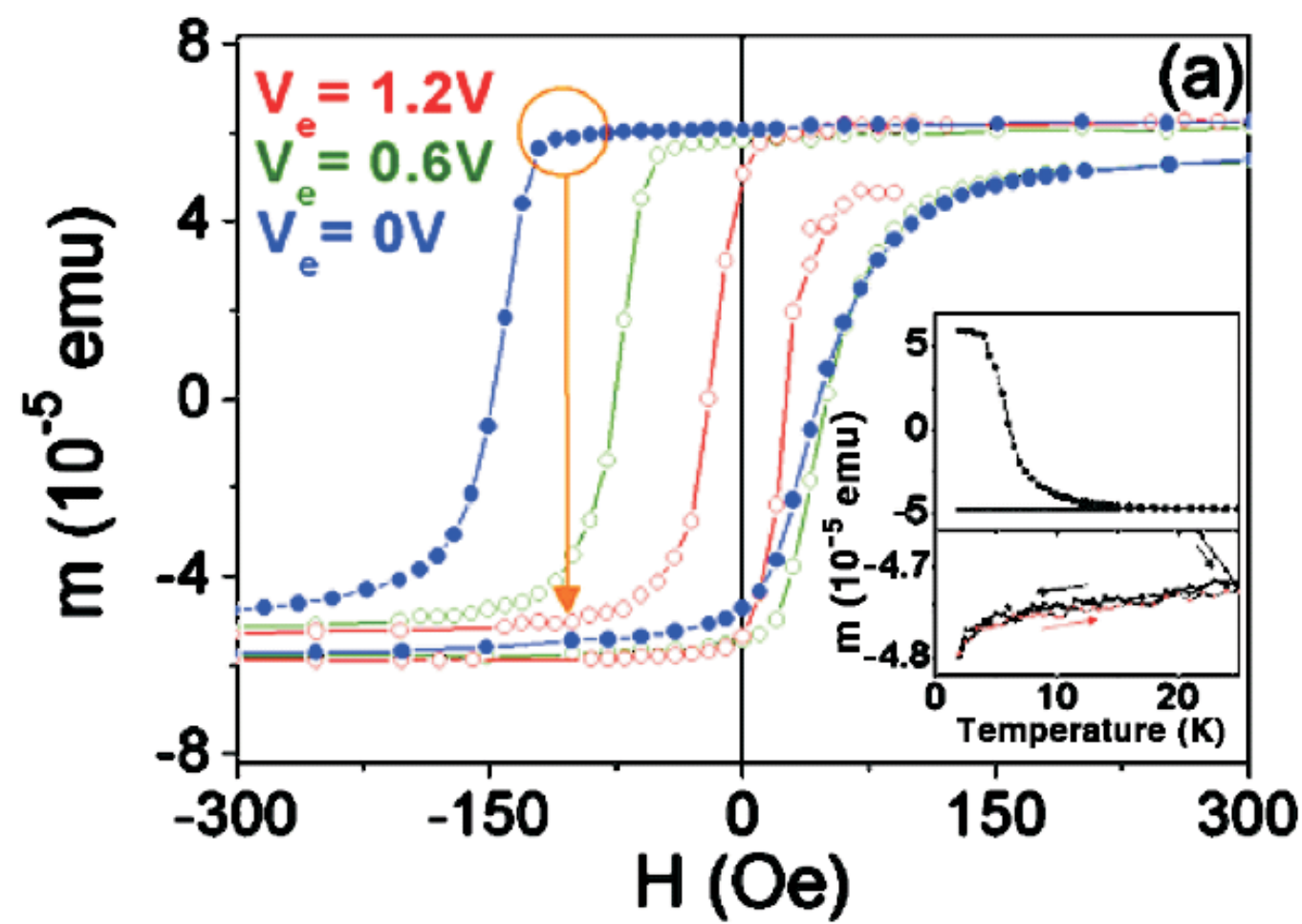

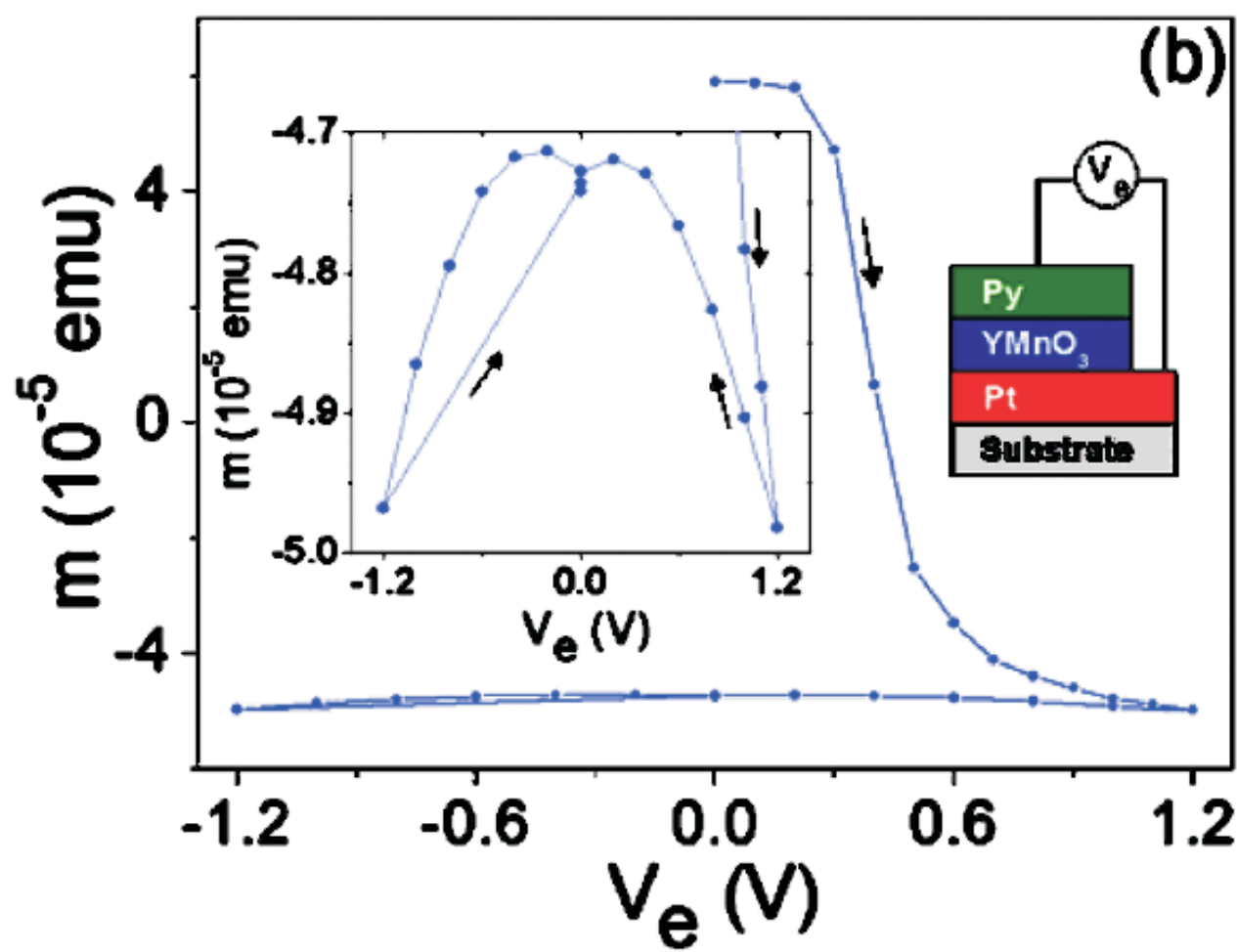

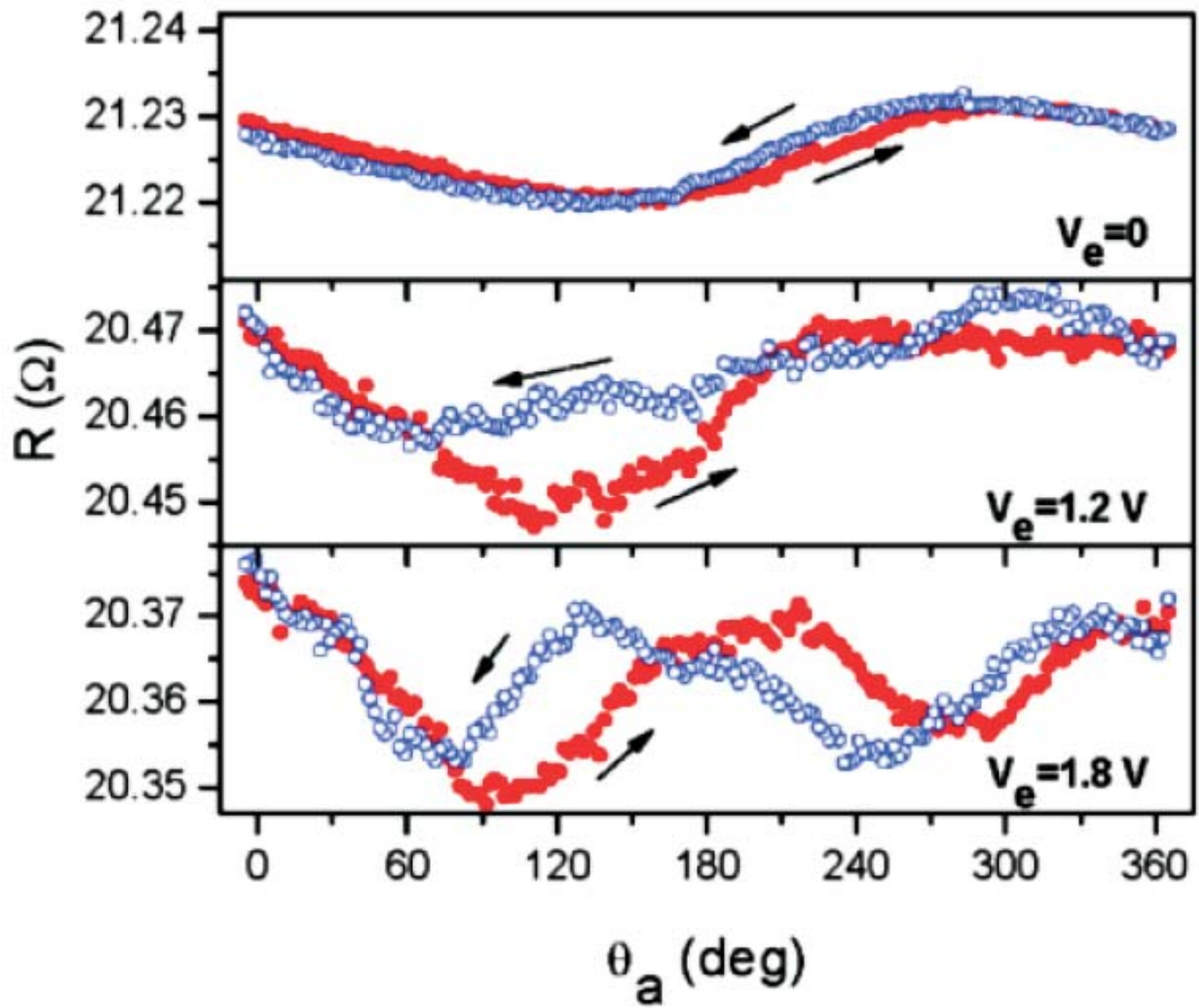

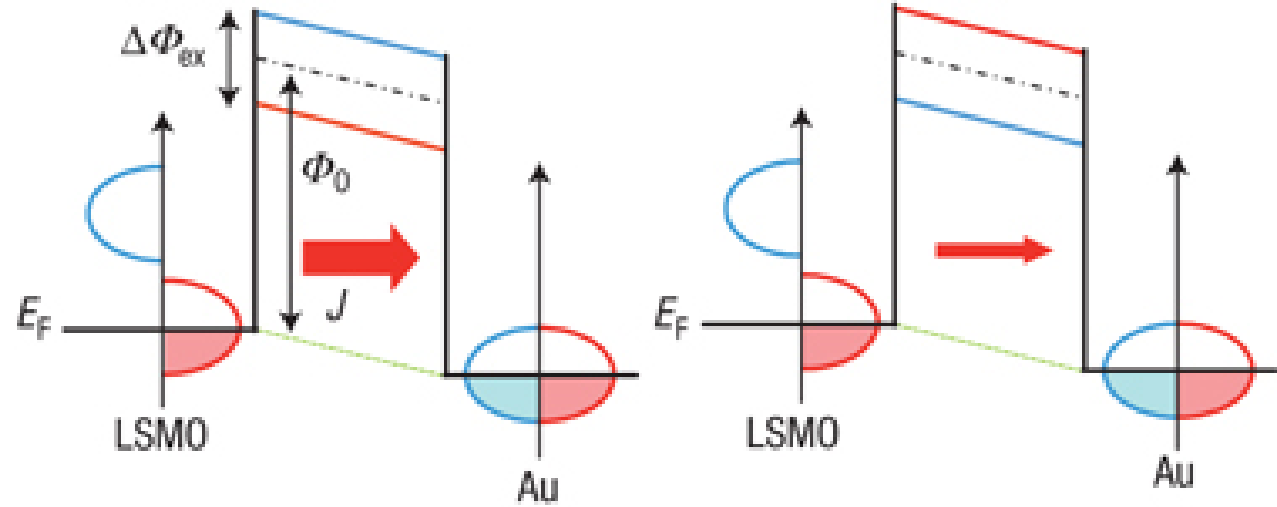

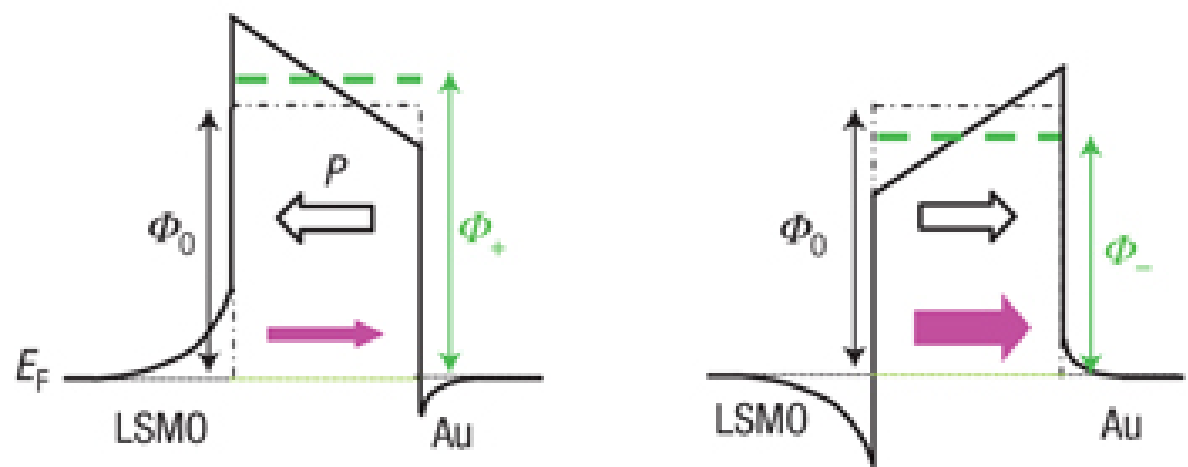

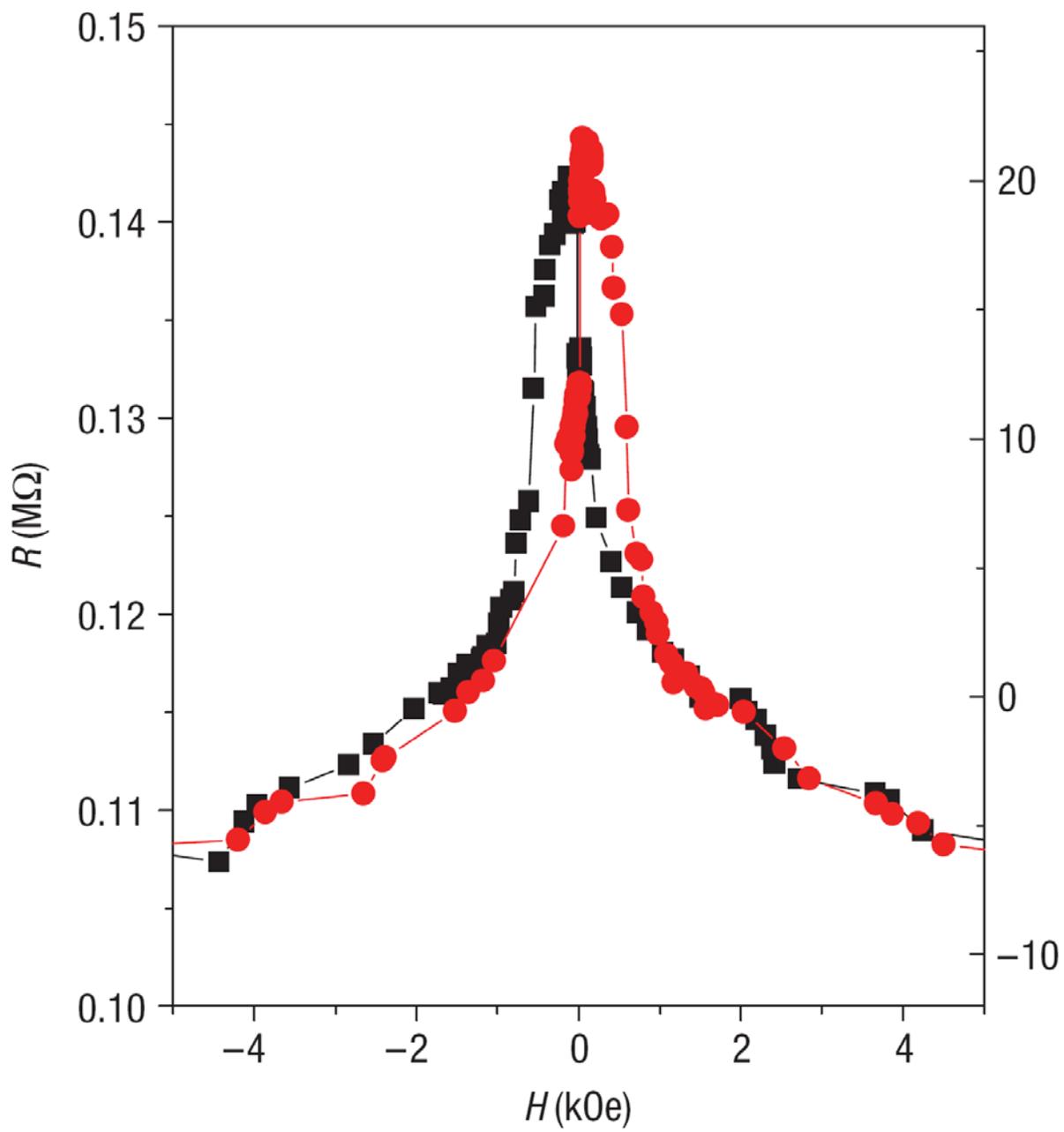

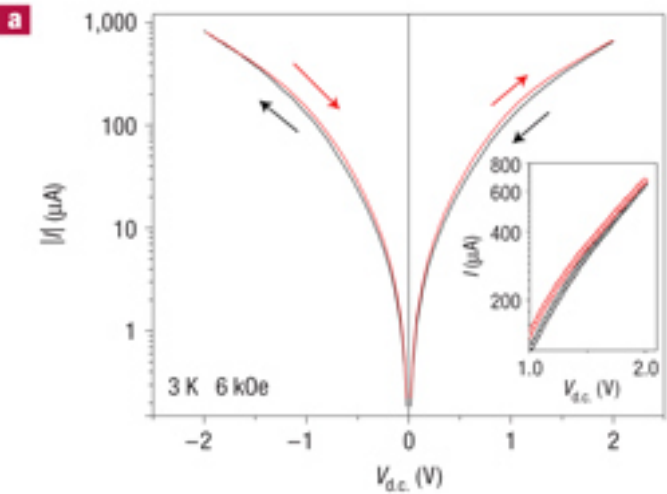

**a**

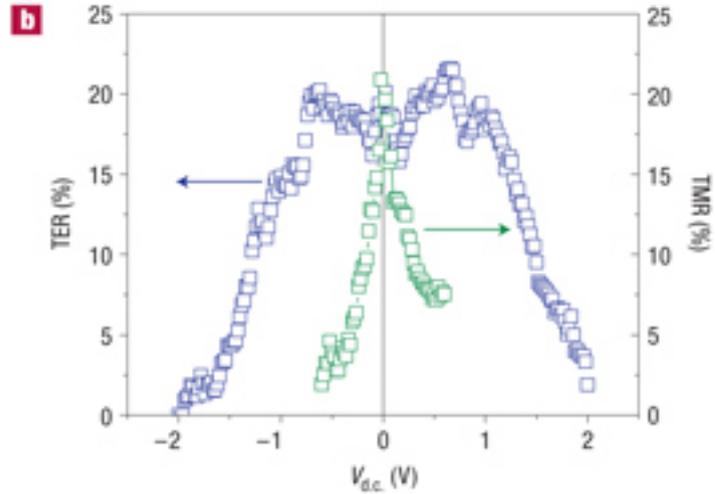

**b**

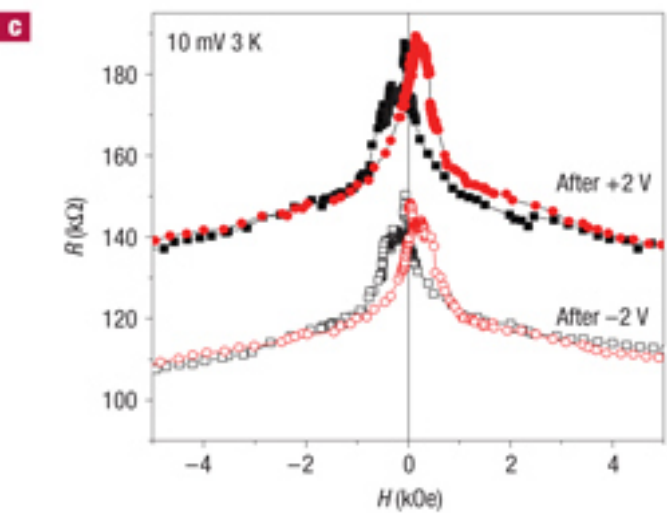

**c**

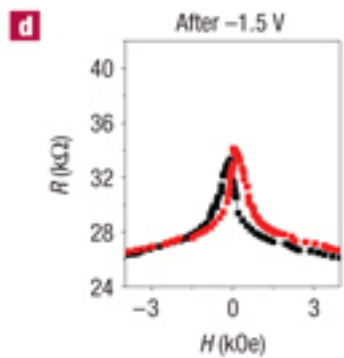

**d** After −1.5 V

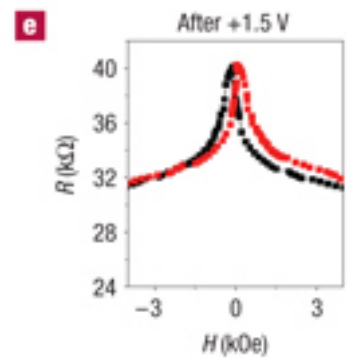

**e** After +1.5 V

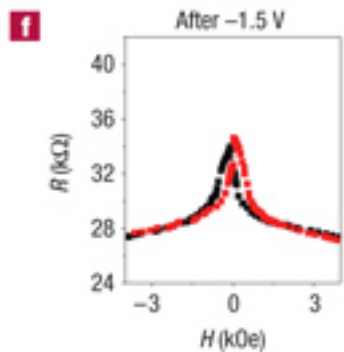

**f** After −1.5 V

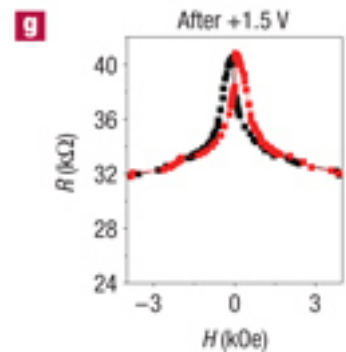

**g** After +1.5 V